\documentclass[a4paper,12pt]{article}
\pdfoutput=1
\usepackage{jheppub}
\usepackage[utf8]{inputenc}
\usepackage{url}
\usepackage{mathrsfs}  
\usepackage{pdflscape}
\usepackage{refcount}
\usepackage{booktabs}
\usepackage{todonotes}
\usepackage{enumitem}
\usepackage{adjustbox}
\usepackage{afterpage}
\usepackage{mathdots}

\usepackage{amsfonts}
\usepackage{bbm}

\usepackage{pdfpages}
\usepackage{caption}
\usepackage{subcaption}

\usepackage{tikz}
\usetikzlibrary{positioning}
\usetikzlibrary{arrows}
\usetikzlibrary{decorations.pathreplacing}
\usetikzlibrary{shapes.geometric}
\usetikzlibrary{calc}
\usetikzlibrary{decorations.pathmorphing}
\usetikzlibrary{shapes.misc}

\definecolor{goodgreen}{RGB}{55,169,49}

\definecolor{darkyellow}{RGB}{230,170,10}

\definecolor{brightyellow}{RGB}{255,240,190}

\tikzset{flavour/.style={draw=none,minimum size=0.3mm,fill=white, regular polygon,regular polygon sides=4,draw}}
\tikzset{gaugeBig/.style={inner sep=1mm,draw=none,fill=white,minimum size=2mm,circle, draw}}
\tikzset{bd/.style={circle, draw=black, inner sep=0pt, fill=black, minimum size=2mm}}
\tikzset{wd/.style={circle, draw=black, inner sep=0pt, fill=white, minimum size=2mm}}
\tikzset{Dynkin/.style={circle, draw=black, inner sep=0pt, fill=white, minimum size=2mm}}
\tikzstyle{ligne}=[draw, very thick] 
\tikzstyle{gridline}=[draw, gray] 
\tikzset{gauge/.style={circle, draw,inner sep=2.5pt}}
\tikzset{gaugeo/.style={circle, draw,inner sep=2.5pt,fill=orange}}
\tikzset{gaugec/.style={circle, draw,inner sep=2.5pt,fill=cyan}}
\tikzset{gauger/.style={circle, draw,inner sep=2.5pt,fill=red}}
\tikzset{gaugeb/.style={circle, draw,inner sep=2.5pt,fill=blue}}
\tikzset{gaugeg/.style={circle, draw,inner sep=2.5pt,fill=green}}
\tikzset{gaugem/.style={circle, draw,inner sep=2.5pt,fill=magenta}}
\tikzset{hasse/.style={circle, fill,inner sep=2pt}}
\tikzset{shrinky/.style={circle, fill,inner sep=1pt}}
\tikzset{sized/.style={circle, draw, inner sep=1.5pt}}
\tikzset{seven/.style={circle, draw,inner sep=3pt}}

\tikzset{dotto/.style={circle, orange, draw,inner sep=1.5pt,fill=orange}}
\tikzset{dottp/.style={circle, purple, draw,inner sep=1.5pt,fill=purple}}
\tikzset{dottc/.style={circle, cyan, draw,inner sep=1.5pt,fill=cyan}}
\tikzset{dottr/.style={circle, red, draw,inner sep=1.5pt,fill=red}}
\tikzset{dottb/.style={circle, blue, draw,inner sep=1.5pt,fill=blue}}
\tikzset{dottg/.style={circle, green, draw,inner sep=1.5pt,fill=green}}
\tikzset{dottm/.style={circle, magenta, draw,inner sep=1.5pt,fill=magenta}}

\DeclareMathOperator{\U}{U}
\DeclareMathOperator{\SU}{SU}
\DeclareMathOperator{\SO}{SO}
\DeclareMathOperator{\SL}{SL}

\DeclareMathOperator{\USp}{USp}
\DeclareMathOperator{\tr}{Tr}

\usepackage{epsfig}
\usepackage{amsmath}
\usepackage{amssymb}
\usepackage{amsfonts}
\usepackage{amsxtra}
\usepackage{amsthm}
\usepackage{mathrsfs}
\usepackage{makeidx}
\usepackage{graphics}
\usepackage{dsfont}
\usepackage{mathtools}
\usepackage{graphicx}
\usepackage{placeins}
\usepackage{bm}
\usepackage[capitalise]{cleveref}
\usepackage{empheq}
\usepackage{colortbl}
\usepackage{xcolor}
\usepackage{titlesec}
\usepackage{longtable}
\usepackage{hhline}
\usepackage{float}
\usepackage{color}
\usepackage{tikz}
\usepackage{xfrac}
\usepackage{footnote}
\usepackage{rotating}
\usepackage{lscape}
\usepackage{makecell}
\usepackage{environ}
\usepackage{tabularx}
\usepackage{subfiles}
\usepackage{ytableau}
\usepackage{tikz-3dplot}
\usepackage{slashed}
\usepackage{pifont}
\usepackage{multirow}
\usepackage{mdframed}
\usepackage{bbm}
\usepackage[normalem]{ulem}
\usepackage{arydshln}
\usepackage{enumitem} 
\usepackage{fancybox}

\usetikzlibrary{positioning,trees,decorations.pathmorphing,decorations.markings,decorations.pathreplacing,calc,shapes,patterns,arrows,chains,arrows.meta,fit,fadings,decorations.markings,graphs,graphs.standard,quotes,plotmarks,calligraphy,decorations.pathreplacing}

\usetikzlibrary{positioning}
\usetikzlibrary{chains}
\usetikzlibrary{arrows, arrows.meta ,fit,decorations.pathreplacing}
\tikzstyle{every picture}+=[remember picture]
\tikzstyle{na} = [baseline]
\tikzstyle{ligne}=[draw, thick]

\usetikzlibrary{arrows, decorations.markings, calc, fadings, decorations.pathreplacing, patterns, decorations.pathmorphing, positioning}
\tikzset{>={Latex[width=1.5mm,length=1.5mm]}}
\tikzset{bd/.style={circle, draw=black, inner sep=0pt, fill=black, minimum size=1.2mm}}
\tikzset{bld/.style={circle, draw=blue, inner sep=0pt, fill=blue, minimum size=1.2mm}}
\tikzset{wd/.style={circle, draw=black, inner sep=0pt, fill=white, minimum size=1.2mm}}
\tikzset{rd/.style={circle, draw=red, inner sep=0pt, fill=red, minimum size=.9mm}}
\tikzset{wrd/.style={circle, draw=red, inner sep=0pt, fill=white, minimum size=.9mm}}
\usetikzlibrary{graphs,graphs.standard,quotes}

\def\node#1#2{\overset{#1}{\underset{#2}{{\color{gray} \bullet}}}}

\def\node#1#2{\overset{#1}{\underset{#2}{\circ}}}

\tikzstyle{every picture}+=[remember picture]
\tikzstyle{na} = [baseline=-.5ex]

\newcommand{\eg}{e.g. }

\newcommand{\ie}{i.e. }

\numberwithin{equation}{section}
\newcommand{\bes}[1]{\begin{equation} \begin{split} #1\end{split} \end{equation}}

\newcommand{\be}{\begin{equation}} \newcommand{\ee}{\end{equation}}
\newcommand{\bea}{\begin{equation} \begin{aligned}} \newcommand{\eea}{\end{aligned} \end{equation}}

\def\tilde{\widetilde}

\def\hat{\widehat}

\def\bar{\overline}

\def\rt2{\sqrt{2}}

\def\det{\mathop{\rm det}}

\def\tr{\mathop{\rm tr}}

\def\CF{{\cal F}}

\def\CI{{\cal I}}

\def\CM{{\cal M}}
\def\CN{{\cal N}}

\def\CP{{\cal P}}

\def\CT{{\cal T}}

\def\CX{{\cal X}}

\def\CZ{{\cal Z}}

\def\1{{\ds 1}}

\newcommand{\fn}{\mathfrak{n}}

\newcommand{\fz}{\mathfrak{z}}

\newcommand{\ID}{\mathds{1}}
\newcommand{\fy}{\mathfrak{y}}
\newcommand{\fq}{\mathfrak{q}}

\def\SO{\mathrm{SO}}
\def\O{\mathrm{O}}
\def\SU{\mathrm{SU}}
\def\SL{\mathrm{SL}}

\def\Spin{\mathrm{Spin}}

\def\su{\mathfrak{su}}

\def\so{\mathfrak{so}}

\def\usp{\mathfrak{usp}}

\def\fN{\mathfrak{N}}
\def\Dih{\mathrm{Dih}}

\def\repa{\raise4pt\hbox{$\square$}\mkern-14mu\raise-4pt\hbox{$\square$}}
\def\repab{\overline{\raise4pt\hbox{$\square$}\mkern-14mu\raise-4pt\hbox{$\square$}\mkern-1mu}}

\def\smileface{\ensuremath{\hbox{\large$\bigcirc$}\mkern-15mu\raise-1pt\hbox{\scriptsize$\smallsmile$}%
\mkern-10mu\raise4pt\hbox{..}\mkern4mu}}
\def\frownface{\ensuremath{\hbox{\large$\bigcirc$}\mkern-15mu\raise-1pt\hbox{\scriptsize$\smallfrown$}%
\mkern-10mu\raise4pt\hbox{..}\mkern4mu}}

\newcommand{\ba}{\begin{array}}
\newcommand{\ea}{\end{array}}
\newcommand{\bi}{\begin{itemize}}
\newcommand{\ei}{\end{itemize}}
\def\vec#1{\bm{#1}}
\def\bea#1\eea{\allowdisplaybreaks \begin{align}#1\end{align}}
 \newcommand{\ben}{\begin{enumerate}}
\newcommand{\een}{\end{enumerate}}
\newcommand{\bean}{\begin{eqnarray*}}
\newcommand{\eean}{\end{eqnarray*}}
\newcommand{\eref}[1]{(\ref{#1})}

\newcommand{\fref}[1]{Figure~\ref{#1}}
\newcommand{\PE}{\mathop{\rm PE}}

\newcommand{\tQ}{\widetilde{Q}}

\newcommand{\BC}{\mathbb{C}}

\newcommand{\BZ}{\mathbb{Z}}

\definecolor{light-gray}{gray}{0.5}
\definecolor{new-green}{rgb}{0,0.7,0.3}

\newcommand{\blue}{\color{blue}}

\newcommand{\red}{\color{red}}
\newcommand{\green}{\color{new-green}}
\newcommand{\violet}{\color{violet}}

\def\aup#1 {\overset{#1}{\uparrow} \, \overset{\tilde{#1}}{\downarrow}}

\tikzset{snake it/.style={decorate, decoration={snake, amplitude=.4mm, segment length=2mm,
                       post length=0mm,pre length=0mm}}}

\newcommand{\MA}{\mathds{A}}
\newcommand{\BM}{\mathds{B}}

\newcommand{\QM}{\mathds{Q}}
\newcommand{\ZM}{\mymathds{0}}

\def\u{\mathfrak{u}}
\DeclareMathAlphabet{\mymathds}{U}{BOONDOX-ds}{m}{n}
\tikzstyle{double_border} = [draw, double, double distance=1pt]

\preprint{\hspace{1cm}}

\title{Wreathing, Discrete Gauging, and Non-invertible Symmetries}
\author[a]{Julius F. Grimminger,}
\author[b,c]{William Harding,}
\author[c,d]{and Noppadol Mekareeya}

\affiliation[a]{Mathematical Institute, University of Oxford,\\Andrew Wiles Building, Woodstock Road, Oxford, OX2 6GG, UK}
\affiliation[b]{Dipartimento di Fisica, Universit\`a di Milano-Bicocca, Piazza della Scienza 3, I-20126 Milano, Italy}
\affiliation[c]{INFN, sezione di Milano-Bicocca,
Piazza della Scienza 3, I-20126 Milano, Italy}
\affiliation[d]{Department of Physics, Faculty of Science, Chulalongkorn University, Phayathai Road, Pathumwan, Bangkok 10330, Thailand}

\emailAdd{julius.grimminger@maths.ox.ac.uk}
\emailAdd{w.harding@campus.unimib.it}
\emailAdd{n.mekareeya@gmail.com}

\abstract{
't Hooft anomalies of discrete global symmetries and gaugings thereof have rich mathematical structures and far-reaching physical consequences. We examine each subgroup $G$, up to automorphisms, of the permutation group $S_4$ that acts on the four legs of the affine $D_4$ quiver diagram, which is mirror dual to the 3d $\mathcal{N}=4$ $\mathrm{SU}(2)$ gauge theory with four flavours. These actions are studied in terms of how each permutation cycle acts on the superconformal index of the theory in question. We present a prescription for refining the index with respect to the fugacities associated with the Abelian discrete symmetries that are subgroups of $G$. This allows us to study sequential gauging of various subgroups of $G$ and construct symmetry webs. We study the effects of 't Hooft anomalies and non-invertible symmetries that arise from discrete gauging on the index. When the whole symmetry $G$ is gauged, our results are in perfect agreement with a type of discrete operations on the quiver, known as wreathing, discussed in the literature. We provide a general prescription for computing the index for any wreathed quivers that contain unitary or special unitary gauge groups. We demonstrate this in an example of the 3d $\mathcal{N}=4$ $\mathrm{U}(N)$ gauge theory with $n$ flavours and compare the results with gauging the charge conjugation symmetry associated with the flavour symmetry of such a theory.
}

\begin{document}
\maketitle

\section{Introduction and summary}
Discrete global symmetries, gaugings thereof, and 't Hooft anomalies have received considerable attention in recent years due to their importance in constraining the dynamics of strongly-coupled quantum field theories and their rich mathematical structures. The central theme of this article is to study such topics in certain simple and well-known 3d $\CN=4$ superconformal field theories (SCFTs). One of the benefits of this class of theories is the availability of exact quantities such as the superconformal index\footnote{We will also refer to it as the {\it index} for brevity.} \cite{Bhattacharya:2008zy,Bhattacharya:2008bja, Kim:2009wb,Imamura:2011su, Kapustin:2011jm, Dimofte:2011py, Aharony:2013dha, Aharony:2013kma} which serves as a main tool for investigating the aforementioned topics in the SCFTs studied in this article. 

The first 3d $\CN=4$ SCFT that we focus on admits a gauge theory description as an $\SU(2)$ vector multiplet coupled to four hypermultiplets in the fundamental representation, also known as $\SU(2)$ SQCD with four flavours. Another description of this SCFT is known, due to 3d mirror symmetry \cite{Intriligator:1996ex}, as the extended $\so(8)$ Dynkin diagram, where the gauge group is $[\U(2) \times \prod_{i=1}^4 \U(1)_i]/\U(1)$, with a bifundamental hypermultiplet under the $\U(2) \times \U(1)_i$ gauge group for each $i=1,2,3,4$. This is depicted below, and will be referred to as the affine $D_4$ quiver. 
\bes{
\vcenter{\hbox{\begin{tikzpicture}
        \node[gauge,label=right:{\footnotesize $2$}] (0) at (0,0) {};
        \node[gauge,label=right:{\footnotesize $1$}] (1) at (1,1) {};
        \node[gauge,label=right:{\footnotesize $1$}] (2) at (1,-1) {};
        \node[gauge,label=left:{\footnotesize $1$}] (3) at (-1,-1) {};
        \node[gauge,label=left:{\footnotesize $1$}] (4) at (-1,1) {};
        \draw (0)--(1);
        \draw (0)--(2);
        \draw (0)--(3);
        \draw (0)--(4);
\end{tikzpicture}}}\quad \text{\footnotesize$/\U(1)$}
}
A feature of this description is that there is a symmetry $S_4$, namely the permutation group of order $24$, that acts by permuting the four legs of the quiver. Gauging this $S_4$ symmetry and subgroup thereof constitutes the first part of this article. In fact, a number of interesting aspects in gauging such symmetries in this particular theory has been explored in \cite{Bourget:2020bxh}. In this reference, the actions of every subgroup of $S_4$ (up to automorphism) on the monopole operators of the affine $D_4$ quiver were studied, and gauging such discrete symmetries was referred to as {\it wreathing} (see also \cite{Hanany:2018cgo, Hanany:2023uzn}). It was proposed that such gauging produces a new theory with a continuous global symmetry being a subgalgebra of $\so(8)$. The Coulomb branches of the resulting theories after gauging were studied very thoroughly by means of the Hilbert series \cite{Cremonesi:2013lqa}, and many interesting aspects of the Higgs branches were also presented.

This article generalises the results of \cite{Bourget:2020bxh} in several directions. First of all, we show that, for the affine $D_4$ quiver, wreathing by any subgroups of $S_4$ that are isomorphic to $\BZ_2$ and $\BZ_2 \times \BZ_2$ can be realised by gauging appropriate charge conjugation symmetries associated with the $\SO(2)^4$ subgroup of the $\SO(8)$ flavour symmetry of $\SU(2)$ SQCD with four flavours. The resulting theories can be realised as mirror theories of the aforementioned wreathed affine $D_4$ quivers.
Secondly, we propose a prescription for computing the superconformal index of wreathed quivers by a general discrete symmetry $G$. In case of the affine $D_4$ quiver, the Higgs and Coulomb branch limits \cite{Razamat:2014pta} of such an index reproduce correctly the Higgs and Coulomb branch Hilbert series reported in \cite{Bourget:2020bxh} for each subgroup $G$ of $S_4$. As an important by-product, we propose the superconformal index of the orginal theory, namely the affine $D_4$ quiver, refined with respect to the fugacities associated with the Abelian subgroups of $G$. This allows us to sequentially gauge various subgroups of $G$ and produce symmetry webs in the same fashion as in \cite{Bhardwaj:2022maz, Bartsch:2022ytj} (see also \cite{Tachikawa:2017gyf, Bhardwaj:2017xup, Buican:2020who, Bhardwaj:2022yxj, Bhardwaj:2022lsg, Bartsch:2022mpm, Bhardwaj:2022kot, Bartsch:2023pzl, Bartsch:2023wvv, Bhardwaj:2023ayw, Schafer-Nameki:2023jdn, Buican:2023bzl, Choi:2024rjm, Bullimore:2024khm}). As pointed out in these references, this way of gauging produces a rich family of symmetries, including interplay between one-form and zero-form symmetries \cite{Gaiotto:2014kfa}, two-group structures, and non-invertible symmetries, which can be summarised in the language of two-categories. As demonstrated in \cite{Mekareeya:2022spm} (see also \cite{Sacchi:2023omn, Nardoni:2024sos}), the superconformal index provides a convenient and efficient way to detect various mixed 't Hooft anomalies, including those involving two discrete zero-form symmetries and a discrete one-form symmetry, where related discussions and further examples were also provided in \cite{Bhardwaj:2022dyt, Bhardwaj:2023zix}. We extend the result of \cite{Mekareeya:2022spm} to study various sequential gaugings of the subgroups $G$ of $S_4$. In particular, we consider gauging various subgroups of $G$ when $G$ is either a semi-direct product of the form $G \cong (\prod_i \BZ_{s_i}) \rtimes \BZ_h$ or when $G$ can be realised as a non-trivial extension of $\BZ_h$ by $\prod_{i} \BZ_{s_i}$. We report the superconformal index and discuss the implications of the 't Hooft anomaly or non-invertible symmetry on the result. Importantly, in case of the semi-direct product, we examine gauging of the normal subgroup $\prod_i \BZ_{s_i}$ of $G$ that results in the split two-group formed by the dual one-form symmetry $\prod_i \BZ^{[1]}_{s_i}$ and the zero-form symmetry $\BZ_h$. One of the new results presented in this article is the prescription for gauging such a two-group in the superconformal index. We show that this procedure actually leads to a correct result in gauging the subgroup $\BZ_h$ of $G$.

In the second part of the article, we consider another 3d $\CN=4$ SCFT described by a $\U(N)$ vector multiplet coupled to $n$ hypermultiplets in the fundamental representation, also known as $\U(N)$ SQCD with $n$ flavours. The mirror theory of this was derived in \cite{Hanany:1996ie} and is depicted in \eref{mirrUnwNflv} in the main text. It turns out that this quiver has a $\BZ_2$ left-right symmetry with respect to the vertical axis. We can therefore perform gauging of this $\BZ_2$ symmetry and compute the index of the resulting theory using the prescription discussed above. At this point, one can pose a natural question: what does wreathing in the mirror theory correspond to in $\U(N)$ SQCD with $n$ flavours? For $n$ odd, we can give a concise answer to this question: it corresponds to gauging the $\BZ_2$ charge conjugation symmetry in the central extension $\SU(n) \rtimes \BZ_2$ of the flavour symmetry (see \eg~ \cite{Bourget:2017tmt, Bourget:2018ond, Arias-Tamargo:2019jyh, Arias-Tamargo:2021ppf, Arias-Tamargo:2022nlf, Arias-Tamargo:2023duo}). We check that the Coulomb branch limit of the wreathed mirror theory indeed coincides with the Higgs branch of $\U(N)$ SQCD with $n$ flavours after the charge conjugation symmetry which is part of the flavour symmetry is gauged. We observe also that, in the latter, the gauging affects the monopole operators and hence the Coulomb branch of SQCD. For even $n$, the answer to the above question is more complicated. First of all, for $N=1$, the wreathing of the mirror theory corresponds to turning the $\U(1)$ gauge theory with $n$ flavours into the $\O(2)$ gauge theory with $n/2$ flavours, whose flavour symmetry is $\usp(n)$ which does not coincide with gauging the $\BZ_2$ charge conjugation symmetry in $\SU(n) \rtimes \BZ_2$ for even $n$. For higher $N$, we only manage to give some sporadic answers to the above question. In fact, we find that, if we try to gauge the charge conjugation symmetry that is part of the flavour symmetry of the $\U(1)$ gauge theory with $n$ flavours, the Higgs branch Hilbert series does not have a palindromic numerator, meaning that the corresponding moduli space is not a symplectic singularity. This leads us to conjecture that, in this case, such a charge conjugation symmetry has an 't Hooft anomaly. We leave the further study in case of even $n$ for future work.

The paper is organised as follows. In Section \ref{sec:discretegaugingSU2w4}, we consider gauging the charge conjugation symmetries $(\BZ_2)_{\chi_i}$ (with $i=1, \ldots, 4$) associated with the $i$-th factor of $\SO(2)$ of the flavour symmetry $\prod_{i=1}^4 \SO(2) \subset \SO(8)$ of $\SU(2)$ SQCD with four flavours, and obtain the mirror theories of the affine $D_4$ quiver wreathed by the subgroups of $S_4$ that are isomorphic to $\BZ_2$ and $\BZ_2 \times \BZ_2$.  In Section \ref{sec:noninv}, we study how permutation cycles act on the index. One of the most important results is that we propose how to refine the index with respect to the Abelian subgroups of $G$.  This allows us to study sequential gauging of the subgroups of $G$ and construct the webs of symmetries. Of course, when all of such Abelian fugacities are gauged, the Higgs and Coulomb branch limits of the index can be matched nicely with the corresponding Hilbert series presented in \cite{Bourget:2020bxh}. Another main result of this paper is to study the consequences of various 't Hooft anomalies and non-invertible symmetries on the superconformal indices. The results are summarised in Section \ref{sec:sumvarious}. In Section \ref{sec:SQCDtildeF}, we examine wreathing of the mirror theory of $\U(N)$ SQCD with $n$ flavours. The Higgs branch of the $\U(1)$ gauge theory with $n$ flavours with the charge conjugation symmetry which is part of the flavour symmetry being gauged is examined in Section \ref{sec:SQEDCCgauged}. In Appendix \ref{sec:wreathedquivers}, we provide the prescription for computing the superconformal index for any wreathed quivers that contain unitary or special unitary gauge groups.
\subsection*{Notations and conventions}
\bi
\item We denote a $p$-form global symmetry $G$ by a superscript $[p]$, \ie $G^{[p]}$.
\item Whenever we want to specify the fugacity $g$ associated with a zero-form global symmetry $G$, we add a subscript $g$, \ie $(G^{[0]})_g$. 
\item Gauging a zero-form finite symmetry $(G^{[0]})_g$ in $d=3$ dimensions gives rise to a dual $(G^{[1]})_{\hat{g}}$ one-form symmetry, where we put $\hat{g}$ to emphasise that this is the dual symmetry.
\item We denote by $\Gamma^{(2)}$ a two-group, formed by a zero-form symmetry $H^{[0]}$ and a dual one-form symmetry $K^{[1]}$ resulting from gauging a discrete zero-form symmetry $K^{[0]}$. Whenever we want to be specific about which symmetries participate in the two-group, we add the subscripts as follows: $\Gamma^{(2)}_{\hat{k}, h}$.
\ei

\section{\texorpdfstring{Discrete gauging of $\SU(2)$ SQCD with four flavours}{Discrete gauging of SU(2) SQCD with four flavours}}\label{sec:discretegaugingSU2w4}
In this section, we study discrete gauging of 3d $\CN = 4$ $\SU(2)$ SQCD with four flavours by the subgroups of $S_4$ and compute the superconformal index. Some aspects of the discrete gauging of such a theory has been explored in \cite{Bourget:2020bxh,Bourget:2020xdz,Giacomelli:2024sex}. Here, we generalise the existing results and present new perspectives on the subject. 
 
Let us denote by $Q_a^i$ and $\widetilde{Q}^a_i$ the chiral fields in the hypermultiplets and with $\Phi^b_a$ the adjoint chiral of the $\SU(2)$ vector multiplet, where $a, b = 1, 2$ are colour indices and $i = 1, 2, 3, 4$ is the flavour index. The superpotential of the theory reads
\bes{ \label{WSU2with4}
\mathcal{W} = \widetilde{Q}^a_i \Phi^b_a Q^i_b~.
}
Since the theory has an $\so(8)$ flavour symmetry, the chirals $Q_a^i$ and $\widetilde{Q}^a_i$ can be combined into the fields $q_a^I$, with $I = 1, \ldots, 8$, transforming under the vector representation of $\so(8)$, as follows:
\bes{ \label{mapQtoq}
q_a^{2i-1}=Q_a^i+\widetilde{Q}_a^i, \quad q_a^{2i}=-i(Q_a^i-\widetilde{Q}_a^i)~, \quad i=1,\ldots,4~,
}
where $\widetilde{Q}_a^i = \epsilon_{ab} \widetilde{Q}^{b,i}$. It follows that the superpotential \eref{WSU2with4} can be rewritten as
\bes{ 
\mathcal{W}=\epsilon^{ab} \epsilon^{cd} \Phi_{ac} q_b^I q_d^J \delta_{IJ}~,
}
where $\Phi_{ab}$ is symmetric. In these variables, it is well known that there are 28 gauge invariant mesons $\epsilon^{ab} q_a^I q_b^J$. It is thus natural to represent the quiver diagram of $\SU(2)$ SQCD with four flavours as
\bes{ \label{USp2SO8chi}
\vcenter{\hbox{\begin{tikzpicture}
        \node[gauge,label=below:{\footnotesize $\USp(2)$}] (1) at (0,0) {};
        \node[flavour,label=below:{\footnotesize $\SO(8)_{\chi}$}] (2) at (2,0) {};
        \draw (1)--(2);
\end{tikzpicture}}}}
where we denote with the subscript $\chi$ the $(\BZ_2^{[0]})_{\chi}$ zero-form change conjugation symmetry of $\SO(8)$.\footnote{Here and throughout this section, we adopt the point of view of \cite[Section 6]{Aharony:2013kma}, whereby, for the $\SO(N)$ group, the fugacity $\chi$ for the charge conjugation symmetry can be turned on. We will later gauge this symmetry by summing over $\chi = \pm 1$. However, for $\SO(8)$, all representations are real and so, strictly speaking, it is not meaningful to speak about the charge conjugation symmetry which acts on them trivially. By charge conjugation symmetry in this context, we mean the outer automorphism that exchanges the two spinor representations $\mathbf{8}_s$ and $\mathbf{8}_c$, whereby the symmetry can be regarded as $\O(8)$. In order to avoid confusion, we adopt the former point of view.} As pointed out in \cite{Giacomelli:2024sex}, \eref{USp2SO8chi} can be rewritten also as
\bes{ \label{USp2SO7SO1}
\vcenter{\hbox{\begin{tikzpicture}
        \node[flavour,label=below:{\footnotesize $\SO(1)_{\chi_\text{I}}$}] (0) at (-2,0) {};
        \node[gauge,label=below:{\footnotesize $\USp(2)$}] (1) at (0,0) {};
        \node[flavour,label=below:{\footnotesize $\SO(7)_{\chi_\text{II}}$}] (2) at (2,0) {};
        \draw (0)--(1);
        \draw (1)--(2);
\end{tikzpicture}}}}
which has the advantage of possessing two charge conjugation symmetries $(\BZ_2^{[0]})_{\chi_\text{I}}$ and $(\BZ_2^{[0]})_{\chi_\text{II}}$, associated with $\SO(1)$ and $\SO(7)$ respectively.  

In this paper, we consider instead another description of \eref{USp2SO8chi}, whose $\so(8)$ flavour symmetry can be split into $\so(2)^4$ as follows:
\bes{ \label{USp2SO2SO2SO2SO2}
\vcenter{\hbox{\begin{tikzpicture}
        \node[gauge,label=right:{\footnotesize $\USp(2)$}] (0) at (0,0) {};
        \node[flavour,label=above:{\footnotesize $\SO(2)_{\chi_1}$}] (1) at (1,1) {};
        \node[flavour,label=below:{\footnotesize $\SO(2)_{\chi_2}$}] (2) at (1,-1) {};
        \node[flavour,label=below:{\footnotesize $\SO(2)_{\chi_3}$}] (3) at (-1,-1) {};
        \node[flavour,label=above:{\footnotesize $\SO(2)_{\chi_4}$}] (4) at (-1,1) {};
        \draw (0)--(1);
        \draw (0)--(2);
        \draw (0)--(3);
        \draw (0)--(4);
\end{tikzpicture}}}}
The advantage of this description is that \eref{USp2SO2SO2SO2SO2} possesses four charge conjugation symmetries $(\BZ_2^{[0]})_{\chi_i}$, with $i=1,\ldots,4$. Its superconformal index can be derived along the lines of \cite{Aharony:2013dha, Aharony:2013kma, Beratto:2021xmn, Mekareeya:2022spm}. For $\chi_i = 1$ $\forall \ i$, it is given by
\bes{ \label{indUSp2SO2SO2SO2SO2}
&\CI_{\eqref{USp2SO2SO2SO2SO2}}(\vec{y}, \vec{m}| a, n_a| \chi_1 = \ldots = \chi_4 = 1; x)  \\ & = \frac{1}{2} \sum_{l \in \BZ} \oint \frac{d z}{2 \pi i z} \CZ^{\USp(2)}_{\text{vec}}\left(z; l; x\right) \prod_{s = {0, \pm 1}} \CZ^{1}_{\chi} \left(z^{2 s} a^{-2}; 2 s l -2 n_a; x\right) \\ 
& \quad \, \, \times \prod_{i = 1}^{4} \prod_{s_1, s_2 = \pm 1} \CZ^{1/2}_{\chi} \left(z^{s_1} y_i^{s_2} a; s_1 l + s_2 m_i + n_a; x\right) ~,
}
where the $\USp(2)$ vector multiplet contribution is
\bes{
&\CZ^{\USp(2)}_{\text{vec}}(z; l; x)
={x^{-{\left|2 l\right|}}} \prod_{{s}={\pm{1}}}{\left({1}-{\left(-{1}\right)^{2 {s}{l}}}{z^{2 s}}{x^{2 \left|{s}{l}\right|}}\right)}
}
and the contribution of the chiral multiplet of $R$-charge $R$ is
\bes{
\CZ^R_{\chi}(z; l;x) = \left( x^{1-R} z^{-1} \right)^{|l|/2} \prod_{j=0}^\infty \frac{1-(-1)^l z^{-1} x^{|l|+2-R+2j}}{1-(-1)^l z\,  x^{|l|+R+2j}}~,
}
where the subscript $\chi$ here stands for {\it chiral} and it should not be confused with the fugacity associated with charge conjugation in \eref{USp2SO8chi}. We denote by $\vec{y} = (y_1, \ldots, y_4)$ the fugacities for the $\so(2)^4 \cong \so(8)$ flavour symmetry, with $\vec{m} = (m_1, \ldots, m_4)$ the corresponding background magnetic fluxes and with $(a, n_a)$ the (fugacity, background magnetic flux) for the axial symmetry. On the other hand, the index with $\chi_j = -1$ and $\chi_i = 1$ $\forall \ i \neq j$ can be obtained from \eref{indUSp2SO2SO2SO2SO2} by setting $y_j = 1$, $y^{-1}_j = -1$ and $m_j = 0$ (see \cite[(6.9)]{Aharony:2013kma}).\footnote{For instance, in the special case $\chi_i = -1$ $\forall \ i$, we set $y_i = 1$, $y^{-1}_i = -1$ and $m_i = 0$ $\forall \ i$.} For our purposes, it is not necessary to turn on the background magnetic fluxes $\vec{m}$ and $n_a$. Hence, from now on, we will set $m_i = n_a = 0$ $\forall \ i$ and drop their dependence from the index. For generic values of $\chi_1, \ldots, \chi_4$, the index of theory \eref{USp2SO2SO2SO2SO2} with $y_1 = \ldots = y_4 = 1$ admits the following expression:
\bes{ \label{indUSp2SO2SO2SO2SO2withchi}
&\CI_{\eqref{USp2SO2SO2SO2SO2}}(y_i=1|a| \chi_1, \chi_2, \chi_3, \chi_4; x) \\ & = 1+ a^2 \left[6 + 4 \sum_{i = 1}^4 \chi_i + \sum_{1 < i < j <4} \chi_i \chi_j\right] x \\ & \quad + \left[a^4 \left(52 + 28 \sum_{i = 1}^4 \chi_i + 17 \sum_{1 < i < j <4} \chi_i \chi_j + 8 \sum_{1 < i < j < k <4} \chi_i \chi_j \chi_k + 2 \prod_{i=1}^4 \chi_i\right) \right. \\ & \quad \left.+ a^{-4} \left(1 + \prod_{i=1}^4 \chi_i\right)-\left(7 + 4 \sum_{i = 1}^4 \chi_i + \sum_{1 < i < j <4} \chi_i \chi_j\right)\right] x^2 + \ldots~.
}
At the level of the index, we can gauge $n$ charge conjugation symmetries of \eref{USp2SO2SO2SO2SO2} individually, with $0 \leq n \leq 4$, as follows:
\bes{ \label{gaugemchi}
\frac{1}{2^n} \sum_{(\chi_1, \chi_2, \chi_3, \chi_4)=\vec r} \CI_{\eqref{USp2SO2SO2SO2SO2}}(\vec{y}| a| \chi_1, \chi_2, \chi_3, \chi_4; x)~,
}
where $\vec r = (r_1, r_2, r_3, r_4)$ has $n$ entries that are summed over $\pm 1$ and $4 - n$ entries which are set to be equal to either $1$ or $-1$. We can also gauge a diagonal combination of two charge conjugation symmetries of \eref{USp2SO2SO2SO2SO2}, say $(\BZ_2^{[0]})_{\chi_1}$ and $(\BZ_2^{[0]})_{\chi_2}$, as follows:
\bes{ \label{gaugediagchi}
\frac{1}{2} \sum_{\chi_1 = \chi_2= \pm 1} \CI_{\eqref{USp2SO2SO2SO2SO2}}(\vec{y}| a| \chi_1, \chi_2, \chi_3, \chi_4; x)~.
}
\subsection{\texorpdfstring{$\BZ_2$ and $\BZ_2 \times \BZ_2$ discrete gauging of quiver \eref{USp2SO2SO2SO2SO2}}{Z2 and Z2 x Z2 discrete gauging of quiver \eref{USp2SO2SO2SO2SO2}}} \label{sec:SU2w4DiscreteGauging}
The description \eref{USp2SO2SO2SO2SO2} of $\SU(2)$ SQCD with four flavours provides a natural interpretation of discrete gauging by all $\BZ_2$ and $\BZ_2 \times \BZ_2$ subgroups of $S_4$. Indeed, theory \eref{USp2SO2SO2SO2SO2}$/\langle (ij) \rangle$, where $\langle (ij) \rangle$ is the $\BZ_2$ group generated by the permutation cycle $(i j)$ of two elements $i$ and $j$, can be implemented by gauging $(\BZ_2^{[0]})_{\chi_j}$ while leaving $\chi_i$ untouched. At the level of the index, using \eref{gaugemchi} with $n = 1$, this corresponds to
\bes{ \label{indUSp2SO2SO2SO2SO2modij}
\eref{USp2SO2SO2SO2SO2}/\langle (i j) \rangle = \frac{1}{2} \sum_{\chi_j=\pm 1} \CI_{\eqref{USp2SO2SO2SO2SO2}}(\vec{y}| a| \chi_i, \chi_j, \chi_l, \chi_m; x)~.
}
Similarly, one can also gauge a $\BZ_2$ group of the type $\langle (ij) (lm) \rangle$. This can be realised by gauging a diagonal combination of $(\BZ_2^{[0]})_{\chi_j}$ and $(\BZ_2^{[0]})_{\chi_m}$, while leaving $(\BZ_2^{[0]})_{\chi_i}$ and $(\BZ_2^{[0]})_{\chi_l}$ untouched. From \eref{gaugediagchi}, the corresponding index reads
\bes{ \label{indUSp2SO2SO2SO2SO2modijkl}
\eref{USp2SO2SO2SO2SO2}/\langle (i j) (l m) \rangle = \frac{1}{2} \sum_{\chi_j=\chi_m=\pm 1} \CI_{\eqref{USp2SO2SO2SO2SO2}}(\vec{y}| a| \chi_i, \chi_j, \chi_l, \chi_m; x)~.
}
Exploiting this prescription, we can now discuss explicitly the discrete gauging of quiver \eref{USp2SO2SO2SO2SO2} by the $\BZ_2$ and $\BZ_2 \times \BZ_2$ subgroups of $S_4$, namely
\bi
\item $\BZ_2 = \langle (12) \rangle$, 
\item double transposition (DT) $= \langle (12)(34) \rangle$, which is also isomorphic to $\BZ_2$, 
\item non-normal Klein (NNK) $= \langle (12),(34) \rangle$, which is isomorphic to $\BZ_2 \times \BZ_2$,
\item normal Klein (NK) $= \langle (12)(34), (13)(24) \rangle$, which is also isomorphic to $\BZ_2 \times \BZ_2$.
\ei
In order to write down more compact expressions, we will use the following notation:
\bes{ \label{USp2SO2SO2SO2SO2mat}
\eref{USp2SO2SO2SO2SO2} \equiv \begin{bmatrix} \chi_4 & \chi_1 \\ \chi_3 & \chi_2 \end{bmatrix}~.}
\subsubsection*{\texorpdfstring{$\BZ_2$ discrete gauging: single transposition}{Z2 discrete gauging: single transposition}}
Let us study the $\BZ_2$ discrete gauging of $\SU(2)$ SQCD with four flavours described by the following action on the elementary fields:
\bes{ \label{Z2onQ}
\begin{cases} Q_{a}^{i}\rightarrow Q_{a}^{i}~, \quad \widetilde{Q}_{a}^{i}\rightarrow\widetilde{Q}_{a}^{i}~, \quad i=1,3,4 \\
Q_{a}^{2}\leftrightarrow\widetilde{Q}_{a}^{2}
\end{cases}~,
}
which, using the map \eref{mapQtoq}, can be equivalently implemented as
\bes{ \label{Z2onq}
\begin{cases}
q_a^I \rightarrow q_a^I~, \quad I \neq 4 \\
q_a^4 \rightarrow -q_a^4
\end{cases}~.
}
As analysed in \cite{Argyres:2016yzz}, if such a transformation is implemented, the resulting Higgs branch is isomorphic to the next-to-minimal orbit closure of $\so(7)$, denoted by $\bar{\mathrm{n.min}\, B_3}$. Indeed, upon gauging this $\BZ_2$ transformation, out of the original 28 gauge invariant mesons, 7 are projected out, namely
\bes{
\epsilon^{ab} q_a^I q_b^4~, \quad I \neq 4~,
}
and the remaining 21 of them survive, transforming in the $[0, 1, 0]$ representation of $\so(7)$. This is in agreement with the branching rule
\bes{ \label{28so8to21so7}
[0,1,0,0]_{\so(8)} \to { [0,1,0]_{\so(7)}} + {\red [1,0,0]_{\so(7)}}~,
}
where the term in red gets removed by the $\BZ_2$ gauging.
It was observed in \cite{Giacomelli:2024sex} that the resulting theory can be realised by gauging $\O(1)$ inside $\so(8)$ in \eref{USp2SO8chi} (see \cite[Section 3.2]{KobakSwann}), or equivalently by gauging $(\BZ_2^{[0]})_{\chi_\text{I}}$ in \eref{USp2SO7SO1}. This leads to the following quiver (see \cite[Appendix D.1]{Cremonesi:2014uva} and \cite{Bourget:2020bxh}):
\bes{ \label{USp2SO7O1}
\eref{USp2SO7SO1}/(\BZ_2^{[0]})_{\chi_\text{I}} = \vcenter{\hbox{\begin{tikzpicture}
        \node[gauge,label=below:{\footnotesize $\O(1)$}] (0) at (-2,0) {};
        \node[gauge,label=below:{\footnotesize $\USp(2)$}] (1) at (0,0) {};
        \node[flavour,label=below:{\footnotesize $\SO(7)_{\chi_\text{II}}$}] (2) at (2,0) {};
        \draw (0)--(1);
        \draw (1)--(2);
\end{tikzpicture}}}}
Using the description \eref{USp2SO2SO2SO2SO2mat}, this theory can also be realised by gauging the $\BZ_2 = \langle (1 2) \rangle$ subgroup of $S_4$ as
\bes{ \label{USp2SO2SO2SO2SO2Z2}
\eref{USp2SO2SO2SO2SO2}/\langle (1 2) \rangle \equiv \eref{USp2SO2SO2SO2SO2}/\BZ_2 = \frac{1}{2} \left\{\begin{bmatrix} + & + \\ + & + \end{bmatrix} + \begin{bmatrix} + & + \\ + & - \end{bmatrix} \right\}~.
}

\subsubsection*{$\BZ_2$ discrete gauging: double transposition}
We are now interested in the $\BZ_2$ transformation which acts on the elementary fields as
\bes{ \label{DTonQ}
\begin{cases} Q_{a}^{i}\rightarrow Q_{a}^{i}~, \quad \widetilde{Q}_{a}^{i}\rightarrow\widetilde{Q}_{a}^{i}~, \quad& i=1,3 \\
Q_{a}^{j}\leftrightarrow\widetilde{Q}_{a}^{j}~, \quad& j=2,4
\end{cases}~,
}
which can be rephrased as
\bes{ \label{DTonq}
\begin{cases}
q_a^I \rightarrow q_a^I~, \quad& I \neq 4, 8 \\
q_a^J \rightarrow -q_a^J~, \quad& J=4, 8
\end{cases}~.
}
The effect of \eref{DTonq} is that of breaking the original $\so(8)$ flavour symmetry into $\su(4) \oplus \u(1)$, where the vector representation of $\so(8)$ gets decomposed as
\bes{ \label{so8tosu4u1}
[1,0,0,0]_{\so(8)} \to { [0,1,0]_{\su(4)}(0)} + {\red [0,0,0]_{\su(4)}(2)} + {\red [0,0,0]_{\su(4)}(-2)}~,
}
where the terms in {\red red} are killed by gauging \eref{DTonq}. The action on the gauge invariant operators is as follows.
\begin{itemize}
\item The mesons $\epsilon^{ab} q_a^I q_b^J$, with $I, J \neq \{4, 8\}$, are invariant.
\item The mesons $\epsilon^{ab} q_a^I q_b^4$, with $I \neq \{4, 8\}$, get mapped to minus themselves.
\item The mesons $\epsilon^{ab} q_a^I q_b^8$, with $I = 1, \ldots, 7$ and $I \neq 4$, also get mapped to minus themselves.
\item The meson $\epsilon^{ab} q_a^4 q_b^8$ is invariant.
\end{itemize}
Therefore, after gauging \eref{DTonq}, we are left with 16 gauge invariant mesons, transforming in the $[1, 0, 1] (0) \oplus [0, 0, 0] (0)$ representation of $\su(4) \oplus \u(1)$, as follows from the branching rule
\bes{ \label{so8tosu4u116}
\scalebox{0.92}{$
[0,1,0,0]_{\so(8)} \to { [1,0,1]_{\su(4)}(0)} + {\red [0,1,0]_{\su(4)}(2)} + {\red [0,1,0]_{\su(4)}(-2)} + { [0,0,0]_{\su(4)}(0)}~.
$}}
Using the description \eref{USp2SO2SO2SO2SO2mat}, the gauging in question can be realised as
\bes{ \label{USp2SO2SO2SO2SO2DT}
\eref{USp2SO2SO2SO2SO2}/\langle (1 2)(3 4) \rangle \equiv \eref{USp2SO2SO2SO2SO2}/\mathrm{DT} = \frac{1}{2} \left\{\begin{bmatrix} + & + \\ + & + \end{bmatrix} + \begin{bmatrix} - & + \\ + & - \end{bmatrix} \right\}~.
}

\subsubsection*{$\BZ_2 \times \BZ_2$ discrete gauging: non-normal Klein subgroup of $S_4$}
We can proceed by studying discrete gauging of $\SU(2)$ SQCD with four flavours by the $\BZ_2 \times \BZ_2$ transformation which combines \eref{Z2onQ} with another $\BZ_2$ transformation, namely
\bes{ \label{NNKonQ}
\eref{Z2onQ}:&\begin{cases} Q_{a}^{i}\rightarrow Q_{a}^{i}~, \quad \widetilde{Q}_{a}^{i}\rightarrow\widetilde{Q}_{a}^{i}~, \quad i=1,3,4 \\
Q_{a}^{2}\leftrightarrow\widetilde{Q}_{a}^{2}
\end{cases} \\ & \qquad \qquad \qquad+ \\
&\begin{cases} Q_{a}^{i}\rightarrow Q_{a}^{i}~, \quad \widetilde{Q}_{a}^{i}\rightarrow\widetilde{Q}_{a}^{i}~, \quad i=1,2,3 \\
Q_{a}^{4}\leftrightarrow\widetilde{Q}_{a}^{4}
\end{cases}~.
}
In the variables \eref{mapQtoq}, this transformation can be expressed by acting separately with both \eref{Z2onq} and the $\BZ_2$ transformation under which $q_a^8$ changes sign:
\bes{ \label{NNKonq}
\eref{Z2onq}:\begin{cases}
q_a^I \rightarrow q_a^I~, \quad I \neq 4 \\
q_a^4 \rightarrow -q_a^4
\end{cases}~ + \quad
\begin{cases}
q_a^I \rightarrow q_a^I~, \quad I \neq 8 \\
q_a^8 \rightarrow -q_a^8
\end{cases}~.
}
This action first breaks the vector representation of $\so(8)$ into the $[1,0,0]$ representation of $\so(7)$ and, then, further reduces the flavour symmetry down to $\su(4)$, according to the branching rule
\bes{
[1,0,0]_{\so(7)} \to { [0,1,0]_{\su(4)}} + {\red [0,0,0]_{\su(4)}}~,
}
where we denoted in {\red red} the term which gets projected out by \eref{NNKonq}. In terms of the meson spectrum, the action of this $\BZ_2 \times \BZ_2$ transformation is the following.
\begin{itemize}
\item The mesons $\epsilon^{ab} q_a^I q_b^J$, with $I, J \neq \{4, 8\}$, are invariant.
\item The mesons $\epsilon^{ab} q_a^I q_b^4$, with $I \neq 4$, get mapped to minus themselves.
\item The mesons $\epsilon^{ab} q_a^I q_b^8$, with $I = 1, \ldots, 7$, also get mapped to minus themselves.
\end{itemize}
This means that the gauge invariant mesons are reduced to 15, corresponding to the $[1,0,1]$ representation of $\su(4)$. Indeed, starting with the original theory, we can exploit the branching rule \eref{28so8to21so7} from $\so(8)$ to $\so(7)$, followed by the branching rule from $\so(7)$ to $\su(4)$
\bes{
[0,1,0]_{\so(7)} \to { [1,0,1]_{\su(4)}} + {\red [0,1,0]_{\su(4)}}~.
}
Using notation \eref{USp2SO2SO2SO2SO2mat}, the resulting gauged theory this time can be expressed as
\bes{ \label{USp2SO2SO2SO2SO2NNK}
\eref{USp2SO2SO2SO2SO2}/\langle (1 2), (3 4) \rangle &\equiv \eref{USp2SO2SO2SO2SO2}/\mathrm{NNK} \\&= \frac{1}{4} \left\{\begin{bmatrix} + & + \\ + & + \end{bmatrix} + \begin{bmatrix} + & + \\ + & - \end{bmatrix} + \begin{bmatrix} - & + \\ + & + \end{bmatrix} + \begin{bmatrix} - & + \\ + & - \end{bmatrix} \right\}~.
}

\subsubsection*{$\BZ_2 \times \BZ_2$ discrete gauging: normal Klein subgroup of $S_4$}
Finally, let us consider the $\BZ_2 \times \BZ_2$ transformation acting via \eref{DTonQ} combined with another $\BZ_2$ transformation as follows:
\bes{  \label{NKonQ}
\eref{DTonQ}:&\begin{cases} Q_{a}^{i}\rightarrow Q_{a}^{i}~, \quad \widetilde{Q}_{a}^{i}\rightarrow\widetilde{Q}_{a}^{i}~, \quad& i=1,3 \\
Q_{a}^{j}\leftrightarrow\widetilde{Q}_{a}^{j}~, \quad& j=2,4
\end{cases} \\ & \qquad \qquad \qquad + \\
&\begin{cases} Q_{a}^{i}\rightarrow Q_{a}^{i}~, \quad \widetilde{Q}_{a}^{i}\rightarrow\widetilde{Q}_{a}^{i}~, \quad& i=1,2 \\
Q_{a}^{j}\leftrightarrow\widetilde{Q}_{a}^{j}~, \quad& j=3,4
\end{cases}~.
}
Equivalently, this transformation can be performed by acting separately with both \eref{DTonq} and another $\BZ_2$ transformation which flips the sign of both $q_a^6$ and $q_a^8$:
\bes{ \label{NKonq}
\eref{DTonQ}:\begin{cases}
q_a^I \rightarrow q_a^I~, \quad& I \neq 4, 8 \\
q_a^J \rightarrow -q_a^J~, \quad& J=4, 8
\end{cases}~ + \quad
\begin{cases}
q_a^I \rightarrow q_a^I~, \quad& I \neq 6, 8 \\
q_a^J \rightarrow -q_a^J~, \quad& J=6, 8
\end{cases}~.
}
Observe that the flavour symmetry now reduces to $\usp(4)$, where the $[0,1,0]$ representation of $\su(4)$ coming from \eref{so8tosu4u1} is decomposed according to the branching rule
\bes{
[0,1,0]_{\su(4)} \to { [0,1]_{\usp(4)}} + {\red [0,0]_{\usp(4)}}
}
and the singlet in {\red red} is projected out. Accordingly, out of the 28 gauge invariant mesons of the original theory, 10 of them are left untouched by this $\BZ_2 \times \BZ_2$ action.
\begin{itemize}
\item The mesons $\epsilon^{ab} q_a^I q_b^J$, with $I, J \neq \{4, 8\}$ and $I, J \neq \{6, 8\}$, are invariant.
\item The mesons $\epsilon^{ab} q_a^I q_b^4$, with $I \neq \{4, 8\}$, get mapped to minus themselves, as well as the mesons $\epsilon^{ab} q_a^I q_b^6$, with $I \neq \{6, 8\}$.
\item The mesons $\epsilon^{ab} q_a^I q_b^8$, with $I = 1, \ldots, 7$ and $I \neq \{4, 6\}$, also get mapped to minus themselves.
\item The mesons $\epsilon^{ab} q_a^I q_b^8$, with $I = \{4, 6\}$ are no longer invariant.
\end{itemize}
Taking into account \eref{so8tosu4u116}, this is consistent with the branching rules
\bes{
[1,0,1]_{\su(4)}&\to[2,0]_{\usp(4)}+{\red [0,1]_{\usp(4)}}~,\\
[0,0,0]_{\su(4)}&\to{\red [0,0]_{\usp(4)}}~.
}
Using notation \eref{USp2SO2SO2SO2SO2mat}, we propose the following description for this $\BZ_2 \times \BZ_2$ gauging:
\bes{ \label{USp2SO2SO2SO2SO2NK}
\eref{USp2SO2SO2SO2SO2}/\langle (1 2) (3 4), (1 3) (2 4) \rangle &\equiv \eref{USp2SO2SO2SO2SO2}/\mathrm{NK} \\&= \frac{1}{4} \left\{\begin{bmatrix} + & + \\ + & + \end{bmatrix} + \begin{bmatrix} - & + \\ + & - \end{bmatrix} + \begin{bmatrix} - & + \\ - & + \end{bmatrix} + \begin{bmatrix} + & + \\ - & - \end{bmatrix} \right\}~.
}

\subsubsection*{Superconformal indices, Higgs branches and Coulomb branches}
A robust check that our prescription for obtaining the $\BZ_2$ and $\BZ_2 \times \BZ_2$ discrete gaugings of $\SU(2)$ SQCD with four flavours discussed above is correct comes from the computation of the superconformal index \eref{indUSp2SO2SO2SO2SO2}, which, as pointed out explicitly in \eref{indUSp2SO2SO2SO2SO2withchi}, can be expressed with all of the four charge conjugation fugacities turned on. Gauging them as explained in \eref{indUSp2SO2SO2SO2SO2modij} and \eref{indUSp2SO2SO2SO2SO2modijkl}, we can therefore derive the indices of \eref{USp2SO2SO2SO2SO2}$/\Gamma$, with $\Gamma$ being either trivial, or one of the $\BZ_2$ and $\BZ_2 \times \BZ_2$ subgroups of $S_4$ discussed above. Importantly, in order for the index to be well-defined, when the charge conjugation $\chi_i$ in the definition of \eref{USp2SO2SO2SO2SO2}$/\Gamma$ takes both values $\pm 1$, we set the corresponding flavour fugacity $y_i = 1$. Without doing so, we would obtain an ill-defined index, whose series expansion in powers of $x$ contains half-integer coefficients.\footnote{For instance, theory \eref{USp2SO2SO2SO2SO2}$/\langle(12)\rangle$ is defined in \eref{USp2SO2SO2SO2SO2Z2} as a sum of two terms, one with $\chi_2 = 1$ and the other with $\chi_2 = -1$. Hence, we have to set $y_2 = 1$ in the corresponding index, otherwise this would take the form
\bes{
1+ a^2 \left( \frac{1}{2} \sum_{i = 1, 3, 4} \, \sum_{s_1, s_2 = \pm 1} y_i^{s_1} y_2^{s_2} + \sum_{i < j, \, i, j \neq 2} \, \sum_{s_1, s_2 = \pm 1} y_i^{s_1} y_j^{s_2} + 3\right) x + \ldots~,
}
with an half-integer coefficient appearing in front of $y_2$.  

This can actually be explained as follows: the discrete symmetries in question are actually part of the Weyl symmetry of $\so(8)$ and they do not commute with all Cartan elements corresponding to fugacities $y_{i}$ (with $i=1,\dots, 4$). In fact, one is allowed to turn on the fugacities associated with commuting symmetries, and so the index cannot be simultaneously refined with respect to such discrete symmetries and all of the variables $y_i$ of the maximal torus of $\so(8)$. 
\label{foot:welldef}}
Upon setting $\vec{y}=(1, \ldots, 1)$, the corresponding indices up to order $x^3$ are reported below.
\bes{ \label{tabSU2w4}
\scalebox{0.93}{
\begin{tabular}{c|l|c}
\hline
$\Gamma$ & \qquad \qquad \quad Index of \eref{USp2SO2SO2SO2SO2}$/\Gamma$ & Flavour  \\
 &  & symmetry  \\
\hline
Trivial & $1+{28 a^2} x+\left(300 a^4+{2}{a^{-4}}-29\right) x^2$ & $\so(8)$\\ & $\, \, \, \,+\left(1925 a^6+{a^{-6}}-649 a^2-{a^{-2}}\right) x^3+\ldots$ & \\
\hline
$\langle (1 2) \rangle$ & $1+{21 a^2 x}+\left(195 a^4+{a^{-4}}-22\right) x^2$ & $\so(7)$\\ = $\BZ_2$ & $\, \, \, \,+\left(1155 a^6-{404}{a^{-2}}\right) x^3+\ldots$ &\\
\hline
$\langle (1 2)(3 4) \rangle$ & $1+{16 a^2 x}+\left(160 a^4+{2 a^{-4}}-17\right) x^2$ & $\su(4) \oplus \u(1)$\\ = DT & $\, \, \, \,+\left(985 a^6+a^{-6}-329 a^2-{a^{-2}}\right) x^3+\ldots$ &\\
\hline
$\langle (1 2),(3 4) \rangle$ & $1+{15 a^2 x}+\left(125 a^4+{a^{-4}}-16\right) x^2$ & $\su(4)$\\ = NNK & $\, \, \, \,+\left(685 a^6-244 a^2\right) x^3+\ldots$ &\\
\hline
$\langle (1 2) (3 4), (1 3) (2 4) \rangle$ & $1+{10 a^2 x}+\left(90 a^4+{2 a^{-4}}-11\right) x^2$ & $\usp(4)$\\ = NK & $\, \, \, \,+\left(515 a^6+a^{-6}-169 a^2-a^{-2}\right) x^3+\ldots$ &\\
\end{tabular}
}
}
In the last column, we point out the global symmetry associated with the 3d $\CN=4$ flavour current multiplet, which can be deduced by looking at the moment map appearing at order $x$ in the index. For instance, in the case $\Gamma =$ trivial, the coefficient 28 in ${28 a^2 x}$ corresponds to the adjoint representation $[0,1,0,0]$ of $\so(8)$.\footnote{Here we keep the fugacity $a$ associated with the $\U(1)$ axial symmetry for the convenience of taking the Higgs branch and Coulomb branch limits of the index, defined in \eref{CBHBlimits}.}

As pointed out in \cite{Razamat:2014pta}, one can take the Higgs and Coulomb branch limits of the index to obtain the Higgs and Coulomb branch Hilbert series of theory \eref{USp2SO2SO2SO2SO2}$/\Gamma$.  We define 
\bes{ \label{CBHBlimits}
&h = x^{1/2} a~, ~ c = x^{1/2} a^{-1}~, \\ 
&\text{or equivalently} \quad  x = h c ~,~ a = (h/c)^{1/2}~,
}
and substitute them in the index.  In the Higgs branch limit we send $c \rightarrow 0$ and keep $h$ fixed, whereas in the Coulomb branch limit we send $h \rightarrow 0$ and keep $c$ fixed. Note that, given a series expansion of the index in terms of $x$, the Higgs ({\it resp.} Coulomb) branch limit can be obtained by reading off the coefficients of the terms $a^{2p} x^p$ ({\it resp.} $a^{-2p} x^p$), with $p$ integral or half-integral, and multiplying each of them to $h^{2p}$ ({\it resp.} $c^{2p}$). Conventionally, the corresponding Hilbert series can be obtained by simply setting $h$ or $c$ in each of the aforementioned limits to $t$.

It turns out that the Higgs branch limit of the index of \eref{USp2SO2SO2SO2SO2}$/\Gamma$ reproduces the following Hilbert series:
\bes{ \label{HBHSSU2w4modGamma}
&\mathrm{HS}\left[\text{HB of \eref{USp2SO2SO2SO2SO2}/$\Gamma$}\right](t,\vec{y}) \\
&=  \oint_{|z|=1} \frac{d z}{2\pi i z} (1-z^2) \times {\blue  \frac{1}{m}\sum_{j=0}^{m-1} f_{\Gamma,j}(z, t, \vec{y}) } \times \PE \left[- (z^2+1+z^{-2}) t^2  \right]~,
}
where we set $t=h$, we denote with $m$ the cardinality of $\Gamma$ and we define
\bes{ \label{fHS}
f_{\Gamma,j}(z, t, \vec{y}) = \frac{1}{\det{\left(\ID - \QM_{\Gamma,j} t \right)}}~,
}
with both $\ID$ and $\QM_{\Gamma,j}$ being $16 \times 16$ matrices. The latter is a block-diagonal matrix of the form
\bes{ \label{QmatGamma}
\QM_{\Gamma,j} = \left(
\begin{array}{c|c}
 \MA_{\Gamma,j} & \ZM \\
 \hline
 \ZM & \BM_{\Gamma,j} \\
\end{array}
\right)~,
}
where $\BM_{\Gamma,j} = \left[\MA_{\Gamma,j}^{-1}\right]^T$ and $\MA_{\Gamma,j}$ depends on the choice of $\Gamma$ and the value of $j$ as follows.
\begin{itemize}
\item When $j = 0$, then $\forall \ \Gamma$ we have
\bes{
\MA_{\Gamma,0} =  \left(
\begin{array}{cccc|cccc}
 \frac{y_1}{z} & 0 & 0 & 0 & 0 & 0 & 0 & 0 \\
 0 & \frac{y_2}{z} & 0 & 0 & 0 & 0 & 0 & 0 \\
 0 & 0 & \frac{y_3}{z} & 0 & 0 & 0 & 0 & 0 \\
 0 & 0 & 0 & \frac{y_4}{z} & 0 & 0 & 0 & 0 \\
 \hline
 0 & 0 & 0 & 0 & \frac{1}{y_1 z} & 0 & 0 & 0 \\
 0 & 0 & 0 & 0 & 0 & \frac{1}{y_2 z} & 0 & 0 \\
 0 & 0 & 0 & 0 & 0 & 0 & \frac{1}{y_3 z} & 0 \\
 0 & 0 & 0 & 0 & 0 & 0 & 0 & \frac{1}{y_4 z} \\
\end{array}
\right)~.
}
\item For $\Gamma = \langle (12) \rangle$, which has cardinality 2, $j$ can be equal to $0$ or $1$. The matrix $\MA_{\langle (12) \rangle,1}$, labelled by $j = 1$, takes the form 
\bes{ \label{AmatZ2}
j=1:  \left(
\begin{array}{cccc|cccc}
 \frac{y_1}{z} & 0 & 0 & 0 & 0 & 0 & 0 & 0 \\
 0 & 0 & 0 & 0 & 0 & \frac{1}{y_2 z} & 0 & 0 \\
 0 & 0 & \frac{y_3}{z} & 0 & 0 & 0 & 0 & 0 \\
 0 & 0 & 0 & \frac{y_4}{z} & 0 & 0 & 0 & 0 \\
 \hline
 0 & 0 & 0 & 0 & \frac{1}{y_1 z} & 0 & 0 & 0 \\
 0 & \frac{y_2}{z} & 0 & 0 & 0 & 0 & 0 & 0 \\
 0 & 0 & 0 & 0 & 0 & 0 & \frac{1}{y_3 z} & 0 \\
 0 & 0 & 0 & 0 & 0 & 0 & 0 & \frac{1}{y_4 z} \\
\end{array}
\right)~,
}
which, together with $\BM_{\langle (12) \rangle,1}$, reproduces the $\BZ_2$ action \eref{Z2onQ}. The action of such matrix projects out the fugacity $y_2$ from \eref{HBHSSU2w4modGamma}.
\item When $\Gamma$ is the double transposition subgroup of $S_4$, which also has cardinality 2, the $\BZ_2$ transformation \eref{DTonQ} is implemented by $\QM_{\langle (1 2) (3 4) \rangle,1}$, with the matrix $\MA_{\langle (1 2) (3 4) \rangle,1}$ given by
\bes{ \label{AmatDT}
j=1: \left(
\begin{array}{cccc|cccc}
 \frac{y_1}{z} & 0 & 0 & 0 & 0 & 0 & 0 & 0 \\
 0 & 0 & 0 & 0 & 0 & \frac{1}{y_2 z} & 0 & 0 \\
 0 & 0 & \frac{y_3}{z} & 0 & 0 & 0 & 0 & 0 \\
 0 & 0 & 0 & 0 & 0 & 0 & 0 & \frac{1}{y_4 z} \\
 \hline
 0 & 0 & 0 & 0 & \frac{1}{y_1 z} & 0 & 0 & 0 \\
 0 & \frac{y_2}{z} & 0 & 0 & 0 & 0 & 0 & 0 \\
 0 & 0 & 0 & 0 & 0 & 0 & \frac{1}{y_3 z} & 0 \\
 0 & 0 & 0 & \frac{y_4}{z} & 0 & 0 & 0 & 0 \\
\end{array}
\right)~.
}
The fugacities $y_2$ and $y_4$ are projected out from \eref{HBHSSU2w4modGamma} upon acting with this matrix.
\item Next, when $\Gamma$ is the non-normal Klein subgroup of $S_4$, which has cardinality 4, $j$ takes values in $\{0,1,2,3\}$. For $j=2$, the matrix $\MA_{\langle (1 2), (3 4) \rangle,2}$ is
\bes{
j=2:  \left(
\begin{array}{cccc|cccc}
 \frac{y_1}{z} & 0 & 0 & 0 & 0 & 0 & 0 & 0 \\
 0 & \frac{y_2}{z} & 0 & 0 & 0 & 0 & 0 & 0 \\
 0 & 0 & \frac{y_3}{z} & 0 & 0 & 0 & 0 & 0 \\
 0 & 0 & 0 & 0 & 0 & 0 & 0 & \frac{1}{y_4 z} \\
 \hline
 0 & 0 & 0 & 0 & \frac{1}{y_1 z} & 0 & 0 & 0 \\
 0 & 0 & 0 & 0 & 0 & \frac{1}{y_2 z} & 0 & 0 \\
 0 & 0 & 0 & 0 & 0 & 0 & \frac{1}{y_3 z} & 0 \\
 0 & 0 & 0 & \frac{y_4}{z} & 0 & 0 & 0 & 0 \\
\end{array}
\right)~,
}
whose action projects out the fugacity $y_4$ from \eref{HBHSSU2w4modGamma}, whereas $\MA_{\langle (1 2), (3 4) \rangle,1}$ and $\MA_{\langle (1 2), (3 4) \rangle,3}$ coincide with \eref{AmatZ2} and \eref{AmatDT} respectively. Observe that the corresponding matrices $\QM_{\langle (1 2), (3 4) \rangle,j}$ reproduce the $\BZ_2 \times \BZ_2$ transformation \eref{NNKonQ}.
\item Finally, let us consider the case in which $\Gamma$ is the normal Klein subgroup of $S_4$, with cardinality 4. For $j = 1$, the matrix $\MA_{\langle (1 2) (3 4), (13) (24) \rangle,1}$ is given by \eref{AmatDT}. Moreover, for $j = 2, 3$, the matrices $\MA_{\langle (1 2) (3 4), (13) (24) \rangle,j}$ take the following form:
\bes{
\scalebox{0.85}{$
j=2:  \left(
\begin{array}{cccc|cccc}
 \frac{y_1}{z} & 0 & 0 & 0 & 0 & 0 & 0 & 0 \\
 0 & \frac{y_2}{z} & 0 & 0 & 0 & 0 & 0 & 0 \\
 0 & 0 & 0 & 0 & 0 & 0 & \frac{1}{y_3 z} & 0 \\
 0 & 0 & 0 & 0 & 0 & 0 & 0 & \frac{1}{y_4 z} \\
 \hline
 0 & 0 & 0 & 0 & \frac{1}{y_1 z} & 0 & 0 & 0 \\
 0 & 0 & 0 & 0 & 0 & \frac{1}{y_2 z} & 0 & 0 \\
 0 & 0 & \frac{y_3}{z} & 0 & 0 & 0 & 0 & 0 \\
 0 & 0 & 0 & \frac{y_4}{z} & 0 & 0 & 0 & 0 \\
\end{array}
\right)~, \quad
j = 3:  \left(
\begin{array}{cccc|cccc}
 \frac{y_1}{z} & 0 & 0 & 0 & 0 & 0 & 0 & 0 \\
 0 & 0 & 0 & 0 & 0 & \frac{1}{y_2 z} & 0 & 0 \\
 0 & 0 & 0 & 0 & 0 & 0 & \frac{1}{y_3 z} & 0 \\
 0 & 0 & 0 & \frac{y_4}{z} & 0 & 0 & 0 & 0 \\
 \hline
 0 & 0 & 0 & 0 & \frac{1}{y_1 z} & 0 & 0 & 0 \\
 0 & \frac{y_2}{z} & 0 & 0 & 0 & 0 & 0 & 0 \\
 0 & 0 & \frac{y_3}{z} & 0 & 0 & 0 & 0 & 0 \\
 0 & 0 & 0 & 0 & 0 & 0 & 0 & \frac{1}{y_4 z} \\
\end{array}
\right)~.
$}
}
Observe that, for $j=2$, the fugacities $y_3$ and $y_4$ are projected out from \eref{HBHSSU2w4modGamma}, whereas, for $j=3$, the associated matrix projects out the fugacities $y_2$ and $y_3$ from \eref{HBHSSU2w4modGamma}. In terms of the matrices $\QM_{\langle (1 2) (3 4), (13) (24) \rangle,j}$, they implement the $\BZ_2 \times \BZ_2$ transformation \eref{NKonQ}.
\end{itemize}
In order for the Hilbert series to be well-defined, if the fugacity $y_i$ is projected out in at least one of the $m$ terms appearing in the sum highlighted in {\blue blue} in \eref{HBHSSU2w4modGamma}, then we set $y_i = 1$ in the terms where $y_i$ is not projected out. This ensures that, upon expanding the Hilbert series in powers of $t$, there are no half-integer coefficients in front of $y_i$ (see Footnote \ref{foot:welldef}). The computation of the Hilbert series \eref{HBHSSU2w4modGamma}, or equivalently of the Higgs branch limit of the index of \eref{USp2SO2SO2SO2SO2}$/\Gamma$ as defined in \eref{CBHBlimits}, once unrefined, \ie we set $\vec{y} = (1, \ldots, 1)$, yields the results reported in the table below.
\bes{ 
\scalebox{0.96}{
\begin{tabular}{c|l}
\hline
$\Gamma$ & \qquad \quad Higgs branch Hilbert series of \eref{USp2SO2SO2SO2SO2}$/\Gamma$  \\
\hline
Trivial & $\frac{\left(1+t^2\right) \left(1+17 t^2+48 t^4+ 17 t^6+t^8\right)}{\left(1-t^2\right)^{10}}$\\ & $=1+{28} t^2+300 t^4+1925 t^6+8918 t^8+32928 t^{10}+\ldots$ \\
\hline
$\langle (1 2) \rangle$ & $\frac{\left(1+t^2\right) \left(1+10 t^2+20 t^4+ 10 t^6+t^8\right)}{\left(1-t^2\right)^{10}}$\\ = $\BZ_2$ & $=1+{21} t^2+195 t^4+1155 t^6+5096 t^8+18228 t^{10}+\ldots$\\
\hline
$\langle (1 2)(3 4) \rangle$ & $\frac{\left(1 + 3 t^2 + 6 t^4 + 3 t^6 + t^8\right) \left(1 + 8 t^2 + 55 t^4 + 64 t^6 + 
   55 t^8 + 8 t^{10} + t^{12}\right)}{\left(1-t^2\right)^{10} \left(1+t^2\right)^5}$\\ = DT & $=1 + 16 t^2 + 160 t^4 + 985 t^6 + 4522 t^8 + 16576  t^{10} +\ldots$\\
\hline
$\langle (1 2),(3 4) \rangle$ & $\frac{1 + 10 t^2 + 55 t^4 + 150 t^6 + 288 t^8 + 336 t^{10} + 288 t^{12} + 
 150 t^{14} + 55 t^{16} + 10 t^{18} + t^{20}}{\left(1-t^2\right)^{10} \left(1+t^2\right)^5}$\\ = NNK & $=1 + 15 t^2 + 125 t^4 + 685 t^6 + 2898 t^8 + 10052  t^{10} +\ldots$\\
\hline
$\langle (1 2) (3 4), (1 3) (2 4) \rangle$ & $\frac{1 + 5 t^2 + 45 t^4 + 130 t^6 + 314 t^8 + 354 t^{10} + 314 t^{12} + 
 130 t^{14} + 45 t^{16} + 5 t^{18} + t^{20}}{\left(1-t^2\right)^{10} \left(1+t^2\right)^5}$\\ = NK & $=1 + 10 t^2 + 90 t^4 + 515 t^6 + 2324 t^8 + 8400  t^{10} +\ldots$\\
\end{tabular}
}
}
On the other hand, the Coulomb branch limit of the index  of \eref{USp2SO2SO2SO2SO2}$/\Gamma$ reproduces either the Hilbert series of $\BC^2/\hat{D}_4$ or $\BC^2/\hat{D}_6$, where $\hat{D}_4$ and $\hat{D}_6$ denote the binary dihedral groups of order 8 and 16. Let us comment these results in more detail.
\begin{itemize}
\item When $\Gamma$ is either trivial or one between $\mathrm{DT}$ and $\mathrm{NK}$, \ie $\langle (1 2) (3 4)\rangle$ and $\langle (1 2) (3 4), (1 3) (2 4) \rangle$ respectively, then the Coulomb branch of \eref{USp2SO2SO2SO2SO2}$/\Gamma$ is isomorphic to $\BC^2/\hat{D}_4$, whose Hilbert series reads
\bes{ \label{HSC2modD4}
\mathrm{HS}\left[\BC^2/\hat{D}_4\right](t) = \PE\left[2 t^4 + t^6 - t^{12}\right]~.
}
There are two generators that contribute to the order $t^4$ of \eref{HSC2modD4}. One of them is the Casimir operator of the $\USp(2)$ gauge group, namely $u \equiv \tr(\varphi^2)$, with $\varphi$ the scalar in the vector multiplet. The other is the monopole operator, denoted by $\mathfrak{M}$. They both come from the coefficients of $a^{-4} x^2$ in \eref{indUSp2SO2SO2SO2SO2withchi}, from which we see that $\mathfrak{M}$ carries the fugacity $\prod_{i=1}^4 \chi_i$, meaning that it is charged under the combination of the charge conjugation symmetries associated with the four $\SO(2)$ gauge groups of \eref{USp2SO2SO2SO2SO2}, whereas $u$ is neutral under these symmetries. There is also a generator $G_6 \equiv \mathfrak{M} \varphi$, corresponding to the term at order $t^6$ in \eref{HSC2modD4}.  These generators satisfy the relation at order $t^{12}$, namely
\bes{
G_6^2 + \mathfrak{M}^2 u = u^3~,
}
which is the defining equation of $\BC^2/\hat{D}_4$.
\item Instead, when $\Gamma$ is either $\BZ_2 = \langle (1 2)\rangle$ or $\mathrm{NNK} = \langle (1 2), (3 4)\rangle$, the Coulomb branch of \eref{USp2SO2SO2SO2SO2}$/\Gamma$ is isomorphic to $\BC^2/\hat{D}_6$, with Hilbert series given by
\bes{ \label{HSC2modD6}
\mathrm{HS}\left[\BC^2/\hat{D}_6\right](t) = \PE\left[t^4 + t^8 + t^{10} - t^{20}\right]~.
}
The operators $\mathfrak{M}$ and $\mathfrak{M} \varphi$ are no longer gauge invariant, hence they do not appear in \eref{HSC2modD6}. The generators contributing at order $t^4$, $t^8$ and $t^{10}$ are $u$, $\mathfrak{M}^2$ and $G_{10} \equiv \mathfrak{M}^2 \varphi$ respectively. They satisfy the relation at order $t^{20}$, namely
\bes{
G_{10}^2 + \mathfrak{M}^4 u = u^5~,
}
which is the defining equation of $\BC^2/\hat{D}_6$.
\end{itemize}

Finally, we remark that the mirror theory of 3d $\CN=4$ $\SU(2)$ SQCD with four flavours is 
\bes{ \label{affineD4a}
\vcenter{\hbox{\begin{tikzpicture}
        \node[gauge,label=right:{\footnotesize $\SU(2)$}] (0) at (0,0) {};
        \node[gauge,label=right:{\footnotesize $1$}] (1) at (1,1) {};
        \node[gauge,label=right:{\footnotesize $1$}] (2) at (1,-1) {};
        \node[gauge,label=left:{\footnotesize $1$}] (3) at (-1,-1) {};
        \node[gauge,label=left:{\footnotesize $1$}] (4) at (-1,1) {};
        \draw (0)--(1);
        \draw (0)--(2);
        \draw (0)--(3);
        \draw (0)--(4);
\end{tikzpicture}}}
\quad \text{\footnotesize$/\BZ_2^{[1]}$}\qquad \quad \text{or} \qquad \quad
\vcenter{\hbox{\begin{tikzpicture}
        \node[gauge,label=right:{\footnotesize $2$}] (0) at (0,0) {};
        \node[gauge,label=right:{\footnotesize $1$}] (1) at (1,1) {};
        \node[gauge,label=right:{\footnotesize $1$}] (2) at (1,-1) {};
        \node[gauge,label=left:{\footnotesize $1$}] (3) at (-1,-1) {};
        \node[gauge,label=left:{\footnotesize $1$}] (4) at (-1,1) {};
        \draw (0)--(1);
        \draw (0)--(2);
        \draw (0)--(3);
        \draw (0)--(4);
\end{tikzpicture}}}\quad \text{\footnotesize$/\U(1)$}
}
We emphasise that, in the left diagram, whose central node is $\SU(2)$, there is an overall $\BZ_2$ that acts trivially on the matter fields, and this needs to be quotiented out. Doing so amounts to gauging the one-form symmetry which we denote by $/\BZ_2^{[1]}$.\footnote{This $\BZ_2^{[1]}$ one-form symmetry is actually a dual symmetry of the $(\BZ_2^{[0]})_\omega$, which will be introduced in Section \ref{sec:Z20omega}. Gauging this $\BZ_2^{[1]}$ symmetry thus leads to $(\BZ_2^{[0]})_\omega$.} Likewise, in the left diagram whose central node is $\U(2)$, there is an overall $\U(1)$ that acts trivially on the matter fields and needs to be modded out. We denote this by $/\U(1)$. 

Discrete gauging of $\SU(2)$ SQCD with four flavours by $\Gamma$ corresponds to wreathing theory \eref{affineD4a} by $\Gamma$. The latter was extensively studied in \cite{{Bourget:2020bxh}}.  In particular, the Higgs branch (resp. Coulomb branch) limits of the index computed above are in agreement with the Coulomb branch (resp. Higgs branch) Hilbert series computed in \cite[Figures 9 and 11]{Bourget:2020bxh}.

\subsection{\texorpdfstring{$(\BZ_2^{[0]})_\omega$ and gauging thereof}{Z2[0]omega and gauging thereof}} \label{sec:Z20omega}
As we mentioned towards the end of the previous subsection, the mirror theory of $\SU(2)$ SQCD with four flavours depicted on the left of \eref{affineD4a} contains gauging of the $\BZ_2$ one-form symmetry. In three dimensions, this leads to a $\BZ_2$ zero-form symmetry, which we shall denote by $(\BZ^{[0]}_2)_\omega$.  In the following, we demonstrate how to turn on the corresponding fugacity $\omega$, with $\omega^2=1$, in the index of $\SU(2)$ SQCD with four flavours.  

As pointed out in \cite{Giacomelli:2024sex}, $\SU(2)$ SQCD with four flavours can also be realised as
\bes{ \label{USp2SO4SO4}
\vcenter{\hbox{\begin{tikzpicture}
        \node[flavour,label=below:{\footnotesize $\SO(4)_{\chi_L}$}] (0) at (-2,0) {};
        \node[gauge,label=below:{\footnotesize $\USp(2)$}] (1) at (0,0) {};
        \node[flavour,label=below:{\footnotesize $\SO(4)_{\chi_R}$}] (2) at (2,0) {};
        \draw (0)--(1);
        \draw[blue,thick] (1) to node[midway,above] {\textcolor{blue}{$(\BZ_2^{[0]})_\omega$}} (2);
\end{tikzpicture}}}}
in which the $\so(8)$ flavour symmetry of \eref{USp2SO8chi} is split into two equal $\so(4)$ sets, where the $(\BZ_2^{[0]})_\omega$ zero-form symmetry acts only on one of the two sets. As explained in detail in \cite{Giacomelli:2024sex}, this $(\BZ_2^{[0]})_\omega$ symmetry indeed corresponds to the zero-form symmetry arising from gauging the $\BZ_2^{[1]}$ one-form symmetry depicted in the left diagram of \eref{affineD4a}.   As in \eref{USp2SO2SO2SO2SO2}, we can further split each $\so(4)$ into $\so(2) \oplus \so(2)$ and introduce $(\BZ_2^{[0]})_\omega$, which acts on two of the four $\so(2)$ legs as follows:
\bes{ \label{USp2SO2SO2SO2SO2omega}
\vcenter{\hbox{\begin{tikzpicture}
        \node[gauge,label=left:{\footnotesize $\USp(2)$},label=right:{\textcolor{blue}{$(\BZ_2^{[0]})_\omega$}}] (0) at (0,0) {};
        \node[flavour,label=above:{\footnotesize $\SO(2)_{\chi_1}$}] (1) at (1,1) {};
        \node[flavour,label=below:{\footnotesize $\SO(2)_{\chi_2}$}] (2) at (1,-1) {};
        \node[flavour,label=below:{\footnotesize $\SO(2)_{\chi_3}$}] (3) at (-1,-1) {};
        \node[flavour,label=above:{\footnotesize $\SO(2)_{\chi_4}$}] (4) at (-1,1) {};
        \draw (0)--(3);
        \draw (0)--(4);
        \draw[blue,thick] (0)--(1);
        \draw[blue,thick] (0)--(2);
\end{tikzpicture}}}}
whose index, for $\chi_i = 1$ $\forall \ i$, reads
\bes{ \label{indUSp2SO2SO2SO2SO2omega}
&\CI_{\eqref{USp2SO2SO2SO2SO2omega}}(\vec{y}| a| \chi_1 = \ldots = \chi_4 = 1| \omega; x)  \\ & = \frac{1}{2} \sum_{l \in \BZ} \oint \frac{d z}{2 \pi i z} \CZ^{\USp(2)}_{\text{vec}}\left(z; l; x\right) \prod_{s = {0, \pm 1}} \CZ^{1}_{\chi} \left(z^{2 s} a^{-2}; 2 s l; x\right) \\ 
& \quad \, \, \times \prod_{i = 1}^{2} \prod_{s_1, s_2 = \pm 1} \CZ^{1/2}_{\chi} \left(\omega z^{s_1} y_i^{s_2} a; s_1 l; x\right) \times \prod_{i = 3}^{4} \prod_{s_1, s_2 = \pm 1} \CZ^{1/2}_{\chi} \left(z^{s_1} y_i^{s_2} a; s_1 l; x\right)~.
}
Observe that, upon setting $\omega = 1$, this coincides with the index \eref{indUSp2SO2SO2SO2SO2}. As instructed in \cite[(6.9)]{Aharony:2013kma}, we can obtain the index with $\chi_j = -1$ and $\chi_i = 1$ $\forall \ i \neq j$ from \eref{indUSp2SO2SO2SO2SO2omega} by setting $y_j = 1$ and $y^{-1}_j = -1$.

Similarly to Section \ref{sec:SU2w4DiscreteGauging}, we introduce the shorthand notation
\bes{ 
\eref{USp2SO2SO2SO2SO2omega} \equiv \begin{bmatrix} \chi_4 & {\blue \chi_1} \\ \chi_3 & {\blue \chi_2} \end{bmatrix}~,
} 
from which the action of the various subgroups of $S_4$ that are isomorphic to $\BZ_2$ and $\BZ_2 \times \BZ_2$ can be computed as follows:
\begin{itemize}
\item Single transposition, \ie~ $\Gamma=\langle (12) \rangle \equiv \BZ_2$:
\bes{
\frac{1}{2} \left\{ \begin{bmatrix} + & {\blue \bm{+}} \\ + & {\blue \bm{+}} \end{bmatrix} + \begin{bmatrix} + & {\blue \bm{+}} \\ + & {\blue \bm{-}} \end{bmatrix}  \right\}~.
}
\item Double transposition, \ie~ $\Gamma =\langle (1 2)(3 4) \rangle \equiv \mathrm{DT}$:
\bes{
\frac{1}{2} \left\{\begin{bmatrix} + & {\blue \bm{+}} \\ + & {\blue \bm{+}} \end{bmatrix} + \begin{bmatrix} - & {\blue \bm{+}} \\ + & {\blue \bm{-}} \end{bmatrix} \right\}~.
}
\item The non-normal Klein subgroup of $S_4$, \ie~ $\Gamma = \langle (1 2), (3 4) \rangle \equiv \mathrm{NNK}$:
\bes{
\frac{1}{4} \left\{\begin{bmatrix} + & {\blue \bm{+}} \\ + & {\blue \bm{+}} \end{bmatrix} + \begin{bmatrix} + & {\blue \bm{+}} \\ + & {\blue \bm{-}} \end{bmatrix} + \begin{bmatrix} - & {\blue \bm{+}} \\ + & {\blue \bm{+}} \end{bmatrix} + \begin{bmatrix} - & {\blue \bm{+}} \\ + & {\blue \bm{-}} \end{bmatrix} \right\}~.
}
\item The normal Klein subgroup of $S_4$, \ie~ $\Gamma = \langle (1 2) (3 4), (1 3) (2 4) \rangle \equiv \mathrm{NK}$:
\bes{ \label{USp2SO2SO2SO2SO2modZ2NK}
\frac{1}{4} \left\{\begin{bmatrix} + & {\blue \bm{+}} \\ + & {\blue \bm{+}} \end{bmatrix} + \begin{bmatrix} - & {\blue \bm{+}} \\ + & {\blue \bm{-}} \end{bmatrix} + \begin{bmatrix} - & {\blue \bm{+}} \\ {\blue \bm{-}} & + \end{bmatrix} + \begin{bmatrix} + & {\blue \bm{+}} \\ {\blue \bm{-}} & - \end{bmatrix} \right\}~.
}
Note that, in the last two terms appearing in the sum \eref{USp2SO2SO2SO2SO2modZ2NK}, $(\BZ_2^{[0]})_\omega$ does not act on the legs associated with $\chi_1$ and $\chi_2$, as described in the prescription \eref{USp2SO2SO2SO2SO2omega}. Instead, it acts on the legs refined with the fugacities $\chi_1$ and $\chi_3$. Indeed, this ensures that the index corresponding to \eref{USp2SO2SO2SO2SO2modZ2NK} is well-defined if we turn on the fugacities $\vec{y}$ as follows: in the first term appearing in the summation \eref{USp2SO2SO2SO2SO2modZ2NK}, we set $y_2$ and $y_4$ to $1$, then we set the remaining fugacities $y_2$ and $y_3$, coming also from the other terms in the summation, to be equal to $y_1$. Concretely:
\bes{ \label{indUSp2SO2SO2SO2SO2modZ2NK}
&\CI_{\eqref{USp2SO2SO2SO2SO2modZ2NK}}(y_1| a; x) \\ &=\frac{1}{4} \left[ \CI_{\eqref{USp2SO2SO2SO2SO2omega}}(y_1, y_3 = y_1, y_2 = y_4 = 1| a| \chi_1 = \ldots = \chi_4 = 1| \omega; x) \right.\\ &\left. \quad \, \, \, + \CI_{\eqref{USp2SO2SO2SO2SO2omega}}(y_1, y_3 = y_1| a| \chi_1 = \chi_3 = 1, \chi_2 = \chi_4 = -1| \omega; x) \right.\\ &\left. \quad \, \, \, +\CI_{\eqref{USp2SO2SO2SO2SO2omega}}(y_1, y_2 = y_1| a| \chi_1 = \chi_2 = 1, \chi_3 = \chi_4 = -1| \omega; x) \right.\\ &\left. \quad \, \, \, +\CI_{\eqref{USp2SO2SO2SO2SO2omega}}(y_1, y_4 = y_1| a| \chi_1 = \chi_4 = 1, \chi_2 = \chi_3 = -1| \omega; x) \right]~.
}
We remark that, if we let $(\BZ_2^{[0]})_\omega$ act on the legs associated with $\chi_1$ and $\chi_2$ also in the last two terms contributing to the sum \eref{USp2SO2SO2SO2SO2modZ2NK}, then \eref{indUSp2SO2SO2SO2SO2modZ2NK} would be an ill-defined index, whose series expansion in powers of $x$ contains half-integer coefficients.
\end{itemize}
Let us summarise the index in each case with $y_1=\ldots=y_4=1$ below.
\bes{ \label{SU2w4refw}
\scalebox{0.85}{
\begin{tabular}{c|l}
\hline
$\Gamma$ & \qquad \qquad \quad Index of theory \eref{USp2SO4SO4} or \eref{USp2SO2SO2SO2SO2omega} \\
\hline
Trivial & $1+(12+16 \omega) a^{2} x + \left[2 a^{-4}+ (156+ 144 \omega) a^{4} -(13 +16 \omega) \right]x^2$ \\
& $\, \, \, \,+  \left[ (949+976 \omega )a^{6}-(329+320 \omega )a^{2}  -a^{-2}+ a^{-6}  \right] x^3 +\ldots$  \\
\hline
$\langle (1 2) \rangle$ & $1+\left(9 + 12 \omega\right) a^{2} x + \left[a^{-4} + \left(103 + 92 \omega \right) a^{4} - \left(10 + 12 \omega \right)\right] x^2$ \\ $=\BZ_2$
&$\, \, \, \,+  [(567+588 \omega )a^{6}-(204+200 \omega)a^{2}] x^3+ \ldots$ \\
\hline
$\langle (1 2)(3 4) \rangle$ & $1+\left(6 + 10 \omega\right) a^{2} x + \left[2 a^{-4} + \left(86 + 74 \omega \right) a^{4} - \left(7 + 10 \omega \right)\right] x^2 $\\ = DT
& $\, \, \, \,+[(479 + 506 \omega])a^{6}  - (169 + 160 \omega)a^{2} -a^{-2} + a^{-6}] x^3 +\ldots $\\ 
\hline
$\langle (1 2),(3 4) \rangle$ & $1 + \left(6 + 9 \omega\right) a^{2} x + \left[a^{-4} + \left(68 + 57 \omega\right) a^{4} - \left(7 + 9 \omega\right) \right] x^2$ \\ = NNK
& $\, \, \, \,+ \left[\left(332 + 353 \omega\right) a^{6} - \left(124 + 120 \omega \right) a^{2}\right] x^3 + \ldots$ \\
\hline
$\langle (1 2) (3 4), (1 3) (2 4) \rangle$ & $1 + \left(3 + 7 \omega\right) a^{2} x + \left[2 a^{-4} + \left(51 + 39 \omega\right) a^{4} - \left(4 + 7 \omega\right) \right] x^2$ \\ = NK
&$\, \, \, \,+ \left[\left(244 + 271 \omega\right) a^{6} - \left(89 + 80 \omega \right) a^{2} - a^{-2}  + a^{-6} \right] x^3 + \ldots$\\
\end{tabular}}
}
We point out that, in each case, the coefficients of the terms $a^{2p} x^p \omega^s$ (with $p=0,1,2,\ldots$ and $s=0,1$) can be obtained from \eref{tabSU2w4} by applying the following branching rules: $\so(8) \rightarrow \su(2)^4$, $\so(7) \rightarrow \su(2)^3$, $\su(4)\oplus \u(1) \rightarrow \su(2)^2 \oplus \u(1)$, $\su(4) \rightarrow \su(2)^2$, and $\usp(4) \rightarrow \su(2)$. For example, the coefficients of the term $a^{2} x$ are given by
\bes{
\mathbf{28} \, &\rightarrow \, [(\mathbf{3},\mathbf{1},\mathbf{1},\mathbf{1}) \oplus (\mathbf{1},\mathbf{3},\mathbf{1},\mathbf{1}) \oplus (\mathbf{1},\mathbf{1},\mathbf{3},\mathbf{1}) \oplus (\mathbf{1},\mathbf{1},\mathbf{1},\mathbf{3}) ] \oplus (\mathbf{2},\mathbf{2},\mathbf{2},\mathbf{2}) \omega~, \\
\mathbf{21} \, &\rightarrow \, [(\mathbf{3},\mathbf{1},\mathbf{1}) \oplus 
(\mathbf{1},\mathbf{3},\mathbf{1}) \oplus
(\mathbf{1},\mathbf{1},\mathbf{3})] \oplus (\mathbf{2} ,\mathbf{2}, \mathbf{3}) \omega~,
\\
\mathbf{16} \, &\rightarrow \, [(\mathbf{3},\mathbf{1})(0) \oplus 
(\mathbf{1},\mathbf{3})(0)] \oplus (\mathbf{3}, \mathbf{3})(0) \omega \oplus (\mathbf{1}, \mathbf{1})(0) \omega~,
\\
\mathbf{15} \, &\rightarrow \, [(\mathbf{3},\mathbf{1}) \oplus 
(\mathbf{1},\mathbf{3})] \oplus (\mathbf{3}, \mathbf{3}) \omega~,
\\
\mathbf{10} \, &\rightarrow \, (\mathbf{3}) \oplus (\mathbf{7}) \omega~.
}

\subsubsection*{Gauging $(\BZ^{[0]}_2)_\omega$}
Let us now gauge $(\BZ^{[0]}_2)_\omega$. We then have the following theory:
\bes{ \label{USp2SO4SO4modZ2}
\vcenter{\hbox{\begin{tikzpicture}
        \node[flavour,label=below:{\footnotesize $\SO(4)_{\chi_L}$}] (0) at (-2,0) {};
        \node[gauge,label=below:{\footnotesize $\USp(2)$}] (1) at (0,0) {};
        \node[flavour,label=below:{\footnotesize $\SO(4)_{\chi_R}$}] (2) at (2,0) {};
        \draw (0)--(1);
        \draw[blue,thick] (1) to node[midway,above] {\textcolor{blue}{$(\BZ_2^{[0]})_\omega$}} (2);
\end{tikzpicture}}}\quad \text{\footnotesize$/(\BZ_2^{[0]})_\omega$}}
or the one with the fugacities $\chi_1, \ldots, \chi_4$ turned on:
\bes{ \label{USp2SO2SO2SO2SO2modZ2}
\vcenter{\hbox{\begin{tikzpicture}
        \node[gauge,label=left:{\footnotesize $\USp(2)$},label=right:{\textcolor{blue}{$(\BZ_2^{[0]})_\omega$}}] (0) at (0,0) {};
        \node[flavour,label=above:{\footnotesize $\SO(2)_{\chi_1}$}] (1) at (1,1) {};
        \node[flavour,label=below:{\footnotesize $\SO(2)_{\chi_2}$}] (2) at (1,-1) {};
        \node[flavour,label=below:{\footnotesize $\SO(2)_{\chi_3}$}] (3) at (-1,-1) {};
        \node[flavour,label=above:{\footnotesize $\SO(2)_{\chi_4}$}] (4) at (-1,1) {};
        \draw (0)--(3);
        \draw (0)--(4);
        \draw[blue,thick] (0)--(1);
        \draw[blue,thick] (0)--(2);
\end{tikzpicture}}}\quad \text{\footnotesize$/(\BZ_2^{[0]})_\omega$}}
whose index can be derived from \eref{indUSp2SO2SO2SO2SO2omega} as
\bes{ \label{indUSp2SO2SO2SO2SO2modZ2}
\CI_{\eqref{USp2SO2SO2SO2SO2modZ2}}(\vec{y}| a| \chi_1, \ldots, \chi_4; x)=\frac{1}{2} \sum_{\omega = \pm 1} \CI_{\eqref{USp2SO2SO2SO2SO2omega}}(\vec{y}| a| \chi_1, \ldots, \chi_4| \omega; x)~.
}
Note that resulting theory possesses a dual $\BZ_2$ one-form symmetry.

Let us report the explicit results of the computation of the indices of theory $\eref{USp2SO2SO2SO2SO2modZ2}/\Gamma$, with $\Gamma$ being one of the $\BZ_2$ or $\BZ_2 \times \BZ_2$ subgroups of $S_4$ discussed above. The indices for the various choices of $\Gamma$ up to order $x^3$ are listed below, where, for simplicity, we set $\vec{y}=(1, \ldots, 1)$. We also report the flavour symmetry of the corresponding theory, which can be read from the moment map appearing at order $x$ of the index.
\bes{ \label{indexUSp2SO2SO2SO2SO2modZ2modGamma}
\scalebox{0.97}{
\begin{tabular}{c|l|c}
\hline
$\Gamma$ & \qquad \qquad \quad Index of \eref{USp2SO2SO2SO2SO2modZ2}$/\Gamma$ & Flavour  \\
 &  & symmetry  \\
\hline
Trivial & $1+{12 a^2} x+\left(156 a^4+{2}{a^{-4}}-13\right) x^2$ & $\su(2)^4$\\ & $\, \, \, \,+\left(949 a^6+{a^{-6}}-329 a^2-{a^{-2}}\right) x^3+\ldots$ & \\
\hline
$\langle (1 2) \rangle$ & $1+{9 a^2 x}+\left(103 a^4+{a^{-4}}-10\right) x^2$ & $\su(2)^3$\\ = $\BZ_2$ & $\, \, \, \,+\left(567 a^6-{204}{a^{-2}}\right) x^3+\ldots$ &\\
\hline
$\langle (1 2)(3 4) \rangle$ & $1+{6 a^2 x}+\left(86 a^4+{2 a^{-4}}-7\right) x^2$ & $\su(2)^2$\\ = DT & $\, \, \, \,+\left(479 a^6+a^{-6}-169 a^2-{a^{-2}}\right) x^3+\ldots$ &\\
\hline
$\langle (1 2),(3 4) \rangle$ & $1+{6 a^2 x}+\left(68 a^4+{a^{-4}}-7\right) x^2$ & $\su(2)^2$\\ = NNK & $\, \, \, \,+\left(332 a^6-124 a^2\right) x^3+\ldots$ &\\
\hline
$\langle (1 2) (3 4), (1 3) (2 4) \rangle$ & $1+{3 a^2 x}+\left(51 a^4+{2 a^{-4}}-4\right) x^2$ & $\su(2)$\\ = NK & $\, \, \, \,+\left(244 a^6+a^{-6}-89 a^2-a^{-2}\right) x^3+\ldots$ &\\
\end{tabular}
}
}
From \eref{SU2w4refw}, we see that $(\BZ_2^{[0]})_\omega$ acts non-trivially on the Higgs branch of theory \eref{USp2SO2SO2SO2SO2omega} but acts trivially on its Coulomb branch, since $\omega$ appears only in the coefficients of the terms $a^{2p} x^p$ but not $a^{-2p} x^p$. Hence, the Coulomb branch of \eref{USp2SO2SO2SO2SO2omega} remains unchanged upon gauging $\Gamma$. The corresponding Hilbert series are given by \eref{HSC2modD4} and \eref{HSC2modD6}. 
 Similarly to \eref{HBHSSU2w4modGamma}, the Higgs branch limit of the index of \eref{USp2SO2SO2SO2SO2modZ2}$/\Gamma$ coincides with the following Hilbert series:
\bes{ \label{HBHSSU2w4modmodGamma}
\scalebox{0.95}{$
\begin{split}
&\mathrm{HS}\left[\text{HB of \eref{USp2SO2SO2SO2SO2modZ2}/$\Gamma$}\right](t,\vec{y}) \\
&=  \frac{1}{2} \sum_{\omega = \pm 1} \oint_{|z|=1} \frac{d z}{2\pi i z} (1-z^2) \times {\blue  \frac{1}{m}\sum_{j=0}^{m-1} \tilde{f}_{\Gamma,j}(z, t, \vec{y}, \omega) } \times \PE \left[- (z^2+1+z^{-2}) t^2  \right]~,
\end{split}
$}
}
where $\tilde{f}_{\Gamma,j}(z, t, \vec{y}, \omega)$ is defined as follows. If $\Gamma$ is either trivial or one among $\BZ_2 = \langle (1 2) \rangle$, $\mathrm{DT} = \langle (1 2)(3 4) \rangle$ and $\mathrm{NNK} = \langle (1 2),(3 4) \rangle$, then
\bes{
\tilde{f}_{\Gamma,j}(z, t, \vec{y}, \omega) = {f}_{\Gamma,j}(z, t, \omega y_1, \omega y_2, y_3, y_4)~,
}
otherwise, for $\Gamma = \mathrm{NK} = \langle (1 2) (3 4), (1 3) (2 4) \rangle$, we have
\bes{ \label{ftildeHSNK}
\scalebox{0.94}{$
\begin{cases}
\tilde{f}_{\Gamma,j}(z, t, \vec{y}, \omega) = {f}_{\Gamma,j}(z, t, \omega y_1, \omega y_2, y_3, y_4)~, \quad& j=0,1 \\
\tilde{f}_{\Gamma,j}(z, t, \vec{y}, \omega) = {f}_{\Gamma,j}(z, t, \omega y_1, y_2, \omega y_3, y_4)~, \quad& j=2,3
\end{cases}~,
$}
}
with ${f}_{\Gamma,j}(z, t, \vec{y})$ defined as in \eref{fHS}. In \eref{ftildeHSNK}, we have taken into account that the Hilbert series \eref{HBHSSU2w4modmodGamma} has to be well-defined, with no half-integer coefficients appearing in its series expansion in the variable $t$. Hence, similarly to the discussion below \eref{USp2SO2SO2SO2SO2modZ2NK}, for $\Gamma = \langle (1 2) (3 4), (1 3) (2 4) \rangle$, the Hilbert series \eref{HBHSSU2w4modmodGamma} can be realised as
\bes{
&\mathrm{HS}\left[\text{HB of \eref{USp2SO2SO2SO2SO2modZ2}/${\Gamma =\langle (1 2) (3 4), (1 3) (2 4) \rangle}$}\right](t,y_1) \\&= \frac{1}{2} \sum_{\omega = \pm 1} \oint_{|z|=1} \frac{d z}{2\pi i z} (1-z^2) \PE \left[- (z^2+1+z^{-2}) t^2  \right] \\ &\qquad \times {\blue  \frac{1}{4} \left[ \tilde{f}_{\Gamma,0}(z, t, y_1, y_3=y_1, y_2 = y_4 = 1, \omega) \right.}{\blue \left. + \tilde{f}_{\Gamma,1}(z, t, y_1, y_3 = y_1, \omega) \right. }\\ &\qquad \quad \, \, \, \, \,  {\blue \left. + \tilde{f}_{\Gamma,2}(z, t, y_1, y_2 = y_1, \omega) \right. }{\blue \left. + \tilde{f}_{\Gamma,3}(z, t, y_1, y_4 = y_1, \omega) \right]}~.
}
Let us report the expression of the unrefined Hilbert series \eref{HBHSSU2w4modmodGamma}, or equivalently the Higgs branch limit of the index of theory \eref{USp2SO2SO2SO2SO2modZ2}$/\Gamma$, for various choices of $\Gamma$ below.
\bes{ \label{HBHSUSp2SO2SO2SO2SO2modZ2modGamma}
\scalebox{0.97}{
\begin{tabular}{c|l}
\hline
$\Gamma$ & \qquad \quad Higgs branch Hilbert series of \eref{USp2SO2SO2SO2SO2modZ2}$/\Gamma$  \\
\hline
Trivial & $\frac{1 + 7 t^2 + 101 t^4 + 244 t^6 + 666 t^8 + 650 t^{10} + 666 t^{12} + 
 244 t^{14} + 101 t^{16} + 7 t^{18} + t^{20}}{\left(1-t^2\right)^{10} \left(1+t^2\right)^5}$\\ & $=1 + 12 t^2 + 156 t^4 + 949 t^6 + 4486 t^8 + 16416  t^{10}+\ldots$ \\
\hline
$\langle (1 2) \rangle$ & $\frac{1 + 4 t^2 + 63 t^4 + 112 t^6 + 352 t^8 + 280 t^{10} + 352 t^{12} + 
 112 t^{14} + 63 t^{16} + 4 t^{18} + t^{20}}{\left(1-t^2\right)^{10} \left(1+t^2\right)^5}$\\ $= \BZ_2$ & $=1 + 9 t^2 + 103 t^4 + 567 t^6 + 2572 t^8 + 9076  t^{10}+\ldots$\\
\hline
$\langle (1 2)(3 4) \rangle$ & $\frac{1 + t^2 + 61 t^4 + 94 t^6 + 378 t^8 + 274 t^{10} + 378 t^{12} + 94 t^{14} + 
 61 t^{16} + t^{18} + t^{20}}{\left(1-t^2\right)^{10} \left(1+t^2\right)^5}$\\ = DT & $=1 + 6 t^2 + 86 t^4 + 479 t^6 + 2288 t^8 + 8240  t^{10} +\ldots$\\
\hline
$\langle (1 2),(3 4) \rangle$ & $\frac{1 + t^2 + 43 t^4 + 37 t^6 + 208 t^8 + 92 t^{10} + 208 t^{12} + 37 t^{14} + 
 43 t^{16} + t^{18} + t^{20}}{\left(1-t^2\right)^{10} \left(1+t^2\right)^5}$\\ = NNK & $=1 + 6 t^2 + 68 t^4 + 332 t^6 + 1473 t^8 + 4988  t^{10} +\ldots$\\
\hline
$\langle (1 2) (3 4), (1 3) (2 4) \rangle$ & $\frac{1 - 2 t^2 + 41 t^4 + 19 t^6 + 234 t^8 + 86 t^{10} + 234 t^{12} + 
 19 t^{14} + 41 t^{16} - 2 t^{18} + t^{20}}{\left(1-t^2\right)^{10} \left(1+t^2\right)^5}$\\ = NK & $=1 + 3 t^2 + 51 t^4 + 244 t^6 + 1189 t^8 + 4152  t^{10} +\ldots$\\
\end{tabular}
}
}
From the left diagram of \eref{affineD4a}, we see that the mirror theory of \eref{USp2SO2SO2SO2SO2modZ2} is simply
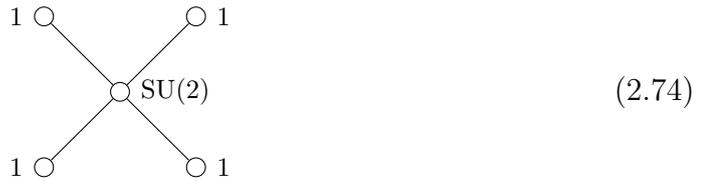
\bes{ \label{affineD4mod}
\vcenter{\hbox{\begin{tikzpicture}
        \node[gauge,label=right:{\footnotesize $\SU(2)$}] (0) at (0,0) {};
        \node[gauge,label=right:{\footnotesize $1$}] (1) at (1,1) {};
        \node[gauge,label=right:{\footnotesize $1$}] (2) at (1,-1) {};
        \node[gauge,label=left:{\footnotesize $1$}] (3) at (-1,-1) {};
        \node[gauge,label=left:{\footnotesize $1$}] (4) at (-1,1) {};
        \draw (0)--(1);
        \draw (0)--(2);
        \draw (0)--(3);
        \draw (0)--(4);
\end{tikzpicture}}}
}
The centre of the $\SU(2)$ gauge group in the middle indeed gives rise to the $\BZ_2^{[1]}$ one-form symmetry depicted in the left diagram of \eref{affineD4a}. Moreover, we find that discrete gauging \eref{USp2SO2SO2SO2SO2modZ2}$/\Gamma$ is equivalent to wreathing \eref{affineD4mod} by $\Gamma$ with the Higgs and Coulomb branches exchanged, as expected from mirror symmetry.  Indeed, the results in this section are perfectly in agreement with those in \cite[Section 4.2]{Giacomelli:2024sex}.

\subsection{Transitions between two discrete gaugings}
Let us conclude this section by describing a recipe for obtaining $\eref{USp2SO2SO2SO2SO2}/\Gamma'$ starting from $\eref{USp2SO2SO2SO2SO2}/\Gamma$ for some $\Gamma'$ and $\Gamma$ that are subgroups of $S_4$. In particular, we use the notation
\bes{
\eref{USp2SO2SO2SO2SO2}/\Gamma \xrightarrow{\begin{bmatrix} \chi_4 & \chi_1 \\ \chi_3 & \chi_2 \end{bmatrix}} \eref{USp2SO2SO2SO2SO2}/\Gamma'~,
}
to mean that
\bes{
\eref{USp2SO2SO2SO2SO2}/\Gamma' = \frac{1}{2} \left\{\eref{USp2SO2SO2SO2SO2}/\Gamma + \begin{bmatrix} \chi_4 & \chi_1 \\ \chi_3 & \chi_2 \end{bmatrix}\right\}~.
}
From the above discussion, we can obtain the following interesting examples.
\bi
\item $\Gamma =$ double transposition and $\Gamma'=$ the non-normal Klein subgroup of $S_4$:
\bes{
\eref{USp2SO2SO2SO2SO2}/\langle (1 2)(3 4) \rangle \xrightarrow{\begin{bmatrix} - & + \\ + & + \end{bmatrix}}\eref{USp2SO2SO2SO2SO2}/\langle (1 2), (3 4) \rangle~,
}
where we use \eref{USp2SO2SO2SO2SO2DT}, \eref{USp2SO2SO2SO2SO2NNK} and the fact that, upon setting $y_2 = y_4 = 1$ in \eref{USp2SO2SO2SO2SO2NNK} in order to obtain a well-defined index, we can identify
\bes{
\begin{bmatrix} - & + \\ + & + \end{bmatrix} = \frac{1}{2} \left\{\begin{bmatrix} + & + \\ + & - \end{bmatrix} + \begin{bmatrix} - & + \\ + & + \end{bmatrix}\right\}~.
}
\item $\Gamma =$ double transposition and $\Gamma'=$ the normal Klein subgroup of $S_4$:
\bes{
\eref{USp2SO2SO2SO2SO2}/\langle (1 2)(3 4) \rangle \xrightarrow{\begin{bmatrix} - & + \\ - & + \end{bmatrix}}\eref{USp2SO2SO2SO2SO2}/\langle (1 2) (3 4), (1 3) (2 4) \rangle ~,
}
where we use \eref{USp2SO2SO2SO2SO2DT} and \eref{USp2SO2SO2SO2SO2NK} and the fact that, upon setting $y_2 = y_4 = 1$ in \eref{USp2SO2SO2SO2SO2NK}, we can identify
\bes{
\begin{bmatrix} - & + \\ - & + \end{bmatrix} = \frac{1}{2} \left\{\begin{bmatrix} - & + \\ - & + \end{bmatrix} + \begin{bmatrix} + & + \\ - & - \end{bmatrix}\right\}~.
}
\ei
Moreover, such transitions can occur between $\eref{USp2SO2SO2SO2SO2}/\Gamma$ and $\eref{USp2SO2SO2SO2SO2}/\Gamma'$, where $\Gamma$ is Abelian and $\Gamma'$ is non-Abelian. For instance, let us take $\Gamma = \BZ_3 = \langle (123) \rangle$ and $\Gamma' = S_3$, whose elements can be represented in terms of cycles as follows:
\bes{
S_3 = \left\{\ID, (12), (13), (23), (123), (132)\right\}~.
}
In particular, we have the following transition:
\bes{
\eref{USp2SO2SO2SO2SO2}/\BZ_3 \xrightarrow{\begin{bmatrix} + & + \\ + & - \end{bmatrix}}\eref{USp2SO2SO2SO2SO2}/S_3~,
}
where we use the fact that, upon setting $y_2 = y_3 = 1$, we can identify
\bes{
\begin{bmatrix} + & + \\ + & - \end{bmatrix} = \frac{1}{3} \left\{\begin{bmatrix} + & + \\ + & - \end{bmatrix} + \begin{bmatrix} + & + \\ - & + \end{bmatrix} + \begin{bmatrix} + & + \\ - & + \end{bmatrix}\right\}~.
}
Up to now, the information presented in this section does not provide a way to obtain neither the index of theory $\eref{USp2SO2SO2SO2SO2}/\BZ_3$, nor that of $\eref{USp2SO2SO2SO2SO2}/S_3$. Nevertheless, we compute such indices in the next section; in particular the results are reported in \eref{indZ3wr} and \eref{indaffineD4/S3} respectively. It can be checked explicitly that such a transition indeed occurs.

\subsection{\texorpdfstring{$\BZ_2$ and $\BZ_2 \times \BZ_2$ actions on the 4d Coulomb branch}{Z2 and Z2xZ2 actions on the 4d Coulomb branch}} 
Hitherto, we have considered 3d $\CN=4$ $\SU(2)$ SQCD with four flavours. In this subsection, we turn to 4d $\CN=2$ $\SU(2)$ SQCD with four flavours and let us briefly discuss how the  $\BZ_2$ and $\BZ_2 \times \BZ_2$ transformations discussed above act on such a theory. We consider the superpotential mass term, which, in the variables $q_a^{I}$ defined in \eref{mapQtoq}, reads
\bes{ 
\sum_{i = 1}^4 \epsilon^{ab} m_i q_a^{2 i -1} q_b^{2 i} ~,
}
from which we can easily analyse the $\BZ_2$ and $\BZ_2 \times \BZ_2$ actions of our interest on the mass parameters.
\bi
\item The $\BZ_2$ transformation \eref{Z2onq} corresponds to a sign flip on one of the four mass parameters, \ie
\bes{ \label{Z2onm}
\eref{Z2onq}:\begin{cases}
q_a^I \rightarrow q_a^I~, \quad I \neq 4 \\
q_a^4 \rightarrow -q_a^4
\end{cases}~ \longleftrightarrow \quad
\begin{cases}
m_i \rightarrow m_i~, \quad i \neq 2 \\
m_2 \rightarrow -m_2
\end{cases}~.
}
This particular $\BZ_2$ discrete gauging was analysed in \cite[Section 2.3]{Giacomelli:2024sex}, where it was pointed out that the resulting Higgs branch is isomorphic to the closure of the next-to-minimal orbit $\bar{\mathrm{n.min}\,B_3}$ of $\so(7)$. The resulting theory after gauging is a rank-one SCFT, with a non-trivial conformal manifold and a Coulomb branch operator of dimension two. As discussed in \cite[(10.4)]{Seiberg:1994aj}, the action \eref{Z2onm} on the mass parameters amounts to exchanging the two spinor representations of $\mathrm{Spin}(8)$, while keeping the vector representation invariant. In terms of the Seiberg-Witten curve \cite[(16.35)]{Seiberg:1994aj}, this corresponds to interchanging  $T_2$ and $T_3$ in \cite[(16.36)]{Seiberg:1994aj}, while leaving $R$, $T_1$ and $N$ invariant. Upon accompanying by the interchange of $e_2$ and $e_3$, the curve \cite[(16.35)]{Seiberg:1994aj} is left invariant. According to \cite[ (16.14)]{Seiberg:1994aj}, the transformation $e_2 \leftrightarrow e_3$ is equivalent to sending $\tau \rightarrow \tau + 1$, where $\tau = \frac{\theta}{\pi}+\frac{8 \pi i}{g^2}$ is defined as in \cite[(16.1)]{Seiberg:1994aj}.\footnote{As pointed out in \cite[(3.2) and Page 15]{Seiberg:1994aj}, the $\BZ_2$ symmetry \eref{Z2onq} is anomalous in general, except if it is accompanied by the map $\tau \rightarrow \tau + 1$.} This shift $\theta \rightarrow \theta + \pi$ of the theta angle can be identified with the $\mathsf{T}$ transformation of $\SL(2,\BZ)$ and acts trivially on the supercharges,\footnote{Under a generic $\SL(2,\BZ)$ transformation which sends $\tau \rightarrow \frac{a \tau + b}{c \tau + d}$, the supercharges transform as
\bes{
Q_{\alpha}^a \rightarrow \left(\frac{c \tau + d}{|c \tau + d|}\right)^{-\frac{1}{2}} Q_{\alpha}^a ~, \quad a=1,2~, \quad \alpha=1,2~, 
}
from which we see that they are left invariant under the action of the $\mathsf{T}$ generator, whereas they pick up a non-trivial phase under the action of the $\mathsf{S}$ generator, where
\bes{
\mathsf{S} = \left(
\begin{array}{c c}
 0 & 1 \\
 -1 & 0 \\
\end{array}
\right)~, \quad
\mathsf{T} = \left(
\begin{array}{c c}
 1 & 1 \\
 0 & 1 \\
\end{array}
\right)~. 
}
}
meaning that there is no need to combine it with a $\U(1)_r$ transformation. Hence, the $\BZ_2$ transformation \eref{Z2onm} acts trivially on the Coulomb branch and the resulting theory after gauging has one Coulomb branch operator of dimension two.
\item Next, we consider the $\BZ_2$ transformation \eref{DTonq}, which corresponds to flipping the sign of two out of the four mass parameters, \ie
\bes{ \label{DTonm}
\eref{DTonq}: \begin{cases}
q_a^I \rightarrow q_a^I~, \quad& I \neq 4, 8 \\
q_a^J \rightarrow -q_a^J~, \quad& J=4, 8
\end{cases}~ \longleftrightarrow \quad
\begin{cases}
m_i \rightarrow m_i~, \quad& i = 1, 3 \\
m_j \rightarrow -m_j~, \quad& j = 2, 4
\end{cases}~.
}
This transformation leaves the parameters $T_{1,2,3}$, $R$ and $N$ in \cite[(16.36)]{Seiberg:1994aj} invariant, and therefore straightforwardly leaves the Seiberg-Witten curve \cite[(16.35)]{Seiberg:1994aj} of the original theory invariant. It thus acts trivially on the Coulomb branch. This results in a rank-one SCFT with $\su(4) \oplus \u(1)$ flavour symmetry.
\item The effect of the $\BZ_2 \times \BZ_2$ transformation \eref{NNKonq} is to act separately with two distinct sign flips on one of the four mass parameters, namely
\bes{ \label{NNKonm}
\eref{Z2onm}:&\begin{cases}
q_a^I \rightarrow q_a^I~, \quad I \neq 4 \\
q_a^4 \rightarrow -q_a^4
\end{cases}~ \longleftrightarrow \quad
\begin{cases}
m_i \rightarrow m_i~, \quad i \neq 2 \\
m_2 \rightarrow -m_2
\end{cases} \\ & \qquad \qquad \qquad \qquad \qquad \, \, \,+ \\
&\begin{cases}
q_a^I \rightarrow q_a^I~, \quad I \neq 8 \\
q_a^8 \rightarrow -q_a^8
\end{cases}~ \longleftrightarrow \quad
\begin{cases}
m_i \rightarrow m_i~, \quad i \neq 4 \\
m_4 \rightarrow -m_4
\end{cases}~.
}
Following the discussion below \eref{Z2onm}, each of the two $\BZ_2$ transformations in \eref{NNKonm} is accompanied by the action of the $\mathsf{T}$ generator of $\SL(2, \BZ)$, which acts trivially on the supercharges and, therefore, on the Coulomb branch. The resulting theory is a rank-one SCFT with $\su(4)$ flavour symmetry.
\item Finally, let us discuss the effect of the $\BZ_2 \times \BZ_2$ transformation \eref{NKonq}. This corresponds to the following two distinct $\BZ_2$ transformations:
\bes{ \label{NKonm}
\eref{DTonm}:&\begin{cases}
q_a^I \rightarrow q_a^I~, \quad& I \neq 4,8 \\
q_a^J \rightarrow -q_a^J~, \quad& J = 4,8
\end{cases} \quad \longleftrightarrow \quad
\begin{cases}
m_i \rightarrow m_i~, \quad& i = 1,3 \\
m_j \rightarrow -m_j~, \quad& j = 2,4
\end{cases} \\ & \qquad \qquad \qquad \qquad \qquad \quad \quad + \\
&\begin{cases}
q_a^I \rightarrow q_a^I~, \quad& I \neq 6,8 \\
q_a^J \rightarrow -q_a^J~, \quad& J = 6,8
\end{cases} \quad \longleftrightarrow \quad
\begin{cases}
m_i \rightarrow m_i~, \quad& i = 1,2 \\
m_j \rightarrow -m_j~, \quad& j = 3,4
\end{cases}~.
}
This transformation acts trivially on the Coulomb branch, as explained below \eref{DTonm}, and gives rise to a rank-one SCFT with $\usp(4)$ flavour symmetry.
\ei

\section{Wreathing and non-invertible symmetries}\label{sec:noninv}
The protagonist of this section is the mirror theory of 3d $\CN=4$ $\SU(2)$ SQCD with four flavours, namely the following affine $D_4$ quiver:
\bes{ \label{affineD4}
\vcenter{\hbox{\begin{tikzpicture}
        \node[gauge,label=right:{\footnotesize $\SU(2)$}] (0) at (0,0) {};
        \node[gauge,label=right:{\footnotesize $1$}] (1) at (1,1) {};
        \node[gauge,label=right:{\footnotesize $1$}] (2) at (1,-1) {};
        \node[gauge,label=left:{\footnotesize $1$}] (3) at (-1,-1) {};
        \node[gauge,label=left:{\footnotesize $1$}] (4) at (-1,1) {};
        \draw (0)--(1);
        \draw (0)--(2);
        \draw (0)--(3);
        \draw (0)--(4);
\end{tikzpicture}}}
\quad \text{\footnotesize$/\BZ_2^{[1]}$}
}
We consider every subgroup $H$ of $S_4$, up to automorphisms, that acts on the four legs of the above quiver. An important result we derive in this section is the superconformal index of theory \eref{affineD4} refined with respect to the fugacities associated with the generators of certain Abelian subgroups of $H$.  This is achieved by considering the action of cycles on the index of \eref{affineD4}, as presented in Section \ref{sec:actioncycles}, and then inserting the fugacities for the discrete Abelian group associated with each type of the cycles. These discrete fugacities are subject to two conditions: (1) upon setting them to unity, we recover the index of \eref{affineD4}, and (2) upon summing all of such fugacities over the allowed values and dividing by the order of $H$, we obtain the index of theory \eref{affineD4} with the whole group $H$ being gauged.  This procedure of obtaining the refined index can, in fact, be applied very generally to any subgroup of a permutation group. We elucidate this procedure explicitly in the examples below. One of the benefits of this refined index is that we can gauge a proper subgroup $K$ of $H$ by summing the corresponding fugacities over the allowed values.

Suppose that the whole group $H$ is gauged. The result coincides with theory \eref{affineD4} wreathed by $H$, and we denote the resulting theory by $\eref{affineD4}/H$ as before. We check that the Higgs and Coulomb branch limits are in agreement with \cite[Figures 9 and 11]{Bourget:2020bxh}. Note that, when $H$ is non-Abelian, gauging $H$ gives rise to a non-invertible symmetry characterised by 2-Rep$(H)$ 
 \cite{Tachikawa:2017gyf} (see also \eg \cite{Bhardwaj:2017xup, Bhardwaj:2022yxj, Bartsch:2022mpm}). 

On the other hand, if we gauge a proper subgroup $K$ of $H$, the result {\it may} or {\it may not} coincide with wreathing theory \eref{affineD4} by $K$. If the resulting theory coincides with \eref{affineD4} wreathed by $K$, we denote it by \eref{affineD4}$/K$ as usual, but if this is not the case, we will specify the wreathing group explicitly. We will discuss several examples in the following subsections. Moreover, the aforementioned refined index also allows us to gauge diverse subgroups of $H$ in various orders. This gives rise to intricate symmetry webs, many of which were studied in \cite{Bhardwaj:2022maz, Bartsch:2022ytj}. We analyse the case in which $H$ can be realised as a semi-direct product of other groups, as well as the case in which $H$ is a non-trivial central extension of other groups. In both cases, the indices manifest interesting features. We summarise these results in Section \ref{sec:sumvarious}.

\subsection{\texorpdfstring{Cycles and their actions on the index of the affine $D_4$ quiver}{Cycles and their actions on the index of the affine D4 quiver}} \label{sec:actioncycles}

Since every element of $S_4$ can be represented as a cycle or a product of disjoint cycles, in this subsection we investigate how it acts on the superconformal index of theory \eref{affineD4}. We  provide only the prescription for computing such indices which will be useful for obtaining the results in the following subsections. The readers are referred to Appendix \ref{sec:wreathedquivers} for arguments and justifications. 

The index of the star-shaped quiver \eref{affineD4} can be obtained by gluing four copies of the $T[\SU(2)]$ theory \cite{Gaiotto:2008ak} together as follows:
\bes{ \label{indSp}
\CI_{\eref{affineD4}}(w_{1,2,3,4}, a| \omega ; x) 
&= \frac{1}{2} \sum_{\epsilon=0}^1 \omega^\epsilon \sum_{m \in \BZ+\frac{\epsilon}{2}} \oint \frac{d z}{2\pi i z}  \,\, \CZ^{\USp(2)}_{\text{vec}} (z;m;x) \\
& \qquad \times \prod_{s = {0, \pm 1}} \CZ^{1}_{\chi} \left(z^{2 s} a^{-2}; 2 s l; x\right) \times \CP~,
}
where $\CP$ is defined as
\bes{
\CP = \prod_{i=1}^4 \CI^\epsilon_{T[\SU(2)]} (w_i,n_i =0|z,m|a, n_a=0;x)~, 
}
and the index of the $T[\SU(2)]$ theory, computed from that of the $\U(1)$ gauge theory with two hypermultiplets of charge one, is
\bes{
&\CI^\epsilon_{T[\SU(2)]}  (w, n| f, m|a, n_a; x) \\
&= \sum_{l \in \BZ + \frac{\epsilon}{2}} (w^2)^l \oint \frac{d\zeta}{2\pi i \zeta}\,\, \zeta^n \,\, \CZ_{\chi}^{1}(a^{-2};-2 n_a; x) \\
& \qquad \times \prod_{s = \pm 1} \CZ_{\chi}^{1/2}( (\zeta f)^s a; s(l+m) + n_a; x)  \CZ_{\chi}^{1/2}( (\zeta^{-1} f)^s a; s(-l+m) + n_a; x)~.
}
Here $w_{1,2,3,4}$ are the fugacities for $\so(8)$.  Upon setting $w_1=\ldots=w_4=1$, we obtain
\bes{ \label{indaffineD4}
\scalebox{0.9}{$
\begin{split}
&\CI_{\eqref{affineD4}}(w_i=1,a| \omega; x)  \\
&= 1+(12+16 \omega) a^{-2} x + \left[2 a^4+ (156+ 144 \omega) a^{-4} -(13 +16 \omega) \right]x^2\\
& \quad \, \, \, \, \,+  \left[ a^6+ (949+976 \omega )a^{-6}-a^2 -(329+320 \omega )a^{-2} \right] x^3 \\
& \quad \, \, \, \, \,+\left[3 a^8 + 184 + 192 \omega + \left(4486+4432 \omega\right) a^{-8}-a^4-\left(2811 + 2832 \omega\right) a^{-4}\right] x^4+\ldots~,
\end{split}
$}
}
which is equal to that of the first line of \eref{SU2w4refw} with $a \leftrightarrow a^{-1}$.

Let us first start by discussing a cycle of length two, say $(12)$. Its action is to identify the first and second legs of \eref{affineD4}. We represent this as 
\bes{ \label{quiver12}
\eref{affineD4}_{(12)} = 
\vcenter{\hbox{\begin{tikzpicture}
        \node[gauge,label=right:{\footnotesize $\SU(2)$}] (0) at (0,0) {};
        \node[gauge,label=right:{\footnotesize $1$}] (1) at (1,1) {};
        \node[gauge,label=left:{\footnotesize $1$}] (2) at (-1,1) {};
        \node[gauge,thick,draw=red,label=right:{\footnotesize $1$}] (3) at (0,-1.4) {};
        \draw (0)--(1);
        \draw (0)--(2);
        \draw[thick, red] (0)--(3) node[midway, left] {\red{\scriptsize $(x,a,z) \rightarrow (x^2, a^2,z^2)$}};
\end{tikzpicture}}}
\quad \text{\footnotesize$/\BZ_2^{[1]}$}
}
where $(x,a,z) \rightarrow (x^2, a^2,z^2)$ denotes the rescaling of the fugacities in the contribution of each red line, with $z$ being the $\SU(2)$ fugacity. Explicitly, the index is given by \eref{indSp} where $\CP$ is defined as
\bes{
\CP &= \CI^\epsilon_{T[\SU(2)]} (w_1 w_2,n_1=n_2 =0|z^2,m|a^2, n_a=0;x^2) \\& \qquad \times \prod_{i=3}^4 \CI^\epsilon_{T[\SU(2)]} (w_i,n_i =0|z,m|a, n_a=0;x)~,
}
with $w_1 = w_2$. Note that, if we set $w_i=1$, the results for any cycle of length two of the form $(i_1 i_2)$ are equal, namely
\bes{ \label{indZ2(12)}
\scalebox{0.99}{$
\begin{split}
&\CI_{\eqref{affineD4}_{(i_1 i_2)}}(w_i=1, a| \omega; x) \\
&= 1+ (6 + 8 \omega) a^{-2} x + \left[(50 + 40 \omega)a^{-4} - (7+8\omega) \right] x^2\\
& \quad \, \, \, \, \,+   \left[a^2 + (185 + 200 \omega)a^{-6} - a^6 - (79 + 80 \omega)a^{-2} \right] x^3\\
& \quad \, \, \, \, \,+\left[a^8+a^4+34+48 \omega + \left(658+616 \omega\right) a^{-8}-\left(405+408 \omega\right) a^{-4}\right] x^4 +\ldots~.
\end{split}
$}
}

Next, we consider the product of two disjoint cycles of length two, say $(12)(34)$. Its action is to identify the first and second legs together, and the third and fourth legs together. This can be represented as
\bes{ \label{quiver12c34}
\eref{affineD4}_{(12)(34)} = 
\vcenter{\hbox{\begin{tikzpicture}
        \node[gauge,label=right:{\footnotesize $\SU(2)$}] (0) at (0,0) {};
        \node[gauge,thick,draw=red,label=right:{\footnotesize $1$}] (1) at (0,1.4) {};
        \node[gauge,thick,draw=red,label=right:{\footnotesize $1$}] (2) at (0,-1.4) {};
        \draw[thick, red] (0)--(1) node[midway, left] {\red{\scriptsize $(x,a,z) \rightarrow (x^2, a^2,z^2)$}};
        \draw[thick, red] (0)--(2) node[midway, left] {\red{\scriptsize $(x,a,z) \rightarrow (x^2, a^2,z^2)$}};
\end{tikzpicture}}}
\quad \text{\footnotesize$/\BZ_2^{[1]}$}
}
Explicitly, the index is given by \eref{indSp} with $\CP$ defined as
\bes{
\CP =&  \CI^\epsilon_{T[\SU(2)]} (w_1 w_2,n_1=n_2 =0|z^2,m|a^2, n_a=0;x^2) \\
\times & \CI^\epsilon_{T[\SU(2)]} (w_3 w_4,n_3=n_4 =0|z^2,m|a^2, n_a=0;x^2)~,
}
where $w_1 = w_2$ and $w_3 = w_4$. If we set $w_i=1$, the results for any product of two disjoint cycles of length two of the form $(i_1 i_2)(i_3 i_4)$ are equal:
\bes{\label{indZ2(12)(34)}
&\CI_{{\eqref{affineD4}_{(i_1 i_2) (i_3 i_4)}}}(w_i=1, a| \omega; x) \\
&= 1+ 4 \omega a^{-2} x + \left[2 a^4 + (16 + 4 \omega)a^{-4} - (1+4\omega)\right] x^2 \\& \quad \, \, \, \, \,+\left[a^6 + \left(9 + 36 \omega\right) a^{-6} - a^2 - 9 a^{-2}\right] x^3\\& \quad \, \, \, \, \,+\left[3 a^8+\left(90+36 \omega\right) a^{-8}-a^4-8-\left(15+36 \omega\right) a^{-4}\right] x^4+\ldots~.
}

We now consider the action of the cycle of length three, say $(123)$, which can be represented graphically as
\bes{ \label{quiver123}
\eref{affineD4}_{(123)} = 
\vcenter{\hbox{\begin{tikzpicture}
        \node[gauge,label=right:{\footnotesize $\SU(2)$}] (0) at (0,0) {};
        \node[gauge,label=right:{\footnotesize $1$}] (1) at (0,1.4) {};
        \node[gauge,thick,draw=red,label=right:{\footnotesize $1$}] (2) at (0,-1.4) {};
        \draw (0)--(1);
        \draw[thick, red] (0)--(2) node[midway, left] {\red{\scriptsize $(x,a,z) \rightarrow (x^3, a^3,z^3)$}};
\end{tikzpicture}}}
\quad \text{\footnotesize$/\BZ_2^{[1]}$}
}
The index is given by \eref{indSp} where $\CP$ defined as
\bes{
\CP =&  \CI^\epsilon_{T[\SU(2)]} (w_1 w_2 w_3,n_1=n_2=n_3 =0|z^3,m|a^3, n_a=0;x^3) \\ \times
& \CI^\epsilon_{T[\SU(2)]} (w_4,n_4 =0|z,m|a, n_a=0;x)~,
}
with $w_1 = w_2 = w_3$. Setting $w_i=1$, we obtain the result for any cycle of length three of the form $(i_1 i_2 i_3)$ as follows:
\bes{ \label{indZ3(123)}
&\CI_{\eref{affineD4}_{(i_1 i_2 i_3)}} (w_i=1, a| \omega; x) \\ &= 1 + \left(3 + 4 \omega \right) a^{-2} x + \left[\left(15 + 12 \omega\right) a^{-4} - a^4 -\left(4 + 4 \omega \right)\right] x^2 \\
&\quad \, \, \, \, \, + \left[a^6 + 2 a^2 + \left(37 + 40 \omega\right) a^{-6} - \left(20 + 20 \omega \right) a^{-2}\right] x^3 \\
&\quad \, \, \, \, \,+\left[7+12 \omega + \left(94+88 \omega\right) a^{-8}-a^4-\left(60+60 \omega \right) a^{-4}\right] x^4+ \ldots~.
}

Finally, the action of the cycle of length four is represented by
\bes{ \label{quiver1234}
\eref{affineD4}_{(1234)} = 
\vcenter{\hbox{\begin{tikzpicture}
        \node[gauge,label=right:{\footnotesize $\SU(2)$}] (0) at (1,0) {};
        \node[gauge,thick,draw=red,label=left:{\footnotesize $1$}] (1) at (-1,0) {};
        \draw[thick, red] (0)--(1) node[midway, above=0.1] {\red{\scriptsize $(x,a,z) \rightarrow (x^4, a^4,z^4)$}};
\end{tikzpicture}}}
\quad \text{\footnotesize$/\BZ_2^{[1]}$}
}
The index is given by \eref{indSp} with $\CP$ defined as
\bes{
\CP &=  \CI^\epsilon_{T[\SU(2)]} (w_1 w_2 w_3 w_4,n_1=n_2=n_3=n_4 =0|z^4,m|a^4, n_a=0;x^4)~,
}
where $w_1 = w_2 = w_3 = w_4$. Setting $w_i=1$, we obtain the result for any cycle of length four of the form $(i_1 i_2 i_3 i_4)$ as follows:
\bes{ \label{ind(1234)}
&\CI_{{\eqref{affineD4}_{(i_1 i_2 i_3 i_4)}}}(w_i=1, a| \omega; x) \\&=
1 + 2 \omega a^{-2} x + [(4+2 \omega)a^{-4} -(1+2 \omega)]x^2 \\
&\quad \, \, \, \, \,+ [a^2 + (3+6\omega)a^{-6} - a^6 -3 a^{-2}]x^3\\
&\quad \, \, \, \, \,+\left[a^8+a^4+\left(12+6 \omega\right) a^{-8}-2-\left(3+6 \omega\right) a^{-4}\right] x^4 +\ldots~.
}

\subsection{\texorpdfstring{$\BZ_2$ wreathing of the affine $D_4$ quiver}{Z2 wreathing of the affine D4 quiver}}
Let us start by analysing the affine $D_4$ quiver wreathed by the $\BZ_2$ subgroups of $S_4$, whose mirror theories have been studied in Section \ref{sec:discretegaugingSU2w4}.

\subsubsection*{Single transposition}
The index associated with the quiver wreathed by $\BZ_2 = \langle (12) \rangle$ can be derived from the relation
\bes{ \label{IndexZ2wr}
\eref{affineD4}/ \BZ_2 = \eref{affineD4}/ \langle (12) \rangle = \frac{1}{2} \left[\eref{affineD4} + \eref{affineD4}_{(12)}\right]~,
}
where the relevant indices are
\begin{subequations}  \label{indaffD4Z2}
\begin{align}
\begin{split} 
&\CI_{\eqref{affineD4}}(a| \omega; x) = \eref{indaffineD4}~, \qquad 
\CI_{{\eqref{affineD4}_{(12)}}}(a| \omega; x) = \eref{indZ2(12)}, 
\end{split} \\
\begin{split} \label{indZ2wr}
& \CI_{{\eqref{affineD4}/ \BZ_2}}(a| \omega; x) \\
&= 1+\left(9 + 12 \omega\right) a^{-2} x + \left[a^4 + \left(103 + 92 \omega \right) a^{-4} - \left(10 + 12 \omega \right)\right] x^2 \\
&\quad \, \, \, \, \,+  [(567+588 \omega )a^{-6}-(204+200 \omega)a^{-2}] x^3 + \left[2 a^8 + 109 + 120 \omega \right. \\
&\quad \, \, \, \, \, \, \, \left.+ \left(2572+2524 \omega\right) a^{-8} - \left(1608+1620 \omega\right) a^{-4}\right] x^4+ \ldots~.
\end{split}
\end{align}
\end{subequations}
It will be useful to obtain the index of the affine $D_4$ quiver \eref{affineD4} refined with a fugacity $b$ associated with the $(\BZ_2^{[0]})_b$ symmetry generated by the element $(12)$ of $\BZ_2$, satisfying $b^2 = 1$. This can be achieved order by order in $x$ using the ansatz
\bes{
\CI_{\eqref{affineD4}}(a| \omega| b; x) &= \CI_{{\eqref{affineD4}/ \BZ_2}}(a| \omega; x) +b \sum_{p=0}^\infty y_p x^p~,
}
where $y_{p} \equiv y_{p}(a, \omega)$ are unknowns that have to satisfy the conditions:
\bes{
\CI_{{\eqref{affineD4}/ \BZ_2}}(a| \omega; x) &= \frac{1}{2} \sum_{b = \pm 1} \CI_{\eqref{affineD4}}(a| \omega | b; x) = \eref{indZ2wr} ~,  \\
\CI_{\eqref{affineD4}}(a| \omega; x) &= \CI_{\eqref{affineD4}}(a| \omega | b=1; x) = \eref{indaffineD4}~.
}
Comparing this with \eref{IndexZ2wr}, we see that they can be computed by solving the following system of equations in the unknown variable $y_p$:
\bes{ \label{systemZ2}
\left[\CI_{{\eqref{affineD4}/ \BZ_2}}(a| \omega; x) \right]_{x^p} + b y_p |_{b = 1} &= \left[ \CI_{\eqref{affineD4}}(a| \omega; x) \right]_{x^p}~, \\
\left[\CI_{{\eqref{affineD4}/ \BZ_2}}(a| \omega; x) \right]_{x^p} + b y_p |_{b = e^{\pi i}} &= \left[\CI_{{\eqref{affineD4}_{(12)}}}(a| \omega; x) \right]_{x^p}~,
}
where the notation $[\CI]_{x^p}$ denotes the coefficient of $x^p$ in the series expansion of $\CI$, and the explicit expressions for $\CI$ in the above are given by \eref{indaffD4Z2}.  As a result, the index of the affine $D_4$ quiver refined with the $\BZ_2$ fugacity $b$ reads
\bes{  \label{indfugZ2}
&\CI_{\eqref{affineD4}}(a| \omega | b; x) \\
&= 1+\left[9 + 12 \omega + \left(3 + 4 \omega\right) b\right] a^{-2} x + \left\{\left(1 + b\right) a^4 \right. \\ & \quad \, \, \, \, \, \left. + \left[103 + 92 \omega + \left(53 + 52 \omega\right) b \right] a^{-4} - \left[10 + 12 \omega + \left(3 + 4 \omega\right) b\right]\right\} x^2 + \ldots~,
}
from which the index of \eref{affineD4} wreathed by $\BZ_2$ can be derived by gauging $(\BZ_2^{[0]})_b$:
\bes{
\CI_{{\eqref{affineD4}/ \BZ_2}}(a| \omega; x) = &\frac{1}{2} \sum_{b = \pm 1} \CI_{\eqref{affineD4}}(a| \omega | b; x) = \eref{indZ2wr}~.
}
Observe that, in agreement with mirror symmetry, upon sending $a \rightarrow a^{-1}$, this coincides with the index of theory $\eref{USp2SO2SO2SO2SO2}/\BZ_2 $ if we set $\omega = 1$, see \eref{tabSU2w4}, and with the index of theory $\eref{USp2SO2SO2SO2SO2modZ2}/\BZ_2$ if we gauge $(\BZ_2^{[0]})_{\omega}$, see \eref{indexUSp2SO2SO2SO2SO2modZ2modGamma}.

\subsubsection*{Double transposition}
Similarly, the theory arising from wreathing \eref{affineD4} by the double transposition (DT) subgroup of $S_4$, which is generated by the $\BZ_2$ element $(12)(34)$, can be derived as
\bes{ \label{IndexDTwr}
\eref{affineD4}/ \mathrm{DT} = \eref{affineD4}/ \langle (12)(34) \rangle = \frac{1}{2} \left[\eref{affineD4} + \eref{affineD4}_{(12)(34)}\right]~,
}
where the relevant indices are
\begin{subequations}
\begin{align}
\begin{split} 
\CI_{\eqref{affineD4}}(a| \omega; x) = \eref{indaffineD4}~, \qquad \CI_{{\eqref{affineD4}_{(12) (34)}}}(a| \omega; x)= \eref{indZ2(12)(34)}~, 
\end{split} \\
\begin{split}  \label{indZ2wrDT}
&\CI_{{\eqref{affineD4}/ \mathrm{DT}}}(a| \omega; x) \\
&= 1+\left(6 + 10 \omega\right) a^{-2} x + \left[2 a^4 + \left(86 + 74 \omega \right) a^{-4} - \left(7 + 10 \omega \right)\right] x^2 \\ & \quad \, \, \, \, \, + \left[a^6 + \left(479+506 \omega\right) a^{-6} - a^2 - \left(169 + 160 \omega\right) a^{-2}\right] x^3 +\left[3 a^8 + 88\right. \\
&\quad \, \, \, \, \, \, \, \left. + 96 \omega +\left(2288+2234 \omega\right) a^{-8} - a^4 -\left(1413+1434 \omega\right) a^{-4}\right] x^4+ \ldots~.
\end{split} 
\end{align}
\end{subequations}
Note that, upon sending $a \rightarrow a^{-1}$ in the last expression, we obtain the result that coincides with the index of its mirror dual $\eref{USp2SO2SO2SO2SO2}/\mathrm{DT}$, reported in \eref{tabSU2w4} for $\omega = 1$. On the other hand, gauging $(\BZ_2^{[0]})_\omega$, the index coincides with the one of the mirror theory $\eref{USp2SO2SO2SO2SO2modZ2}/\mathrm{DT}$, listed in \eref{indexUSp2SO2SO2SO2SO2modZ2modGamma}.

The index of the affine $D_4$ quiver \eref{affineD4} can be refined with a fugacity $c$ associated with the $(\BZ_2^{[0]})_c$ symmetry corresponding to the $\BZ_2$ element $(12)(34)$, with $c^2 = 1$. We use the ansatz
\bes{ \label{indexDTref}
\CI_{\eqref{affineD4}}(a| \omega| c; x) = \CI_{{\eqref{affineD4}/ \mathrm{DT}}}(a| \omega; x) +c \sum_{p=0}^\infty y_p x^p~,
}
where $y_p \equiv y_p(a, \omega)$ are unknowns that satisfy
\bes{
\CI_{{\eqref{affineD4}/ \mathrm{DT}}}(a| \omega; x) &= \frac{1}{2} \sum_{c = \pm 1} \CI_{\eqref{affineD4}}(a| \omega | c; x) = \eref{indZ2wrDT} ~,  \\
\CI_{\eqref{affineD4}}(a| \omega; x) &= \CI_{\eqref{affineD4}}(a| \omega | c=1; x) = \eref{indaffineD4}~.
}
From \eref{IndexDTwr}, we can obtain $y_p$ by solving the following system of equations:
\bes{ \label{systemDT}
\CF_p( c=1)  &= \left[ \CI_{\eqref{affineD4}}(a| \omega; x) = \eref{indaffineD4} \right]_{x^p}~, \\ 
\CF_p( c=-1) &= \left[\CI_{\eqref{affineD4}_{(12)(34)}}(a| \omega; x) = \eref{indZ2(12)(34)} \right]_{x^p}~,
}
where we define
\bes{
\CF_p( c) \equiv  \left[\CI_{{\eqref{affineD4}/ \mathrm{DT}}}(a| \omega; x) = \eref{indZ2wrDT} \right]_{x^p} + c y_p
}
and the notation $[\CI]_{x^p}$ denotes the coefficient of $x^p$ in the series expansion of $\CI$. Solving \eref{systemDT}, it follows that, explicitly, the index \eref{indexDTref} is given by
\bes{ 
\scalebox{0.95}{$
\begin{split}
&\CI_{\eqref{affineD4}}(a| \omega | c; x) \\
&= 1+\left[6 + 10 \omega + \left(6 + 6 \omega\right) c \right] a^{-2} x \\ & \quad \, \, \, \, \, + \left\{2 a^4 + \left[86 + 74 \omega + \left(70 + 70 \omega\right) c \right] a^{-4} - \left[7 + 10 \omega + \left(6 + 6 \omega\right) c\right]\right\} x^2 + \ldots~.
\end{split}
$}
}
Indeed, the index of \eref{affineD4} wreathed by the double transposition subgroup of $S_4$ can be obtained by gauging $(\BZ_2^{[0]})_c$:
\bes{ \label{inedxaffD4DTwr}
\CI_{{\eqref{affineD4}/ \mathrm{DT}}}(a| \omega; x) = &\frac{1}{2} \sum_{c = \pm 1} \CI_{\eqref{affineD4}}(a| \omega | c; x) = \eref{indZ2wrDT}.
}

\subsection{\texorpdfstring{$\BZ_2 \times \BZ_2$ wreathing of the affine $D_4$ quiver}{Z2xZ2 wreathing of the affine D4 quiver}}
We can now proceed by examining the effect of wreathing the affine $D_4$ quiver \eref{affineD4} by the $\BZ_2 \times \BZ_2$ subgroups of $S_4$, whose mirror theories have also been studied in Section \ref{sec:discretegaugingSU2w4}.

\subsubsection*{Non-normal Klein subgroup of $S_4$}

The theory arising from wreathing the affine $D_4$ quiver by the non-normal Klein subgroup of $S_4$ can be expressed as
\bes{ \label{IndexNNKwr}
\eref{affineD4}/\text{NNK} &\equiv \eref{affineD4}/{ \langle (12), (34) \rangle} \\
&= \frac{1}{4} \left[\eref{affineD4} + \eref{affineD4}_{(12)} + \eref{affineD4}_{(34)} + \eref{affineD4}_{(12) (34)}\right]~,
}
where the indices of $\eref{affineD4}_{(12)}$ and $\eref{affineD4}_{(34)}$ are given by \eref{indZ2(12)},\footnote{Strictly speaking, the non-normal Klein group elements $(12)$ and $(34)$ act on different legs of quiver \eref{affineD4}. However, since we are interested in computing the index unrefined with respect to the flavour fugacities associated with each leg of the affine $D_4$ quiver, the indices of $\eref{affineD4}_{(12)}$ and $\eref{affineD4}_{(34)}$ are consequently the same, given by \eref{indZ2(12)}. Hence, we are going to identify both $\eref{affineD4}_{(12)}$ and $\eref{affineD4}_{(34)}$ with \eref{quiver12}, even if the legs which are altered by the group action in the two cases are not the same. We will use a similar argument throughout this section.} and the index of $\eref{affineD4}_{(12) (34)}$ is given by \eref{indZ2(12)(34)}.  As a result, we obtain
\bes{ \label{indmod(12),(34)}
&\CI_{\eref{affineD4}/{ \mathrm{NNK}}}\\ 
&= 1 + \left(6 + 9 \omega\right) a^{-2} x + \left[a^4 + \left(68 + 57 \omega\right) a^{-4} - \left(7 + 9 \omega\right) \right] x^2 \\
& \quad \, \, \, \, \, + \left[\left(332 + 353 \omega\right) a^{-6} - \left(124 + 120 \omega \right) a^{-2}\right] x^3 \\
& \quad \, \, \, \, \,+ \left[2 a^8 + 61 + 72 \omega + \left(1473+1425 \omega\right) a^{-8} - \left(909+921 \omega\right) a^{-4}\right] x^4+ \ldots~.
}

We can refine the index of \eref{affineD4} with two fugacities $b_1$ and $b_2$, satisfying $b_1^2 = b_2^2 = 1$, associated with the $(\BZ_2^{[0]})_{b_1}$ and $(\BZ_2^{[0]})_{b_2}$ symmetries generated by $(12)$ and $(34)$ respectively as follows. Let us consider the ansatz
\bes{
&\CI_{\eqref{affineD4}}(a| \omega| b_1,b_2; x)  \\
&=  \CI_{{\eqref{affineD4}/ \mathrm{NNK}}}(a| \omega; x) + (b_1+b_2) \sum_{p=0}^\infty y_{1,p} x^p + b_1 b_2 \sum_{p=0}^\infty y_{2,p} x^p~,
}
where there are two sets of the unknown variables $y_{1,p} \equiv y_{1,p}(a, \omega)$ and $y_{2,p} \equiv y_{2,p}(a, \omega)$. In this ansatz, the fugacities associated with the elements of the same cycle structure appear with the same set of the unknowns, \ie~ $b_{1}$ and $b_2$ associated with $(12)$ and $(34)$ appear with $y_{1,p}$, whereas $b_1b_2$ associated with $(12)(34)$ appears with $y_{2,p}$.  The above ansatz must satisfy the conditions:
\bes{
\CI_{{\eqref{affineD4}/ \mathrm{NNK}}}(a| \omega; x) &= \frac{1}{4} \sum_{b_1,b_2 = \pm 1} \CI_{\eqref{affineD4}}(a| \omega | b_1,b_2; x) = \eref{indmod(12),(34)} ~,  \\
\CI_{\eqref{affineD4}}(a| \omega; x) &= \CI_{\eqref{affineD4}}(a| \omega | b_1=1, b_2=1; x) = \eref{indaffineD4}~. 
}
Using \eref{IndexNNKwr}, we can determine $y_{1,p}$ and $y_{2,p}$ by solving the following system of equations:
\bes{ \label{systemNNK}
\left[ \CI_{\eqref{affineD4}}(a| \omega; x) = \eref{indaffineD4}\right]_{x^p} &= \CF_p(b_1=1,b_2=1)~, \\
\left[ \CI_{{\eqref{affineD4}_{(12)\, \text{or}\, (34)}}}(a| \omega; x) = \eref{indZ2(12)}\right]_{x^p} &= \frac{1}{2} \Big[ \CF_p(b_1=1,b_2=-1)\\
& \quad \, \, +\CF_p(b_1=-1,b_2=1)\Big]~,  \\
\left[\CI_{{\eqref{affineD4}_{(12) (34)}}}(a| \omega; x)= \eref{indZ2(12)(34)}\right]_{x^p}  &= \CF_p(b_1=-1,b_2=-1) ~,
}
where we define
\bes{
\CF_p(b_1,b_2) &= \left[\CI_{{\eqref{affineD4}/ \mathrm{NNK}}}(a| \omega; x) = \eref{indmod(12),(34)}\right]_{x^p} + \left(b_1 + b_2\right) y_{1,p} + b_1 b_2 y_{2,p}
}
and the notation $[\CI]_{x^p}$ denotes the coefficient of $x^p$ in the series expansion of $\CI$. As a result, the index of quiver \eref{affineD4} refined with respect to the fugacities $b_1$ and $b_2$ reads
\bes{ \label{indexNNKref}
&\CI_{\eqref{affineD4}}(a| \omega| b_1, b_2; x) \\ 
&= 1+\left[6 + 9 \omega + \left(3 + 3 \omega\right) \left(b_1 + b_2\right) + \omega b_1 b_2 \right] a^{-2} x + \left\{\left(1 + b_1 b_2\right) a^4 \right. \\ & \quad \, \, \, \, \, \left. + \left[68 + 57 \omega + \left(35 + 35 \omega\right) \left(b_1 + b_2\right) + 
 \left(18 + 17 \omega\right) b_1 b_2 \right] a^{-4} \right. \\ & \quad \, \, \, \, \, \left.  - \left[7 + 9 \omega + \left(3 + 3 \omega\right) \left(b_1 + b_2\right) + \omega b_1 b_2\right]\right\} x^2 + \ldots~.
}
Note that the coefficients of the fugacities associated with the elements with the same cycle structure are equal. 

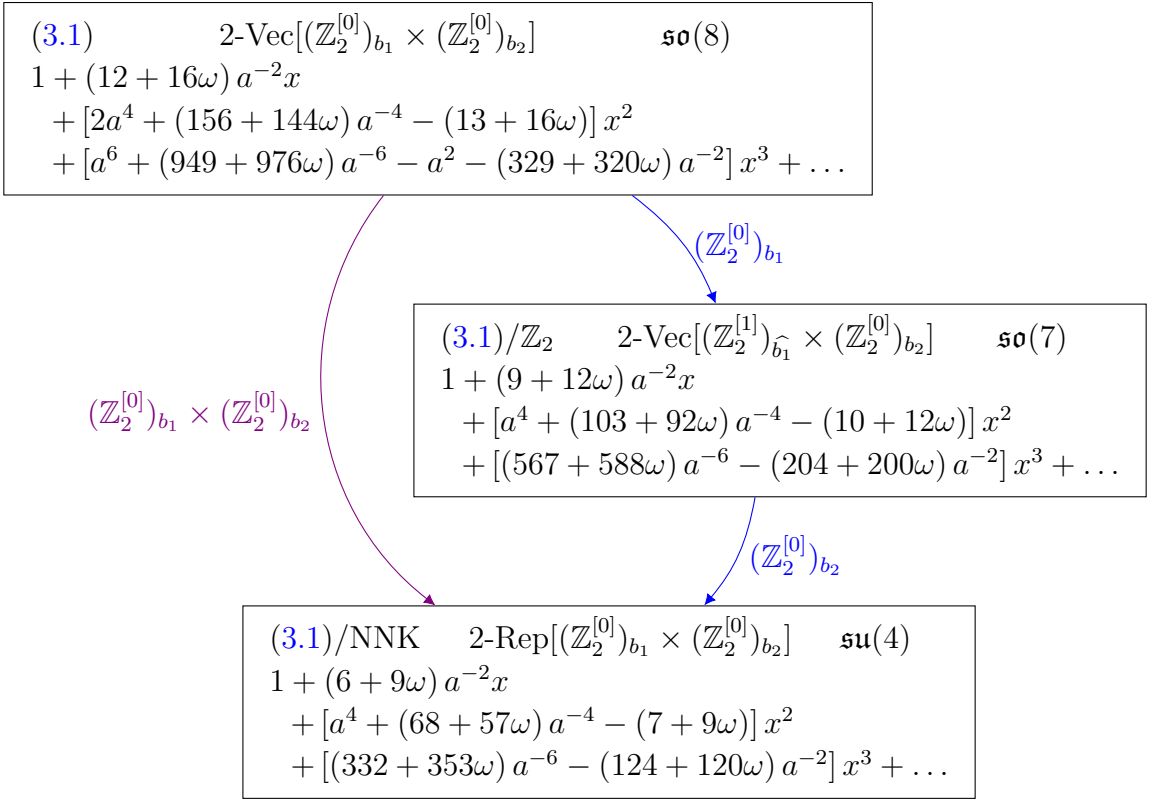
\begin{figure}
    \centering
\begin{tikzpicture}
			\node[draw] (2VecZ2Z2) at (-2.25,4) {\begin{tabular}{l}
			 \eref{affineD4}   \hfill 2-Vec$[(\BZ_2^{[0]})_{b_1} \times (\BZ_2^{[0]})_{b_2}]$ \hfill $\mathfrak{so}(8)$ \\$1 + \left(12 + 16 \omega\right) a^{-2} x $ \\$\, \, \, \, + \left[2 a^4 + \left(156 + 144 \omega\right) a^{-4} - \left(13 + 16 \omega\right) \right] x^2$ \\ $\, \, \, \, + \left[a^6 + \left(949 + 976 \omega\right) a^{-6} - a^2 - \left(329 + 320 \omega \right) a^{-2}\right] x^3 + \ldots$ \end{tabular} }; 
			\node[draw] (2VecZ2Z21) at (2.25,0) {\begin{tabular}{l}
			$\eref{affineD4}/\BZ_2$ \hfill 2-Vec$[(\BZ_2^{[1]})_{\hat{b_1}} \times (\BZ_2^{[0]})_{b_2}]$ \hfill $\mathfrak{so}(7)$ \\$1 + \left(9 + 12 \omega\right) a^{-2} x $ \\$\, \, \, \, + \left[a^4 + \left(103 + 92 \omega\right) a^{-4} - \left(10 + 12 \omega\right) \right] x^2$ \\ $\, \, \, \, + \left[\left(567 + 588 \omega\right) a^{-6} - \left(204 + 200 \omega \right) a^{-2}\right] x^3 + \ldots$	\end{tabular} }; 
			\node[draw] (2RepZ2Z2) at (0,-4) {\begin{tabular}{l}
			    \eref{affineD4}/NNK \hfill 2-Rep$[(\BZ_2^{[0]})_{b_1} \times (\BZ_2^{[0]})_{b_2}]$ \hfill $\mathfrak{su}(4)$ \\$1 + \left(6 + 9 \omega\right) a^{-2} x $ \\$\, \, \, \, + \left[a^4 + \left(68 + 57 \omega\right) a^{-4} - \left(7 + 9 \omega\right) \right] x^2$ \\ $\, \, \, \, + \left[\left(332 + 353 \omega\right) a^{-6} - \left(124 + 120 \omega \right) a^{-2}\right] x^3 + \ldots$	\end{tabular} };  
            \draw[->,blue] (2VecZ2Z2) to [bend left=15] node[midway, right] {\blue $(\BZ_2^{[0]})_{b_1}$} (2VecZ2Z21);
            \draw[->,blue] (2VecZ2Z21) to [bend left=15] node[midway, right] {\blue $(\BZ_2^{[0]})_{b_2}$} (2RepZ2Z2);
            \draw[->,violet] (2VecZ2Z2) to [bend right=45] node[midway, left] {\violet $(\BZ_2^{[0]})_{b_1} \times (\BZ_2^{[0]})_{b_2}$} (2RepZ2Z2);
\end{tikzpicture}
    \caption[NNK web]{The affine $D_4$ quiver \eref{affineD4} with global $(\BZ_2^{[0]})_{b_1} \times (\BZ_2^{[0]})_{b_2} \equiv \langle (12), (34) \rangle$ zero-form symmetry and gauging of various subgroups. Each arrow with label $F$ connecting two boxes denotes the gauging of the zero-form symmetry $F$. In each box, which is associated with a theory arising from the affine $D_4$ quiver via discrete gauging, we report the corresponding symmetry category on the left, its flavour algebra on the right and, below, the index refined with the fugacities $a$ and $\omega$ (we set the fugacities $b_1$ and $b_2$ to unity after gauging). Since the original theory is symmetric in $b_1$ and $b_2$, see \eref{indexNNKref}, the sequence of {\blue blue} arrows on the right-hand side can be taken to be equivalently either gauging $(\BZ_2^{[0]})_{b_1}$ first, followed by gauging $(\BZ_2^{[0]})_{b_2}$, or the opposite.}
    \label{figNNKweb}
\end{figure}

Having derived the index \eref{indexNNKref} refined with $b_{1}$ and $b_2$, we can now study sequential gauging. Starting from theory \eref{affineD4} with the global $(\BZ_2^{[0]})_{b_1} \times (\BZ_2^{[0]})_{b_2}$ zero-form symmetry, whose symmetry category is 2-Vec$[(\BZ_2^{[0]})_{b_1} \times (\BZ_2^{[0]})_{b_2}]$, we can gauge $(\BZ_2^{[0]})_{b_1}$ (or, equivalently, $(\BZ_2^{[0]})_{b_2}$) in order to land on theory $\eref{affineD4}/\BZ_2 = \eref{affineD4}/{ \langle (12) \rangle}$. Since gauging $(\BZ_2^{[0]})_{b_1}$ leads to a dual $(\BZ_2^{[1]})_{\hat{b_1}}$ one-form symmetry, the corresponding symmetry category is 2-Vec$[(\BZ_2^{[1]})_{\hat{b_1}} \times (\BZ_2^{[0]})_{b_2}]$. Subsequently, one can gauge $(\BZ_2^{[0]})_{b_2}$ and obtain the wreathed theory $\eref{affineD4}/{ \mathrm{NNK}}$, whose associated symmetry category is 2-Rep$[(\BZ_2^{[0]})_{b_1} \times (\BZ_2^{[0]})_{b_2}]$. The same end result can also be achieved by directly gauging the whole $(\BZ_2^{[0]})_{b_1} \times (\BZ_2^{[0]})_{b_2}$ symmetry. The whole process of gauging this $(\BZ_2^{[0]})_{b_1} \times (\BZ_2^{[0]})_{b_2}$ symmetry and subgroups thereof is described in Figure \ref{figNNKweb}. In particular, the index of the wreathed theory ${\eqref{affineD4}}/{ \mathrm{NNK}}$ can be computed by gauging the whole $(\BZ_2^{[0]})_{b_1} \times (\BZ_2^{[0]})_{b_2}$ symmetry starting from \eref{indexNNKref} as
\bes{
\CI_{{\eqref{affineD4}/ \mathrm{NNK}}}(a| \omega; x) = \frac{1}{4} \sum_{b_1, b_2 = \pm 1} \left[\CI_{\eqref{affineD4}}(a| \omega | b_1, b_2; x) =  \eref{indexNNKref}\right] = \eref{indmod(12),(34)}~.
}
The corresponding expression is reported in Figure \ref{figNNKweb}, where we point out that the flavour symmetry algebra is $\su(4)$. Once again, in agreement with mirror symmetry, we have
\bes{
\CI_{{\eqref{affineD4}/ \mathrm{NNK}}}(a| \omega = 1; x) &= \CI_{\eref{USp2SO2SO2SO2SO2}/\mathrm{NNK}}(a^{-1}; x)~, \\
\CI_{{{\eqref{affineD4}}/{ \mathrm{NNK}}}/(\BZ_2^{[0]})_{\omega}}(a; x) &= \frac{1}{2} \sum_{\omega = \pm 1} \CI_{{\eqref{affineD4}/ \mathrm{NNK}}}(a| \omega; x) \\ &= \CI_{\eref{USp2SO2SO2SO2SO2modZ2}/\mathrm{NNK}}(a^{-1}; x)~.
}
where the indices of the mirror theories are reported in \eref{tabSU2w4} and \eref{indexUSp2SO2SO2SO2SO2modZ2modGamma}.

\subsubsection*{Normal Klein subgroup of $S_4$}

Let us now analyse the affine $D_4$ quiver \eref{affineD4} wreathed by the normal Klein (NK) subgroup of $S_4$ generated by $(12) (34)$ and $(13) (24)$, \ie $\mathrm{NK}= \langle (12) (34), (13) (24) \rangle$:
\bes{ \label{IndexNKwr}
\eref{affineD4}/\mathrm{NK} &= \frac{1}{4} \left[\eref{affineD4} + \eref{affineD4}_{(12) (34)} + \eref{affineD4}_{(13) (24)} + \eref{affineD4}_{(14) (23)}\right]~,
}
where $\eref{affineD4}_{(12) (34)}$,$\eref{affineD4}_{(13) (24)}$ and $\eref{affineD4}_{(14) (23)}$ take the form \eref{quiver12c34}, and the contribution of each of these terms to the index is equal, given by \eref{indZ2(12)(34)}. 

We can refine the index of \eref{affineD4} with respect to the two fugacities $c_1$ and $c_2$, such that $c_1^2 = c_2^2 = 1$, associated with the $(\BZ_2^{[0]})_{c_1}$ and $(\BZ_2^{[0]})_{c_2}$ symmetries generated by $(12) (34)$ and $(13) (24)$ respectively. As before we consider the ansatz:
\bes{ \label{indexaffD4c1c2}
\CI_{\eqref{affineD4}}(a| \omega| c_1, c_2; x) = \CI_{\eqref{affineD4}/ \mathrm{NK}}(a| \omega; x) +\left(c_1 + c_2 + c_1 c_2\right) \sum_{p = 1}^{\infty}  y_p  x^p~,
}
where it satisfies the conditions that, upon setting $c_1=c_2=1$, we recover the index $\CI_{{\eqref{affineD4}}}(a| \omega; x)$, and upon summing over $c_1, c_2 = \pm 1$ and dividing by 4, we obtain $\CI_{\eqref{affineD4}/\mathrm{NK}}(a| \omega; x)$.  In particular, $y_p \equiv y_p(a, \omega)$ can be determined the following system of equations:
\bes{ \label{systemNK}
\left[\CI_{\eqref{affineD4}}(a| \omega; x) = \eref{indaffineD4}\right]_{x^p} &= \CF_p(c_1=1, c_2=1)~,  \\
\left[\CI_{{\eqref{affineD4}_{(i_1 i_2)(i_3 i_4)}}}(a| \omega; x) = \eref{indZ2(12)(34)}\right]_{x^p}  &= \frac{1}{3} \Big[\CF_p(c_1=-1, c_2=1) \\
& \quad + \CF_p(c_1=1, c_2=-1) \\
& \quad +\CF_p(c_1=-1, c_2=-1)\Big]~,
}
where we define
\bes{
\CF_p(c_1, c_2) =\left[ \CI_{{\eqref{affineD4}/ \mathrm{NK}}}(a| \omega; x)\right]_{x^p} + \left(c_1 + c_2 + c_1 c_2\right) y_p~,
}
where $[\CI]_{x^p}$ denotes the coefficient of $x^p$ in the series expansion of $\CI$. Solving \eref{systemNK} for the unknown variable $y_p$, we obtain that the index \eref{indexaffD4c1c2} of the affine $D_4$ quiver refined with the fugacities $c_1$ and $c_2$ is given by
\begin{figure}
    \centering
\begin{tikzpicture}
			\node[draw] (2VecZ2Z2) at (-1.9,4) {\begin{tabular}{l}
			\eref{affineD4}    \hfill 2-Vec$[(\BZ_2^{[0]})_{c_1} \times (\BZ_2^{[0]})_{c_2}]$ \hfill $\mathfrak{so}(8)$ \\$1 + \left(12 + 16 \omega\right) a^{-2} x $ \\$\, \, \, \, + \left[2 a^4 + \left(156 + 144 \omega\right) a^{-4} - \left(13 + 16 \omega\right) \right] x^2$ \\ $\, \, \, \, + \left[a^6 + \left(949 + 976 \omega\right) a^{-6} - a^2 - \left(329 + 320 \omega \right) a^{-2}\right] x^3 + \ldots$ \end{tabular} }; 
			\node[draw] (2VecZ2Z21) at (1.9,0) {\begin{tabular}{l}
			    \eref{affineD4}/DT \hfill 2-Vec$[(\BZ_2^{[1]})_{\hat{c_1}} \times (\BZ_2^{[0]})_{c_2}]$ \hfill $\mathfrak{su}(4) \oplus \u(1)$ \\$1 + \left(6 + 10 \omega\right) a^{-2} x $ \\$\, \, \, \, + \left[2 a^4 + \left(86 + 74 \omega\right) a^{-4} - \left(7 + 10 \omega\right) \right] x^2$ \\ $\, \, \, \, + \left[a^6 + \left(479 + 506 \omega\right) a^{-6} - a^2 \left(169 + 160 \omega \right) a^{-2}\right] x^3 + \ldots$	\end{tabular} }; 
			\node[draw] (2RepZ2Z2) at (0,-4) {\begin{tabular}{l}
			    \eref{affineD4}/NK  \hfill 2-Rep$[(\BZ_2^{[0]})_{c_1} \times (\BZ_2^{[0]})_{c_2}]$ \hfill $\mathfrak{usp}(4)$ \\$1 + \left(3 + 7 \omega\right) a^{-2} x $ \\$\, \, \, \, + \left[2 a^4 + \left(51 + 39 \omega\right) a^{-4} - \left(4 + 7 \omega\right) \right] x^2$ \\ $\, \, \, \, + \left[a^6 + \left(244 + 271 \omega\right) a^{-6} - a^2 - \left(89 + 80 \omega \right) a^{-2}\right] x^3 + \ldots$	\end{tabular} };  
            \draw[->,blue] (2VecZ2Z2) to [bend left=15] node[midway, right] {\blue $(\BZ_2^{[0]})_{c_1}$} (2VecZ2Z21);
            \draw[->,blue] (2VecZ2Z21) to [bend left=15] node[midway, right] {\blue $(\BZ_2^{[0]})_{c_2}$} (2RepZ2Z2);
            \draw[->,violet] (2VecZ2Z2) to [bend right=52] node[midway, left] {\violet $(\BZ_2^{[0]})_{c_1} \times (\BZ_2^{[0]})_{c_2}$} (2RepZ2Z2);
\end{tikzpicture}
    \caption[NK web]{The affine $D_4$ quiver \eref{affineD4} with global $(\BZ_2^{[0]})_{c_1} \times (\BZ_2^{[0]})_{c_2} \equiv \langle (12) (34), (13) (24) \rangle$ zero-form symmetry and discrete gaugings thereof. The notation for arrows and boxes is as described in Figure \ref{figNNKweb}.}
    \label{figNKweb}
\end{figure}
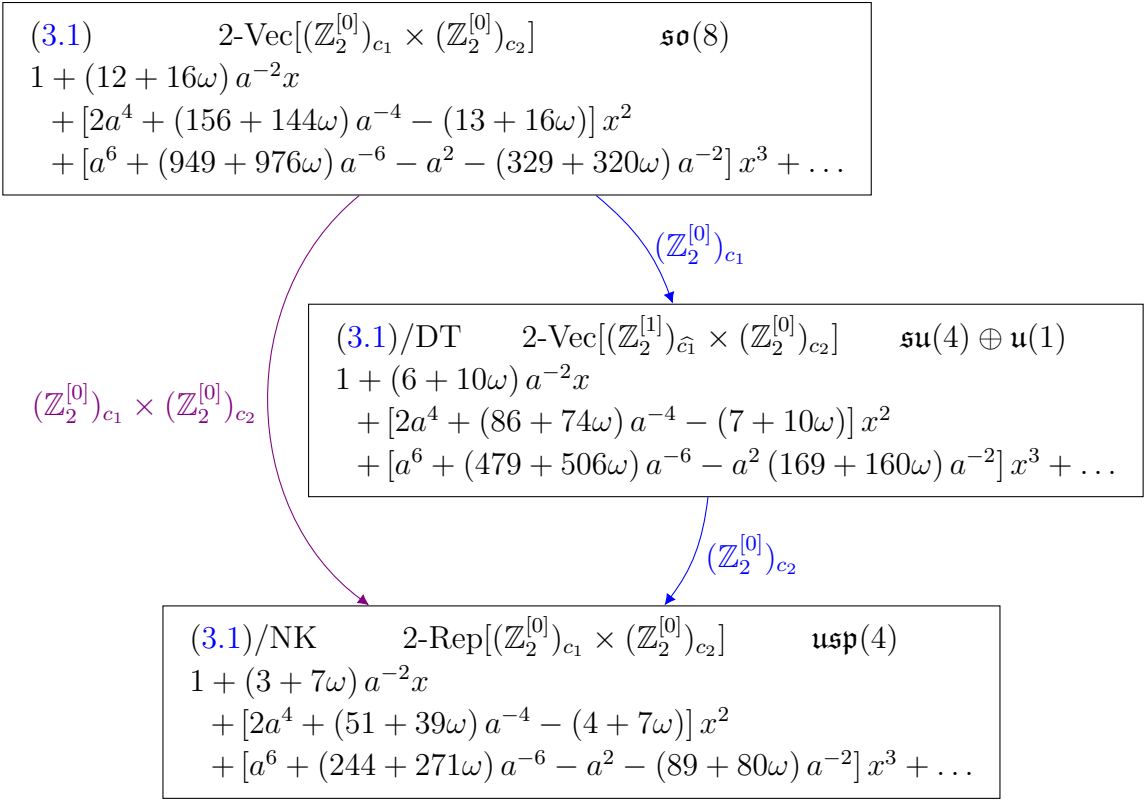
\bes{ \label{indexNKref}
&\CI_{\eqref{affineD4}}(a| \omega| c_1, c_2; x) \\
&= 1+\left[3 + 7 \omega + \left(3 + 3 \omega\right) \left(c_1 + c_2 + c_1 c_2\right) \right] a^{-2} x \\ & \quad \, \, \, \, \, + \left\{2 a^4 + \left[51 + 39 \omega + \left(35 + 35 \omega\right) \left(c_1 + c_2 + c_1 c_2\right) \right] a^{-4} \right. \\ & \quad \, \, \, \, \, \left.  - \left[4 + 7 \omega + \left(3 + 3 \omega\right) \left(c_1 + c_2 + c_1 c_2\right)\right]\right\} x^2 + \ldots~.
}
Note that the result is invariant under the interchange of $c_1$ and $c_2$.

We can now sequentially gauge the subgroups of $(\BZ_2^{[0]})_{c_1} \times (\BZ_2^{[0]})_{c_2}$, as summarised in Figure \ref{figNKweb}, where the corresponding indices are also presented. Gauging $(\BZ_2^{[0]})_{c_1}$ (resp. $(\BZ_2^{[0]})_{c_2}$) results in theory $\eref{affineD4}/{ \langle (12) (34) \rangle}$, \ie ~ \eref{affineD4} wreathed by the double transposition $\langle (12) (34) \rangle$. We can then further gauge the $(\BZ_2^{[0]})_{c_2}$ (resp. $(\BZ_2^{[0]})_{c_1}$) symmetry to obtain the wreathed theory $\eref{affineD4}/\mathrm{NK}$. This latter can be indeed implemented as
\bes{ \label{indNKwr}
\scalebox{0.94}{$
\begin{split}
&\CI_{{\eqref{affineD4}/ \mathrm{NK}}}(a| \omega; x) \\
&= \frac{1}{4} \sum_{c_1, c_2 = \pm 1} \CI_{\eqref{affineD4}}(a| \omega | c_1, c_2; x)  \\
&= 1 + \left(3 + 7 \omega\right) a^{-2} x + \left[2 a^4 + \left(51 + 39 \omega\right) a^{-4} - \left(4 + 7 \omega\right) \right] x^2 \\
& \quad \, \, \, \, \, + \left[a^6 + \left(244 + 271 \omega\right) a^{-6} - a^2 - \left(89 + 80 \omega \right) a^{-2}\right] x^3 \\
& \quad \, \, \, \, \,+ \left[3 a^8 + 40 + 48 \omega + \left(1189+1135 \omega\right) a^{-8}-a^4-\left(714+735 \omega\right) a^{-4}\right] x^4+ \ldots~,
\end{split}
$}
}
which coincides with the index of its mirror dual $\eref{USp2SO2SO2SO2SO2}/\mathrm{NK}$ upon sending $a \rightarrow a^{-1}$, compare the results in Figure \ref{figNKweb} and in \eref{tabSU2w4} for $\omega = 1$. It also follows that, gauging $(\BZ_2^{[0]})_{\omega}$ in theory ${\eqref{affineD4}}/{ \mathrm{NK}}$, the corresponding mirror theory is $\eref{USp2SO2SO2SO2SO2modZ2}/\mathrm{NK}$, whose index can be found in \eref{indexUSp2SO2SO2SO2SO2modZ2modGamma}.

\subsection{\texorpdfstring{$\BZ_3$ wreathing of the affine $D_4$ quiver}{Z3 wreathing of the affine D4 quiver}}
We can now examine the $\BZ_3 = \langle (123) \rangle$ wreathing of the affine $D_4$ quiver, for which, to the best of our knowledge, the mirror theory is not known. The $\BZ_3$ wreathed quiver is given by
\bes{ \label{IndexZ3wr}
\eref{affineD4}/{ \BZ_3} = \frac{1}{3} \left[\eref{affineD4} + \eref{affineD4}_{(123)} + \eref{affineD4}_{(132)}\right]~,
}
where the index for a cycle of length three is given by $\CI_{\eref{affineD4}_{(i_1 i_2 i_3)}} (a| \omega; x) = \eref{indZ3(123)}$. The index wreathed by $\BZ_3$, as computed using \eref{IndexZ3wr}, is therefore
\bes{ \label{indZ3wr}
\scalebox{0.95}{$
\begin{split}
&\CI_{{\eqref{affineD4}/ \BZ_3}}(a| \omega; x)  \\
&=1 + \left(6 + 8 \omega\right) a^{-2} x + \left[\left(62 + 56 \omega\right) a^{-4} - \left(7 + 8 \omega\right) \right] x^2 \\ & \quad \, \, \, \, \,+ \left[a^6 + a^2 + \left(341 + 352 \omega\right) a^{-6} - \left(123 + 120 \omega \right) a^{-2}\right] x^3 \\ & \quad \, \, \, \, \,+ \left[a^8 + 66 + 72 \omega + \left(1558 + 1536 \omega\right) a^{-8} - a^4 - \left(977 + 984 \omega \right) a^{-4}\right] x^4 + \ldots~.
\end{split}
$}
}

Let us now refine the index of \eref{affineD4} with respect to a fugacity $u$ associated with the $(\BZ_3^{[0]})_u$ symmetry generated by $(123)$, such that $u^3 = 1$.  We consider the following ansatz:
\bes{ \label{indexaffD4d}
\CI_{\eqref{affineD4}}(a| \omega| u; x) = \CI_{{\eqref{affineD4}/ \BZ_3}}(a| \omega; x)  + (u+u^2) \sum_{p = 1}^{\infty}  y_p x^p~,
}
where $y_p \equiv y_p(a, \omega)$ are unknowns that can be determined from the following system of equations:
\bes{ \label{systemZ3}
\CF_p(u=1) &= \left[\CI_{{\eqref{affineD4}}}(a| \omega; x) = \eref{indaffineD4}\right]_{x^p}~, \\
\CF_p(u=e^{\frac{2\pi i}{3}}) &= \left[\CI_{\eref{affineD4}_{(i_1 i_2 i_3)}} (a| \omega; x) 
 = \eref{indZ3(123)}\right]_{x^p}~,
}
where we define
\bes{
\CF_p(u) &= \left[\CI_{{\eqref{affineD4}/ \BZ_3}}(a| \omega; x) = \eref{indZ3wr}\right]_{x^p} + (u+u^2) y_p
}
and the notation $[\CI]_{x^p}$ denotes the coefficient of $x^p$ in the series expansion of $\CI$. The logic of \eref{systemZ3} can be explained as follows: we associate $u=1$ with the identity element of $\BZ_3$, $u=e^{\frac{2\pi i}{3}}$ with the element $(123) \in \BZ_3$ and $u=e^{\frac{4\pi i}{3}}$ with the element $(132) \in \BZ_3$; the latter two contribute the same to $\CF_p(u)$.

As a result, the index of theory \eref{affineD4} refined with the fugacity $u$ is
\bes{ \label{indexZ3}
&\CI_{\eqref{affineD4}}(a| \omega | u; x) \\
&= 1+\left[6 + 8 \omega + \left(3 + 4 \omega\right) \left(u + u^2\right) \right] a^{-2} x \\ & \quad \, \, \, \, \, + \left\{\left(u + u^2\right) a^4 + \left[62 + 56 \omega + \left(47 + 44 \omega\right) \left(u + u^2\right) \right] a^{-4} \right. \\ & \quad \, \, \, \, \, \left.- \left[7 + 8 \omega + \left(3 + 4 \omega\right) \left(u + u^2\right)\right]\right\} x^2 + \ldots~.
}
Note that, if we gauge the $(\BZ_3^{[0]})_u$ symmetry of theory \eref{affineD4}, the result is indeed that of the wreathed quiver ${\eqref{affineD4}}/{ \BZ_3}$:
\bes{ \label{indexZ3wreathing}
&\CI_{{\eqref{affineD4}/ \BZ_3}}(a| \omega; x) = \frac{1}{3} \sum_{j = 0, 1, 2} \CI_{\eqref{affineD4}}(a| \omega | u = e^{\frac{2 \pi i j}{3}}; x) = \eref{indZ3wr}~.
}
Taking the Coulomb and Higgs branch limits of the index $\eref{indexZ3wreathing} |_{\omega=1}$ as described in \eqref{CBHBlimits}, we obtain the same results as the ones reported in \cite[Figures 9 and 11]{Bourget:2020bxh}.
\subsection{\texorpdfstring{$\BZ_4$ wreathing of the affine $D_4$ quiver}{Z4 wreathing of the affine D4 quiver}}
Left us focus on the global zero-form $\BZ_4 \subset S_4$ symmetry of the affine $D_4$ quiver \eref{affineD4}, where we denote the elements of $\BZ_4$ as
\bes{
\BZ_4 = \left\{1, D, D^2, D^3 \right\}~, \qquad D^4=1~.
}
The theory arising from the $\BZ_4$ wreathing of \eref{affineD4} is given by
\bes{ \label{IndexZ4wr}
\eref{affineD4}/{ \BZ_4} = \frac{1}{4} \left[\eref{affineD4} + \eref{affineD4}_{(1234)} + \eref{affineD4}_{(13) (24)} + \eref{affineD4}_{(1432)}\right]~,
}
where the index of $\eref{affineD4}_{(13) (24)}$ is given by \eref{indZ2(12)(34)}, while those for both $\eref{affineD4}_{(1234)}$ and $\eref{affineD4}_{(1432)}$ are given by \eref{ind(1234)}. The index of the wreathed quiver is therefore
\bes{
\scalebox{0.99}{$
\begin{split}
&\CI_{{\eqref{affineD4}/\BZ_4}}(a| \omega; x) \\ &= 1 + \left(3 + 6 \omega\right) a^{-2} x + \left[a^4 + \left(45 + 38 \omega\right) a^{-4} - \left(4 + 6 \omega\right) \right] x^2 \\ 
& \qquad + \left[\left(241 + 256 \omega\right) a^{-6} - \left(86 + 80 \omega \right) a^{-2}\right] x^3 \\ 
& \qquad+ \left[2 a^8 + 43 + 48 \omega + \left(1150 + 1120 \omega\right) a^{-8} - \left(708 + 720 \omega\right) a^{-4}\right] x^4 + \ldots~.
\end{split}
$}
}
Note that the Higgs and Coulomb branch limits \eref{CBHBlimits} of this index are in agreement with \cite[Figure 9 and Figure 11]{Bourget:2020bxh}.

The index of the affine $D_4$ quiver \eref{affineD4} can be refined with a fugacity $d$ associated with the $(\BZ_4^{[0]})_d$ symmetry generated by $D = (1234)$, such that $d^4 = 1$. We consider the following ansatz:
\bes{ \label{indexaffD4dZ4}
\CI_{\eqref{affineD4}}(a| \omega| d; x) = \CI_{{\eqref{affineD4}/\BZ_4}}(a| \omega; x) + \left(d + d^3\right) \sum_{p = 1}^{\infty} y_{1,p} x^p  + d^2 \sum_{p = 1}^{\infty} y_{2,p} x^p~,
}
where $y_{1,p} \equiv y_{1,p}(a, \omega)$ and $y_{2,p} \equiv y_{2,p}(a, \omega)$ are unknown that can be determined by solving the following system of equations:
\bes{ \label{systemZ4}
\CF_p(d=1) &= \left[\CI_{\eqref{affineD4}}(a| \omega; x) = \eref{indaffineD4}\right]_{x^p}~, \\
\CF_p(d=e^{i \pi/2}) &= \left[\CI_{\eqref{affineD4}_{(i_1i_2i_3i_4)}}(a| \omega; x) = \eref{ind(1234)}\right]_{x^p}~,  \\
\CF_p(d=-1) &= \left[\CI_{\eqref{affineD4}_{(12)(34)}}(a| \omega; x) = \eref{indZ2(12)(34)}\right]_{x^p}~,
}
where we define
\bes{ \label{FaffD4dZ4}
\CF_p(d) = [\CI_{{\eqref{affineD4}/\BZ_4}}(a| \omega; x)]_{x^d} + \left(d + d^3\right) y_{1,p} + d^2 y_{2,p}
}
and the notation $[\CI]_{x^p}$ denotes the coefficient of $x^p$ in the series expansion of $\CI$. The logic of \eref{systemZ4} can be explained as follows. We associate the element $D^p \in \BZ_4$ to the value of the fugacity $d=e^\frac{2\pi i p}{4}$. The first line corresponds to the identity element $(d=1)$, the second line corresponds to $D$ $(d=i)$ and $D^3$ $(d=-i)$, where both contribute the same to $\CF_p(d)$, and the third line corresponds to $D^2$ $(d=-1)$. We thus obtain
\bes{ \label{indexZ4}
&\CI_{\eqref{affineD4}}(a| \omega | d; x) \\
&= 1+\left[3 + 6 \omega + \left(3 + 3 \omega\right) \left(d + d^3\right) + \left(3 + 4 \omega\right) d^2 \right] a^{-2} x + \left\{\left(1 + d^2\right) a^4 \right. \\ & \quad \, \, \, \, \, \left. + \left[45 + 38 \omega + \left(35 + 35 \omega\right) \left(d + d^3\right) + \left(41 + 36 \omega\right) d^2 \right] a^{-4} \right. \\ & \quad \, \, \, \, \, \left.- \left[4 + 6 \omega + \left(3 + 3 \omega\right) \left(d + d^3\right) + \left(3 + 4 \omega\right) d^2\right]\right\} x^2 + \ldots~.
}

\subsubsection{\texorpdfstring{Sequentially gauging subgroups of $\BZ_4$}{Gauging sequentially subgroups of Z4}}
\begin{figure}[h]
    \centering
\hspace*{-1.5cm}
\scalebox{0.7}{
\begin{tikzpicture}
			\node[draw] (2RepZ4) at (0,-5) {\begin{tabular}{l}
			     $\eref{affineD4}/\BZ_4$ \hfill 2-Rep$(\BZ_4)$ \hfill $\mathfrak{su}(3) \oplus \mathfrak{u}(1)$ \\$1 + \left(3 + 6 \omega\right) a^{-2} x $ \\$\, \, \, \, + \left[a^4 + \left(45 + 38 \omega\right) a^{-4} - \left(4 + 6 \omega\right) \right] x^2$ \\ $\, \, \, \, + \left[\left(241 + 256 \omega\right) a^{-6} - \left(86 + 80 \omega \right) a^{-2}\right] x^3 + \ldots$ \end{tabular} }; 
			\node[draw] (2VecZ2Z2) at (7.5,0) {\begin{tabular}{l}
			     \hfill \hfill $\mathfrak{su}(4)*$ \\$1 + \left(6 + 9 \omega\right) a^{-2} x $ \\$\, \, \, \, + \left[a^4 + \left(80 + 73 \omega\right) a^{-4} - \left(7 + 9 \omega\right) \right] x^2$ \\ $\, \, \, \, + \left[\left(476 + 491 \omega\right) a^{-6} - \left(166 + 160 \omega \right) a^{-2}\right] x^3 + \ldots$	\end{tabular} }; 
			\node[draw] (2RepZ2Z2) at (-7.5,0) {\begin{tabular}{l}
			     $\eref{affineD4}/\mathrm{DT}$ \hfill 2-Vec$^{\alpha}[(\BZ_2^{[1]})_{\hat{\fn}} \times (\BZ_2^{[0]})_{\fq}]$ \hfill $\mathfrak{su}(4) \oplus \mathfrak{u}(1)$ \\$1 + \left(6 + 10 \omega\right) a^{-2} x $ \\$\, \, \, \, + \left[2 a^4 + \left(86 + 74 \omega\right) a^{-4} - \left(7 + 10 \omega\right) \right] x^2$ \\ $\, \, \, \, + \left[a^6 + \left(479 + 506 \omega\right) a^{-6} - a^2 - \left(169 + 160 \omega \right) a^{-2}\right] x^3 + \ldots$	\end{tabular} }; 
			\node[draw] (2VecZ4) at (0,5) {\begin{tabular}{l}
			     \eref{affineD4} \hfill 2-Vec$(\BZ_4)$ \hfill $\so(8)$ \\$1 + \left(12 + 16 \omega\right) a^{-2} x $ \\$\, \, \, \, + \left[2 a^4 + \left(156 + 144 \omega\right) a^{-4} - \left(13 + 16 \omega\right) \right] x^2$ \\ $\, \, \, \, + \left[a^6 + \left(949 + 976 \omega\right) a^{-6} - a^2 - \left(329 + 320 \omega \right) a^{-2}\right] x^3 + \ldots$ \end{tabular} }; 
            \draw[->,blue] (2VecZ4) to [bend left=15] node[midway, right=0.2] {\blue $(\BZ_2^{[0]})_{\fq}$} (2VecZ2Z2);
            \draw[->,blue] (2VecZ2Z2) to [bend left=15] node[midway, right] {\blue $(\BZ_2^{[0]})_{\fn}$} (2RepZ4);
            \draw[->,blue] (2VecZ4) to [bend right=15] node[midway, left=0.2] {\blue $(\BZ_2^{[0]})_{\fn}$} (2RepZ2Z2);
            \draw[->,blue] (2RepZ2Z2) to [bend right=15] node[midway, left] {\blue $(\BZ_2^{[0]})_{\fq}$} (2RepZ4);
            \draw[->,violet] (2VecZ4) to node[midway, right] {\violet $(\BZ_4^{[0]})_{d}$} (2RepZ4);
\end{tikzpicture}}
    \caption[Z4 web]{$\BZ_4$ symmetry of the affine $D_4$ quiver and gauging of various subgroups. Each arrow with label $F$ connecting two boxes denotes the gauging of the zero-form symmetry $F$. In each box, which is associated with a theory arising from the affine $D_4$ quiver via discrete gauging, we report the corresponding symmetry category on the left, its flavour algebra on the right and, below, the index refined with the fugacities $a$ and $\omega$ (we set the fugacity $d$ to unity and the variables $\fq$ and $\fn$ to zero after gauging). This notation will be used throughout this section.}
    \label{figZ4web}
\end{figure}
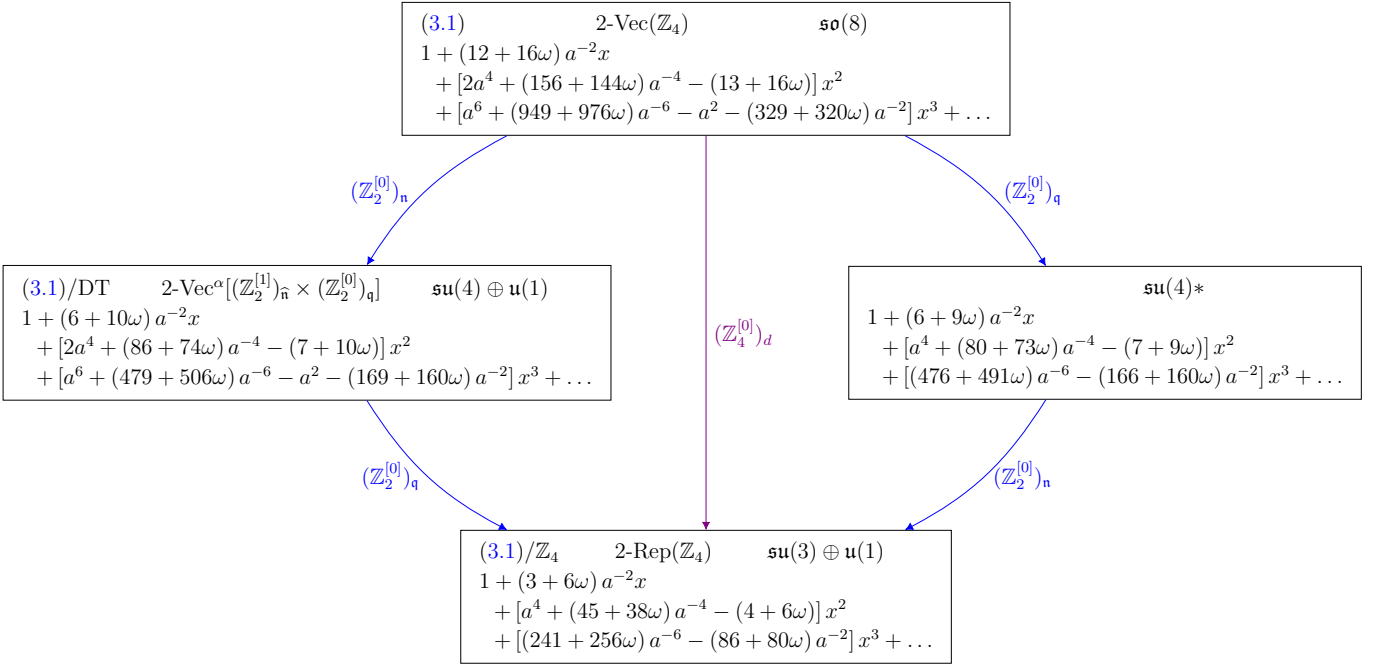

Let us now gauge the subgroups of $\BZ_4$ sequentially, in a similar fashion to that studied in \cite{Bhardwaj:2022maz} and \cite{Bartsch:2022ytj}. To make it precise, we shall denote this $\BZ_4$ by $(\BZ_4^{[0]})_d$ where the subscript $d$ is the fugacity introduced earlier, and the superscript $[0]$ emphasises that this is a zero-form symmetry. Note that $(\BZ_4^{[0]})_d$ can be viewed as a non-trivial extension between $(\BZ_2^{[0]})_{\fn}$ and $(\BZ_2^{[0]})_{\fq}$:
\bes{ \label{sequenceZ4}
0 \longrightarrow (\BZ_2^{[0]})_{\fn} \longrightarrow (\BZ_4^{[0]})_{d} \longrightarrow (\BZ_2^{[0]})_{\fq} \longrightarrow 0 ~,
}
where $(\BZ_2^{[0]})_{\fn} = \left\{1, D^2\right\} \equiv \langle D^2 \rangle$ is a normal subgroup of $(\BZ_4^{[0]})_d$, and in terms of cycles $D^2= (13)(24)$. Here $(\BZ_2^{[0]})_{\fq}$ is the quotient group $(\BZ_2^{[0]})_{\fq} \cong (\BZ^{[0]}_4)_d/(\BZ^{[0]}_2)_{\fn}$, which can be represented in terms of distinct cosets as $\{\langle D^2 \rangle, D\langle D^2 \rangle \}$. The subscripts $\fn$ and $\fq$ denote the variables associated with the normal subgroup and the quotient group respectively, which can be introduced by expressing the $(\BZ_4^{[0]})_d$ fugacity $d$ as
\bes{ \label{dintermsofqandn}
d = \exp \left[ {\frac{2 \pi i}{4} (\fq + 2 \fn)} \right]~, \quad \text{with} \quad \fq = 0,1~, \quad \fn = 0, 1~.
}
 Starting from the index \eref{indexZ4} of the affine $D_4$ quiver \eref{affineD4}, we see that there are three ways of gauging as depicted in Figure \ref{figZ4web}:
 \ben
 \item we can gauge directly the whole $(\BZ_4^{[0]})_d$ symmetry to arrive at the wreathed quiver $\eref{affineD4}/{ \BZ_4}$ with 2-Rep$(\BZ_4)$ symmetry,
 \item we can first gauge $(\BZ_2^{[0]})_{\fq}$ and then gauge $(\BZ_2^{[0]})_{\fn}$ to arrive at $\eref{affineD4}/{ \BZ_4}$, or
 \item we can first gauge $(\BZ_2^{[0]})_{\fn}$ and then gauge $(\BZ_2^{[0]})_{\fq}$ to arrive at $\eref{affineD4}/{ \BZ_4}$.
 \een
The relevant indices are given by\footnote{We thank Gabi Zafrir for pointing out to us the parametrisation \eref{dintermsofqandn}, which leads to these expressions in gauging.}
\bes{ \label{indseqgaugeZ4}
\CI_{\eqref{affineD4}/(\BZ_4^{[0]})_{d}}(a| \omega; x) =& \frac{1}{4} \sum_{\fn, \fq = 0, 1} \CI_{\eqref{affineD4}}(a| \omega | \fq, \fn; x)~, \\
\CI_{\eqref{affineD4}/(\BZ_2^{[0]})_{\fq}}(a| \omega | \fn; x) =& \frac{1}{2} \sum_{\fq = 0, 1} \CI_{\eqref{affineD4}}(a| \omega | \fq, \fn; x)~, \\
\CI_{\eqref{affineD4}/(\BZ_2^{[0]})_{\fn}}(a| \omega | \fq; x) =& \frac{1}{2} \sum_{\fn = 0, 1} \CI_{\eqref{affineD4}}(a| \omega | \fq, \fn; x)~,
}
where we define
\bes{
\CI_{\eqref{affineD4}}(a| \omega | \fq, \fn; x) = \CI_{\eqref{affineD4}}(a| \omega | d = e^{\frac{2 \pi i}{4} (\fq + 2 \fn)}; x)~.
}
Note that gauging the normal subgroup $(\BZ_2^{[0]})_{\fn}$ and the quotient group $(\BZ_2^{[0]})_{\fq}$ is equivalent to setting
\bes{ \label{Gaugeqn}
\begin{array}{lll}
\text{gauge $(\BZ_2^{[0]})_{\fn}$}:& \quad d + d^3 = 0~, & \quad d^2 = (-1)^{\fq}~, \\
\text{gauge $(\BZ_2^{[0]})_{\fq}$}:& \quad d + d^3 = (-1)^{\fn}~, & \quad d^2 = 0
\end{array}
}
in the index \eref{indexZ4}. The various indices can then be computed using this prescription and their explicit expressions are given in \fref{figZ4web}. Let us comment on the results we obtain.
\bi
\item Gauging the normal subgroup $(\BZ_2^{[0]})_{\fn}$ by summing $\fn$ over $0, 1$ and setting $\fq = 0$ is equivalent to the following manipulation of the indices:
\bes{ \label{gaugeZ2Q}
&\CI_{\eqref{affineD4}/(\BZ^{[0]}_2)_{\fn}}(a| \omega | \fq = 0; x)\\
&\overset{\eref{FaffD4dZ4}}{=} \frac{1}{2} \sum_{d=\pm 1} \sum_{p=0}^\infty \CF_p(d) x^p \\
&\overset{\eref{systemZ4}}{=} \frac{1}{2} \left [\CI_{{\eqref{affineD4}}}(a| \omega; x) + \CI_{{\eqref{affineD4}_{(13)(24)}}}(a| \omega; x)\right] \\
&\overset{\eref{IndexDTwr}}{=} \CI_{{\eqref{affineD4}/ \langle (13)(24)\rangle}}(a| \omega; x) =  \CI_{\eqref{affineD4}/ \mathrm{DT}}(a| \omega; x)~.
} 
We emphasise that the index of theory $\eqref{affineD4}/(\BZ_2^{[0]})_{\fn}$ depends on an extra $\BZ_2$ variable $\fq$ in comparison with that of $\eqref{affineD4}/ \mathrm{DT}$. They are precisely equal if we set $\fq=0$ in the former. On the other hand, we can also analyse the index of theory $\eqref{affineD4}/(\BZ_2^{[0]})_{\fn}$ without turning off the variable $\fq$. Due to the non-trivial extension in \eref{sequenceZ4}, gauging the normal subgroup $(\BZ_2^{[0]})_{\fn}$ of $(\BZ_4^{[0]})_{d}$ leads to a {\bf mixed anomaly} between the dual one-form symmetry $(\BZ^{[1]}_2)_{\hat{\fn}}$ and the remaining $(\BZ_2^{[0]})_{\fq}$ zero-form symmetry \cite{Tachikawa:2017gyf}.  The resulting symmetry is denoted by 2-Vec$^{\alpha}(\Gamma^{(2)}_{\hat{\fn}, \fq})$, with $\Gamma^{(2)}_{\hat{\fn}, \fq}$ the two-group $(\BZ_2^{[1]})_{\hat{\fn}} \times (\BZ_2^{[0]})_{\fq}$, where the ’t Hooft anomaly is specified by $\alpha \in H^4(\Gamma^{(2)}_{\hat{\fn}, \fq},\U(1))$, 
and is characterised by the following anomaly theory (see \eg~\cite[(2.19)]{Cordova:2017vab}, \cite[(D1),(D2)]{Kaidi:2021xfk} and \cite[(2.47)]{Bartsch:2022ytj}):
\bes{ \label{anomZ4}
&\pi \int_{M_4} B^{(2)}_{\fn} \cup A^{(1)}_{\fq} \cup A^{(1)}_{\fq} \\
=&\pi \int_{M_4} B^{(2)}_{\fn} \cup \mathrm{Bock}(A^{(1)}_{\fq}) = \pi \int_{M_4} \mathrm{Bock} (B^{(2)}_{\fn}) \cup A^{(1)}_{\fq}~,
}
where $\mathrm{Bock}$ is the obstruction to lifting $B^{(2)}_{\fn}$ from a $\BZ_2$ gauge field to a $\BZ_4$ gauge field. As discussed in \cite[Page 21]{Cordova:2017vab}, if we further gauge $(\BZ_2^{[0]})_{\fq}$, the resulting one-form global symmetry is extended to $\BZ^{[1]}_4$. Similarly, if we gauge the one-form symmetry $(\BZ_2^{[1]})_{\hat{\fn}}$, the mixed anomaly \eref{anomZ4} forces the resulting zero-form symmetry to be extended to $\BZ^{[0]}_4$.

The index of theory $\eqref{affineD4}/(\BZ_2^{[0]})_{\fn}$, with the variable $\fq$ turned on, reads
\bes{ \label{IndexmodZ2nfy}
&\CI_{\eqref{affineD4}/(\BZ^{[0]}_2)_{\fn}}(a| \omega | \fq; x) \\&= 1 + \left(6 \fy + 2 \omega + 8 \fy \omega \right) a^{-2} x \\ & \quad \, \, \, \, \, + \left[2 \fy a^4 + \left(4 + 82 \fy + 2 \omega + 72 \fy \omega\right) a^{-4} - \left(1 + 6 \fy + 2 \omega + 8 \fy \omega\right)\right] x^2 \\ & \quad \, \, \, \, \, + \left[\left(-1 + 2 \fy\right) a^6 + \left(3 + 476 \fy + 6 \omega + 500 \fy \omega\right) a^{-6} \right. \\& \, \, \, \, \, \, \, \, \, \, \, \, \, \left.- \left(-1 + 2 \fy \right) a^2 - \left(3 + 166 \fy + 160 \fy \omega\right) a^{-2}\right] x^3 + \ldots~,
}
where we define the fugacity $\fy$ as
\bes{
\fy \equiv \frac{1 + (-1)^{\fq}}{2}~.
}
We remark that, upon setting $\fq = 0$, or equivalently $\fy = 1$, the index \eref{IndexmodZ2nfy} is well-defined and reproduces the index of $\eqref{affineD4}/ \mathrm{DT}$, as shown in \eref{gaugeZ2Q}. 

On the other hand, if we set $\fq = 1$ (\ie $\fy = 0$), which amounts to turning on the background field for the $(\BZ^{[0]}_2)_\fq$ symmetry, we obtain instead
\bes{ \label{IndexmodZ2nfq1}
\scalebox{0.98}{$
\begin{split}
\CI_{\eqref{affineD4}/(\BZ^{[0]}_2)_{\fn}}(a| \omega | \fq = 1; x) = 1 &+ 2 \omega a^{-2} x + \left[\left(4 + 2 \omega\right) a^{-4} - \left(1 + 2 \omega\right)\right] x^2 \\&+ \left[a^2 + \left(3 + 6 \omega\right) a^{-6} {\red - a^6} - 3 a^{-2}\right] x^3 + \ldots~,
\end{split}
$}
}
which is not a well-defined index for the following reason. If we take the Higgs branch limit of the index as described in \eref{CBHBlimits}, \ie we read the coefficients of the terms contributing as $a^{2 p} x^p$, with $p \ge 0$, we see that the Higgs branch Hilbert series contains a negative term, which is highlighted in {\red red} in the index above. Given that such term cannot be associated with gauge invariant operators parametrising the Higgs branch, we refer to \eref{IndexmodZ2nfq1} as an {\it invalid index}. 

Nevertheless, if we further gauge the $(\BZ_2^{[0]})_{\fq}$ symmetry in theory $\eqref{affineD4}/(\BZ^{[0]}_2)_{\fn}$ to obtain $\eqref{affineD4}/(\BZ^{[0]}_2)_{\fn}/(\BZ_2^{[0]})_{\fq}$ by summing $\fq$ over $0, 1$ and dividing by $2$, or equivalently by setting $\fy = 1/2$, in \eref{IndexmodZ2nfy}, we reach the wreathed quiver $\eref{affineD4}/\BZ_4$. The latter has a $\BZ_4$ one-form symmetry. This is in accordance with the discussion around \eref{anomZ4}.

\item We may consider an attempt to gauge the quotient group $(\BZ_2^{[0]})_{\fq}$ in theory \eref{affineD4}. In this case, the corresponding index has an interesting inconsistency that deserves a detailed discussion. Explicitly, it reads
\bes{ \label{IndexmodZ2nfz}
\scalebox{0.95}{$
\begin{split}
&\CI_{\eqref{affineD4}/(\BZ^{[0]}_2)_{\fq}}(a| \omega | \fn; x) \\&= 1 + \left(6 \fz + 3 \omega + 6 \fz \omega \right) a^{-2} x \\ & \quad \, \, \, \, \, + \left[a^4 + \left(10 + 70 \fz + 3 \omega + 70 \fz \omega\right) a^{-4} - \left(1 + 6 \fz + 3 \omega + 6 \fz \omega\right)\right] x^2 \\ & \quad \, \, \, \, \, + \left[\left(6 + 470 \fz + 21 \omega + 470 \fz \omega\right) a^{-6} - \left(6 + 160 \fz + 160 \fz \omega\right) a^{-2}\right] x^3 + \ldots~,
\end{split}
$}
}
where we define the fugacity $\fz$ as
\bes{
\fz \equiv \frac{1 + (-1)^{\fn}}{2}~.
}
In contrast to \eref{IndexmodZ2nfq1}, if we set $\fn = 1$ (\ie $\fz = 0$) in \eref{IndexmodZ2nfz}, we do not obtain an invalid index anymore: 
\bes{ \label{IndexmodZ2nfn1}
\scalebox{1}{$
\begin{split}
\CI_{\eqref{affineD4}/(\BZ^{[0]}_2)_{\fq}}(a| \omega | \fn = 1; x) &= 1 + 3 \omega a^{-2} x \\
& \quad \, \, \, \, \, + \left[a^4+ \left(10 + 3 \omega\right) a^{-4} - \left(1 + 3 \omega\right)\right] x^2 \\
& \quad \, \, \, \, \, + \left[\left(6 + 21 \omega\right) a^{-6}  - 6 a^{-2}\right] x^3 + \ldots~.
\end{split}
$}
}

Let us now examine what happens to the above index if we turn off the fugacity $\fn$ by setting $\fn = 0$ (\ie~ $\fz = 1$) instead. The corresponding expression is reported in Figure \ref{figZ4web}. First of all, the terms $(6+9 \omega)$ at order $x$ implies that the moment map operators transform in the adjoint representation $\mathbf{15}$ of $\su(4)$.  Indeed, using \eref{Gaugeqn} with $\fn=0$, we see that the terms $(3+3 \omega)+ (3+4 \omega) = (6+7 \omega)$, associated with the representation $\mathbf{6} \oplus \mathbf{6} \oplus \mathbf{1}$ of $\su(4)$, at order $x$ in \eref{indexZ4}, are removed upon gauging $(\BZ_2^{[0]})_{\fq}$. This is consistent with the branching rule
\bes{ \label{br28tosu4u1}
\mathbf{28} \rightarrow (\mathbf{15})(0)  \oplus (\mathbf{6})(2) \oplus (\mathbf{6})(-2) \oplus (\mathbf{1})(0)
}
of $\so(8) \rightarrow \su(4) \oplus \u(1)$. However, upon examining the coefficient of $a^{-4} x^2$ in the index of \eref{affineD4}, this term corresponds to the irreducible representation $\mathbf{300}$ of $\so(8)$ \cite{Benvenuti:2010pq}. The branching rule of $\so(8) \rightarrow \su(4) \oplus \u(1)$ gives 
\bes{ \label{br300tosu4u1}
\mathbf{300} &\rightarrow (\mathbf{84})(0) \oplus (\mathbf{20}')(0) \oplus (\mathbf{15})(0) \oplus (\mathbf{1})(0) \oplus (\mathbf{64})(-2) \oplus (\mathbf{6})(-2) \\
& \qquad \oplus (\mathbf{64})(2) \oplus (\mathbf{6})(2) \oplus (\mathbf{20}')(-4) \oplus (\mathbf{20}')(4)~.
}
As can be seen by using the second line of \eref{Gaugeqn} with $\fn=0$, upon gauging $(\BZ_2^{[0]})_{\fq}$ in \eref{indexZ4}, the term $\left(35 + 35 \omega\right) + \left(41 + 36 \omega\right) = (76+71 \omega)$ at order $a^{-4} x^2$ is removed. However, we cannot subtract a $147$-dimensional representation out of the right-hand side of the branching rule \eref{br300tosu4u1}, since the equation
\bes{
&84 x_{\mathbf{84}} + 64 (2 x_{\mathbf{64}}) + 20 (3 x_{\mathbf{20'}}) + 15 x_{\mathbf{15}} + 6 (2 x_{\mathbf{6}}) + x_{\mathbf{1}} = 147~, \\
&\qquad \qquad \text{with}~ 0 \leq x_i \leq 1,
}
does not have a non-negative integer solution.\footnote{Even if we consider instead the equation $84 x_{\mathbf{84}} + 64 x_{\mathbf{64}} + 20 x_{\mathbf{20'}} + 15 x_{\mathbf{15}} + 6  x_{\mathbf{6}} + x_{\mathbf{1}} = 147$, with $0 \leq x_{\mathbf{84}} \leq 1, ~ 0 \leq x_{\mathbf{64}} \leq 2, ~ 0 \leq x_{\mathbf{20'}} \leq 3, ~ 0 \leq x_{\mathbf{15}} \leq 1, ~ 0 \leq x_{\mathbf{6}} \leq 2,~ 0\leq x_{\mathbf{1}} \leq 1$, there is still no non-negative integer solution.}  
We observe that this inconsistency with the branching rule is a feature whenever the quotient group involved in a non-intrivial extension is gauged. Due to the inconsistency with the latter branching rule, we put $*$ in $\su(4)*$ in Figure \ref{figZ4web} to denote this phenomenon.

It is instructive to compare this with $\eref{affineD4}/\mathrm{NNK}$, whose moment map operators also transform in $\mathbf{15}$ of $\su(4)$. Indeed, from \eref{indexNNKref}, we see that gauging $(\BZ^{[0]}_2)_{b_1}$ and $(\BZ^{[0]}_2)_{b_2}$ removes the terms $2 \times (35+35 \omega) + (18 + 17 \omega)$, corresponding to the representation $2 \times (\mathbf{64} \oplus \mathbf{6}) \oplus \mathbf{20'} \oplus \mathbf{15}$ of $\su(4)$, at order $a^{-4} x^2$. This is perfectly consistent with the above branching rule. Upon computing the difference between the two indices $\CI_{\eqref{affineD4}/(\BZ_2^{[0]})_{\fq}}(a| \omega | \fn=0; x)$ given by the rightmost box in \fref{figZ4web} and $\CI_{\eqref{affineD4}/\mathrm{NNK}}(a|\omega, x)$ given by \eref{IndexNNKwr}, we obtain
\bes{ \label{anomdiff}
&\CI_{\eqref{affineD4}/(\BZ_2^{[0]})_{\fq}}(a| \omega | \fn = 0; x) -\CI_{\eqref{affineD4}/\mathrm{NNK}}(a|\omega, x)   \\
&=  (12+16 \omega) a^{-4} x^2 + \Big[(144+138 \omega) a^{-6} -(42+40\omega) a^{-2}  \Big] x^3 + \ldots~.
}
We will see these terms also in the case of the dihedral group of order eight $(\BZ_2 \times \BZ_2) \rtimes \BZ_2$.
\ei

\subsection{\texorpdfstring{$S_3$ wreathing of the affine $D_4$ quiver}{S3 wreathing of the affine D4 quiver}}
Let us study the non-Abelian global zero-form $S_3 \subset S_4$ symmetry of the affine $D_4$ quiver, where $S_3$ is defined as
\bes{
S_3 = \left\{1, G, G^2, H, G H, G^2 H\right\}~,
}
such that
\bes{
G^3 = H^2 =1~,\quad H G H = G^2~.
}
It is convenient to represent $G$ and $H$ in terms of cycles, namely $G=(123)$ and $H=(12)$.  
Note that $S_3$ is a semidirect product $S_3 \cong \BZ_3 \rtimes \BZ_2$, where $\BZ_3$ is generated by $G$ and $\BZ_2$ is generated by $H$, and $\BZ_2$ exchanges the elements $G$ and $G^{-1} =G^2$ of $\BZ_3$ by conjugation. We therefore have the following {\it split} exact sequence:
\bes{
1 \,\, \longrightarrow \,\, \BZ_3 \,\, \longrightarrow \,\, S_3 \,\, \longrightarrow \,\, \BZ_2 \,\, \longrightarrow \,\, 1~,
}
where $\BZ_3$ is a normal subgroup of $S_3$.

 The $S_3$ wreathing of the affine $D_4$ quiver \eref{affineD4} is given by
\bes{ \label{IndexS3wr}
\scalebox{0.98}{$
\eref{affineD4}/{ S_3} = \frac{1}{6} \left[\eref{affineD4} + \eref{affineD4}_{(12)} + \eref{affineD4}_{(23)} + \eref{affineD4}_{(13)} + \eref{affineD4}_{(123)} + \eref{affineD4}_{(132)}\right]~,
$}
}
where $\eref{affineD4}_{(12)}$, $\eref{affineD4}_{(23)}$ and $\eref{affineD4}_{(13)}$ take the form \eref{quiver12} and the explicit expression for the index is given by \eref{indZ2(12)}, whereas both $\eref{affineD4}_{(123)}$ and $\eref{affineD4}_{(132)}$ take the form \eref{quiver123} and the expression for the index is given by \eref{indZ3(123)}.  The index of the affine $D_4$ quiver wreathed by $S_3$ is therefore
\bes{ \label{indaffineD4/S3}
& \CI_{{\eqref{affineD4}/ S_3}}(a| \omega; x) \\
&= 1 + \left(6 + 8 \omega\right) a^{-2} x + \left[\left(56 + 48 \omega\right) a^{-4} - \left(7 + 8 \omega\right) \right] x^2 \\ 
& \quad \, \, \, \, \,+ \left[a^2 + \left(263 + 276 \omega\right) a^{-6} - \left(101 + 100 \omega \right) a^{-2}\right] x^3 \\ 
& \quad \, \, \, \, \,+ \left[a^8 + 50 + 60 \omega +\left(1108 + 1076 \omega \right) a^{-8} - \left(691 + 696 \omega\right) a^{-4}\right] x^4 + \ldots~,
}
where the Higgs and Coulomb branch limits \eref{CBHBlimits} of this index are in agreement with \cite[Figure 9 and Figure 11]{Bourget:2020bxh}.

Let us denote by $g$ and $h$ the fugacities associated with the $(\BZ_3^{[0]})_g$ and $(\BZ_2^{[0]})_h$ zero-form symmetries generated by $G$ and $H$ respectively, such that $g^3=h^2=1$.  We can turn on $g$ and $h$ in the index of the affine $D_4$ quiver as follows. We consider the ansatz:
 \bes{ \label{indexaffD4gh}
&\CI_{\eqref{affineD4}}(a| \omega| g, h; x) \\
&= \CI_{{\eqref{affineD4}/ S_3}}(a| \omega; x)+   \left(h + g h + g^2 h\right) \sum_{p=0}^\infty y_{1,p} x^p + \left(g + g^2\right) \sum_{p=0}^\infty y_{2,p}  x^p~,
}
where the fugacities associated with the elements with the same cycle structure are grouped together. Here $y_{1,p} \equiv y_{1,p}(a, \omega)$ and $y_{2,p} \equiv y_{2,p}(a, \omega)$ are unknowns that can be determined from the following system of equations:
\bes{ \label{systemS3}
\CF_p(g=1,h=1) &= [\CI_{\eqref{affineD4}}(a| \omega; x) = \eref{indaffineD4}]_{x^p}~, \\
\CF_p(g= e^{\frac{2 \pi i}{3}}, h = 1) &= [\CI_{{\eqref{affineD4}_{(i_1i_2i_3)}}}(a| \omega; x) = \eref{indZ3(123)}]_{x^p}~,  \\
\CF_p(g= e^{\frac{2 \pi i}{3}}, h = e^{\pi i}) &= [\CI_{{\eqref{affineD4}_{(i_1i_2)}}}(a| \omega; x) = \eref{indZ2(12)}]_{x^p}~,
}
where the notation $[\CI]_{x^p}$ denotes the coefficient of $x^p$ in the series expansion of $\CI$, and
\bes{ \label{FaffD4gh}
\CF_p(g,h) = [\CI_{{\eqref{affineD4}/ S_3}}(a| \omega; x)]_{x^p} + \left(h + g h + g^2 h\right) y_{1,p} + \left(g + g^2\right) y_{2,p}~.
}
The logic of \eref{systemS3} can be explained as follows: we associate the element $G^{p_1} H^{p_2} \in S_3$ to the values of the fugacities $g= \exp\left(\frac{2\pi i p_1}{3}\right)$ and $h= \exp\left(\frac{2\pi i p_2}{2}\right)$. The first line corresponds to the identity element $(g=h=1)$ of $S_3$, the second line corresponds to $G$ $(g= e^{\frac{2 \pi i}{3}}, h = 1)$ and $G^{-1} =G^2$, where both contribute the same to $\CF_p(g,h)$, and the third line corresponds to $H$, $GH$ $(g= e^{\frac{2 \pi i}{3}}, h = e^{\pi i})$ and $G^2H$, where all of them contribute the same to $\CF_p(g,h)$. The required result is therefore
\bes{ \label{indexS3}
&\CI_{\eqref{affineD4}}(a| \omega | g, h; x) \\
&= 1+\left[6 + 8 \omega + \left(3 + 4 \omega\right) \left(g + g^2\right)\right] a^{-2} x + \left\{\left(g + g^2\right) a^4 \right. \\ & \quad \, \, \, \, \, \left. + \left[56 + 48 \omega + \left(6 + 8 \omega\right) (h + g h + g^2 h) + \left(41 + 36 \omega\right) (g + g^2) \right] a^{-4} \right. \\ & \quad \, \, \, \, \, \left.- \left[7 + 8 \omega + \left(3 + 4 \omega\right) \left(g + g^2\right)\right]\right\} x^2 + \ldots~.
}
We remark that the group elements $G$ and $H$ obey the relation $GH= HG^2$, but not $GH=HG$; in other words, $G$ and $H$ are not commuting elements. In fact, the associated fugacities $g$ and $h$ should not be refined simultaneously in the index. However, we have done so in order to keep track of the representations of operators in the index. Indeed, we will see in the following discussion that gauging $(\BZ_2^{[0]})_h$ leads to an inconsistent result.\footnote{We acknowledge the JHEP referee for this comment.}

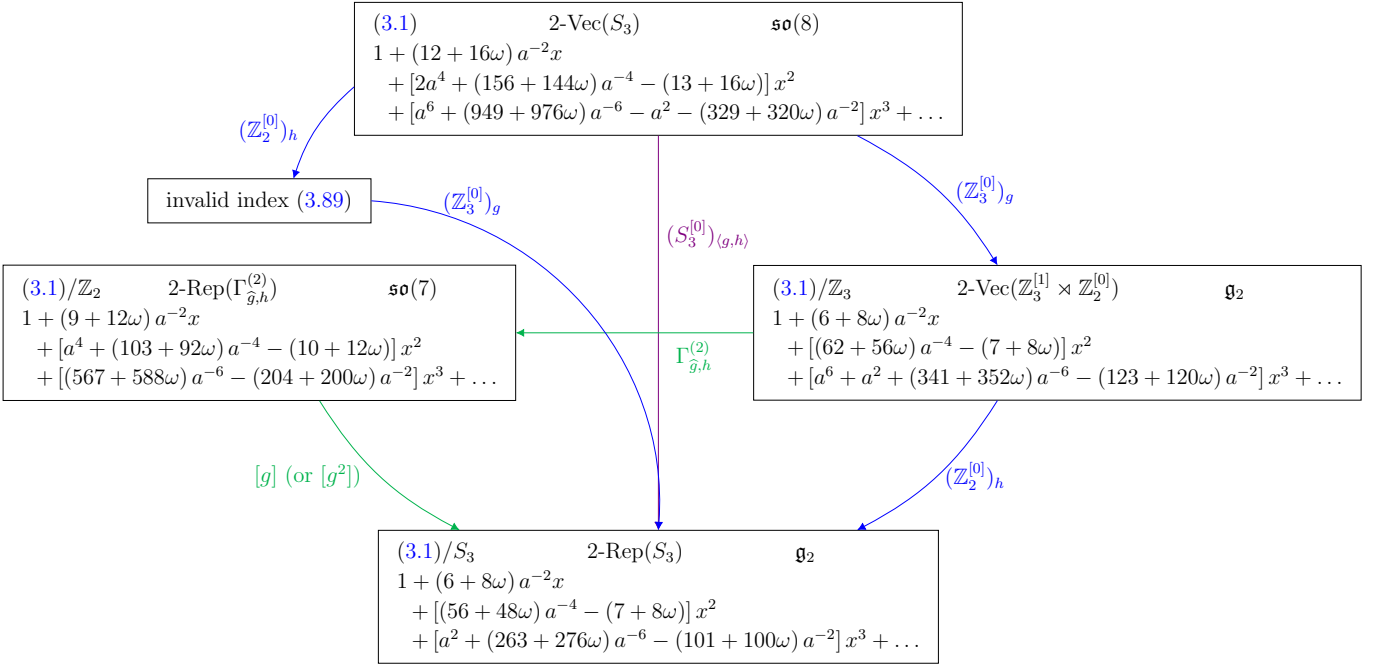
\begin{figure}[h]
    \centering
\hspace*{-1.5cm}
\scalebox{0.7}{
\begin{tikzpicture}
			\node[draw] (2RepS3) at (0,-5) {\begin{tabular}{l}
			    $\eref{affineD4}/S_3$ \hfill 2-Rep$(S_3)$ \hfill $\mathfrak{g}_2$ \\$1 + \left(6 + 8 \omega\right) a^{-2} x $ \\$\, \, \, \, + \left[\left(56 + 48 \omega\right) a^{-4} - \left(7 + 8 \omega\right) \right] x^2$ \\ $\, \, \, \, + \left[a^2 + \left(263 + 276 \omega\right) a^{-6} - \left(101 + 100 \omega \right) a^{-2}\right] x^3 + \ldots$ \end{tabular} }; 
			\node[draw] (2VecZ3Z2) at (7.5,0) {\begin{tabular}{l}
			     $\eref{affineD4}/\BZ_3$ \hfill 2-Vec$(\BZ_3^{[1]} \rtimes \BZ_2^{[0]})$ \hfill $\mathfrak{g}_2$ \\$1 + \left(6 + 8 \omega\right) a^{-2} x $ \\$\, \, \, \, + \left[\left(62 + 56 \omega\right) a^{-4} - \left(7 + 8 \omega\right) \right] x^2$ \\ $\, \, \, \, + \left[a^6 + a^2 + \left(341 + 352 \omega\right) a^{-6} - \left(123 + 120 \omega \right) a^{-2}\right] x^3 + \ldots$	\end{tabular} }; 
                \node[draw] (nv) at (-7.5,2.5) {\begin{tabular} {l}
			     \hfill invalid index \eref{IndexS3gaugeZ2}\end{tabular}};
                \node[draw] (2RepL) at (-7.5,0) {\begin{tabular}{l}
                $\eref{affineD4}/\BZ_2$ \hfill 2-Rep$(\Gamma^{(2)}_{\hat{g}, h})$ \hfill \qquad $\so(7)$
                \\$1 + \left(9 + 12 \omega\right) a^{-2} x $ \\$\, \, \, \, + \left[a^4 + \left(103 + 92 \omega\right) a^{-4} - \left(10 + 12 \omega\right) \right] x^2$ \\ $\, \, \, \, + \left[\left(567 + 588 \omega\right) a^{-6} - \left(204 + 200 \omega \right) a^{-2}\right] x^3 + \ldots$	
                \end{tabular} }; 
			\node[draw] (2VecS3) at (0,5) {\begin{tabular}{l}
			    \eref{affineD4} \hfill 2-Vec$(S_3)$ \hfill $\so(8)$ \\$1 + \left(12 + 16 \omega\right) a^{-2} x $ \\$\, \, \, \, + \left[2 a^4 + \left(156 + 144 \omega\right) a^{-4} - \left(13 + 16 \omega\right) \right] x^2$ \\ $\, \, \, \, + \left[a^6 + \left(949 + 976 \omega\right) a^{-6} - a^2 - \left(329 + 320 \omega \right) a^{-2}\right] x^3 + \ldots$ \end{tabular} }; 
            \draw[->,blue] (2VecS3) to [bend left=15] node[midway, right=0.2] {\blue $(\BZ_3^{[0]})_g$} (2VecZ3Z2);
            \draw[->,blue] (2VecZ3Z2) to [bend left=15] node[midway, right] {\blue $(\BZ_2^{[0]})_{h}$} (2RepS3);
            \draw[->,blue] (2VecS3) to [bend right=15] node[midway, left=0.2] {\blue $(\BZ_2^{[0]})_{h}$} (nv);
            \draw[->,violet] (2VecS3) to node[right, near start] {\violet $(S_3^{[0]})_{\langle g, h \rangle}$} (2RepS3);
            \draw[->,new-green] (2VecZ3Z2) to node[near start, below] {\green $\Gamma^{(2)}_{\hat{g}, h}$} (2RepL);
            \draw[->,new-green] (2RepL) to [bend right=15] node[midway, left = 0.2] {\green $[g]$ (or $[g^2]$)} (2RepS3);
            \draw[->,blue] (nv) to [bend left=45] node[at start, right=1.2] {\blue $(\BZ_3^{[0]})_{g}$} (2RepS3);
\end{tikzpicture} }
    \caption[S3 web]{$S_3$ symmetry of the affine $D_4$ quiver and gauging of various subgroups. Here $\Gamma^{(2)}_{\hat{g},h}$ denotes the two-group formed by the one-form symmetry that is dual to $(\BZ_3^{[0]})_g$ and the zero-form symmetry $(\BZ_2^{[0]})_h$.}
    \label{WebS3}
\end{figure}

We can now study the sequential gauging by various subgroups of $S_3$ in a similar way to \cite{Bhardwaj:2022maz} and \cite{Bartsch:2022mpm, Bartsch:2022ytj} as shown in \fref{WebS3}.  We comment on the results as follows.
\bi
\item Let us start with the index \eref{indexS3} and gauge $(\BZ_3^{[0]})_g$ by summing $g$ over the third roots of unity, with $h$ being set to $1$. As a result, we obtain the index of theory \eref{affineD4} wreathed by the $\BZ_3$ subgroup of $S_4$. This can be seen as follows:
\bes{ \label{gaugeZ3g}
\scalebox{0.94}{$
\begin{split}
&\CI_{\eqref{affineD4}/(\BZ_3^{[0]})_{g}}(a| \omega | h = 1; x) \\
&\overset{\eref{FaffD4gh}}{=} \frac{1}{3} \sum_{j = 0}^2 \sum_{p=0}^\infty \CF_p(g = e^{\frac{2 \pi i j}{3}}, h = 1) x^p \\
&\overset{\eref{systemS3}}{=} \frac{1}{3} \left\{\CI_{{\eqref{affineD4}}}(a| \omega; x) + \CI_{{\eqref{affineD4}_{(123)}}}(a| \omega; x) + \CI_{{\eqref{affineD4}_{(132)}}}(a| \omega; x)\right\} \\
&\overset{\eref{IndexZ3wr}}{=} \CI_{{\eqref{affineD4}/ \BZ_3}}(a| \omega; x) = \eref{indZ3wr} ~,
\end{split}
$}
}
where, in the third line, we used the fact that the expression \eref{indexaffD4gh} evaluated in $g = \exp({2 \pi i/3})$ and in $g = \exp({4 \pi i/3})$ yields the same result.  Note that in the above we set $h=1$ for the sake of convenience in comparing with $\CI_{{\eqref{affineD4}/ \BZ_3}}(a| \omega; x)$. The zero-form symmetry $(\BZ^{[0]}_2)_h$ is, in fact, still there and acts by interchanging the elements $G$ and $G^2$ of the dual $(\BZ^{[1]}_3)_{\hat{g}}$ one-form symmetry.  In other words, the dual $(\BZ^{[1]}_3)_{\hat{g}}$ one-form symmetry forms a {\it split two-group}, denoted by $\Gamma^{(2)}_{\hat{g}, h} = (\BZ^{[1]}_3)_{\hat{g}} \rtimes (\BZ^{[0]}_2)_h$, with the $(\BZ^{[0]}_2)_h$ zero-form symmetry. The corresponding symmetry is denoted by 2-Vec$(\Gamma^{(2)}_{\hat{g}, h})$.

\item Let us start with the index \eref{indexS3} and gauge $(\BZ_2^{[0]})_h$ by summing $h$ over $\pm 1$, with $g$ being set to $1$.  The result is the following index:
\bes{ \label{IndexS3gaugeZ2}
&\CI_{\eqref{affineD4}/(\BZ_2^{[0]})_h}(a| \omega | g = 1; x) = \eref{indexS3}/{(\BZ_2^{[0]})_h}\big|_{g = 1} \\&=
1 + \left(12 + 16 \omega\right) a^{-2} x + \left[2 a^4 + \left(138 + 120 \omega\right) a^{-4} - \left(13 + 16 \omega\right) \right] x^2 \\ & \quad \, \, \, \, \,+ \left[ \left(715 + 748 \omega\right) a^{-6} {\red - 2 a^6} - a^2 - \left(263 + 260 \omega \right) a^{-2}\right] x^3 + \ldots
}
This index is in fact problematic for the following reason. Upon taking the Higgs branch limit which amounts to reading off the coefficients of the terms $a^{2p} x^p$ (with $p\geq 0$), we obtain the Higgs branch Hilbert series with a negative term due to the presence of the term highlighted in red in \eref{IndexS3gaugeZ2}. We therefore refer to \eref{IndexS3gaugeZ2} as an {\it invalid index}.  We will mention another method, namely by gauging the two-group, that leads to a valid index below. An interpretation of \eref{IndexS3gaugeZ2} will be shortly provided in the next two bullet points.

Nevertheless, if we turn on $g$ in \eref{IndexS3gaugeZ2}, which is equal to $\eref{indexS3} |_{h=0}$, and sum $g$ over the cube root of unity, we recover the index wreathed by $S_3$, namely \eref{indaffineD4/S3}, as expected.

\item One way to explain the invalidity of the index \eref{IndexS3gaugeZ2} is as follows. There are three irreducible representations of the group $S_3$: (1) the trivial representation, (2) the one-dimensional sign representation, and (3) the two-dimensional standard representation. Note that, under the sign representation, the even permutations (\eg~ a cycle of length 3) get mapped to $1$ and the odd permutations (\eg~ a cycle of length 2) get mapped to $-1$. On the other hand, the standard representation acts on a non-trivial permutation cycle in a non-trivial way.  The fugacities $g$ and $h$ in the index keep track of how the operators transform under such representations; in particular, the first, second and third terms in \eref{FaffD4gh} keep track of the trivial, sign and standard representations, respectively. 

Due to the action discussed above, the sign representation transforms trivially under the $(\BZ_3^{[0]})_g$ normal subgroup of $S_3$, and it is actually a non-trivial representation of the $(\BZ_2^{[0]})_h$ quotient group. Upon gauging $(\BZ_3^{[0]})_g$, we are therefore left with the trivial and non-trivial representation of $(\BZ_2^{[0]})_h$, as expected. This explains why we obtained a sensible result in \eref{gaugeZ3g}.

On the other hand, both the sign and the standard representations transform non-trivially under  $(\BZ_2^{[0]})_h$. As a consequence, projecting only the second term in \eref{FaffD4gh}, corresponding to summing over $h=\pm 1$, does not correspond to gauging the group $(\BZ_2^{[0]})_h$. This explains the invalid index \eref{IndexS3gaugeZ2}.\footnote{We acknowledge the JHEP referee for the comment in this bullet point.}

\item Let us further comment on gauging a subgroup $H$ of the original group $K$. As pointed out in \cite[(1.4)]{Cordova:2017kue}, the global symmetry after this gauging is given by the normaliser $N(H, K)$ of the subgroup $H$ that we gauge in the original group $K$ quotiented by the subgroup $H$ that we gauge,\footnote{We thank Gabi Zafrir for pointing this out to us.} \ie~ 
\bes{ \label{normalizerquotient}
N(H,K)/H~.
}

In the case of gauging $H=(\BZ^{[0]}_2)_h$, the normaliser of $H=(\BZ^{[0]}_2)_h$ in $K=S_3$ is $N(H, K)=(\BZ^{[0]}_2)_h$, and so the quotient $N(H,K)/H$ is trivial.  One interpretation for this is that, after gauging, the whole symmetry becomes non-invertible described by 2-$\mathrm{Rep}(\Gamma^{(2)}_{\hat{g},h})$ = 2-$\mathrm{Rep}(\BZ^{[1]}_3 \rtimes \BZ_2)$, as pointed out in \cite[(4.34)]{Bhardwaj:2022maz}. The consequence of this is that, if we naively gauged $(\BZ^{[0]}_2)_h$, we would obtain the {\it invalid} index as in \eref{IndexS3gaugeZ2}. In fact, the valid index obtained by gauging the two-group will be discussed in the next bullet point.

In the case of gauging $H=(\BZ^{[0]}_3)_g$, the normaliser of $H=(\BZ^{[0]}_3)_g$ in $K=S_3$ is the original group itself, \ie~ $N(H, K)=K=S_3$. Thus,  $N(H,K)/H$ is simply the original group $S_3$ quotiented by $\BZ_3$ (which is isomorphic to $\BZ_2$). This explains \eref{gaugeZ3g}, which is the original theory \eref{affineD4} wreathed by $\BZ_3$.

\item {\bf Gauging the two-group $\Gamma^{(2)}_{\hat{g}, h}$.}  Starting from the index of $\eref{affineD4}/(\BZ^{[0]}_3)_g$ with symmetry 2-Vec$(\BZ_3^{[1]} \rtimes \BZ_2^{[0]})$ = 2-Vec$(\Gamma^{(2)}_{\hat{g},h})$, we can gauge the two-group as follows. We first gauge $(\BZ_2^{[0]})_h$ in the index of theory $\eqref{affineD4}/(\BZ_3^{[0]})_{g}$ and, subsequently, we restore all of the terms in {\it either} $[g]=\{g, gh \}$ {\it or} $[g^2]=\{g^2, g^2h \}$.  For definiteness, let us choose the former option. We then obtain the index for $\eref{affineD4}/(\BZ^{[0]}_3)_g/\Gamma^{(2)}_{\hat{g},h} \cong \eref{affineD4}/(\BZ^{[0]}_2)_h$ as follows:
\bes{ \label{ind/Z3/Gamma}
&\CI_{\eref{affineD4}/(\BZ^{[0]}_3)_g/\Gamma^{(2)}_{\hat{g},h}}(a| \omega | [g]; x)  \\
&=1+\left[6 + 8 \omega + \left(3 + 4 \omega\right) g \right] a^{-2} x \\
&\quad \, \, \, \, \, + \left\{g  a^4  + \left[56 + 48 \omega + \left(6 + 8 \omega\right) g h + \left(41 + 36 \omega\right)g  \right] a^{-4} \right. \\ 
& \quad \, \, \, \, \, \left.- \left[7 + 8 \omega + \left(3 + 4 \omega\right)g \right]\right\} x^2 + \ldots~.
}
A feature of this index is that $g$ and $g^2$ no longer appear together, since $(\BZ^{[0]}_2)_h$ that exchanges $G$ and $G^2$ has been gauged.\footnote{Upon gauging $(\BZ_2^{[0]})_h$, we obtain the dual one-form symmetry $(\BZ_2^{[1]})_{\hat{h}}$. However, the latter can no longer swap the elements $G$ and $G^{-1} = G^2$ of the zero-form symmetry $(\BZ_3^{[0]})_g$ by conjugation, due to the fact that the symmetry generator of a one-form symmetry is of codimension two (\ie~ a line in three dimensions) whereas that of the zero-form symmetry is of codimension one (\ie~ a surface in three dimensions). \label{foot:codim}}  Setting the fugacities in $[g]=\{g, gh\}$ to unity, we obtain
\bes{
\CI_{\eref{affineD4}/(\BZ^{[0]}_3)_g/\Gamma^{(2)}_{\hat{g},h}}(a| \omega | [g] = 1; x) = \CI_{{\eqref{affineD4}/ \BZ_2}}(a| \omega; x) = \eref{indZ2wr}~.
}
We indeed obtain a valid index as mentioned in the previous bullet point, and this is in agreement with \cite[(4.33)]{Bhardwaj:2022maz}.

Starting from \eref{ind/Z3/Gamma}, we can further sum $g$ over the three cube roots of unity and sum $gh$ over $\pm 1$. As a result, we obtain the index of $\eref{affineD4}/S_3$ given by \eref{indaffineD4/S3} as expected.
\ei

\subsection{\texorpdfstring{$\Dih_8$ wreathing of the affine $D_4$ quiver}{Dih8 wreathing of the affine D4 quiver}}
Let us analyse the global zero-form $\Dih_8 \subset S_4$ symmetry of the affine $D_4$ quiver, where $\Dih_8$ is the dihedral group of order eight
\bes{ \label{Dih8asZ4Z2}
\Dih_8 = \left\{1, R , R^2, R^3, S, R S, R^2 S, R^3 S \right\}~,
}
such that
\bes{ \label{Dih8conjswap}
R^4 = S^2 =1~,\quad  S R S = R^3 ~.
}
It is convenient to represent $R=(1234)$ and $S=(13)$. In this way, $\Dih_8$ is viewed a semidirect product $\Dih_8 \cong \BZ_4 \rtimes \BZ_2$, where $\BZ_4$ is generated by $R$ and $\BZ_2$ is generated by $S$, where $\BZ_2$ exchanges the elements $R$ and $R^{-1} = R^3$ of $\BZ_4$ by conjugation. As a result, we have a {\it split} exact sequence:
\bes{
1 \,\, \longrightarrow \,\, \BZ_4 \,\, \longrightarrow \,\, \Dih_8 \,\, \longrightarrow \,\, \BZ_2 \,\,  \longrightarrow \,\, 1~.
}
In the following we write the zero-form symmetries associated with the above Abelian groups as $(\BZ_4^{[0]})_r$ and $(\BZ_2^{[0]})_s$, where $r$ and $s$ are the fugacities associated with $R$ and $S$, such that $r^4=s^2=1$, respectively. 

Let us consider the theory arising from the $\Dih_8$ wreathing of the affine $D_4$ quiver \eref{affineD4}, which can be derived as
\bes{ \label{IndexDih8wr}
\eref{affineD4}/{ \Dih_8} =& \frac{1}{8} \left[\eref{affineD4} + \eref{affineD4}_{(13)} + \eref{affineD4}_{(24)} + \eref{affineD4}_{(13)(24)} + \eref{affineD4}_{(12)(34)} \right.\\& \qquad \, \, \, \, \, \, \, \,+\left. \eref{affineD4}_{(14)(23)} + \eref{affineD4}_{(1234)} + \eref{affineD4}_{(1432)}\right]~,
}
where the corresponding indices are
\bes{
\begin{array}{ll}
\CI_{\eqref{affineD4}}(a| \omega; x) = \eref{indaffineD4} ~, &\quad \CI_{{\eqref{affineD4}_{(i_1 i_2)}}}(a| \omega; x) = \eref{indZ2(12)} ~, \\  \CI_{{\eqref{affineD4}_{(i_1 i_2) (i_3 i_4)}}}(a| \omega; x) = \eref{indZ2(12)(34)} ~, & \quad \CI_{{\eqref{affineD4}_{(i_1 i_2 i_3 i_4)}}}(a| \omega; x) = \eref{ind(1234)}~.
\end{array}
}
We then obtain the index of theory \eref{affineD4} wreathed by $\Dih_8$ as follows:
\bes{ \label{indmodDih8}
&\CI_{\eqref{affineD4}/\Dih_8}(a| \omega; x) \\&= 1 + \left(3 + 6 \omega\right) a^{-2} x + \left[a^4 + \left(39 + 30 \omega\right) a^{-4} - \left(4 + 6 \omega\right) \right] x^2\\ & \quad \, \, \, \, \, + \left[\left(169 + 187 \omega\right) a^{-6} - \left(65 + 60 \omega \right) a^{-2}\right] x^3 \\ & \quad \, \, \, \, \,+ \left[ 2 a^8 + 28 + 36 \omega + \left(762 + 723 \omega\right) a^{-8} - \left(459 + 471 \omega\right) a^{-4}\right] x^4 + \ldots~,
}
whose Higgs and Coulomb branch limits \eref{CBHBlimits} are in agreement with \cite[Figure 9 and Figure 11]{Bourget:2020bxh}.

The index of the affine $D_4$ quiver can be refined with respect to the fugacities $r$ and $s$ as follows:
\bes{ \label{indexaffD4rs}
&\CI_{\eqref{affineD4}}(a| \omega| r, s; x) = \CI_{{\eqref{affineD4}/ \Dih_8}}(a| \omega; x) \\
& \qquad + \sum_{p = 1}^{\infty} \left[ y_{1,p} \left(s + r^2 s\right) + y_{2,p} \left(r^2 + r s + r^3 s\right) + y_{3,p} \left(r + r^3\right)\right] x^p~,
}
where the fugacities associated with the elements that have the same cycle structure are grouped together, and $y_{1,p} \equiv y_{1,p}(a, \omega)$, $y_{2,p} \equiv y_{2,p}(a, \omega)$ and $y_{3,p} \equiv y_{3,p}(a, \omega)$ are unknown that can be determined from the following system of equations:
\bes{ \label{systemDih8}
\scalebox{0.96}{$
\begin{split}
\CF_p(r =1, s = 1) &= \left[\CI_{\eqref{affineD4}}(a| \omega; x)\right]_{x^p}~, \\
\CF_p(r = e^{\frac{\pi i}{2}}, s = 1) &= \left[\CI_{\eqref{affineD4}_{(i_1 i_2 i_3 i_4)}}(a| \omega; x)\right]_{x^p}~, \\
\frac{1}{2} \left[\CF_p(r = 1, s = e^{\pi i}) + \CF_p(r =e^{\pi i}, s=e^{\pi i})\right] &= \left[\CI_{\eqref{affineD4}_{(i_1 i_2)}}(a| \omega; x)\right]_{x^p}~, \\
\frac{1}{3} \left[\CF_p(r = e^{\pi i}, s = 1) + 2 \times \CF_p(r = e^{\frac{\pi i}{2}}, s = e^{\pi i})\right] &= \left[\CI_{\eqref{affineD4}_{(i_1 i_2)(i_3i_4)}}(a| \omega; x)\right]_{x^p}~,
\end{split}
$}
}
with $[\CI]_{x^p}$ the coefficient of $x^p$ in the series expansion of $\CI$, and
\bes{ \label{FpDih8}
\CF_p(r,s) &= [\CI_{{\eqref{affineD4}/ \Dih_8}}(a| \omega; x)]_{x^p} \\
& \quad + y_{1,p} \left(s + r^2 s\right) + y_{2,p} \left(r^2 + r s + r^3 s\right) + y_{3,p} \left(r + r^3\right)~. 
}
Each line in \eref{systemDih8} can be explained as follows. We associate the element $R^{p_1} S^{p_2} \in \Dih_8$ to the following values of the fugacites: $r= \exp\left(\frac{2\pi i p_1}{4}\right)$, $s=\exp\left(\frac{2\pi i p_2}{2}\right)$. The first line corresponds to the identity element $(r=s=1)$ of $\Dih_8$. The second line corresponds to $R$, \ie~ $r= i , s=1$. The third line corresponds to $S$ $(r=1,\, s=-1)$ and $R^2S$ $(r=-1,\, s=-1)$. The fourth line corresponds to $R^2$ $(r=-1,\, s=1)$, $RS$ $(r=i,\, s=-1)$ and $R^3S$ $(r=-i, \, s=-1)$, where the latter two contribute the same to \eref{FpDih8}.

As a result, we obtain the index for \eref{affineD4} refined with the fugacities $r$ and $s$ as follows:
\bes{ \label{indexDih8}
\scalebox{0.91}{$
\begin{split}
&\CI_{\eqref{affineD4}}(a| \omega | r, s; x) \\
&= 1+\left[3 + 6 \omega + 
  \omega \left(s + r^2 s\right) + \left(3 + 4 \omega\right) \left(r^2 + r s + r^3 s\right) - 2 \omega \left(r + r^3\right)\right] a^{-2} x \\ & \quad \, \, \, \, \, +\left \{\left[1 + s + r^2 s + r^2 + r s + r^3 s - 2 \left(r + r^3\right)\right] a^4 + \left[39 + 30 \omega \right.\right. \\ &\quad \, \, \, \,  \left.\left.+ \left(12 + 
    9 \omega\right) \left(s + r^2 s\right) + \left(35 + 
    28 \omega\right) \left(r^2 + r s + r^3 s\right) + \left(-6 + 6 \omega\right) \left(r + r^3\right) \right] a^{-4} \right. \\ & \quad \, \, \, \, \left.- \left[4 + 6 \omega + 
  \omega \left(s + r^2 s\right) + \left(3 + 4 \omega\right) \left(r^2 + r s + r^3 s\right) - 2 \omega \left(r + r^3\right)\right]\right\} x^2 + \ldots~.
  \end{split}
$}
}
\subsubsection{\texorpdfstring{Sequentially gauging subgroups of $\Dih_8 \cong \BZ_4 \rtimes \BZ_2$}{Sequentially gauging subgroups of Dih8 = Z4xZ2}}
Starting from the index \eref{indexDih8}, we can now sequentially gauge the $(\BZ_4^{[0]})_r$ and $(\BZ_2^{[0]})_s$ symmetries associated with the fugacities $r$ and $s$. The various possibilities arising from gauging such symmetries in different order are depicted in Figure \ref{figDih8web}.
\ben
 \item Theory $\eref{affineD4}/{\Dih_8}$ can be obtained by first gauging the $(\BZ_4^{[0]})_{r}$ symmetry and, subsequently, the remaining $(\BZ_2^{[0]})_{s}$ symmetry.
  \item We can also first gauge $(\BZ_2^{[0]})_{s}$ and then gauge $(\BZ_4^{[0]})_{r}$ to arrive at theory $\eref{affineD4}/{\Dih_8}$.
  \item Observe that theory $\eref{affineD4}/(\BZ_4^{[0]})_{r}$ possesses a dual $(\BZ_4^{[1]})_{\hat{r}}$ one-form symmetry, which forms a two-group together with the $(\BZ_2^{[0]})_{s}$ symmetry. Gauging this two-group in theory $\eref{affineD4}/(\BZ_4^{[0]})_{r}$ leads to the wreathed quiver $\eref{affineD4}/\BZ_2$, as we will explain below.
 \een
\begin{figure}
\hspace*{-1.5cm}
\scalebox{0.72}{
\begin{tikzpicture} 
			\node[draw] (2RepD8) at (0,-5) {\begin{tabular}{l}
			$\eref{affineD4}/\Dih_8$    \hfill 2-Rep$(\Dih_8)$ \hfill $\mathfrak{su}(3) \oplus \u(1)$ \\$1 + \left(3 + 6 \omega\right) a^{-2} x $ \\$\, \, \, \, + \left[a^4 + \left(39 + 30 \omega\right) a^{-4} - \left(4 + 6 \omega\right) \right] x^2$ \\ $\, \, \, \, + \left[\left(169 + 187 \omega\right) a^{-6} - \left(65 + 60 \omega \right) a^{-2}\right] x^3 + \ldots$ \end{tabular} }; 
			\node[draw] (2VecZ4Z2) at (7.5,0) {\begin{tabular}{l}
			    \eref{affineD4} wreathed by NK \hfill 2-Vec$[(\BZ_4^{[1]})_{\hat{r}} \rtimes (\BZ_2^{[0]})_s]$ \hfill $\mathfrak{usp}(4)$ \\$1 + \left(3 + 7 \omega\right) a^{-2} x $ \\$\, \, \, \, + \left[2 a^4 + \left(51 + 39 \omega\right) a^{-4} - \left(4 + 7 \omega\right) \right] x^2$ \\ $\, \, \, \, + \left[a^6 + \left(244 + 271 \omega\right) a^{-6} - a^2 - \left(89 + 80 \omega \right) a^{-2}\right] x^3 + \ldots$	\end{tabular} }; 
			\node[draw] (2VecD8) at (0,5) {\begin{tabular}{l}
			    $\eref{affineD4}$ \hfill 2-Vec$(\Dih_8)$ \hfill $\so(8)$ \\$1 + \left(12 + 16 \omega\right) a^{-2} x $ \\$\, \, \, \, + \left[2 a^4 + \left(156 + 144 \omega\right) a^{-4} - \left(13 + 16 \omega\right) \right] x^2$ \\ $\, \, \, \, + \left[a^6 + \left(949 + 976 \omega\right) a^{-6} - a^2 - \left(329 + 320 \omega \right) a^{-2}\right] x^3 + \ldots$ \end{tabular} }; 
                \node[draw] (nv) at (-7.5,2.5) {\begin{tabular} {l}
			     \hfill invalid index \eref{IndexD8gaugeZ2}\end{tabular}};
                \node[draw] (2RepL) at (-7.5,0) {\begin{tabular}{l}
                $\eref{affineD4}/\BZ_2$ \hfill 2-Rep$(\Gamma^{(2)}_{\hat{r}, s})$ \hfill \qquad $\so(7)$
                \\$1 + \left(9 + 12 \omega\right) a^{-2} x $ \\$\, \, \, \, + \left[a^4 + \left(103 + 92 \omega\right) a^{-4} - \left(10 + 12 \omega\right) \right] x^2$ \\ $\, \, \, \, + \left[\left(567 + 588 \omega\right) a^{-6} - \left(204 + 200 \omega \right) a^{-2}\right] x^3 + \ldots$	
                \end{tabular} };
            \draw[->,blue] (2VecD8) to [bend right=15] node[midway, left=0.2] {\blue $(\BZ_2^{[0]})_s$} (nv);
            \draw[->,blue] (2VecZ4Z2) to [bend left=15] node[midway, right] {\blue $(\BZ_2^{[0]})_s$} (2RepD8);     
            \draw[->,blue] (2VecD8) to [bend left=15] node[midway, right=0.2] {\blue $(\BZ_4^{[0]})_r$} (2VecZ4Z2);
            \draw[->,violet] (2VecD8) to node[right, near start] {\violet $(\Dih_8^{[0]})_{\langle r, s \rangle}$} (2RepD8);
            \draw[->,new-green] (2VecZ4Z2) to node[near start, below] {\green $\Gamma^{(2)}_{\hat{r}, s}$} (2RepL);
            \draw[->,new-green] (2RepL) to [bend right=15] node[midway, left = 0.2] {\green $[r]$ or $[r^3]$} (2RepD8);
            \draw[->,blue] (nv) to [bend left=45] node[at start, right=1.2] {\blue $(\BZ_4^{[0]})_{r}$} (2RepD8);
\end{tikzpicture}} 
    \caption[Dih8 web]{The $\Dih_8 \cong \BZ_4 \rtimes \BZ_2$ symmetry of the affine $D_4$ quiver and gauging of various subgroups. Here $\Gamma^{(2)}_{\hat{r},s}$ denotes the two-group formed by the one-form symmetry that is dual to $(\BZ_4^{[0]})_r$ and the zero-form symmetry $(\BZ_2^{[0]})_s$.} \label{figDih8web}
\end{figure}

 Let us now study such possible gaugings in more detail.
\bi
\item Let us consider theory \eref{affineD4} with $\Dih_8$ zero-form symmetry, whose symmetry category is 2-Vec$(\Dih_8)$. If we gauge $(\BZ_4^{[0]})_r$, we land on theory \eref{affineD4} wreathed by the normal Klein (NK) subgroup of $S_4$.  The corresponding symmetry category is 2-Vec$(\Gamma^{(2)}_{\hat{r},s})$ where $\Gamma^{(2)}_{\hat{r},s}$ is the two-group $\Gamma^{(2)}_{\hat{r},s}$ formed by the dual $(\BZ_4^{[1]})_{\hat{r}}$ one-form symmetry and the $(\BZ_2^{[0]})_s$ zero-form symmetry. Using the index, this can be shown as follows:
\bes{ \label{gaugeZ4r}
&\CI_{\eqref{affineD4}/(\BZ_4^{[0]})_{r}}(a| \omega | s = 1; x)
\\&\overset{\eref{FpDih8}}{=} \frac{1}{4} \sum_{j = 0, \ldots, 3} \CF_p(r = e^{\frac{2 \pi i j}{4}}, s = 1) x^p
\\
&\overset{\eref{systemDih8}}{=} \frac{1}{4} \left[\CI_{\eqref{affineD4}}(a| \omega; x) + 3 \times \CI_{\eqref{affineD4}_{(i_1 i_2) (i_3 i_4)}}(a| \omega; x) \right] \\ &\overset{\eref{IndexNKwr}}{=} \CI_{{\eqref{affineD4}/ \text{NK}}}(a| \omega; x) = \eref{indNKwr}~,
}
where we set $s = 1$ and, in the third line, we used the following properties:
\bes{ \label{solveZ4}
\scalebox{0.91}{$
\begin{split}
&\CF_p(r = e^{\frac{\pi i}{2}}, s = 1) = \CF_p(r = e^{\frac{3 \pi i}{2}}, s = 1) = \CI_{\eqref{affineD4}_{(i_1 i_2 i_3 i_4)}}(a| \omega; x)~, \\
&\CF_p(r = e^{\pi i}, s = 1) \overset{\eref{systemDih8}}{=} 3 \times \CI_{\eqref{affineD4}_{(i_1 i_2) (i_3 i_4)}}(a| \omega; x) - 2 \times \CI_{\eqref{affineD4}_{(i_1 i_2 i_3 i_4)}}(a| \omega; x) ~.
\end{split}
$}
}
Note that, in this case, the wreathing group, namely the normal Klein subgroup of $S_4$, does not coincide with the group that we gauge, namely $(\BZ_4^{[0]})_r$.  This is due to the mixed anomaly \eref{anomZ4}, which makes the Klein four-group $\BZ_2 \times \BZ_2$ extended to $\BZ_4$.\footnote{This is completely analogous to the discussion in \cite[Footnote 20]{Cordova:2017vab} in the following way. In $(2+1)$ dimensions, we can go from the $\Spin(6)$ gauge theory, which has a $\BZ_4^{[1]}$ one-form symmetry, to the $\SO(6)/\BZ_2$ gauge theory, which has a $\BZ_4^{[0]}$ zero-form magnetic symmetry, in two steps. The first one is to gauge a $\BZ^{[1]}_2$ subgroup of $\BZ_4^{[1]}$ to obtain the $\SO(6)$ gauge theory, which has a $\BZ^{[1]}_2$ one-form symmetry and a dual $\BZ^{[0]}_2$ zero-form magnetic symmetry. The second step is to gauge $\BZ^{[1]}_2$ of the $\SO(6)$ gauge theory to obtain the $\SO(6)/\BZ_2$ gauge theory. In the latter, the original $\BZ^{[0]}_2$ zero-form magnetic symmetry of the $\SO(6)$ gauge theory, together with the dual $\BZ_2$ zero-form symmetry arising from gauging, forms a $\BZ_4$ zero-form magnetic symmetry.}

Moreover, if we turn on the fugacity $s$, we can further gauge the $(\BZ_2^{[0]})_s$ symmetry and we reach theory $\eref{affineD4}/\Dih_8$.
\item On the other hand, if we gauge $(\BZ_2^{[0]})_s$ first in \eref{indexDih8}, we obtain an index, whose explicit expression reads
\bes{ \label{IndexD8gaugeZ2}
&\CI_{\eqref{affineD4}/(\BZ_2^{[0]})_{s}}(a| \omega | r = 1; x) \\&=1 + \left(6 + 6 \omega\right) a^{-2} x + \left[\left(62 + 70 \omega\right) a^{-4} {\red - 2 a^4} - \left(7 + 6 \omega\right) \right] x^2 \\ & \quad \, \, \, \, \,+ \left[3 a^2 + \left(467 + 446 \omega\right) a^{-6} {\red - 3 a^6} - \left(157 + 160 \omega \right) a^{-2}\right] x^3 + \ldots~.
}
If we take the Higgs branch limit of the index as described in \eref{CBHBlimits}, \ie we read the coefficients of the terms contributing as $a^{2 p} x^p$, with $p \ge 0$, we see that the Higgs branch Hilbert series contains negative terms, which are highlighted in {\red red} in the index above. Given that such terms cannot be associated with gauge invariant operators parametrising the Higgs branch, we refer to \eref{IndexD8gaugeZ2} as an {\it invalid index}. 

The reason for this invalid index is similar to that described around \eref{normalizerquotient}. The normaliser of $H= (\BZ^{[0]}_2)_s = \langle S \rangle$ in $K= \Dih_8 \cong (\BZ^{[0]}_4)_r \rtimes (\BZ^{[0]}_2)_s  = \langle R, S \rangle$ is the Klein four-group $N(H,K) = \{1, S, R^2, R^2 S \}$. Therefore, after gauging $H$, we are left with the $\BZ_2$ global symmetry described by \cite[(1.4)]{Cordova:2017kue}
\bes{ \label{normalizergaugeS} N(H,K)/H \cong \{1 , R^2 \}~.}
Upon gauging the symmetry associated with $s$ in \eref{indexDih8} by summing it over $\pm 1$ and dividing by two, we are left with the fugacities $r$ and $r^3$, but these correspond to the $\BZ_4$ symmetry which is not really there after gauging $H$. Upon setting them to unity, we then obtain the invalid index. For example, the term $-2a^4 x^2$, highlighted in red in \eref{IndexD8gaugeZ2}, originates from $[(1+r^2)-2(r+r^3)] a^4 x^2$ in \eref{indexDih8} with $r$ set to $1$.

Another point of view on this is as follows. Gauging the non-normal subgroup $(\BZ^{[0]}_2)_s$ in $\Dih_8$ makes the whole symmetry become non-invertible, described by
\bes{ \label{2RepZ4Z2}
\text{2-}\mathrm{Rep}(\BZ_4^{[1]} \rtimes \BZ_2) \cong \text{2-}\mathrm{Rep}[(\BZ_2^{[1]} \times \BZ_2^{[1]}) \rtimes \BZ_2]  \cong \text{2-} \mathrm{Rep}(\Gamma^{(2)}_{\hat{r}, s})~,
}
as pointed out in \cite[(5.83)]{Bhardwaj:2022maz} and \cite[(3.64)]{Bartsch:2022ytj}.

The invalid index, due to gauging $H$ in such a naive manner, as discussed above, reflects the presence of this non-invertible symmetry. We will shortly describe the procedure of getting a valid index in \eref{indgaugeZ4rGammawithr}.


Nevertheless, if we turn on the fugacity $r$ in \eref{IndexD8gaugeZ2} and gauge the corresponding $(\BZ_4^{[0]})_r$ symmetry, we reach the wreathed quiver $\eref{affineD4}/\Dih_8$.
\item {\bf Gauging the two-group $\Gamma^{(2)}_{\hat{r}, s}$.} Starting from theory $\eref{affineD4}/(\BZ_4^{[0]})_r$, whose symmetry category is 2-Vec$(\Gamma^{(2)}_{\hat{r},s})$, we can reach the theory associated with the 2-Rep$(\Gamma^{(2)}_{\hat{r},s})$ symmetry category by gauging the two-group. This can be achieved at the level of the index by gauging the $(\BZ_2^{[0]})_s$ symmetry in the index of $\eref{affineD4}/(\BZ_4^{[0]})_r$ first and, subsequently, restore all of the terms containing {\it either} $[r] = \{r^2, r, r s\}$ {\it or} $[r^3] = \{r^2, r^3, r^3 s\}$.\footnote{The reason for this can be understood by considering the map \eref{mapRStoX1X2}, in which case $S$ acts by swapping $X_1$ and $X_2$, see \eref{Dih8SX1X2}. Using this parametrisation, $[r]$ and $[r^3]$ coincide respectively with $[x_1] = \{x_1, x_1 s, x_1 x_2\}$ and $[x_2] = \{x_2, x_2 s, x_1 x_2\}$, where $x_1$ and $x_2$ denote the fugacities associated with $X_1$ and $X_2$. Since, in our case, $(\BZ_2^{[0]})_s$ has been gauged, the dual $(\BZ_2^{[1]})_{\hat{s}}$ one-form symmetry can no longer exchange $X_1$ and $X_2$ (see the reason in Footnote \ref{foot:codim}), meaning that we can restore all of the terms containing {\it either} $[x_1]$ {\it or} $[x_2]$, but not both, in the index.} For instance, if we choose the former option, we obtain a valid index for theory $[\eref{affineD4}/(\BZ_4^{[0]})_r]/\Gamma^{(2)}_{\hat{r},s}$, whose expression reads
\bes{ \label{indgaugeZ4rGammawithr}
&\CI_{[\eqref{affineD4}/(\BZ_4^{[0]})_{r}]/\Gamma^{(2)}_{\hat{r}s}}(a| \omega | [r] ; x) \\& = 1+\left[3 + 6 \omega + \left(3 + 4 \omega\right) \left(r^2 + r s\right) - 2 \omega r\right] a^{-2} x \\ & \quad \, \, \, \, \, +\left \{\left[1 + r^2 + r s - 2 r\right] a^4 + \left[39 + 30 \omega \right.\right. \\ & \quad \, \, \, \, \, +\left.\left. \left(35 + 
    28 \omega\right) \left(r^2 + r s\right) + \left(-6 + 6 \omega\right) r \right] a^{-4} \right. \\ & \quad \, \, \, \, \, -\left. \left[4 + 6 \omega + \left(3 + 4 \omega\right) \left(r^2 + r s\right) - 2 \omega r\right]\right\} x^2 + \ldots~.
}
Upon setting $[r] = \{r^2, r, r s\}$ to unity, we obtain the index of \eref{affineD4} wreathed by $\BZ_2$:
\bes{ \label{gaugeZ4andthentwogroup}
\CI_{[\eqref{affineD4}/(\BZ_4^{[0]})_{r}]/\Gamma^{(2)}_{\hat{r}s}}(a| \omega | [r] = 1 ; x) = \CI_{\eqref{affineD4}/\BZ_2}(a| \omega; x) = \eref{indZ2wr}~.
}
This is in agreement with $[\eqref{affineD4}/(\BZ_4^{[0]})_{r}]/\Gamma^{(2)}_{\hat{r}s} \cong \eqref{affineD4}/\BZ_2$.

As a final step, starting from \eref{indgaugeZ4rGammawithr}, we can further sum $r s$ and $r^2$ over $\pm 1$ and sum $r$ over the four fourth roots of unity. Doing so, we recover the index of $\eref{affineD4}/\Dih_8$ given by \eref{indmodDih8}, as expected.
\ei

\subsubsection{$\Dih_8$ as a semi-direct product $(\BZ_2 \times \BZ_2) \rtimes \BZ_2$}
Instead of \eref{Dih8asZ4Z2}, we can alternatively represent the dihedral group of order eight as 
\bes{
\Dih_8 = \left\{1, X_1 S, X_1 X_2, X_2 S, S, X_1, X_1 X_2 S, X_2\right\}~,
}
with
\bes{ \label{Dih8SX1X2}
S^2 = X_1^2 = X_2^2 = 1,\quad  X_1 X_2 = X_2 X_1,\quad S X_1 S = X_2~.
}
To recover \eref{Dih8asZ4Z2}, we can set 
\bes{ \label{mapRStoX1X2}
X_1 = R S \quad \text{and} \quad X_2 = S R = R^3 S~.
}
The group $\Dih_8$ can be realised as a semi-direct product $\Dih_8 \cong (\BZ_2 \times \BZ_2) \rtimes \BZ_2$ characterised by the following {\it split} exact sequence:
\bes{
1 \, \longrightarrow \, \BZ_2 \times \BZ_2 \, \longrightarrow \, \Dih_8 \,  \longrightarrow \,\BZ_2 \, \longrightarrow \, 1~.
}
We denote the zero-form symmetry corresponding to the subgroup $\BZ_2 \times \BZ_2$ of $\Dih_8$ as $(\BZ_2^{[0]})_{x_1} \times (\BZ_2^{[0]})_{x_2}$, where each factor is generated by $X_1$ and $X_2$ respectively.  We denote by $(\BZ_2^{[0]})_{s}$ the zero-form symmetry associated with the last $\BZ_2$, generated by $S$, in the above exact sequence. As can be seen from the last equality of \eref{Dih8SX1X2}, $(\BZ_2^{[0]})_s$ swaps $(\BZ_2^{[0]})_{x_1}$ and $(\BZ_2^{[0]})_{x_2}$ by conjugation. 

Let us consider the index of theory \eref{affineD4} refined with respect to the fugacities $s$, $x_1$ and $x_2$ associated with the elements $S$, $X_1$ and $X_2$ in \eref{Dih8SX1X2} respectively. It can be derived by using the fugacity maps $x_1= r s$ and $x_2 = r^3 s$ in \eref{indexDih8}. The required result is
\bes{ \label{indsyz}
\scalebox{0.93}{$
\begin{split}
&\CI_{\eqref{affineD4}}(a| \omega | s, \{y, z\}; x) \\
&= 1+\left[3 + 6 \omega + 
  \omega \left(s + y z s\right) + \left(3 + 4 \omega\right) \left(y z + y + z\right) - 2 \omega \left(y s + z s\right)\right] a^{-2} x \\ & \quad \, \, \, \, \, + \left\{\left[1 + s + yzs + y z + y + z - 2 \left(y s + z s\right)\right] a^4 + \left[39 + 30 \omega \right.\right. \\ & \quad \, \, \, \, \, \left.+\left. \left(12 + 
    9 \omega\right) \left(s + y z s\right) + \left(35 + 
    28 \omega\right) \left(y z + y + z\right) + \left(-6 + 6 \omega\right) \left(y s + z s\right)\right] a^{-4} \right. \\ & \quad \, \, \, \, \, \left.- \left[4 + 6 \omega + 
  \omega \left(s + y z s\right) + \left(3 + 4 \omega\right) \left(y z +y + z\right) - 2 \omega \left(y s + z s\right)\right]\right\} x^2 + \ldots~,
  \end{split}
$}}
where, for the sake of conciseness, we write $y=x_1$ and $z=x_2$.  This index satisfies
\bes{ \label{swapsymm}
\CI_{\eqref{affineD4}}(a| \omega | s, \{y, z\}; x) = \CI_{\eqref{affineD4}}(a| \omega | s, \{z, y\}; x)~,
}
namely it is invariant under swapping $y \leftrightarrow z$. Upon gauging $(\BZ_2^{[0]})_{x_2}$, the resulting theory possesses a dual $(\BZ_2^{[1]})_{\hat{x_2}}$ one-form symmetry and its index is given by
\bes{
\CI_{\eqref{affineD4}/{(\BZ_2^{[0]})_{x_2}}}(a| \omega | s, \{y, \cdot \}; x) = \frac{1}{2} \sum_{z = \pm 1} \CI_{\eqref{affineD4}}(a| \omega | s, \{y, z\}; x)~,
}
where we put a dot in the curly brackets to indicate that, since $(\BZ_2^{[0]})_{x_2}$ is gauged, the corresponding fugacity disappears from the index. If we now gauge the symmetry $(\BZ_2^{[0]})_{x_1} \times (\BZ_2^{[1]})_{\hat{x_2}}$ of theory $\CT \equiv \eref{affineD4}/{(\BZ_2^{[0]})_{x_2}}$, at the level of the index we obtain a similar result to \cite[(A.12)]{Choi:2024rjm}:
\bes{
\CI_{\CT/[(\BZ_2^{[0]})_{x_1} \times (\BZ_2^{[1]})_{\hat{x_2}}]}(a| \omega | s, \{\cdot, z\}; x)  &= \frac{1}{2} \sum_{y = \pm 1} \CI_{\CT/(\BZ_2^{[1]})_{\hat{x_2}}}(a| \omega | s, \{y, z\}; x) \\ &= \frac{1}{2} \sum_{y = \pm 1} \CI_{\eqref{affineD4}}(a| \omega | s, \{y, z\}; x) \\ &\overset{\eref{swapsymm}}{=} \frac{1}{2} \sum_{y = \pm 1} \CI_{\eqref{affineD4}}(a| \omega | s, \{z, y\}; x) \\ &= \CI_{\CT}(a| \omega | s, \{z, \cdot\}; x)~,
}
where, in the first line, we used the fact that, upon gauging the $(\BZ_2^{[1]})_{\hat{x_2}}$ symmetry of theory $\CT$, the fugacity associated with the dual $(\BZ_2^{[0]})_{x_2}$ symmetry is restored in the index. As pointed out in \cite[Appendix A]{Choi:2024rjm}, this means that theory $\CT$ is self-dual under gauging its $(\BZ_2^{[0]})_{x_1} \times (\BZ_2^{[1]})_{\hat{x_2}}$ symmetry, which implies that the $(\BZ_2^{[0]})_s$ symmetry of theory \eref{affineD4} becomes a non-invertible symmetry in theory $\CT$. The corresponding non-invertible defect implements the self-duality property \cite{Choi:2021kmx, Kaidi:2021xfk} and defines the non-invertible symmetry forming the fusion two-category 2-Rep$[(\BZ_2^{[1]})_{\hat{x_1}} \times (\BZ_2^{[1]})_{\hat{x_2}} \rtimes (\BZ_2^{[0]})_{s}]$ \cite{Bhardwaj:2022maz, Bartsch:2022ytj}.\footnote{We can see the origin of the notation $(\BZ_2^{[1]})_{\hat{x_1}} \times (\BZ_2^{[1]})_{\hat{x_2}} \rtimes (\BZ_2^{[0]})_{s}$ as follows. Note that we can form a topological interface between the dual theories $\CT$ and $\CT/[(\BZ_2^{[0]})_{x_1} \times (\BZ_2^{[1]})_{\hat{x_2}}]$, where the Dirichlet boundary condition is imposed on the $(\BZ_2^{[0]})_{x_1} \times (\BZ_2^{[1]})_{\hat{x_2}}$ gauge field. Since the two theories are dual to each other, this creates a topological defect in the single theory $\CT$. In theory $\CT$ we have the symmetry $(\BZ_2^{[1]})_{\hat{x_2}}$, whereas in theory $\CT/[(\BZ_2^{[0]})_{x_1} \times (\BZ_2^{[1]})_{\hat{x_2}}]$ we have the symmetry $(\BZ_2^{[1]})_{\hat{x_1}}$. As we move across the topological interface, these two symmetries get interchanged by the action of the non-invertible symmetry that arises from $(\BZ_2^{[0]})_{s}$. This therefore explains the aforementioned notation.} Note that we have encountered this symmetry in \eref{2RepZ4Z2}.

\subsubsection*{Sequential gauging of subgroups of $\Dih_8$ via \eref{indsyz}}
\begin{figure}
    \centering
\vspace*{-1.5cm}\hspace*{-1.5cm}
\scalebox{0.7}{
\begin{tikzpicture}
			\node[draw] (2RepD8) at (0,-6.75) {\begin{tabular}{l}
		 $\eref{affineD4}/\Dih_8$    \hfill 2-Rep$(\Dih_8)$ \hfill $\mathfrak{su}(3) \oplus \u(1)$ \\$1 + \left(3 + 6 \omega\right) a^{-2} x $ \\$\, \, \, \, + \left[a^4 + \left(39 + 30 \omega\right) a^{-4} - \left(4 + 6 \omega\right) \right] x^2$ \\ $\, \, \, \, + \left[\left(169 + 187 \omega\right) a^{-6} - \left(65 + 60 \omega \right) a^{-2}\right] x^3 + \ldots$ \end{tabular} }; 
			\node[draw] (2VecZ4Z2) at (7,-2.25) {\begin{tabular}{l}
			    $\eref{affineD4}/\mathrm{NK}$ \hfill 2-Vec$\{[(\BZ_2^{[1]})_{\hat{x_1}} \times (\BZ_2^{[1]})_{\hat{x_2}}] \rtimes \BZ_2^{[0]}\}$ \hfill $\mathfrak{usp}(4)$ \\$1 + \left(3 + 7 \omega\right) a^{-2} x $ \\$\, \, \, \, + \left[2 a^4 + \left(51 + 39 \omega\right) a^{-4} - \left(4 + 7 \omega\right) \right] x^2$ \\ $\, \, \, \, + \left[a^6 + \left(244 + 271 \omega\right) a^{-6} - a^2 - \left(89 + 80 \omega \right) a^{-2}\right] x^3 + \ldots$	\end{tabular} }; 
			\node[draw] (2Vecw) at (7,2.25) {\begin{tabular}{l}
			     \hfill 2-Vec$^{\alpha}$$\{[(\BZ_2^{[1]})_{\hat{x_1}} \times (\BZ_2^{[0]})_{x_2}] \rtimes \BZ_2^{[0]}\}$ \hfill $\su(4)*$ \\$1 + \left(6 + 9 \omega\right) a^{-2} x $ \\$\, \, \, \, + \left[a^4 + \left(80 + 73 \omega\right) a^{-4} - \left(7 + 9 \omega\right) \right] x^2$ \\ $\, \, \, \, + \left[\left(476 + 491 \omega\right) a^{-6} - \left(166 + 160 \omega \right) a^{-2}\right] x^3 + \ldots$	\end{tabular} }; 
			\node[draw] (2VecD8) at (0,6.75) {\begin{tabular}{l}
			    \eref{affineD4} \hfill 2-Vec$(\Dih_8)$ \hfill $\so(8)$ \\$1 + \left(12 + 16 \omega\right) a^{-2} x $ \\$\, \, \, \, + \left[2 a^4 + \left(156 + 144 \omega\right) a^{-4} - \left(13 + 16 \omega\right) \right] x^2$ \\ $\, \, \, \, + \left[a^6 + \left(949 + 976 \omega\right) a^{-6} - a^2 - \left(329 + 320 \omega \right) a^{-2}\right] x^3 + \ldots$ \end{tabular} }; 
                \node[draw] (2RepL) at (-7,2.25) {\begin{tabular}{l}
			      \hfill $\so(8)\oplus \u(1) \oplus \u(1) \dagger$ \\$1 + \left(12 + 18 \omega\right) a^{-2} x $ \\$\, \, \, \, + \left[4 a^4 + \left(144 + 114 \omega\right) a^{-4} - \left(13 + 18 \omega\right) \right] x^2$ \\ $\, \, \, \, + \left[3 a^6 + \left(667 + 730 \omega\right) a^{-6} - 3 a^2 - \left(251 + 240 \omega \right) a^{-2}\right] x^3 + \ldots$	\end{tabular} };
			\node[draw] (2VecL) at (-7,-2.25) {\begin{tabular}{l}
			     \hfill $\su(4) \oplus \u(1)**$  \\$1 + \left(6 + 10 \omega\right) a^{-2} x $ \\$\, \, \, \, + \left[2 a^4 +\left(74 + 58 \omega\right) a^{-4}- \left(7 + 10 \omega\right) \right] x^2$ \\ $\, \, \, \, + \left[a^6 + \left(335 + 368 \omega\right) a^{-6} -a^2 - \left(127 + 120 \omega \right) a^{-2}\right] x^3 + \ldots$ \end{tabular} };
            \draw[->,blue] (2VecD8) to [bend right=15] node[midway, left=0.2] {\blue $(\BZ_2^{[0]})_s$} (2RepL);
            \draw[->,blue] (2VecD8) to [bend left=15] node[midway, right=0.2] {\blue $(\BZ_2^{[0]})_{x_1}$} (2Vecw);
            \draw[->,blue] (2RepL) to [bend right=15] node[midway, left] {\blue $(\BZ_2^{[0]})_{x_1}$} (2VecL);
            \draw[->,blue] (2Vecw) to [bend left=15] node[midway, right] {\blue $(\BZ_2^{[0]})_{x_2}$} (2VecZ4Z2);
            \draw[->,blue] (2VecL) to [bend right=15] node[midway, left] {\blue $(\BZ_2^{[0]})_{x_2}$} (2RepD8);
            \draw[->,blue] (2VecZ4Z2) to [bend left=15] node[midway, right] {\blue $(\BZ_2^{[0]})_s$} (2RepD8);     
            \draw[->,blue] (2Vecw) to [bend left=5] node[near end, below] {\blue $(\BZ_2^{[0]})_s$} (2VecL);
            \draw[->,violet] (2VecD8) to [bend right=35] node[midway, left] {\violet $(\BZ_4^{[0]})_r$} (2VecZ4Z2);
\end{tikzpicture} }
    \caption[Dih8 web]{The $\Dih_8 \cong (\BZ_2 \times \BZ_2) \rtimes \BZ_2$ symmetry of the affine $D_4$ quiver and gauging of various subgroups.} 
    \label{figDih8syz}
\end{figure}
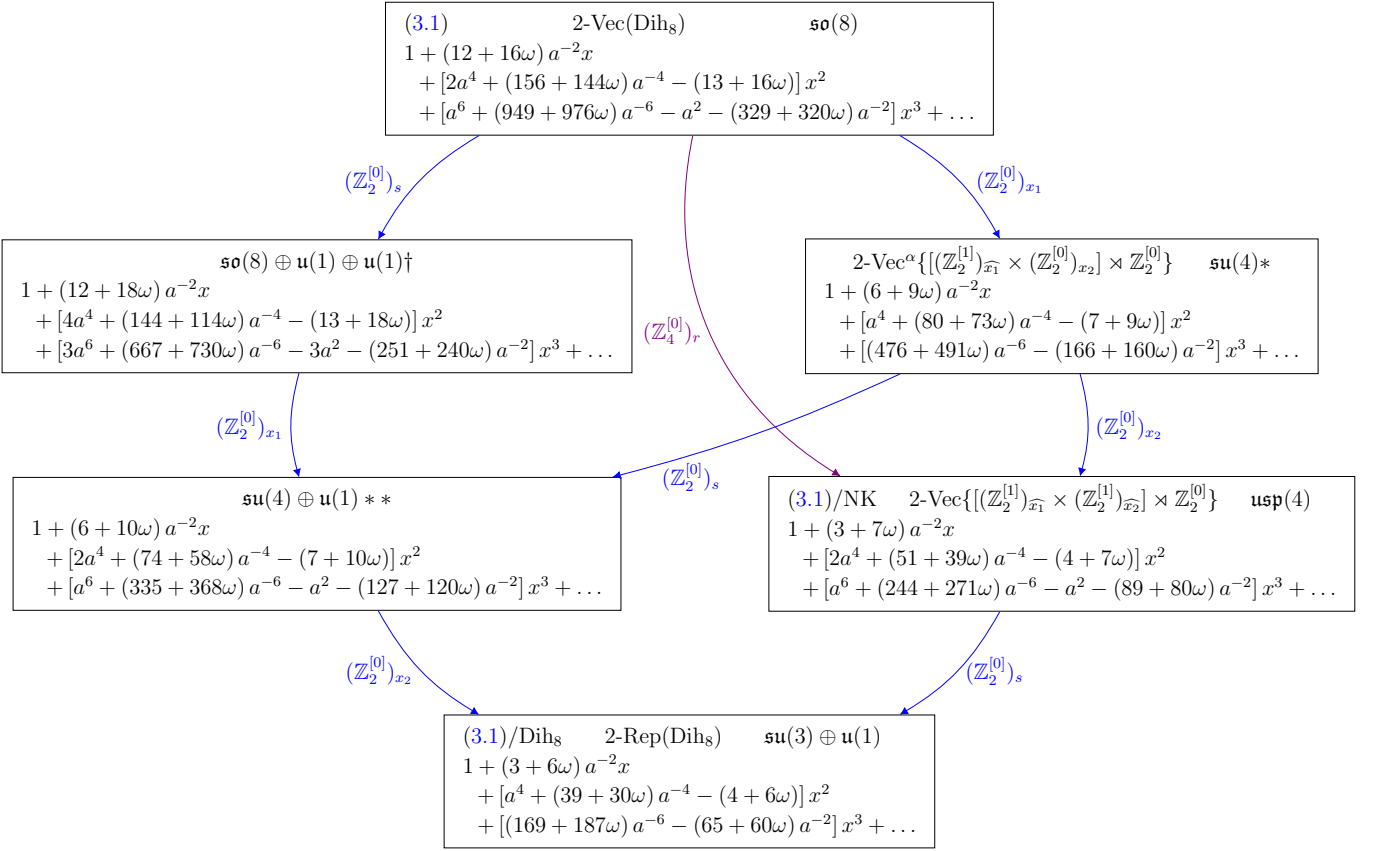

We can now study the sequential gauging of subgroups of $\Dih_8$ starting from \eref{indsyz}. The results are summarised in \fref{figDih8syz}. Let us comment on each of them as follows.  We start by considering the right half of \fref{figDih8syz}, and then discuss the left half of the diagram.
\bi
\item  Starting from \eref{indsyz}, if we gauge the whole normal subgroup $(\BZ^{[0]}_2)_{x_1} \times (\BZ^{[0]}_2)_{x_1}$ of $\Dih_8$, we obtain the index for the normal Klein wreathing of \eref{affineD4}, in a very similar fashion to \eref{gaugeZ4r} where the $(\BZ_4^{[0]})_r$ normal subgroup of $\Dih_8$ is gauged. The result is shown in the box marked by the symmetry 2-Vec$((\BZ_2^{[1]})^2 \rtimes \BZ_2^{[0]})$.  

Upon further gauging $(\BZ_2^{[0]})_s$, we obtain the quiver \eref{affineD4} wreathed by $\Dih_8$, whose symmetry is 2-Rep$(\Dih_8)$, as required.

\item On the other hand, if we start from \eref{indsyz} and gauge only $(\BZ^{[0]}_2)_{x_1}$, we have the dual one-form symmetry $(\BZ^{[1]}_2)_{\hat{x}_1}$ and the zero-form symmetry $(\BZ^{[0]}_2)_{x_2}$ with $(\BZ^{[0]}_2)_s$ trying to exchange $X_1$ in the one-form symmetry with $X_2$ in the zero-form symmetry. This suggests that $[(\BZ^{[1]}_2)_{\hat{x}_1} \times (\BZ^{[0]}_2)_{x_2}] \rtimes (\BZ^{[0]}_2)_s$ forms a two-group, denoted by $\Gamma^{(2)}_{\hat{x}_1, x_2, s}$, such that its 't Hooft anomaly is characterised by $\alpha \in H^4(\Gamma^{(2)}_{\hat{x}_1, x_2, s}, \U(1))$.  The correponding symmetry is denoted by 2-Vec$^{\alpha}$$\{[(\BZ_2^{[1]})_{\hat{x_1}} \times (\BZ_2^{[0]})_{x_2}] \rtimes \BZ_2^{[0]}\}$. The result is similar to that discussed around \eref{br28tosu4u1}--\eref{anomdiff}, and we shall not repeat the argument here. As before, we denote the global symmetry in this case by $\su(4)*$. We emphasise again that if we compute the difference between this index and that of the quiver wreathed by the non-normal Klein subgroup of $S_4$ $\eref{affineD4}/\mathrm{NNK}$ given by \eref{indmod(12),(34)}, we obtain precisely \eref{anomdiff}, which indicates the 't Hooft anomaly of the two-group symmetry in question.
\item Let us now move to the left part of \fref{figDih8syz}.  Starting from \eref{indsyz} and gauging $(\BZ^{[0]}_2)_{s}$, we expect to get an {\it invalid} index, as discussed in \eref{IndexS3gaugeZ2} and \eref{IndexD8gaugeZ2}, for the reason provided around \eref{normalizergaugeS}. Note that here we use a different parametrization in the index \eref{indsyz} from that described in \eref{indexDih8}, and so it is natural to expect to see the different invalidity. 

Indeed, we see that gauging $(\BZ^{[0]}_2)_{s}$ leads to two extra moment map operators associated with the term $2\omega$ at order $x$, and so the global symmetry is $\so(8) \oplus \u(1) \oplus \u(1)$. This is because the terms with negative sign at order $x$ in \eref{indsyz} are removed. This seems to be physically unreasonable since the global symmetry gets larger than the original theory \eref{affineD4}, which is $\so(8)$, after gauging $(\BZ^{[0]}_2)_{s}$. For this reason, we put $\dagger$ to denote this feature. The invalid index, obtained by gauing $(\BZ^{[0]}_2)_s$ naively in this way, reflects the presence of the non-invertible symmetry \eref{2RepZ4Z2}.  In fact, we remind the readers that the valid index for $\text{2-Rep}[(\BZ_2^{[1]})^2 \rtimes \BZ_2^{[0]}]$ has already been obtained in \eref{indgaugeZ4rGammawithr} by means of gauging the appropriate two-group.

Nevertheless, we can press on and further gauge $(\BZ_2^{[0]})_{x_1}$. This gets rid of the term $(6+6\omega)$ at order $x$.  This amounts to removing a 12-dimensional representation from the right-hand side of the branching rule \eref{br28tosu4u1}. This can indeed be achieved by removing the terms $(\mathbf{6})(2) \oplus (\mathbf{6})(-2)$.  Furthermore, we need to remove the term $(156 + 144 \omega) - (74 + 58 \omega) = 82+86\omega$ at order $a^{-4}x^2$. However, we cannot remove a 168-dimensional representation of $\su(4) \oplus \u(1)$ out of the irreducible representations on the right-hand side of \eref{br300tosu4u1}, since the equation
\bes{ \label{inconsistencysu4u1}
&84 x_{\mathbf{84}} + 64 (2 x_{\mathbf{64}}) + 20 (3 x_{\mathbf{20'}}) + 15 x_{\mathbf{15}} + 6 (2 x_{\mathbf{6}}) + x_{\mathbf{1}} = 168~, \\
&\qquad \qquad \text{with}~ 0 \leq x_i \leq 1,
}
does not have a non-negative integer solution.\footnote{We can consider instead the equation $84 x_{\mathbf{84}} + 64 x_{\mathbf{64}} + 20 x_{\mathbf{20'}} + 15 x_{\mathbf{15}} + 6  x_{\mathbf{6}} + x_{\mathbf{1}} = 168$, with $0 \leq x_{\mathbf{84}} \leq 1, ~ 0 \leq x_{\mathbf{64}} \leq 2, ~ 0 \leq x_{\mathbf{20'}} \leq 3, ~ 0 \leq x_{\mathbf{15}} \leq 1, ~ 0 \leq x_{\mathbf{6}} \leq 2,~ 0\leq x_{\mathbf{1}} \leq 1$, and find two non-negative integer solutions: (i) $x_{\mathbf{64}}=x_{\mathbf{20'}}=2$ and the others $x_i=0$, and (ii) $x_{\mathbf{84}}=x_{\mathbf{64}}=x_{\mathbf{20'}}=1$ and the others $x_i=0$. In this case, we need to take the $\u(1)$ charge into account, since the symmetry of the resulting theory is $\su(4) \oplus \u(1)$. In solution (ii), there is no reason why we should remove one of the $\mathbf{64}(\pm 2)$ representations but not the others. In solution (i), there is no reason why we should remove two of the $\mathbf{20'}$ representations but not the other.}

Due to the inconsistency with this branching rule, we thus denote the global symmetry of the gauged theory by $\su(4) \oplus \u(1)**$. 

Moreover, the difference between the index of \eref{affineD4} wreathed by the double transposition subgroup of $S_4$, given by \eref{indZ2wrDT}, and the index in the box labelled by $\su(4)\oplus\u(1)**$ is precisely equal to \eref{anomdiff}.  This seems to indicate an 't Hooft anomaly of the two-group formed by the two dual symmetries $(\BZ_2^{[1]})_{\hat{x}_1}$, $(\BZ_2^{[1]})_{\hat{s}}$, and the zero-form symmetry $(\BZ_2^{[0]})_{x_2}$.

Finally, if we further gauge $(\BZ_2^{[0]})_{x_2}$, we arrive at $\eref{affineD4}/\Dih_8$ as required.
\ei

\subsubsection{Gauging the centre $\langle R^2 \rangle$ of $\Dih_8$ and the Type III anomaly}
The dihedral group $\Dih_8$ of order eight has a centre $Z(\Dih_8)$ generated by $R^2$. In other words, $Z(\Dih_8)=\langle R^2 \rangle \cong \BZ_2$.
The quotient of $\Dih_8$ by $Z(\Dih_8)$ is the inner automorphism group: $\mathrm{Inn}(\Dih_8) \cong \Dih_8/Z(\Dih_8) \cong \BZ_2 \times \BZ_2$ by $Z(\Dih_8)$. We can describe $\Dih_8$ as a non-trivial extension of $\mathrm{Inn}(\Dih_8)$ by $Z(\Dih_8)$, characterised by
\bes{ \label{extensionD8}
0 \, \longrightarrow \, Z(\Dih_8) \, \longrightarrow\, \Dih_8 \, \longrightarrow \, \mathrm{Inn}(\Dih_8) \, \longrightarrow \, 0~.
}
Explicitly, $\mathrm{Inn}(\Dih_8)$ is generated by $\sigma_R$ and $\sigma_S$, with the action given by $\sigma_g(a) = g ag^{-1}$. Its element can be viewed in terms of the corresponding cosets of the centre, whereby the identity element of $\mathrm{Inn}(\Dih_8)$ corresponds to $\{1, R^2\}$. We can identify $\sigma_R$ with $\{R, R^3 \}$, $\sigma_S$ with $\{S, R^2S \}$, and $\sigma_{R}\sigma_{S} =\sigma_{S}\sigma_{R}$ with $\{RS, R^3S\}$.

As pointed out in \cite{Tachikawa:2017gyf}, \cite[(5.5)]{Bhardwaj:2022maz} and \cite{Choi:2024rjm}, gauging the zero-form symmetry $Z(\Dih_8)$ leads to a mixed 't Hooft anomaly between the one-form symmetry $\BZ_2^{[1]}$ dual to $Z(\Dih_8)$, $(\BZ_2^{[0]})_{\sigma_R}$ and $(\BZ_2^{[0]})_{\sigma_S}$. Let $B^{(2)}$, $A^{(1)}_{\sigma_R}$ and $A^{(1)}_{\sigma_S}$ be the background fields for $\BZ_2^{[1]}$, $(\BZ_2^{[0]})_{\sigma_R}$ and $(\BZ_2^{[0]})_{\sigma_S}$ respectively.  The aforementioned mixed anomaly is sometimes referred to as the Type III anomaly in the literature, and is characterised by (see \eg \cite[(3.5)]{Gaiotto:2017yup}, \cite[(D3)]{Kaidi:2021xfk}, \cite[(2.55)]{Bartsch:2022ytj} and \cite[(B.7)]{Choi:2024rjm})
\bes{ \label{TypeIIIanom}
\int_{M_4} B^{(2)} \cup A^{(1)}_{\sigma_R} \cup A^{(1)}_{\sigma_S}~,
}
where $M_4$ is a four manifold whose boundary is the three dimensional spacetime in which our theory lives. In this section, we examine the manifestation of this anomaly in the index.

In fact, we have seen a similar situation for $\BZ_4$ in \eref{sequenceZ4}. In the same way as in \eref{dintermsofqandn}, we parametrise the fugacity $r$ associated with the group element $R$ by
\bes{ \label{rintermsofqandn}
r = \exp \left[ {\frac{2 \pi i}{4} (\fq + 2 \fn)} \right]~, \quad \text{with} \quad \fq = 0,1~, \quad \fn = 0, 1~,
} 
where $\fn$ is the variable associated with $Z(\Dih_8)$, and $\fq$ is associated with $(\BZ^{[0]}_2)_{\sigma_R}$.

\subsubsection*{Gauging $Z(\Dih_8)$ in the index}
Let us first gauge the centre $Z(\Dih_8)$ by substituting \eref{rintermsofqandn} into \eref{indexDih8}, summing $\fn$ over $0$ and $1$, and dividing by two. The result is
\bes{ \label{IndexmodZ2nfyDih8}
\scalebox{0.99}{$
\begin{split}
&\CI_{\eqref{affineD4}/Z(\Dih_8)}(a| \omega | \fq, s; x) \\&= \CI_{\eqref{affineD4}/(\BZ^{[0]}_2)_{\fn}}(a| \omega | \fq; x)~\text{given by}~ \eref{IndexmodZ2nfy} \\
&\quad + (2 \fy s \omega ) a^{-2} x + \left[(-12 \fy + 24 \fy s -16 \fy \omega + 18 \fy s \omega) a^{-4} + 2 \fy s a^4  - 2 \fy s \omega \right] x^2 \\
& \quad + \left[ (-144 \fy + 150 \fy s -138 \fy \omega + 168 \fy s \omega) a^{-6} \right. \\
& \quad \, \, \, \, \left.+ (42 \fy - 48 \fy s + 40 \fy \omega - 40 \fy s \omega) a^{-2}  -2 \fy s  a^2 + 2 \fy  s  a^6 \right] x^3+\ldots~,
\end{split}
$}
}
where, as before, we define the fugacity $\fy$ as
\bes{
\fy \equiv \frac{1 + (-1)^{\fq}}{2}~.
}
We emphasise that the fugacity $s$ always appears together with $\fy$ as the combination $\fy s$ in the index. Upon turning on the background field for $(\BZ_2^{[0]})_{\sigma_R}$, which amounts to setting $\fq=1$ (\ie~ $\fy=0$), the fugacity $s$ disappears completely from the index. This indicates that the operators in the $(\BZ_2^{[0]})_{\sigma_R}$-twisted sector of theory $\eqref{affineD4}/Z(\Dih_8)$ transform trivially under $(\BZ_2^{[0]})_{\sigma_s}$.


Let us report the above index with $\fq=1$ (\ie~$\fy=0$) and $s=\pm 1$:
\bes{ \label{gaugecentreq1}
&\CI_{\eqref{affineD4}/Z(\Dih_8)}(a| \omega | \fq=1, s = \pm 1; x) \\
&= 1+ 2 \omega a^{-2} x + [(4+2 \omega) a^{-4} -(1+ 2 \omega) ] x^2 \\
& \qquad + [(3+6\omega)a^{-6} -3 a^{-2} {\red - a^6}+ a^2] x^3 +\ldots~.
}
Note that this is an {\it invalid index} for the same reason as discussed above, namely the negative term highlighted in {\red red} gives rise to a negative term in the Higgs branch Hilbert series when the Higgs branch limit is taken.  

It is also interesting to consider the index \eref{IndexmodZ2nfyDih8} when $\fq=0$ (\ie $\fy=1$), namely the background field associated with $(\BZ_2^{[0]})_{\sigma_R}$ is turned off. We report the result with $s=1$ below:
\bes{ \label{gaugecentreq0}
&\CI_{\eqref{affineD4}/Z(\Dih_8)}(a| \omega | \fq=0, s=1; x) \\
&= 1+\left(6+12 \omega\right) a^{-2} x + [(98+76 \omega)a^{-4}+4 a^4-(7+12 \omega)] x^2  \\
& \qquad + [(485 + 536 \omega) a^{-6} - (175 + 160 \omega) a^2 -3  a^2 + 3  a^6] x^3 +\ldots~.
}
Here we see that the moment map operators indicate that the global symmetry is $[\su(4) \oplus \u(1)]\oplus \u(1)\oplus \u(1)$, where from \eref{IndexmodZ2nfyDih8} we see the $\su(4) \oplus \u(1)$ comes from \eref{IndexmodZ2nfy} and the last two $\u(1)$ factors come from the term $2 \fy s \omega$. 

Note that, if we start from $\CI_{\eqref{affineD4}/Z(\Dih_8)}(a| \omega | \fq=0, s; x)$ and gauge the symmetry $(\BZ_2^{[0]})_{\sigma_S}$ by summing $s$ over $\pm 1$ and dividing by $2$, we obtain precisely the index indicated in Figure \ref{figDih8syz} labelled by $\su(4) \oplus \u(1)**$. The notation $**$ indicates the inconsistency with the branching rule discussed around \eref{inconsistencysu4u1}. We encounter this situation again due to the mixed 't Hooft anomaly \eref{TypeIIIanom}.  We point out again that the difference between the index of \eref{affineD4} wreathed by the double transposition given by \eref{indZ2(12)(34)} and the index of \eqref{affineD4}$/Z(\Dih_8)/(\BZ^{[0]}_2)_{\sigma_S}$ is precisely equal to \eref{anomdiff}.  As before, this indicates the 't Hooft anomaly of the two-group formed by the one-form symmetries, that are dual to $Z(\Dih_8)$ and $(\BZ^{[0]}_2)_{\sigma_S}$, and the zero-form symmetry $(\BZ^{[0]}_2)_{\sigma_R}$.

Summing $\eref{gaugecentreq1}$ and $\eref{gaugecentreq0}$ and dividing by 2, which amounts to gauging $Z(\Dih_8)$ and $(\BZ^{[0]}_2)_{\sigma_R}$, we obtain theory \eref{affineD4} wreathed by the normal Klein subgroup of $S_4$, \ie~$\eref{affineD4}/\mathrm{NK}$, whose index is depicted in Figure \ref{figDih8syz} labelled by $\usp(4)$.

\subsubsection*{Gauging $(\BZ^{[0]}_2)_{\sigma_R}$ in the index}
On the other hand, we can consider gauging $(\BZ^{[0]}_2)_{\sigma_R}$ by substituting \eref{rintermsofqandn} into \eref{indexDih8}, summing $\fq$ over $0$ and $1$, and dividing by two. As a result, we have
\bes{ \label{gaugesigmaR}
&\CI_{\eqref{affineD4}/(\BZ^{[0]}_2)_{\sigma_R}}(a| \omega | \fn, s; x) \\
&=  1+\left[(3 - 3 s + 6 \fz s) + (8 - 4 \fz - 3 s + 8 \fz s)\omega\right] a^{-2} x \\
& \quad \, \, \, \, \,+ \left\{\left[(45 - 12 \fz - 23 s + 
   70 \fz s) + (24 + 12 \fz - 19 s + 56 \fz s)  \omega \right] a^{-4} \right. \\
& \quad \, \, \, \, \, \, \, \, \, \left. + \left[(-4 + 3 s - 6 \fz s) + (-8 + 4 \fz + 3 s - 8 \fz s)  \omega \right] + (3 - 4 \fz + 2 \fz s) a^4  \right\} x^2\\
& \quad \, \, \, \, \, \,+ \left\{\left[(103 + 132 \fz - 91 s + 
   332 \fz s) + (148 + 78 \fz - 97 s + 362 \fz s)  \omega\right] a^{-6} \right.\\
& \quad  \, \, \, \, \, \, \, \,\, \, + \left. \left[(-50 - 30 \fz + 38 s - 
   124 \fz s) + (-40 - 40 \fz + 40 s - 120 \fz s)  \omega \right]a^{-2} \right.\\
& \quad \, \, \, \, \, \, \, \, \, \, \left. +(-2 + 4 \fz - 2 \fz s)  a^2 + (2 - 4 \fz + 2 \fz s)  a^6\right\} x^3 +\ldots~,
}
where, as before, we define the fugacity $\fz$ as
\bes{
\fz \equiv \frac{1 + (-1)^{\fn}}{2}~.
}
We have the following results:
\bi
\item Upon setting $\fn=0$ (\ie~ $\fz=1$) in \eref{gaugesigmaR}, we obtain the index depicted in Figure \ref{figDih8syz} labelled by $\su(4)*$.  This is the same result as gauging theory \eref{affineD4} by $(\BZ^{[0]}_2)_{x_1}$ or $(\BZ^{[0]}_2)_{x_2}$.
\item Starting from \eref{gaugesigmaR}, further gauging $Z(\Dih_8)$ by summing $\fn$ over $0$ and $1$, and setting $s=1$, we obtain the index of $\eref{affineD4}/\mathrm{NK}$, depicted in Figure \ref{figDih8syz} labelled by $\usp(4)$. As expected, this is the same result as gauging theory \eref{affineD4} by $(\BZ^{[0]}_2)_{x_1}$ and $(\BZ^{[0]}_2)_{x_2}$.
\item However, starting from \eref{gaugesigmaR}, further gauging $(\BZ^{[0]}_2)_{\sigma_S}$ by summing $s$ over $\pm 1$, and setting $\fn=0$ (\ie~ $\fz=1$), we obtain an {\it invalid} index
\bes{
1+(3+4\omega)a^{-2} x +[(33+36 \omega)a^{-4} {\red - a^4} - (4+4\omega)] x^2+\ldots ~,
}
where the {\red red} term gives rise to a negative term in the Higgs branch Hilbert series upon taking the Higgs branch limit.
\ei

\subsection{\texorpdfstring{$A_4$ wreathing of the affine $D_4$ quiver}{A4 wreathing of the affine D4 quiver}}
We can now examine the global zero-form $A_4 \subset S_4$ symmetry of the affine $D_4$ quiver \eref{affineD4}, where $A_4$ is the alternating group on four objects defined as
\bes{
A_4 = \left\{1, Q, Q^2, M, N, M N, M Q, N Q, M Q^2, N Q^2, M N Q, M N Q^2 \right\}~.
}
with
\bes{ \label{A4def}
\scalebox{0.97}{$
Q^3 = M^2 = N^2 =1,~ Q M Q^2 = M N = N M,~ Q N Q^2 = M~,
$}
}
It can be realised as a semi-direct product $A_4 \cong (\BZ_2 \times \BZ_2) \rtimes \BZ_3$ characterised by the {\it split} exact sequence:
\bes{
1 \, \longrightarrow\, \BZ_2 \times \BZ_2 \, \longrightarrow \, A_4 \, \longrightarrow \, \BZ_3\,  \longrightarrow \, 1~.
}
We denote the zero-form symmetry associated with the normal subgroup $\BZ_2 \times \BZ_2$ by $(\BZ_2^{[0]})_m \times (\BZ_2^{[0]})_n$, where
$m$ and $n$ are the fugacities associated $M$ and $N$ such that $m^2=n^2=1$, and denote by $(\BZ^{[0]}_3)_q$ the zero-form symmetry associated with $\BZ_3$ in the exact sequence generated by $Q$. The corresponding fugacity $q$ is such that $q^3=1$.

The theory arising from the $A_4$ wreathing of the affine $D_4$ quiver \eref{affineD4} can be derived as
\bes{ \label{IndexA4wr}
\eref{affineD4}/{ A_4} =& \frac{1}{12} \left[\eref{affineD4} + \eref{affineD4}_{(12)(34)} + \eref{affineD4}_{(13)(24)} + \eref{affineD4}_{(14)(23)} \right.\\& \qquad \, \, \, \, \, \, \, \, \, \, \,+\left. \eref{affineD4}_{(123)} + \eref{affineD4}_{(132)} + \eref{affineD4}_{(134)} + \eref{affineD4}_{(243)} \right.\\& \qquad \, \, \, \, \, \, \, \, \, \, \,+\left.  \eref{affineD4}_{(234)} + \eref{affineD4}_{(124)} + \eref{affineD4}_{(142)} + \eref{affineD4}_{(143)}\right]~,
}
where $\eref{affineD4}_{(12)(34)}$, $\eref{affineD4}_{(13)(24)}$ and $\eref{affineD4}_{(14)(23)}$ take the form \eref{quiver12c34}, whereas the eight terms of type $\eref{affineD4}_{(i_1 i_2 i_3)}$ are given by \eref{quiver123}. Explicitly, the expressions for the associated indices are given by \eref{indZ2(12)(34)} and \eref{indZ3(123)} respectively, from which we obtain 
\bes{ \label{indmodA4}
\scalebox{0.96}{$
\begin{split}
&\CI_{\eqref{affineD4}/A_4}(a| \omega; x) \\ & =1 + \left(3 + 5 \omega\right) a^{-2} x + \left[\left(27 + 21 \omega\right) a^{-4} - \left(4 + 5 \omega\right) \right] x^2 \\ & \quad \, \, \, \, \, + \left[a^6 + a^2 + \left(106 + 117 \omega\right) a^{-6} - \left(43 + 40 \omega \right) a^{-2}\right] x^3 \\ & \quad \, \, \, \, \, + \left[a^8 + 18 + 24 \omega + \left(459 + 437 \omega \right) a^{-8} - a^4 - \left(278 + 285 \omega \right) a^{-4}\right] x^4+ \ldots~,
\end{split}
$}
}
whose Higgs and Coulomb branch limits \eref{CBHBlimits} reproduce the results in \cite[Figure 9 and Figure 11]{Bourget:2020bxh}.

The index of theory \eref{affineD4} can be refined with the fugacities $m$, $n$ and $q$ by considering the follwing ansatz:
\bes{ \label{indexaffD4mnq}
\scalebox{0.91}{$
\begin{split}
\CI_{\eqref{affineD4}}(a| \omega| m, n, q; x) &= \CI_{\eqref{affineD4}/A_4}(a| \omega; x) + \left(m + n + m n\right) \sum_{p = 0}^{\infty} y_{1,p} x^p \\& +\left(q + q^2 + m q + n q + m q^2 + n q^2 + m n q + m n q^2\right) \sum_{p = 0}^{\infty} y_{2,p} x^p~,
\end{split}
$}
}
which must satisfy the conditions
\bes{
\CI_{\eqref{affineD4}/ A_4}(a| \omega; x) &= \frac{1}{12} \sum_{m,n = \pm 1} \, \sum_{j = 0, 1 ,2} \CI_{\eqref{affineD4}}(a| \omega | m, n, q = e^{\frac{2 \pi i j}{3}}; x) = \eref{indmodA4} ~,  \\
\CI_{\eqref{affineD4}}(a| \omega; x) &= \CI_{\eqref{affineD4}}(a| \omega | m= n = q = 1; x) = \eref{indaffineD4}~,
}
where $y_{1,p} \equiv y_{1,p}(a, \omega)$ and $y_{2,p} \equiv y_{2,p}(a, \omega)$ are two sets of unknown variables. We remark that the fugacities associated with the elements of the same cycle structure appear with the same set of unknowns in the ansatz \eref{indexaffD4mnq}, \ie the fugacities appearing with $y_{1,p}$ correspond to the elements $(i_1 i_2) (i_3 i_4)$, whereas the fugacities appearing with $y_{2,p}$ are associated with the elements $(i_1 i_2 i_3)$. From \eref{IndexA4wr}, the unknowns $y_{1,p}$ and $y_{2,p}$ can be determined by solving the following system of equations:
\bes{  \label{systemA4}
\left[ \CI_{\eqref{affineD4}}(a| \omega; x) = \eref{indaffineD4}\right]_{x^p} =& \CF_p(m = n = q =1)~, \\
\left[ \CI_{{\eqref{affineD4}_{(i_1 i_2)(i_3 i_4)}}}(a| \omega; x) = \eref{indZ2(12)(34)}\right]_{x^p} =& \frac{1}{3} \Big[ \CF_p(m=-1, n = q=1)\\
&\quad +\CF_p(n=-1, m = q =1) \\
& \quad + \CF_p(m = n = -1, = q =1)\Big]~,  \\
\left[\CI_{{\eqref{affineD4}_{(i_1 i_2 i_3)}}}(a| \omega; x)= \eref{indZ3(123)}\right]_{x^p}  =& \frac{1}{8} \sum_{m, n \pm 1} \, \sum_{j = 1, 2} \CF_p(m, n, q=e^{\frac{2 \pi i j}{3}})~,
}
where $[\CI]_{x^p}$ is the coefficient of $x^p$ in the series expansion of $\CI$ and we define
\bes{ \label{Fmnq}
\CF_p(m, n, q) &= \left[\CI_{\eqref{affineD4}/ A_4}(a| \omega; x) = \eref{indmodA4}\right]_{x^p} + \left(m + n + m n\right) y_{1,p} \\ &+ \left(q + q^2 + m q + n q + m q^2 + n q^2 + m n q + m n q^2\right) y_{2,p}~.
}
We therefore find the following series expansion of the index of \eref{affineD4} refined with the fugacities $m$, $n$ and $q$:
\bes{ \label{indexA4}
\scalebox{0.91}{$
\begin{split}
&\CI_{\eqref{affineD4}}(a| \omega | m, n, q; x) \\
&= 1+\left[3 + 5 \omega + 
  \left(3 + \omega\right) \left(m + n + m n\right) \right. \\ & \, \, \, \, \, \, \, \, \, \, \, \, \, \, \,  \left.+ \omega \left(q + q^2 + m q + n q + m q^2 + n q^2 + m n q + m n q^2\right)\right] a^{-2} x \\ & \quad \, \, \, \, \, +\left \{\left[-2 \left(m + n + m n\right) + \left(q + q^2 + m q + n q + m q^2 + n q^2 + m n q + m n q^2\right)\right] a^4 \right. \\ & \, \, \, \, \, \, \, \, \, \, \, \, \, \, \, \left.+ \left[27 + 21 \omega + \left(11 + 17 \omega\right) \left(m + n + m n\right) \right.\right. \\ & \, \, \, \, \, \, \, \, \, \, \, \, \, \, \, \left.+\left. \left(12 + 9 \omega\right) \left(q + q^2 + m q + n q + m q^2 + n q^2 + m n q + m n q^2\right) \right] a^{-4} \right. \\ & \, \, \, \, \, \, \, \, \, \, \, \, \, \, \, \left.- \left[4 + 5 \omega + 
  \left(3 + \omega\right) \left(m + n + m n\right) \right.\right. \\ & \, \, \, \, \, \, \, \, \, \, \, \, \, \, \, \left.+\left. \omega \left(q + q^2 + m q + n q + m q^2 + n q^2 + m n q + m n q^2\right)\right]\right\} x^2 + \ldots~.
    \end{split}
$}
}
\subsubsection{\texorpdfstring{Sequentially gauging subgroups of $A_4$}{Sequentially gauging subgroups of A4}}
\begin{figure}
    \centering
\hspace*{-2.3cm}
\scalebox{0.7}{
\begin{tikzpicture}
			\node[draw] (2RepA4) at (0,-6.75) {\begin{tabular}{l}
			    $\eref{affineD4}/A_4$ \hfill 2-Rep$(A_4)$ \hfill $\mathfrak{su}(3)$ \\$1 + \left(3 + 5 \omega\right) a^{-2} x $ \\$\, \, \, \, + \left[\left(27 + 21 \omega\right) a^{-4} - \left(4 + 5 \omega\right) \right] x^2$ \\ $\, \, \, \, + \left[a^6 + a^2 + \left(106 + 117 \omega\right) a^{-6} - \left(43 + 40 \omega \right) a^{-2}\right] x^3 + \ldots$ \end{tabular} }; 
			\node[draw] (2VecZ21Z21Z3) at (8,-2.25) {\begin{tabular}{l}
			    $\eref{affineD4}/\mathrm{NK}$ \hfill 2-Vec$(\Gamma^{(2)}_{\hat{m}, \hat{n}, q})$ \hfill $\mathfrak{usp}(4)$ \\$1 + \left(3 + 7 \omega\right) a^{-2} x $ \\$\, \, \, \, + \left[2 a^4 + \left(51 + 39 \omega\right) a^{-4} - \left(4 + 7 \omega\right) \right] x^2$ \\ $\, \, \, \, + \left[a^6 + \left(244 + 271 \omega\right) a^{-6} - a^2 - \left(89 + 80 \omega \right) a^{-2}\right] x^3 + \ldots$	\end{tabular} }; 
			\node[draw] (2VecZ21Z2Z3) at (8,2.25) {\begin{tabular}{l}
			    $\eref{affineD4}/\mathrm{DT}$ \hfill 2-Vec$[(\BZ_2^{[1]})_{\hat{n}} \times (\BZ_2^{[0]})_m \rtimes (\BZ_3^{[0]})_q]$ \hfill $\su(4) \oplus \u(1)$ \\$1 + \left(6 + 10 \omega\right) a^{-2} x $ \\$\, \, \, \, + \left[2 a^4 + \left(86 + 74 \omega\right) a^{-4} - \left(7 + 10 \omega\right) \right] x^2$ \\ $\, \, \, \, + \left[a^6 + \left(479 + 506 \omega\right) a^{-6} - a^2 - \left(169 + 160 \omega \right) a^{-2}\right] x^3 + \ldots$	\end{tabular} }; 
			\node[draw] (2VecA4) at (0,6.75) {\begin{tabular}{l}
			    \eref{affineD4} \hfill 2-Vec$(A_4)$ \hfill $\so(8)$ \\$1 + \left(12 + 16 \omega\right) a^{-2} x $ \\$\, \, \, \, + \left[2 a^4 + \left(156 + 144 \omega\right) a^{-4} - \left(13 + 16 \omega\right) \right] x^2$ \\ $\, \, \, \, + \left[a^6 + \left(949 + 976 \omega\right) a^{-6} - a^2 - \left(329 + 320 \omega \right) a^{-2}\right] x^3 + \ldots$ \end{tabular} }; 
			\node[draw] (2RepL) at (-8,-2.25) {\begin{tabular}{l}
                $\eref{affineD4}/\BZ_3$ \hfill 2-Rep$(\Gamma^{(2)}_{\hat{m}, \hat{n}, q})$ \hfill \qquad $\mathfrak{g}_2$
                \\$1 + \left(6 + 8 \omega\right) a^{-2} x $ \\$\, \, \, \, + \left[\left(62 + 56 \omega\right) a^{-4} - \left(7 + 8 \omega\right) \right] x^2$ \\ $\, \, \, \, + \left[a^6 + a^2 + \left(341 + 352 \omega\right) a^{-6} - \left(123 + 120 \omega \right) a^{-2}\right] x^3 + \ldots$	
                \end{tabular} }; 
                \node[draw] (nv1) at (-8,3.25) {\begin{tabular} {l}
			     \hfill non-valid index \eref{IndexA4gaugeZ3}\end{tabular}};
                \node[draw] (nv2) at (-8,1) {\begin{tabular} {l}
			     \hfill non-valid index \eref{IndexA4gaugeZ3gaugeZ2}\end{tabular}};
            \draw[->,blue] (2VecA4) to [bend right=15] node[midway, left=0.2] {\blue $(\BZ_3^{[0]})_q$} (nv1);
            \draw[->,blue] (nv1) to [bend right=15] node[midway, left] {\blue $(\BZ_2^{[0]})_{m\,  \text{(resp. $n$)}}$} (nv2);
            \draw[->,blue] (nv2) to [bend left=45] node[at start, right=1.2] {\blue $(\BZ_2^{[0]})_{n\,  \text{(resp. $m$)}}$} (2RepA4);
            \draw[->,blue] (2VecA4) to [bend left=15] node[midway, right=0.2] {\blue $(\BZ_2^{[0]})_{m\,  \text{(resp. $n$)}}$} (2VecZ21Z2Z3);
            \draw[->,blue] (2VecZ21Z2Z3) to [bend left=15] node[midway, right] {\blue $(\BZ_2^{[0]})_{n\,  \text{(resp. $m$)}}$} (2VecZ21Z21Z3);
            \draw[->,blue] (2VecZ21Z21Z3) to [bend left=15] node[midway, right] {\blue $(\BZ_3^{[0]})_q$} (2RepA4);     
            \draw[->,new-green] (2VecZ21Z21Z3) to node[near start, below] {\green $\Gamma^{(2)}_{\hat{m}, \hat{n}, q}$} (2RepL);
            \draw[->,new-green] (2RepL) to [bend right=15] node[midway, left = 0.2] {\green $[m]$ (or $[n]$ or $[m n]$)} (2RepA4);
            \draw[->,violet] (2VecA4) to node[near start, left] {\violet $(A_4^{[0]})_{\langle m, n, q\rangle}$} (2RepA4);
\end{tikzpicture} }
    \caption[A4 web]{The $A_4$ symmetry of the affine $D_4$ quiver and gauging of subgroups.}
    \label{figA4}
\end{figure}
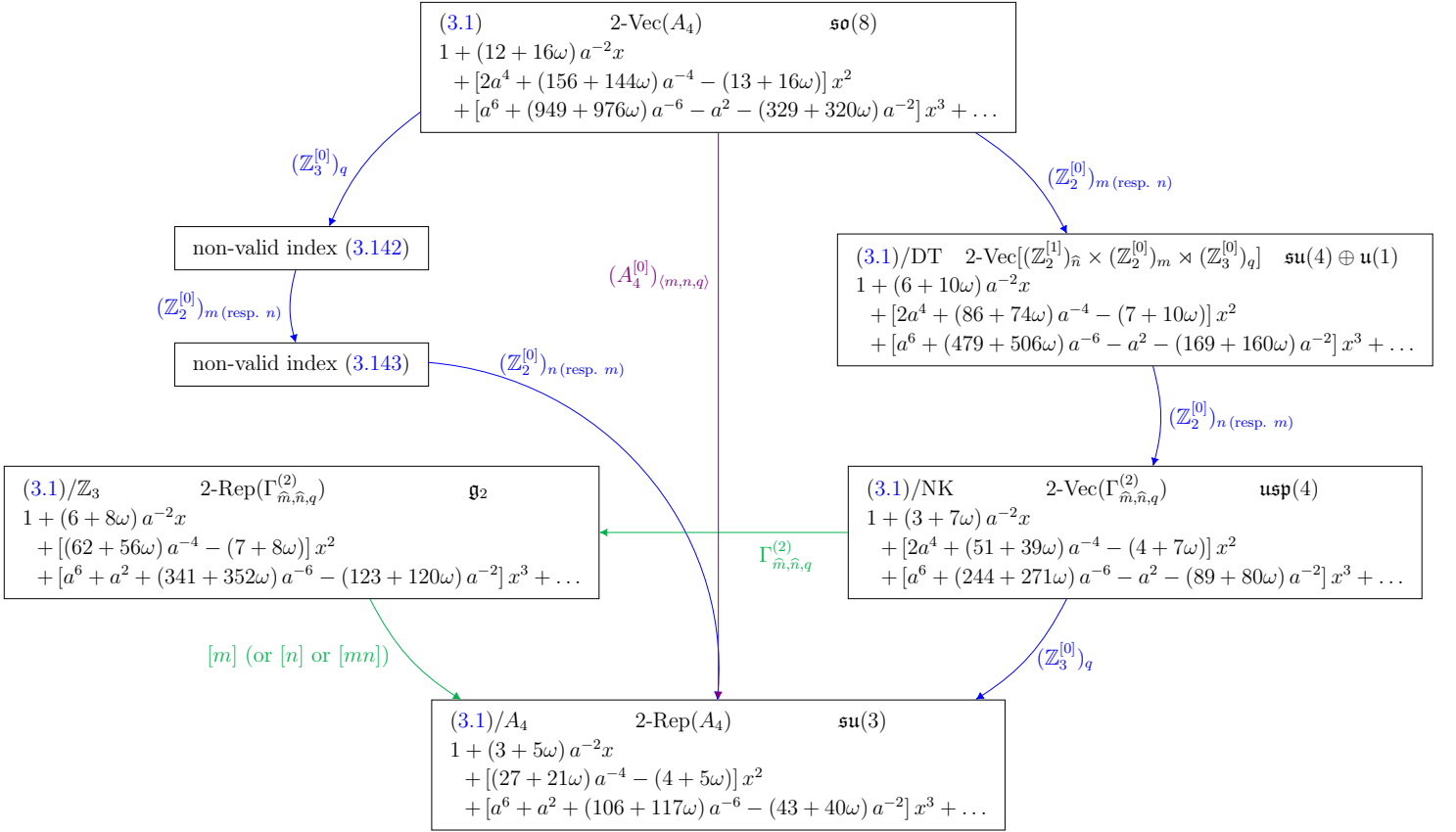

Starting from the above expression of the index, we can now sequentially gauge the $(\BZ_2^{[0]})_m$, $(\BZ_2^{[0]})_n$ and $(\BZ_3^{[0]})_q$ symmetries associated with the fugacities $m$, $n$ and $q$ in various ways, as summarised in Figure \ref{figA4}. Let us discuss the following options.
\ben
 \item We can first gauge the $(\BZ_2^{[0]})_{m} \times (\BZ_2^{[0]})_{n}$ symmetry and, subsequently, we can gauge the remaining $(\BZ_3^{[0]})_{q}$ symmetry to arrive at theory $\eref{affineD4}/{A_4}$.
  \item We can also reach theory $\eref{affineD4}/{A_4}$ by first gauging $(\BZ_3^{[0]})_{q}$ and, subsequently, gauging the remaining $(\BZ_2^{[0]})_{m} \times (\BZ_2^{[0]})_{n}$ symmetry.
 \item If we gauge only the $(\BZ_2^{[0]})_{m} \times (\BZ_2^{[0]})_{n}$ subgroup of the $A_4$ zero-form symmetry, the theory possesses a dual $(\BZ_2^{[1]})_{\hat{m}} \times (\BZ_2^{[1]})_{\hat{n}}$ one-form symmetry. This forms a two-group with the $(\BZ_3^{[0]})_q$ zero-form symmetry. We will see that, upon gauging such a two-group symmetry, we obtain the wreathed quiver $\eref{affineD4}/\BZ_3$.
 \een
 Let us now analyse in detail the various options presented in Figure \ref{figA4}.

 \bi
\item Given theory \eref{affineD4} with the global $A_4$ zero-form symmetry, whose symmetry category is 2-Vec$(A_4)$, we can gauge $(\BZ_2^{[0]})_m$, or, equivalently, $(\BZ_2^{[0]})_n$, and we land on theory $\eref{affineD4}/\mathrm{DT}$, with associated symmetry category 2-Vec$[(\BZ_2^{[1]})_{\hat{m}} \times (\BZ_2^{[0]})_n \rtimes (\BZ_3^{[0]})_q]$. At the level of the index, this can be shown as follows:
\bes{ \label{gaugeZ2m}
\scalebox{0.95}{$
\begin{split}
\CI_{\eqref{affineD4}/(\BZ_2^{[0]})_{m}}(a| \omega | n = q = 1; x) 
&\overset{\eref{indexaffD4mnq} + \eref{Fmnq}}{=} \frac{1}{2} \sum_{m = \pm 1} \sum_{p = 0}^{\infty} \CF_p(m, n = q = 1) x^p \\&\overset{\eref{systemA4}}{=} \frac{1}{2} \left[\CI_{{\eqref{affineD4}}}(a| \omega; x) + \CI_{{\eqref{affineD4}_{(12)(34)}}}(a| \omega; x)\right] \\ &\overset{\eref{IndexDTwr}}{=} \CI_{{\eqref{affineD4}/ \mathrm{DT}}}(a| \omega; x) =  \eref{indZ2wrDT}~,
    \end{split}
$}
}
where we set $m = q = 1$ and, in the second line, we used the fact that
\bes{
\CF_p(m = -1, n = q = 1) &= \CF_p(n = -1, m = q = 1) \\&= \CF_p(m = n = -1, q = 1) ~.
}
After gauging $(\BZ_2^{[0]})_m$, if we turn on the fugacity $n$ in \eref{gaugeZ2m}, then also $(\BZ_2^{[0]})_n$ can be gauged. This amounts to gauging the zero-form $(\BZ_2^{[0]})_m \times (\BZ_2^{[0]})_n \subset A_4$ symmetry of theory \eref{affineD4}, which, at the level of the index, can be performed as
\bes{ \label{gaugeZ2mZ2n}
&\CI_{\eqref{affineD4}/[(\BZ_2^{[0]})_{m} \times (\BZ_2^{[0]})_{n}]}(a| \omega | q = 1; x) 
\\&\overset{\eref{indexaffD4mnq} + \eref{Fmnq}}{=} \frac{1}{4} \sum_{m, n = \pm 1} \sum_{p = 0}^{\infty} \CF_p(m, n, q = 1) x^p \\&\overset{\eref{systemA4}}{=} \frac{1}{4} \left[\CI_{{\eqref{affineD4}}}(a| \omega; x) + 3 \times \CI_{{\eqref{affineD4}_{(i_1 i_2)(i_3 i_4)}}}(a| \omega; x)\right] \\&\overset{\eref{IndexNKwr}}{=} \CI_{{\eqref{affineD4}/ \mathrm{NK}}}(a| \omega; x) = \eref{indNKwr}~.
}
This shows that, upon gauging $(\BZ_2^{[0]})_m \times (\BZ_2^{[0]})_n$, the resulting theory is $\eref{affineD4}/\mathrm{NK}$, with symmetry category 2-Vec$(\Gamma^{(2)}_{\hat{m}, \hat{n}, q})$, where $\Gamma^{(2)}_{\hat{m}, \hat{n}, q}$ is the two-group formed by the $(\BZ_2^{[1]})_{\hat{m}} \times (\BZ_2^{[1]})_{\hat{n}}$ one-form symmetry and the $(\BZ_3^{[0]})_q$ zero-form symmetry. Finally, if we turn on the fugacity $q$ in \eref{gaugeZ2mZ2n} and we further gauge the remaining $(\BZ_3^{[0]})_q$ symmetry, we arrive at the wreathed quiver $\eref{affineD4}/A_4$.
\item On the other hand, gauging $(\BZ_3^{[0]})_q$ first in theory \eref{affineD4} leads to an {\it invalid} index, in the same way as discussed around \eref{IndexS3gaugeZ2} and \eref{IndexD8gaugeZ2}, whose explicit expression reads
\bes{ \label{IndexA4gaugeZ3}
&\CI_{\eqref{affineD4}/(\BZ_3^{[0]})_q}(a| \omega | m = n = 1; x) = \eref{indexA4}/{(\BZ_3^{[0]})_q}\big|_{m = n = 1} \\&= 1 + \left(12 + 8 \omega\right) a^{-2} x + \left[\left(60 + 72 \omega\right) a^{-4} {\red - 6 a^4} - \left(13 + 8 \omega\right) \right] x^2 + \ldots~,
}
where the term highlighted in {\red red} signals that the Higgs branch limit admits a series expansion containing negative terms, which cannot describe gauge invariant quantities parametrising the Higgs branch of theory $\eqref{affineD4}/(\BZ_3^{[0]})_q$.

The invalidity of the index follows from the argument presented in \cite[(1.4)]{Cordova:2017kue} which is also described around \eref{normalizerquotient}. Indeed, the normaliser of $H = (\BZ_3^{[0]})_q$ in $K = A_4$ is $N(H,K) = (\BZ_3^{[0]})_q$, meaning that the quotient $N(H,K)/H$ is trivial. However, if we naively gauge the $(\BZ_3^{[0]})_q$ symmetry, we are left with the fugacities $m$ and $n$ associated with the $(\BZ_2^{[0]})_m$ and $(\BZ_2^{[0]})_n$ symmetries respectively, but these are not really there after gauging $(\BZ_3^{[0]})_q$. As a consequence, upon setting such fugacities to unity, we obtain the invalid index. The interpretation is that, after gauging $(\BZ_3^{[0]})_q$, the whole symmetry becomes non-invertible and is described by $\text{2-Rep}(\Gamma^{(2)}_{\hat{m}, \hat{n}, q}) = \text{2-Rep}\{[(\BZ_2^{[1]})_{\hat{m}} \times (\BZ_2^{[1]})_{\hat{n}}] \rtimes (\BZ_3^{[0]})_q\}$. Therefore, the invalid index, obtained by naively gauging $(\BZ_3^{[0]})_q$, signals the presence of such non-invertible symmetry.  We will describe another way, namely gauging the two-group, that leads to a valid index below.

The same problem persists if we turn on the fugacity $m$ (resp. $n$) in \eref{IndexA4gaugeZ3} and we further gauge $(\BZ_2^{[0]})_{m}$ (resp. $(\BZ_2^{[0]})_{n}$) in theory $\eqref{affineD4}/(\BZ_3^{[0]})_q$, for which the index reads
\bes{ \label{IndexA4gaugeZ3gaugeZ2}
&\CI_{[\eqref{affineD4}/(\BZ_3^{[0]})_q]/(\BZ_2^{[0]})_{m}}(a| \omega | n = 1; x) = [\eref{indexA4}/{(\BZ_3^{[0]})_q}]/(\BZ_2^{[0]})_{m}\big|_{n = 1}\\&= 1 + \left(6 + 6 \omega\right) a^{-2} x + \left[\left(38 + 38 \omega\right) a^{-4} {\red - 2 a^4} - \left(7 + 6 \omega\right) \right] x^2 + \ldots~.
}
Nonetheless, we can turn on the remaining fugacity $n$ (resp. $m$) in \eref{IndexA4gaugeZ3gaugeZ2} and gauge the associated $(\BZ_2^{[0]})_{n}$ (resp. $(\BZ_2^{[0]})_{m}$) symmetry, from which we reach the wreathed quiver $\eref{affineD4}/A_4$.
\item {\bf Gauging the two-group $\Gamma^{(2)}_{\hat{m}, \hat{n}, q}$.} Let us start from the index of theory $\eref{affineD4}/[(\BZ_2^{[0]})_{m} \times (\BZ_2^{[0]})_{n}]$, whose symmetry category is 2-Vec$(\Gamma^{(2)}_{\hat{m}, \hat{n}, q})$, and let us gauge the two-group as follows. First, we gauge $(\BZ_3^{[0]})_q$ in the index of theory $\eqref{affineD4}/[(\BZ_2^{[0]})_{m} \times (\BZ_2^{[0]})_{n}]$ and, subsequently, we restore all of the terms which appear with a combination of discrete fugacities containing {\it only one} among $m$, $n$, or their diagonal combination $m n$. More explicitly, these three options amount to restore all fugacities contained in either $[m]$, $[n]$, or $[m n]$, where
\bes{
\scalebox{0.97}{$
[m] \equiv \left\{m, m q, m q^2\right\} ~,~ [n] \equiv \left\{n, n q, n q^2\right\} ~,~ [m n] \equiv \left\{m n, m n q, m n q^2\right\} ~.
$}
}
For definiteness, let us start with theory $\eqref{affineD4}/[(\BZ_2^{[0]})_{m} \times (\BZ_2^{[0]})_{n}]$, then gauge $(\BZ_3^{[0]})_q$ and, subsequently, restore only the terms appearing with $[m]$. From \eref{indexaffD4mnq}, the required index is of the form
\bes{ \label{gaugeZ2mZ2nGammawithm}
\CI_{\eqref{affineD4}/A_4}(a| \omega; x) + m \sum_{p = 0}^{\infty} y_{1,p} x^p +\left(m q + m q^2\right) \sum_{p = 0}^{\infty} y_{2,p} x^p~,
}
with $y_{1,p}$ and $y_{2,p}$ solutions of \eref{systemA4}. Explicitly, we obtain
\bes{ \label{indgaugeZ2mZ2nGammawithm}
&\CI_{\{\eqref{affineD4}/[(\BZ_2^{[0]})_{m} \times (\BZ_2^{[0]})_{n}]\}/\Gamma^{(2)}_{\hat{m}, \hat{n}, q}}(a| \omega | [m] ; x) \\& = 1 + \left[3 + 5 \omega + \left( 3 + \omega\right) m + \omega \left(m q + m q^2\right) \right] a^{-2}  x \\ & \quad \, \, \, \, \, + \left\{ \left[- 2 m + \left(m q + m q^2\right) \right] a^4 \right. \\ & \quad \, \, \, \, \, \, \, \, \, \left.+ \left[27 + 21 \omega + \left(11 + 17 \omega\right) m + \left(12 + 9 \omega\right) \left(m q + m q^2\right) \right] a^{-4} \right. \\ & \quad \, \, \, \, \, \, \, \, \, \left.- \left[4 + 5 \omega + \left( 3 + \omega\right) m + \omega \left(m q + m q^2\right) \right]\right\} x^2~.
}
Since there is no longer a zero-form symmetry which swap $M$, $N$ and $M N$ among them, we cannot restore terms appearing with two or more combinations among $[m]$, $[n]$, or $[m n]$.\footnote{From \eref{A4def}, $Q$ swaps $M$ with $M N$, $M N$ with $N$ and $N$ with $M$ by conjugation. Gauging $(\BZ_3^{[0]})_q = \langle Q \rangle$ leads to the dual one-form symmetry $(\BZ_3^{[1]})_{\hat{q}}$, but the latter can no longer act on the elements of the zero-form symmetry by conjugation as discussed in Footnote \ref{foot:codim}, due to the fact that the symmetry defect of the one-form symmetry is of higher codimension than that of the zero-form symmetry.} The theory arising from this procedure of gauging the two-group turns out to be equivalent to the theory arising from the $\BZ_3$ wreathing of \eref{affineD4}. Indeed, at the level of the index, the following equality holds:
\bes{ \label{equalitygauge2group}
\scalebox{0.99}{$
\CI_{\{\eqref{affineD4}/[(\BZ_2^{[0]})_{m} \times (\BZ_2^{[0]})_{n}]\}/\Gamma^{(2)}_{\hat{m}, \hat{n}, q}}(a| \omega | [m] = 1; x) = \CI_{{\eqref{affineD4}/ \BZ_3}}(a| \omega; x) = \eref{indZ3wr}~,
$}
}
where the notation $[m] = 1$ means that we set $m$, $m q$ and $m q^2$ to unity. The validity of this equality can be shown as follows. First, observe that, given $y_{1,p}$ and $y_{2,p}$ solutions of \eref{systemA4}, we have
\bes{
&m \sum_{p = 0}^{\infty} y_{1,p} x^p +\left(m q + m q^2\right) \sum_{p = 0}^{\infty} y_{2,p} x^p \\ &= \CI_{\eqref{affineD4}/(\BZ_2^{[0]})_{n}}(a| \omega | m, q; x) - \CI_{\eqref{affineD4}/[(\BZ_2^{[0]})_{m} \times (\BZ_2^{[0]})_{n}]}(a| \omega | q; x)~,
}
where \eref{gaugeZ2m} and \eref{gaugeZ2mZ2n} imply that
\bes{
&\CI_{\eqref{affineD4}/(\BZ_2^{[0]})_{n}}(a| \omega | m = q = 1; x) - \CI_{\eqref{affineD4}/[(\BZ_2^{[0]})_{m} \times (\BZ_2^{[0]})_{n}]}(a| \omega | q = 1; x) \\ &= \CI_{{\eqref{affineD4}/ \mathrm{DT}}}(a| \omega; x) - \CI_{{\eqref{affineD4}/ \mathrm{NK}}}(a| \omega; x)~,
}
meaning that the expression \eref{gaugeZ2mZ2nGammawithm} for $[m] = 1$ can be written as
\bes{ \label{gaugeZ2mZ2nGamma}
&\CI_{\{\eqref{affineD4}/[(\BZ_2^{[0]})_{m} \times (\BZ_2^{[0]})_{n}]\}/\Gamma^{(2)}_{\hat{m}, \hat{n}, q}}(a| \omega | [m] = 1; x) \\ & = \CI_{{\eqref{affineD4}/ A_4}}(a| \omega; x) + \CI_{{\eqref{affineD4}/ \mathrm{DT}}}(a| \omega; x) - \CI_{{\eqref{affineD4}/ \mathrm{NK}}}(a| \omega; x)~.
}
Next, we can use \eref{IndexDTwr} and \eref{IndexNKwr} to rewrite \eref{gaugeZ2mZ2nGamma} in the form
\bes{
\scalebox{0.99}{$
\begin{split}
\eref{gaugeZ2mZ2nGamma}&=\CI_{{\eqref{affineD4}/ A_4}}(a| \omega; x) + \frac{1}{2} \left[\CI_{{\eqref{affineD4}}}(a| \omega; x) + \CI_{{\eqref{affineD4}_{(i_1 i_2)(i_3 i_4)}}}(a| \omega; x)\right] \\ & \qquad \qquad \qquad \, \, \, \, \, \, \, \, \, \, \, - \frac{1}{4} \left[\CI_{{\eqref{affineD4}}}(a| \omega; x) + 3 \CI_{{\eqref{affineD4}_{(i_1 i_2)(i_3 i_4)}}}(a| \omega; x)\right] \\ &= \CI_{{\eqref{affineD4}/ A_4}}(a| \omega; x) + \frac{1}{4} \left[\CI_{{\eqref{affineD4}}}(a| \omega; x) - \CI_{{\eqref{affineD4}_{(i_1 i_2)(i_3 i_4)}}}(a| \omega; x)\right]~.
\end{split}
$}
}
Finally, using \eref{IndexA4wr}, we can replace $\CI_{{\eqref{affineD4}_{(i_1 i_2)(i_3 i_4)}}}(a| \omega; x)$ in the above expression with the following linear combination of indices:
\bes{ 
\scalebox{0.96}{$
\begin{split}
&\CI_{{\eqref{affineD4}_{(i_1 i_2)(i_3 i_4)}}}(a| \omega; x) \\ &= \frac{1}{3} \times \left[12 \times \CI_{\eqref{affineD4}/A_4}(a| \omega; x) - \CI_{\eqref{affineD4}}(a| \omega; x) - 8 \times \CI_{{\eqref{affineD4}_{(i_1 i_2 i_3)}}}(a| \omega; x) \right]~.
\end{split}
$}
}
Doing so, we end up with
\bes{
\scalebox{0.95}{$
\eref{gaugeZ2mZ2nGamma} = \frac{1}{3} \left[\CI_{\eqref{affineD4}}(a| \omega; x) + 2 \times \CI_{{\eqref{affineD4}_{(i_1 i_2 i_3)}}}(a| \omega; x) \right]\overset{\eref{IndexZ3wr}}{=}\CI_{{\eqref{affineD4}/ \BZ_3}}(a| \omega; x)~,
$}
}
which is precisely the equality \eref{equalitygauge2group}.

Starting from \eref{indgaugeZ2mZ2nGammawithm}, we can further sum $m q$ and $m q^2$ over the three cube roots of unity and sum $m$ over $\pm 1$, from which we obtain the index of $\eref{affineD4}/A_4$ given by \eref{indmodA4}.
 \ei
\subsection{$S_4$ wreathing of the affine $D_4$ quiver}
Hitherto, we have discussed wreathing of theory \eref{affineD4} by every subgroup of $S_4$, up to automorphisms.  Let us now analyse the wreathing by the group $S_4$ itself. This is given by a sum over 24 terms, each of them corresponding to an element of $S_4$, divided by the order of the group, \ie 24, namely
\bes{ \label{IndexS4wr}
\eref{affineD4}/{S_4} =& \frac{1}{24} \left[\eref{affineD4} + \eref{affineD4}_{(12)} + \eref{affineD4}_{(13)} + \eref{affineD4}_{(14)} + \eref{affineD4}_{(23)} + \eref{affineD4}_{(24)} \right.\\& \qquad \, \, \, \, \, \, \, \, \, \, \,+\left. \eref{affineD4}_{(34)} + \eref{affineD4}_{(12)(34)} + \eref{affineD4}_{(13)(24)} + \eref{affineD4}_{(14)(23)} \right.\\& \qquad \, \, \, \, \, \, \, \, \, \, \,+\left. \eref{affineD4}_{(123)} + \eref{affineD4}_{(124)} + \eref{affineD4}_{(132)} + \eref{affineD4}_{(134)} \right.\\& \qquad \, \, \, \, \, \, \, \, \, \, \,+\left. \eref{affineD4}_{(142)} + \eref{affineD4}_{(143)} + \eref{affineD4}_{(234)} +\eref{affineD4}_{(243)} \right.\\& \qquad \, \, \, \, \, \, \, \, \, \, \,+\left. \eref{affineD4}_{(1234)} + \eref{affineD4}_{(1243)} + \eref{affineD4}_{(1324)} \right.\\& \qquad \, \, \, \, \, \, \, \, \, \, \,+\left. \eref{affineD4}_{(1342)} + \eref{affineD4}_{(1423)} + \eref{affineD4}_{(1432)}\right]~,
}
where the six terms $\eref{affineD4}_{(i_1 i_2)}$ are of the form \eref{quiver12}, the three terms $\eref{affineD4}_{(i_1 i_2)(i_3 i_4)}$ are given by \eref{quiver12c34}, the eight terms $\eref{affineD4}_{(i_1 i_2 i_3)}$ take the form \eref{quiver123} and, finally, the six terms $\eref{affineD4}_{(i_1 i_2 i_3 i_4)}$ can be depicted as \eref{quiver1234}. Explicitly, their indices are given by
\bes{
\begin{array}{ll}
\CI_{{\eqref{affineD4}_{(i_1 i_2)}}}(a| \omega; x) = \eref{indZ2(12)} ~, &\quad \CI_{{\eqref{affineD4}_{(i_1 i_2) (i_3 i_4)}}}(a| \omega; x) = \eref{indZ2(12)(34)} ~, \\  \CI_{{\eqref{affineD4}_{(i_1 i_2 i_3)}}}(a| \omega; x) = \eref{indZ3(123)} ~, & \quad \CI_{{\eqref{affineD4}_{(i_1 i_2 i_3 i_4)}}}(a| \omega; x) = \eref{ind(1234)}~,
\end{array}
}
whereas the index of \eref{affineD4} is given by \eref{indaffineD4}. Using \eref{IndexS4wr}, we can compute the index of theory $\eref{affineD4}/S_4$, which possesses 2-Rep$(S_4)$ symmetry. This reads
\bes{ \label{indmodS4}
&\CI_{\eqref{affineD4}/S_4}(a| \omega; x) \\ & =1 + \left(3 + 5 \omega\right) a^{-2} x + \left[\left(27 + 21 \omega\right) a^{-4} - \left(4 + 5 \omega\right) \right] x^2 \\ & \quad \, \, \, \, \, + \left[a^2 + \left(100 + 110 \omega\right) a^{-6} - \left(42 + 40 \omega \right) a^{-2}\right] x^3 \\ & \quad \, \, \, \, \, + \left[a^8 + 17 + 24 \omega + \left(397 + 374 \omega\right) a^{-8} - \left(241 + 246 \omega \right) a^{-4}\right] x^4 + \ldots~.
}
Note that the Coulomb branch moment map operators, which contribute to the term $a^{-2} x$ in the above index, indicate that the flavour symmetry of the wreathed quiver $\eref{affineD4}/S_4$ is $\su(3)$. Taking the Coulomb and Higgs branch limits of the index $\eref{indmodS4} |_{\omega=1}$, we obtain the same results as the ones reported in \cite[Figures 9 and 11]{Bourget:2020bxh}. 

\subsection{Summary: semi-direct products versus non-trivial extensions} \label{sec:sumvarious}
We have discussed several examples concerning gauging symmetries involved in a semi-direct product and those involved in a non-trivial central extension of groups. Let us summarise the features of the indices in each case.

Let $\CT$ be a theory of our interest with a continuous non-Abelian symmetry $\CX$, and a zero-form discrete finite symmetry $G$ that is manifest in a certain quiver description of $\CT$. Here $G$ or its subgroup is a symmetry of the quiver description of theory $\CT$. In the preceding subsection, we take $\CT$ to be as described by the affine $D_4$ quiver \eref{affineD4} with $\CX=\so(8)$ and $G$ is a subgroup of $S_4$.

Let us first consider the case in which $G \cong \fN \rtimes \BZ_h$, where $\fN$ is a normal subgroup of $G$. For simplicity and definiteness of the discussion, we assume that $\fN$ is a product of Abelian factors $\fN = \prod_i \BZ_{s_i}$. We can refine the index $\CI_\CT$ of theory $\CT$ with respect to the fugacities associated with the generators of $\BZ_h$ and each factor $\BZ_{s_i}$ in $\fN$. Here are the important points:
\bi
\item We observe that gauging $\fN$ leads to the index of theory $\CT$ wreathed by $\Gamma_w$, namely $\CT/\Gamma_w$, where $\Gamma_w$ is such that 
\bes{
\BZ_h \cong G/\Gamma_w
}
where we have discussed the following examples in the precedent subsections:
\bes{
&(G, \, \fN, \,  \BZ_h, \, \Gamma_w) \\&= \left\{(S_3, \, \BZ_3, \, \BZ_2, \, \BZ_3 )~,~(\Dih_8, \, \BZ_4,\, \BZ_2, \, \mathrm{NK})~,~(A_4,\, \mathrm{NK}, \,\BZ_3,\, \mathrm{NK})\right\}~,
}
where, as before, $\text{NK}$ denotes the normal Klein subgroup of $S_4$.  Note that, in the second case, the wreathing group $\Gamma_w = \text{NK}$ does not coincide with the group that is gauged, namely $\fN = \BZ_4$. The reason for the latter is due to a mixed 't Hooft anomaly, as explained below \eref{solveZ4}.  

We remark that gauging the symmetry $\fN$ of theory $\CT$ leads to a split two-group $\Gamma^{(2)} \cong \prod_i \BZ^{[1]}_{s_i} \rtimes \BZ_h$ formed by the dual one-form symmetry $\prod_{i} \BZ^{[1]}_{s_i}$ and the remaining zero-form symmetry $\BZ_h$. When $G$ is a semi-direct product, the two-group $\Gamma^{(2)}$ does not have an 't Hooft anomaly and can be gauged; see \eg \cite{Bhardwaj:2017xup, Tachikawa:2017gyf, Bhardwaj:2022maz, Bartsch:2022ytj}.
\item On the other hand, an attempt to directly gauge the quotient group $\BZ_h \cong G/\fN$ from the index $\CI_\CT$ leads to the so-called {\it invalid} index. Such an index has certain undesirable properties. For example, upon taking the Higgs or Coulomb branch limit to obtain the Higgs or Coulomb branch Hilbert series, the result contains terms with negative coefficients. These are not reasonable since the Hilbert series counts gauge invariant operators on the Higgs or Coulomb branch. We have provided various reasons for this, for example, below \eref{indexS3} and the second bullet point on Page 47 for the case of $S_3$. We also observe that the invalid index is related to the fact that gauging $\BZ_h$ makes $G$ become a non-invertible symmetry, whose fusion category is $\text{2-Rep}(\Gamma^{(2)})$; see \eg \cite{Bhardwaj:2017xup, Tachikawa:2017gyf, Bhardwaj:2022lsg, Bhardwaj:2022maz, Bartsch:2022mpm, Bartsch:2022ytj}.
\item Such an invalid index can be ``cured'' as follows. Recall that theory $\CT$ with $\fN$ gauged has a split two-group $\Gamma^{(2)}$ which is anomaly-free. Now we can gauge the two-group $\Gamma^{(2)}$ in this theory to obtain theory $\CT$ gauged by $\BZ_h$, \ie~ $\CT/\BZ_h$, via the relation
\bes{
(\CT/\fN)/\Gamma^{(2)} \cong \CT/\BZ_h~.
}
The resulting index turns out to be a valid one, namely that of theory $\CT$ wreathed by $\BZ_h$, as indicated by the right-hand side. In terms of the index, the gauging of the two-group $\Gamma^{(2)}$ can be performed by first summing over the fugacities of $\BZ_h$ (as well as dividing by $h$), and then restoring an appropriate combination of fugacities of $\fN$.
\ei

Let us now suppose that theory $\CT$ has a zero-form discrete finite symmetry $G$ which is a non-trivial extension of $\fN = \prod_i \BZ_{s_i}$ and $\BZ_h$, with $\fN$ a normal subgroup of $G$ as before. In this case, upon gauging $\fN$, the resulting two-group $\Gamma^{(2)}$ formed by the dual symmetry of $\fN$ and $\BZ_h$ has an 't Hooft anomaly. Furthermore, let us suppose that the continuous symmetry $\CX$ of $\CT$ becomes $\CX'$ which is a subalgebra of $\CX$ after gauging $\BZ_h$.  The symmetry $\CX'$ is determined from the moment map operators that transform in the adjoint representation of $\CX'$.  However, upon counting the marginal operators after gauging $\BZ_h$, we see that such a number does not fit into the representations appearing in the branching rule of $\CX$ to $\CX'$. Moreover, we observe a certain universal result upon computing the difference between the index of theory $\CT$ with $\BZ_h$ gauged and that of the wreathed quiver with global symmetry $\CX'$. For the affine $D_4$ quiver \eref{affineD4} for which $\CX=\so(8)$, we denote $\CX'$ in such cases by $\su(4)*$ and $\su(4)\oplus \u(1)**$. The corresponding indices are presented in Figures \ref{figZ4web} and \ref{figDih8syz}, and such a universal result is reported in \eref{anomdiff}. It would be nice to understand the physical origin of this result in future work.

\subsection{\texorpdfstring{$\BZ_6$ discrete gauging: a combination of $\BZ_3$ wreathing and $(\BZ_2^{[0]})_\omega$}{Z6 discrete gauging: a combination of Z3 wreathing and Z2omega}}
Let us now discuss a $\BZ_6$ discrete gauging of theory \eref{affineD4} which is realised in a different fashion from those presented in the precedent subsections. Given the index \eref{indexS3}, we can combine the fugacities $g$ and $\omega$ associated with the $(\BZ_3^{[0]})_g$ and the $(\BZ_2^{[0]})_{\omega}$ symmetries into a fugacity $t \equiv \omega g$ associated with a $(\BZ_6^{[0]})_t$ symmetry. Written in terms of the fugacities $t$ and $h$, the index \eref{indexS3} reads
\bes{ \label{indexZ6}
&\CI_{\eqref{affineD4}}(a| t, h; x) \\
&= 1+\left(6 + 4 t + 3 t^2 + 8 t^3 + 3 t^4 + 4 t^5\right) a^{-2} x \\ & \quad \, \, \, \, \, + \left\{\left(t^2 + t^4\right) a^4 + \left[56 + 36 t + 41 t^2 + 48 t^3 + 41 t^4 + 36 t^5 \right. \right. \\ & \quad \, \, \, \, \, \, \, \, \left. \left.+ \left(6 + 8 t + 6 t^2 + 8 t^3 + 6 t^4 + 8 t^5\right) h \right] a^{-4} \right. \\ & \quad \, \, \, \, \, \, \, \, \left.- \left[7 + 4 t + 3 t^2 + 8 t^3 + 3 t^4 + 4 t^5\right]\right\} x^2 + \ldots~,
}
from which we can gauge the $(\BZ_6^{[0]})_t$ symmetry and obtain the index of theory $\eref{affineD4}/(\BZ_6^{[0]})_t$:
\bes{ \label{indexmodZ6}
\CI_{\eqref{affineD4}/(\BZ_6^{[0]})_t}(a| h; x) = 1 + 6 a^{-2} x + \left[\left(56 + 6 h \right) a^{-4} - 7\right] x^2 + \ldots~.
}
The same index can be obtained equivalently starting from \eref{indexS3} and gauging the $(\BZ_3^{[0]})_g$ and the $(\BZ_2^{[0]})_{\omega}$ symmetries sequentially, as shown in Figure \ref{WebZ6}. This discretely gauged theory will also be considered in the upcoming work \cite{Grimminger:2024doq}.
\begin{figure}[h]
    \centering
\hspace*{-1.5cm}
\scalebox{0.7}{
\begin{tikzpicture}
			\node[draw] (D4modZ6) at (0,-5) {\begin{tabular}{l}
			    $\eref{affineD4}/(\BZ_6^{[0]})_t$ \hfill $\su(2)^2$ \\$1 + 6 a^{-2} x + \left(62 a^{-4} - 7 \right) x^2$ \\ $\, \, \, \, + \left(a^6 + a^2 + 341 a^{-6} - 123 a^{-2}\right) x^3 + \ldots$ \end{tabular} }; 
			\node[draw] (D4modZ3) at (-7.5,0) {\begin{tabular}{l}
			     $\eref{affineD4}/\BZ_3$ \hfill $\mathfrak{g}_2$ \\$1 + \left(6 + 8 \omega\right) a^{-2} x $ \\$\, \, \, \, + \left[\left(62 + 56 \omega\right) a^{-4} - \left(7 + 8 \omega\right) \right] x^2$ \\ $\, \, \, \, + \left[a^6 + a^2 + \left(341 + 352 \omega\right) a^{-6} - \left(123 + 120 \omega \right) a^{-2}\right] x^3 + \ldots$	\end{tabular} }; 
                \node[draw] (D4modZ2) at (7.5,0) {\begin{tabular}{l}
                $\eref{affineD4}/(\BZ_2^{[0]})_{\omega}$ \hfill $\su(2)^4$
                \\$1 + 12 a^{-2} x + \left(2 a^4 + 156 a^{-4} - 13 \right) x^2$ \\ $\, \, \, \, + \left(a^6 + 949 a^{-6} - a^2 - 329 a^{-2}\right) x^3 + \ldots$	
                \end{tabular} }; 
			\node[draw] (D4) at (0,5) {\begin{tabular}{l}
			    \eref{affineD4} \hfill $\so(8)$ \\$1 + \left(12 + 16 \omega\right) a^{-2} x $ \\$\, \, \, \, + \left[2 a^4 + \left(156 + 144 \omega\right) a^{-4} - \left(13 + 16 \omega\right) \right] x^2$ \\ $\, \, \, \, + \left[a^6 + \left(949 + 976 \omega\right) a^{-6} - a^2 - \left(329 + 320 \omega \right) a^{-2}\right] x^3 + \ldots$ \end{tabular} }; 
            \draw[->,blue] (D4) to [bend right=15] node[midway, left = 0.2] {\blue $(\BZ_3^{[0]})_g$} (D4modZ3);
            \draw[->,blue] (D4) to [bend left=15] node[midway, right = 0.2] {\blue $(\BZ_2^{[0]})_\omega$} (D4modZ2);
            \draw[->,blue] (D4modZ3) to [bend right=15] node[midway, left] {\blue $(\BZ_2^{[0]})_{\omega}$} (D4modZ6);
            \draw[->,violet] (D4) to node[right] {\violet $(\BZ_6^{[0]})_t$} (D4modZ6);
            \draw[->,blue] (D4modZ2) to [bend left=15] node[midway, right = 0.1] {\blue $(\BZ_3^{[0]})_g$} (D4modZ6);
\end{tikzpicture} }
    \caption[Z6 web]{$\BZ_6$ discrete gauging of the affine $D_4$ quiver via a combination of $(\BZ^{[0]}_3)_g$ and $(\BZ^{[0]}_2)_\omega$.}
    \label{WebZ6}
\end{figure}

\section{Charge conjugation and flavour symmetry of SQCD}\label{sec:SQCDtildeF}
In this section, we consider gauging the charge conjugation symmetry associated with the flavour symmetry of the 3d $\CN=4$ $\mathrm{U}(N)$ gauge theory with $n$ flavours.\footnote{We will refer to this theory as SQCD for general $N$ and as SQED for $N=1$.} In the following, we study such gauging from two perspectives: the first one involves superconformal indices of the wreathed mirror theory of SQCD, and the second one involves the chiral ring and Hilbert series of the Higgs branch of SQED.

\subsection{Index of the wreathed mirror theory}
The mirror theory of the $\U(N)$ gauge theory with $n\geq2N$ flavours is
\bes{ \label{mirrUnwNflv}
\vcenter{\hbox{\begin{tikzpicture}
        \node[gauge,label=below:{\footnotesize $1$}] (1L) at (-6,0) {};
        \node[gauge,label=below:{\footnotesize $2$}] (2L) at (-5,0) {};
        \node[label=below:{}] (3L) at (-4,0) {$\ldots$};
        \node[gauge,label=below:{\footnotesize $N-1$}] (Nm1L) at (-3,0) {};
        \node[gauge,label=below:
        {\footnotesize $N$}] (NL) at (-2,0) {};
        \node[flavour,label=above:{\footnotesize $1$}] (fL) at (-2,1) {};
        \node[gauge,label=below:
        {\footnotesize $N$}] (N1) at (-1,0) {};
        \node[label=below:{}] (N2) at (0,0) {$\ldots$};
        \node[gauge,label=below:
        {\footnotesize $N$}] (N3) at (1,0) {};
        \node[gauge,label=below:
        {\footnotesize $N$}] (NR) at (2,0) {};
        \node[flavour,label=above:{\footnotesize $1$}] (fR) at (2,1) {};
        \node[gauge,label=below:{\footnotesize $N-1$}] (Nm1R) at (3,0) {};
        \node[gauge,label=below:{\footnotesize $1$}] (1R) at (6,0) {};
        \node[gauge,label=below:{\footnotesize $2$}] (2R) at (5,0) {};
        \node[label=below:{}] (3R) at (4,0) {$\ldots$};
        \draw (1L)--(2L)--(3L)--(Nm1L)--(NL)--(N1)--(N2)--(N3)--(NR)--(Nm1R)--(3R)--(2R)--(1R);
        \draw (NL)--(fL);
        \draw (NR)--(fR);
        \draw[decorate,decoration={brace, amplitude=6pt, mirror}] (-2.1,-0.6) -- (2.1,-0.6) node[midway, below=0.2cm] {$n-2N+1$};
\end{tikzpicture}}}}
This quiver has a left-right symmetry with respect to the vertical axis. For $n$ odd, this axis of symmetry passes through the bifundamental hypermultiplet in the middle of the quiver, whereas, for $n$ even, it passes through the gauge node with label $N$ in the middle of the quiver. The prescription for computing the index of quiver \eref{mirrUnwNflv} wreathed by this $\BZ_2$ symmetry is therefore different in these two cases. We refer the reader to Appendix \ref{sec:Z2wreathedquivers} for a detailed analysis. For simplicity, we focus on the cases of $N=1$ and $N=2$. We report the results for $n$ odd and $n$ even separately below.

\subsubsection*{The cases with odd $n$}
Let us focus on the cases with odd $n$. As explained in Appendix \ref{sec:Z2wreathedquivers}, quiver \eref{mirrUnwNflv} with $n$ odd wreathed by $\BZ_2$ can be realised using \eref{Z2wrQnomiddlenode}, \eref{quivernomiddlenode}, \eref{quivernomiddlenode12}. The corresponding index can be obtained from equations \eref{indexquivernomiddlenode}--\eref{indexquivernomiddlenodechir12}. Their explicit expressions, up to order $x^3$, are given below.
\bes{ \label{indexwreathedmirrU1wnodd}
\begin{tabular}{c|l}
\hline
$N=1$ & \qquad \qquad \quad Index of the $\BZ_2$ wreathed \eref{mirrUnwNflv}  \\
\hline
$n=3$ & $1+\left(a^2+{5}{a^{-2}}\right) x+a^3 x^{3/2}+\left(a^4+{15}{a^{-4}}-4\right) x^2$ \\
& $\, \, \, \,+\left(a^5-a\right) x^{5/2}+\left(2 a^6+{34}{a^{-6}}-{22}{a^{-2}}\right) x^3+\ldots$ \\
\hline
$n=5$ & $1+\left(a^2+{14}{a^{-2}}\right) x+\left(a^4+{105}{a^{-4}}-11\right) x^2$  \\
& $\, \, \, \,+a^5 x^{5/2}+\left(a^6+{510}{a^{-6}}-{221}{a^{-2}}\right) x^3+\ldots$ \\
\hline
$n=7$ & $1+\left(a^2+{27}{a^{-2}}\right) x + \left(a^4+{378}{a^{-4}}-22\right) x^2$\\ &$\, \, \, \,+\left(a^6+{3164}{a^{-6}}-{897}{a^{-2}}\right) x^3+a^7 x^{7/2}+\ldots$ \\
\hline
\hline
$N=2$ & \qquad \qquad \quad Index of the $\BZ_2$ wreathed \eref{mirrUnwNflv}  \\
\hline
$n=3$ & Equal to as that of $N=1$ and $n=3$ \\
\hline
$n=5$ & $1+\left(a^2+{14}{a^{-2}}\right) x+a^3 x^{3/2}+\left(2 a^4+{145}{a^{-4}}+3\right) x^2$ \\
& $\, \, \, \,+\left(2 a^5+23 a\right) x^{5/2}+\left(4 a^6+3 a^2+{1030}{a^{-6}}-{175}{a^{-2}}\right) x^3+\ldots$\\
\hline
$n=7$ & $1 +\left(a^2+{27}{a^{-2}}\right) x+\left(2 a^4+{581}{a^{-4}}+5\right) x^2$ \\
& $\, \, \, \,+a^5 x^{5/2}+\left(2 a^6+5 a^2+{8316}{a^{-6}}-{756}{a^{-2}}\right) x^3+\ldots$\\
\hline
\end{tabular}
}
 It is interesting to consider the Coulomb branch moment maps, which contribute to the term $a^{-2} x$ in the index. The coefficients $5$, $14$, and $27$ that appear above are precisely the dimensions of the principal extension $\tilde{\SU}(n)_{\text{I}} = \SU(n) \rtimes \BZ_2$ of $\SU(n)$ \cite{Bourget:2018ond, Arias-Tamargo:2019jyh, Arias-Tamargo:2021ppf} with the $\BZ_2$ charge conjugation symmetry being gauged . Let us explain this statement as follows. According to \cite[Table 2]{Arias-Tamargo:2021ppf}, such a principal extension has a representation which is known as $\mathrm{Adj} \otimes \chi$, whose character is\footnote{To be consistent with the rest of the paper, we denote by $\chi$ the fugacity associated with the charge conjugation symmetry. Note, however, that in \cite{Arias-Tamargo:2021ppf} this is denoted by $\epsilon$.}
\bes{
\chi^{\SU(n) \rtimes \BZ_2}_{\mathrm{Adj} \otimes \chi} (\vec z, \chi)  &= \left(\frac{1+\chi}{2}\right) \chi^{\SU(n)}_{\text{Adj}} (\vec z) - (1-n) \left(\frac{1-\chi}{2}\right) \\
&= \left(\frac{1+\chi}{2}\right) \left[ (n-1) + \sum_{1\leq i\neq j \leq n} \frac{z_i}{z_j} \right] - (1-n) \left(\frac{1-\chi}{2}\right)~.
}
Observe that upon setting $\chi = 1$, we indeed obtain the character $\chi^{\SU(n)}_{\text{Adj}} (\vec z)$ of the adjoint representation of $\SU(n)$. On the other hand, we can gauge the charge conjugation symmetry by summing over $\chi= \pm 1$ and obtain the following character\footnote{Note that we can perform an analogous computation for $\SO(n) \rtimes \BZ_2$, where the character of the adjoint representation is given by \cite[(2.11)]{Arias-Tamargo:2021ppf}. Upon gauging the charge conjugation symmetry, we obtain the character of the adjoint representation of $\USp(2n-2)$.
}
\bes{ \label{charadjSUntilde}
\frac{1}{2} \sum_{\chi=\pm 1} \chi^{\SU(n) \rtimes \BZ_2}_{\mathrm{Adj} \otimes \chi} (\vec z, \chi) = \frac{1}{2} \left(\chi^{\SU(n)}_{\text{Adj}}(\vec z) +n-1 \right)~.
}
Setting $z_i=1$, we obtain the dimension of the corresponding representation
\bes{ \label{dimadjSUntilde}
\frac{1}{2}(n^2+n-2) = \frac{1}{2} (n-1) (n+2)~.
}
This is precisely the coefficient of the term $a^{-2}x$ that appears in the index. In other words, this confirms that wreathing of the mirror theory \eref{mirrUnwNflv} for $n$ odd indeed reproduces gauging of the charge conjugation symmetry associated with the flavour symmetry of the $\U(N)$ gauge theory with $n$ flavours.\footnote{For the special case of $N=1$ and $n=3$, we find that the coefficients of the $a^{-2p} x^p$ terms in the index is given by $\frac{1}{2} (p+1) \left(p^2+2 p+2\right)$. These terms consitute the Coulomb branch Hilbert series of the wreathed mirror theory \eref{mirrUnwNflv}, or equivalently the Higgs branch Hilbert series of SQED with three flavours with the charge conjugation symmetry associated with the flavour symmetry ibeing gauged. Such a coefficient can also be written as $\dim \, [p,p]_{\su(3)} - 3 \dim \, [p-1,0]_{\usp(4)}$.}

Let us examine the Higgs branch of the wreathed mirror theory \eref{mirrUnwNflv}. The corresponding Hilbert series can be constructed from the coefficients $c_p$ of the terms $a^{2p} x^p$ in the index, namely $\sum_{p \in \frac{1}{2}\BZ} c_p t^{2p}$. For $N=1$, we find that the explicit expression of the Higgs branch Hilbert series of the wreathed theory in question is
\bes{
\PE[ t^2 +t^n]~.
}
It is instructive to compare this with the Higgs branch Hilbert series of the original quiver \eref{mirrUnwNflv} without wreathing, or equivalently the Coulomb branch Hilbert series of the $\U(1)$ gauge theory (SQED) with $n$ flavours, namely 
\bes{
\mathrm{HS}[\BC^2/\BZ_n](t, w) = \PE[ t^2+ (w+w^{-1}) t^n -t^{2n}]~.
}
We see that upon gauging the charge conjugation symmetry associated with the flavour symmetry of SQED, precisely one of the elementary monopole operators is projected out. As a result, the relation that contributes to the term $t^{2n}$ in the Hilbert series also disappears. We observe a similar phenomenon for the case of $N=2$.\footnote{As an example, for $N=2$ and $n=5$, we find that the Hilbert series in question is $\PE[t^2 + t^3 + t^4 + t^5 + t^6 - t^{16}]$, whereas the Coulomb branch Hilbert series of $\U(2)$ SQCD with five flavours is $\PE\left[ t^2 + (w+w^{-1}) t^3 + t^4 + (w+w^{-1}) t^5 - t^8 - t^{10} \right]$.}

\subsubsection*{The cases with even $n$}
Let us now turn to the cases with even $n$. As reported in Appendix \ref{sec:Z2wreathedquivers}, quiver \eref{mirrUnwNflv} with $n$ even wreathed by $\BZ_2$ can be obtained from \eref{Z2wrQmiddlenode}, \eref{quivermiddlenode}, \eref{quivermiddlenode12}. The corresponding index can be derived using equations \eref{indexquivermiddlenode}--\eref{indexquivermiddlenodechir12}. Explicitly, the expressions of the indices up to order $x^3$ are given below.
\bes{ \label{indexwreathedmirrU1wneven}
\begin{tabular}{c|l}
\hline
$N=1$ & \qquad Index of the $\BZ_2$ wreathed \eref{mirrUnwNflv}  \\
\hline
$n=4$ & $1+{10}{a^{-2}} x+\left(2 a^4+{49}{a^{-4}}-11\right) x^2$ \\ & $\, \, \, \,+\left(a^6+{165}{a^{-6}}-a^2-{93}{a^{-2}}\right) x^3+\ldots$ \\
\hline
$n=6$ & $1+{21}{a^{-2}} x+\left(a^4+{216}{a^{-4}}-22\right) x^2$ \\ & $\, \, \, \,+\left(a^6+{1386}{a^{-6}}-{495}{a^{-2}}\right) x^3+\ldots$\\
\hline \hline
$N=2$ & \qquad Index of the $\BZ_2$ wreathed \eref{mirrUnwNflv}  \\
\hline
$n=4$& $1+\left(3 a^2+{10}{a^{-2}}\right) x+\left(6 a^4+{64}{a^{-4}}+16\right) x^2$ \\ &$\, \, \, \,+\left(10 a^6+12 a^2+{270}{a^{-6}}-{a^{-2}}\right) x^3+\ldots$\\
\hline
$n=6$ & $1+{21}{a^{-2}} x +\left(3 a^4+{321}{a^{-4}}-8\right) x^2$ \\ & $\, \, \, \,+\left(2 a^6+40 a^2+{3276}{a^{-6}}-{446}{a^{-2}}\right) x^3+\ldots$\\
\end{tabular}
}
Observe that the coefficient of the term $a^{-2} x$, which is the contribution of the Coulomb branch moment map, is the dimension of the adjoint representation $\USp(n)$. This means that the Coulomb branch symmetry of the wreathed mirror theory \eref{mirrUnwNflv} is $\USp(n)$.  

Let us consider the case of $N=1$, namely SQED with $n$ flavours. Wreathing quiver \eref{mirrUnwNflv} gives the mirror dual to the $\mathrm{O}(2)$ gauge theory with $n/2$ flavours:
\bes{ \label{O2wn/2flv}
\vcenter{\hbox{\begin{tikzpicture}
        \node[gauge,label=below:{\footnotesize $\mathrm{O}(2)$}] (1L) at (0,0) {};
        \node[flavour,label=below:{\footnotesize $\USp(n)$}] (2L) at (2,0) {};
        \draw (1L)--(2L);
\end{tikzpicture}}}}     
where it can be checked that the indices of both theories are equal upon transforming $a \leftrightarrow a^{-1}$.  The Higgs branch of \eref{O2wn/2flv} is the closure of next to the minimal nilpotent orbit of $\usp(n)$, namely $\bar{\text{n.min}\, C_{n/2}}$, and the Coulomb branch of \eref{O2wn/2flv} is isomorphic to $\BC^2/\hat{D}_n$. We see that the Hilbert series of these are equal to the Coulomb branch and Higgs branch limits of the index given by the coefficients of the terms $a^{-2p} x^p$ and $a^{2p} x^p$, respectively.  We see that in this case wreathing is equivalent to gauging the charge conjugation symmetry that turns the $\U(1) \cong \SO(2)$ gauge group of SQED into the $\O(2)$ gauge group. Moreover, it is worth discussing the case of $N=1$ and $n=4$ in detail. Note that SQED with four flavours is dual to the $\USp(2)$ gauge theory with three flavours:
\bes{ \label{USp2wn/3flv}
\vcenter{\hbox{\begin{tikzpicture}
        \node[gauge,label=below:{\footnotesize $\USp(2)$}] (1L) at (0,0) {};
        \node[flavour,label=below:{\footnotesize $\SO(6)$}] (2L) at (2,0) {};
        \draw (1L)--(2L);
\end{tikzpicture}}} \qquad
= \qquad 
\vcenter{\hbox{\begin{tikzpicture}
\node[flavour,label=below:{\footnotesize $\SO(1)$}] (0L) at (-2,0) {};
        \node[gauge,label=below:{\footnotesize $\USp(2)$}] (1L) at (0,0) {};
        \node[flavour,label=below:{\footnotesize $\SO(5)$}] (2L) at (2,0) {};
        \draw (0L)--(1L)--(2L);
\end{tikzpicture}}}
}
Wreathing corresponds to gauging the charge conjugation symmetry associated with the $\SO(1)$ flavour node which turns it into an $\O(1)$
 gauge group. We therefore arrive at the following theory:
\bes{ \label{O1USp2SO5}
\vcenter{\hbox{\begin{tikzpicture}
\node[gauge,label=below:{\footnotesize $\O(1)$}] (0L) at (-2,0) {};
        \node[gauge,label=below:{\footnotesize $\USp(2)$}] (1L) at (0,0) {};
        \node[flavour,label=below:{\footnotesize $\SO(5)$}] (2L) at (2,0) {};
        \draw (0L)--(1L)--(2L);
\end{tikzpicture}}}
}
It can be checked using the index that theory \eref{O1USp2SO5} is indeed dual to theory $\eref{O2wn/2flv}_{n=4}$.

Let us turn to the case of $N=2$ and $n=4$. The wreathed theory discussed above is related to the following theory by mirror symmetry:
\bes{ \label{dualwreathedN2n4}
\vcenter{\hbox{\begin{tikzpicture}
\node[gauge,label=below:{\footnotesize $\SO(2)$}] (0L) at (-2,0) {};
        \node[gauge,label=below:{\footnotesize $\USp(2)$}] (1L) at (0,0) {};
        \node[flavour,label=below:{\footnotesize $\SO(5)$}] (2L) at (2,0) {};
        \node[gauge,label=right:{\footnotesize $\O(1)$}] (1B) at (0,1) {};
        \draw (0L)--(1L)--(2L);
        \draw (1L)--(1B);
\end{tikzpicture}}}
}
where the index of the wreathed theory reported above is related to that of \eref{dualwreathedN2n4} by exchanging $a \leftrightarrow a^{-1}$.  Let us provide an argument why these two theories are mirror dual to each other. $\U(2)$ SQCD with four flavours flows to an SCFT known as $T_{(2^2)}[\SU(4)]$ discussed in \cite{Gaiotto:2008ak}. Due to the isomorphism between $\so(6)$ and $\su(4)$, this turns out to be the same SCFT as $T_{(3,1^3)}[\SO(6)]$, which is described by
\bes{ \label{SO2USp2SO6}
\vcenter{\hbox{\begin{tikzpicture}
\node[gauge,label=below:{\footnotesize $\SO(2)$}] (0L) at (-2,0) {};
        \node[gauge,label=below:{\footnotesize $\USp(2)$}] (1L) at (0,0) {};
        \node[flavour,label=below:{\footnotesize $\SO(6)$}] (2L) at (2,0) {};
        \draw (0L)--(1L)--(2L);
\end{tikzpicture}}}
\qquad = \qquad 
\vcenter{\hbox{\begin{tikzpicture}
\node[gauge,label=below:{\footnotesize $\SO(2)$}] (0L) at (-2,0) {};
        \node[gauge,label=below:{\footnotesize $\USp(2)$}] (1L) at (0,0) {};
        \node[flavour,label=below:{\footnotesize $\SO(5)$}] (2L) at (2,0) {};
        \node[flavour,label=right:{\footnotesize $\SO(1)$}] (1B) at (0,1) {};
        \draw (0L)--(1L)--(2L);
        \draw (1L)--(1B);
\end{tikzpicture}}}
}
Gauging the charge conjugation symmetry associated with the $\SO(1)$ flavour node turns the latter into an $\O(1)$ gauge group.  We thus arrive at \eref{dualwreathedN2n4} as expected.

\subsection{Higgs branch of SQED with charge conjugation symmetry gauged} \label{sec:SQEDCCgauged}
In this section, we study the Higgs branch chiral ring of the $\U(1)$ gauge theory with $n$ flavours and the action of the charge conjugation associated with the flavour symmetry. We will demonstrate that, for odd $n$, gauging the latter reproduces the result of wreathing the mirror theory \eref{mirrUnwNflv} discussed in the precedent subsection. On the other hand, for even $n$, such a procedure does not produce the hyperK\"ahler cone, whose Hilbert series does not have a palindromic numerator. In other words, the aforementioned procedure does not reproduce the result of wreathing for even $n$. In any case, for both odd and even $n$, we see that the Higgs branch symmetry of the discretely gauged SQED is given by \eref{charadjSUntilde} and \eref{dimadjSUntilde}.

Let us briefly summarise the Higgs branch of the SQED with $n$ flavours. We denote the quarks and antiquarks by $Q_i$ and $\tQ^i$ with $i,j,k=1, \ldots,n$. The $F$-term condition reads
\bes{ \label{FtermSQED}
F= \sum_{i=1}^n Q_i \tQ^i =0~.
}
We also denote the gauge invariant mesons by $M^j_i = \tQ^j Q_i$, satisfying the conditions $\mathrm{tr}(M)=M^i_i=0$ and $(M^2)^i_k = M^i_j M^j_k =0$.  We now discuss gauging of the charge conjugation symmetry associated with the flavour symmetry separately for odd and even $n$.

\subsubsection*{The cases with odd $n$}
The $\BZ_2$ charge conjugation symmetry associated with the flavour symmetry in question acts on the quarks and antiquarks as\footnote{Under \eref{act1}, the $\su(n)$ flavour symmetry is broken to $\BZ_2^{n-1}$. To see this, we consider the flavour fugacities $f_i$ (with $i=1, \ldots, n$) such that $f_1 f_2 \ldots f_n=1$. We see that \eref{act1} imposes the conditions $f_i = f_i^{-1}$. For simplicity, we will not refine the Hilbert series with respect to such a $\BZ_2^{n-1}$ symmetry. Alternatively to \eref{act1}, we can choose the action to be 
\bes{ \label{act2}
Q_i \leftrightarrow \tQ^{n+1-i}~, \quad i=1, \ldots, n~.
}
Under \eref{act2}, some combinations of the Cartan elements of $\su(n)$ are preserved, since the action imposes the conditions $f_1 f_{n}= f_2 f_{n-1} = f_3 f_{n-2} =\ldots =1$. Since \eref{act2} interchanges $Q_{(n+1)/2} \leftrightarrow \tQ^{(n+1)/2}$, it follows that $f_{(n+1)/2}= f^{-1}_{(n+1)/2}$ and the corresponding Cartan element is broken to $\BZ_2$.
}
\bes{ \label{act1}
Q_i \leftrightarrow \tQ^i~, \quad i=1, \ldots, n~.
}
We see that the following components of the mesons are invariants under \eref{act1}: $M^1_1$, $M^2_2$, \ldots, $M^n_n$. Taking into account of \eref{FtermSQED}, we have $\sum_{i=1}^n M^i_i =0$.  However, when $i\neq j$, $M^i_j$ is mapped to $M^j_i$. Therefore the combination $S^i_j \equiv M^i_j+M^j_i$, with $1\leq i < j \leq n$, is invariant under \eref{act1}.  We see that the total number of invariants, subject to the $F$-term relation, is
\bes{
(n-1)+\frac{1}{2}n(n-1)= \frac{1}{2} (n-1) (n+2)~.
}
This is precisely equal to \eref{dimadjSUntilde}, which is the dimension of the principal extension of $\SU(n)$ with the charge conjugation symmetry being gauged. This indeed justified the action proposed in \eref{act1}. For definiteness, we take $M^i_i$ with $i=1, \ldots, n-1$, and $S^i_j$ with $1\leq i < j \leq n$ to be the generators of the moduli space. Let us now discuss the relation they satisfy.

For $n=3$, the generators satisfy precisely one relation, namely 
\bes{ \label{detM3by3}
\det \, \mathfrak{M}_{3 \times 3} =\frac{1}{3!} \epsilon^{i_1 i_2 i_3} \epsilon^{j_1 j_2 j_3} (\mathfrak{M}_{3 \times 3})_{i_1 j_1}(\mathfrak{M}_{3 \times 3})_{i_2 j_2}(\mathfrak{M}_{3 \times 3})_{i_3 j_3} =0~, 
}
where
\bes{
\mathfrak{M}_{3 \times 3} = \begin{pmatrix} 
2 M^1_1 & S^1_2 & S^1_3 \\
S^1_2 & 2 M^2_2 & S^2_3 \\
S^1_3 &  S^2_3 & - 2 M^1_1-2 M^2_2
\end{pmatrix}~.
}
Given that there are five generators subject to one relation, the Higgs branch Hilbert series for $n=3$ is therefore
\bes{
\PE[ 5t^2-t^6] &= \frac{1+t^2+t^4}{\left(1-t^2\right)^4} \\
&= 1 + 5 t^2 + 15 t^4 + 34 t^6 + 65 t^8 + 111  t^{10}+\ldots~.
}
We see that the Higgs branch is a two quarternionic dimensional complete intersection, whose defining relation is given by \eref{detM3by3}. Note that the above Hilbert series agree with the coefficients of the terms $a^{-2p}x^p$ in the index reported in \eref{indexwreathedmirrU1wnodd}.

For general $n$, the Higgs branch Hilbert series of the discretely gauged SQED by \eref{act1} can be computed as follows:
\bes{ \label{HBSQEDmodZ2}
H[\text{SQED}/\BZ_2^\chi](t) =\frac{1}{2} \left[ H_+(t, f_i=1)  + H_-(t)\right]
}
where $H_+$ is the usual Higgs branch Hilbert series of SQED with $n$ flavours:
\bes{
H_+(t, \vec f) &= \oint_{|z|=1} \frac{dz}{2\pi i z} \PE \left[ t z^{-1} \sum_{i=1}^n f_i + t z \sum_{i=1}^n f^{-1}_i -t^2\right] \\
&= \sum_{k=1}^n \chi^{\su(n)}_{[k,0,\ldots,0,k]}(\vec f) t^{2k}
}
 and $H_-$ can be computed using the method described in \cite{Bourget:2018ond, Arias-Tamargo:2019jyh} (see also \cite[(4.15)]{Hanany:2012dm}):\footnote{Note that if one considers instead the action \eref{act2}, one should replace the anti-diagonal elements $\left(\frac{z}{f_n}, \ldots, \frac{z}{f_2}, \frac{z}{f_1}\right)$ in the lower left block of $\sigma^-_Q$ by $\left(\frac{z}{f_1}, \ldots, \frac{z}{f_{n-1}}, \frac{z}{f_n}\right)$. The refined Hilbert series $H_-$ is then
\bes{ \label{HplusSQED}
H_-(t, \vec f) = \PE \left[t^2 \left( \sum_{i=1}^n \frac{f_i}{f_{n+1-i}} \right) -t^2 \right] = \PE \left[t^2 \sum_{i \neq \frac{n+1}{2}} \frac{f_i}{f_{n+1-i}} \right]~. 
}
Upon setting $f_i=1$, we obtain the unrefined Hilbert series $(1-t^2)^{-(n-1)}$, agreeing with \eref{HminusSQED}.
}
\bes{ \label{HminusSQED}
H_-(t) = \frac{1-t^2}{\det(\mathbf{1}_{2n\times 2n} - t \sigma^-_Q) }= \frac{1-t^2}{(1-t^2)^n} = \frac{1}{(1-t^2)^{n-1}}
}
where $\sigma^-_Q$ implements the action \eref{act1}, namely
\bes{ \label{sigmaQminus}
\sigma^-_Q = 
\left(
\begin{array}{cccc|cccc}
 \frac{f_1}{z} & 0 & 0 & 0 & 0 & 0 & 0 & 0 \\
 0 & \frac{f_2}{z} & 0 & 0 & 0 & 0 & 0 & 0 \\
 0 & \ldots & \ddots & \ldots & 0 & 0 & 0 & 0 \\
 0 & 0 & 0 & \frac{f_n}{z} & 0 & 0 & 0 & 0 \\
 \hline
 0 & 0 & 0 & 0 & \frac{z}{f_n} & 0 & 0 & 0 \\
 0 & 0 & 0 & 0 & \ldots & \ddots & \ldots & 0 \\
 0 & 0 & 0 & 0 & 0 & 0 & \frac{z}{f_2} & 0 \\
 0 & 0 & 0 & 0 & 0 & 0 & 0 & \frac{z}{f_1} \\
\end{array}
\right)
\times \CP
=  \left(
\begin{array}{cccc|cccc}
 0 & 0 & 0 & 0 & 0 & 0 & 0 & \frac{f_1}{z} \\
 0 & \ddots & 0 & 0 & 0 & 0 & \frac{f_2}{z} & 0 \\
 0 & 0 & 0 & 0 & 0 & \iddots & 0 & 0 \\
 0 & 0 & 0 & 0 & \frac{f_n}{z} & 0 & 0 & 0 \\
 \hline
 0 & 0 & 0 & \frac{z}{f_n} & 0 & 0 & 0 & 0 \\
 0 & 0 & \iddots & 0 & 0 & \ddots & 0 & 0 \\
 0 & \frac{z}{f_2} & 0 & 0 & 0 & 0 & 0 & 0 \\
 \frac{z}{f_1} & 0 & 0 & 0 & 0 & 0 & 0 & 0 \\
\end{array}
\right)~,
}
with
\bes{
\CP = \left(
\begin{array}{cccc|cccc}
 0 & 0 & 0 & 0 & 0 & 0 & 0 & 1 \\
 0 & \ddots & 0 & 0 & 0 & 0 & 1 & 0 \\
 0 & 0 & 0 & 0 & 0 & \iddots & 0 & 0 \\
 0 & 0 & 0 & 0 & 1 & 0 & 0 & 0 \\
 \hline
 0 & 0 & 0 & 1 & 0 & 0 & 0 & 0 \\
 0 & 0 & \iddots & 0 & 0 & \ddots & 0 & 0 \\
 0 & 1 & 0 & 0 & 0 & 0 & 0 & 0 \\
 1 & 0 & 0 & 0 & 0 & 0 & 0 & 0 \\
\end{array}
\right)~,
}
and the numerator $(1-t^2)$ in the first and second equalities denote the contribution of the $F$-term relation. Observe that, although $\sigma^-_Q$ depends on the gauge fugacity $z$ and the flavour fugacities $f_i$, the Hilbert series $H_-$ does not.
As a result of the computation, we find that the Hilbert series up to order $t^6$ takes the form
\bes{
&H[\text{SQED}/\BZ_2^\chi](t) \\
&= \PE \left[\frac{1}{2} (n-1) (n+2) t^2 - \frac{1}{144} (n-2) (n-1)^2 n^2 (n+1) t^6 +\ldots \right]~.
}
The coefficient of $t^6$ inside $\PE$ indicates the number of relations that the generators satisfy.  For $n > 3$, the Higgs branch is no longer a complete intersection.

\subsubsection*{The cases with even $n$}
For $n$ even, one can analyse the Higgs branch in the same way as the cases with $n$ odd.  However, we point out that the Hilbert series \eref{HminusSQED} has odd order of the pole at $t=1$, whereas the Higgs branch Hilbert series $H_+(t, f_i=1)$ of SQED given by \eref{HplusSQED} has even order of the pole at $t=1$ due to the fact that the Higgs branch is hyperK\"ahler and so its complex dimension is even.  As a result, the Hilbert series of $H[\text{SQED}/\BZ_2^\chi](t)$ given by \eref{HBSQEDmodZ2} does not have a palindromic numerator for $n\geq 4$.  

For the special case of $n=2$, the Higgs branch is freely generated by $M^1_1$ and $S^1_2 := M^1_2+M^2_1$. Note that $M^2_2 = -M^1_1$ by the $F$-term relation. There is no relation between $M^1_1$ and $S^1_2$. The Hilbert series is therefore $\PE[2t^2] = \frac{1}{(1-t^2)^2}$. We remark that the number of moment map operators, which is 2, is equal to the dimension of the representation $\mathrm{Adj} \otimes \chi$ of the principal extension $\SU(2) \rtimes \BZ_2$ with the charge conjugation symmetry being gauged. 

Let us now discuss the case of $n=4$ in detail, in which case the Higgs branch Hilbert series computed using \eref{HBSQEDmodZ2} is\footnote{Note that we obtain the same Hilbert series as in \eref{HBSQEDn4modZ2} even if we consider Type II of the principal extension of $\SU(4)$ by taking the left matrix in the first equality of \eref{sigmaQminus} to be as in \cite[(5.15)]{Arias-Tamargo:2019jyh}, namely taking the antidiagonal elements in the upper right block to be $(1,-1,1,-1)$ from right to left, and those in the lower left block to be $(-1,1,-1,1)$ from right to left.}
\bes{ \label{HBSQEDn4modZ2}
\frac{1 + 3 t^2 + 6 t^4}{(1-t^2)^6} &= 1 + 9 t^2 + 45 t^4 + 155 t^6 + 420  t^8+\ldots\\
&= \PE[ 9 t^2 - 10 t^6 + 15 t^8+\ldots]~.
}
The 9 generators of the moduli space are $M^1_1$, $M^2_2$, $M^3_3$, and $S^i_j := M^i_j + M^j_i$ for $1\leq i<j \leq 4$. Note that $M^4_4 = -\sum_{i=1}^3 M^i_i$ by the $F$-term condition.  The 10 relations at order $t^6$ are the $3\times 3$ {\bf minors} of the following matrix\footnote{It can be explicitly shown using {\tt Macaulay2} \cite{M2} that the Hilbert series of $R/I$, where $R$ is the ring of the polynomials in $M^1_1$, $M^2_2$, $M^3_3$, $S^i_j$ ($1\leq i < j \leq 4$), and $I$ is the ideal consisting of the $3 \times 3$ minors of $\mathfrak{M}_{4\times 4}$, indeed reproduces the Hilbert series \eref{HBSQEDn4modZ2}.}
\bes{
\mathfrak{M}_{4\times 4} = \left(
\begin{array}{cccc}
 2 M^1_1 & S^1_2 & S^1_3 & S^1_4 \\
 S^1_2 & 2 M^2_2 & S^2_3 & S^2_4 \\
 S^1_3 & S^2_3 & 2 M^3_3 & S^3_4 \\
 S^1_4 & S^2_4 & S^3_4 & -2 M^1_1-2 M^2_2-2 M^3_3 \\
\end{array}
\right)~.
}
We remark again that the number of moment map operators, which is 9, is equal to the dimension of the representation $\mathrm{Adj} \otimes \chi$ of the principal extension $\SU(4) \rtimes \BZ_2$ with the charge conjugation symmetry being gauged. 

Due to the non-palindromic numerator in the Hilbert series for $n\geq 4$, we see that the gauging the $\BZ_2^\chi$ charge conjugation symmetry associated with the flavour symmetry is incompatible with the symplectic singularity structure of the moduli space. This leads us to conjecture that $\BZ_2^\chi$ has an 't Hooft anomaly. It would be nice to understand this point further in future work.


In order to obtain the Hilbert series that agrees with the one obtained by wreathing, we have to modify the $F$-term condition. In particular, it has to be taken as
\bes{ \label{Fterm2}
\sum_{i,j=1}^n Q_i \tilde{Q}^j J^i_j =0~,
}
with $J$ the $n \times n$ symplectic matrix $\mathbf{1} \otimes i \sigma_2$.  The difference between \eref{FtermSQED} and \eref{Fterm2} is in the contraction of the flavour indices by the Kronecker delta and the symplectic matrix.  The $\frac{1}{2}n(n+1)$ generators, which are the moment map operators, in the adjoint representation of $\USp(n)$ are
\bes{ \label{gennminCn}
&M^1_1 = 2Q^1 \tQ_1~, \,\, M^2_2 = 2Q^2 \tQ_2~, \,\, \ldots~, \,\, M^n_n = 2Q^n \tQ_n, \\
&S^i_j = Q^i \tQ_j + Q^j \tQ_i \,\, (1 \leq i<j\leq n)~.
}
As an example, for $n=4$, there are 10 Higgs branch moment map operators transforming in the adjoint representation of $\USp(4)$, not $9$ as discussed above.  Let us define an $n \times n$ {\it symmetric} matrix $\CM$ such that the diagonal entries are $M^i_i$ with $i=1, \ldots, n$ and the off-diagonal entries are $S^i_j$. The Higgs branch Hilbert series of the discretely gauged theory can be computed, using \eg~ {\tt Macaulay2}, from that of the ring of polynomials of \eref{gennminCn} quotiented by the ideals consisting of the relations arising from the $F$-term condition \eref{Fterm2} as well as the conditions $\CM^i_j \CM^{j'}_k J^j_{j'} = 0$. The result is indeed the Hilbert series of $\bar{\mathrm{n.min}\, C_{n/2}}$ as required.

\section*{Acknowledgements}

We would like to thank Gabi Zafrir for insightful discussions that lead to several improvements of the manuscript. We also thank the JHEP referee for a number of valuable comments, especially those below Eq. \eref{indexS3} and in the second bullet point on Page 47. We also express our gratitude to Alessandro Mininno and Simone Giacomelli for a close collaboration on wreathed quivers, and to Antoine Bourget, Riccardo Comi and Sara Pasquetti for several useful conversations. JFG is supported by the EPSRC Open Fellowship (Schafer-Nameki) EP/X01276X/1 and the ``Simons Collaboration on Special Holonomy in Geometry, Analysis and Physics''. JFG and NM gratefully acknowledge support from the Simons Center for Geometry and Physics, Stony Brook University, at which part of this research project was conducted during the Simons Physics Summer Workshop 2024.
NM and WH are partially supported by the MUR-PRIN grant No. 2022NY2MXY.

\appendix

\section{Prescription for computing indices of wreathed quivers} \label{sec:wreathedquivers}
In this appendix, we consider 3d $\CN = 4$ quiver gauge theories consisting of $n$ identical subquivers with (special) unitary nodes, which naturally give rise to an $S_n$ symmetry. Gauging such an $S_n$ symmetry, or a subgroup thereof, leads to the original quivers wreathed by (a subgroup of) $S_n$. 

The main result we present in this appendix is the derivation of a novel prescription which can be adopted to obtain the superconformal index of (special) unitary wreathed quivers. In particular, we use patterns of the Hilbert series of wreathed quivers as a guideline to come up with the prescription for the indices of such theories. This prescription passes many consistency checks, including the expected Higgs and Coulomb branch limits, defined as in \eref{CBHBlimits}. Indeed, it generalises the methods presented in \cite{Hanany:2018cgo} and \cite[Section 3.6]{Bourget:2020bxh}, which are recovered as special cases by taking the Coulomb and Higgs branch limits of the index respectively.

In the following, we start by describing the simplest case, that is $\BZ_2$ wreathing, and then generalise the results obtained to the more complicated case of quivers wreathed by generic subgroups of $S_n$. The prescription we present here has been tested extensively throughout Sections \ref{sec:noninv} and \ref{sec:SQCDtildeF}.
\subsection{\texorpdfstring{Quivers wreathed by $\BZ_2$}{Quivers wreathed by Z2}}\label{sec:Z2wreathedquivers}
Let us consider a 3d $\CN=4$ quiver gauge theory $\mathsf{Q}$, containing (special) unitary gauge and flavour nodes, which possesses a $\BZ_2$ symmetry acting on two identical subsets of $\mathsf{Q}$. In the following, we consider the wreathed quiver $\mathsf{Q}/{ \BZ_2}$ and we provide a systematic way of defining it which works at the level of the index. Let us deal with two simple cases, which can then be generalised to more complicated quivers. First of all, let us consider a quiver possessing a left-right symmetry with respect to its vertical axis. This $\BZ_2$ symmetry can be gauged, leading to a wreathed quiver of the following type:
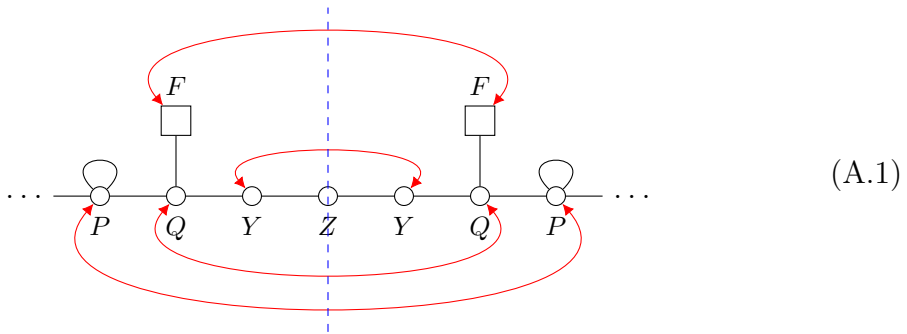
\bes{ \label{Z2wrmiddlenode}
\vcenter{\hbox{\begin{tikzpicture}
        \node[label=below:{}] (3L) at (-4,0) {$\ldots$};
        \node[gauge,label=below:{\footnotesize $P$}] (AL) at (-3,0) {};
        \node[gauge,label=below:
        {\footnotesize $Q$}] (BL) at (-2,0) {};
        \node[flavour,label=above:{\footnotesize $F$}] (FL) at (-2,1) {};
        \node[gauge,label=below:
        {\footnotesize $Y$}] (CL) at (-1,0) {};
        \node[gauge,label=below:
        {\footnotesize $Z$}] (D) at (0,0) {};
        \node[gauge,label=below:
        {\footnotesize $Y$}] (CR) at (1,0) {};
        \node[gauge,label=below:
        {\footnotesize $Q$}] (BR) at (2,0) {};
        \node[flavour,label=above:{\footnotesize $F$}] (FR) at (2,1) {};
        \node[gauge,label=below:{\footnotesize $P$}] (AR) at (3,0) {};
        \node[label=below:{}] (3R) at (4,0) {$\ldots$};
        \path (AL) edge [out=45,in=135,looseness=10] (AL);
        \path (AR) edge [out=45,in=135,looseness=10] (AR);
        \draw (3L)--(AL)--(BL)--(CL)--(D)--(CR)--(BR)--(AR)--(3R);
        \draw (BL)--(FL);
        \draw (BR)--(FR);
        \draw [<->,red] (FL) to [out=130,in=50,looseness=1] (FR);
        \draw [<->,red] (CL) to [out=130,in=50,looseness=1] (CR);
        \draw [<->,red] (BL) to [out=-130,in=-50,looseness=1] (BR);
        \draw [<->,red] (AL) to [out=-130,in=-50,looseness=1] (AR);
        \draw [dashed,blue] (0,-1.8)--(0,2.5);
\end{tikzpicture}}}}
where the dashed {\blue blue} line denotes the vertical axis of symmetry and the {\red red} arrows show the action of the $\BZ_2$ discrete gauging on the nodes of the quiver. Observe that, in theory \eref{Z2wrmiddlenode}, the axis of symmetry passes through the gauge node labelled by $Z$. We can also encounter the case in which the axis of symmetry does not pass through a gauge node and, instead, cuts the bifundamental hypermultiplet connecting two nodes. Upon gauging the left-right $\BZ_2$ symmetry of the theory, this case gives rise to a wreathed quiver of the following type:
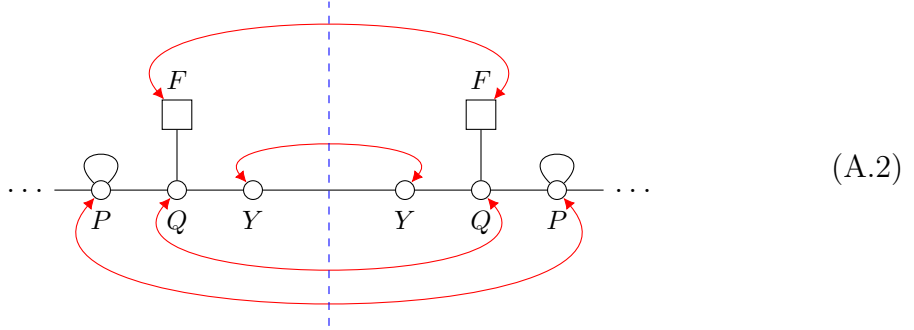
\bes{ \label{Z2wrnomiddlenode}
\vcenter{\hbox{\begin{tikzpicture}
        \node[label=below:{}] (3L) at (-4,0) {$\ldots$};
        \node[gauge,label=below:{\footnotesize $P$}] (AL) at (-3,0) {};
        \node[gauge,label=below:
        {\footnotesize $Q$}] (BL) at (-2,0) {};
        \node[flavour,label=above:{\footnotesize $F$}] (FL) at (-2,1) {};
        \node[gauge,label=below:
        {\footnotesize $Y$}] (CL) at (-1,0) {};
        \node[gauge,label=below:
        {\footnotesize $Y$}] (CR) at (1,0) {};
        \node[gauge,label=below:
        {\footnotesize $Q$}] (BR) at (2,0) {};
        \node[flavour,label=above:{\footnotesize $F$}] (FR) at (2,1) {};
        \node[gauge,label=below:{\footnotesize $P$}] (AR) at (3,0) {};
        \node[label=below:{}] (3R) at (4,0) {$\ldots$};
        \path (AL) edge [out=45,in=135,looseness=10] (AL);
        \path (AR) edge [out=45,in=135,looseness=10] (AR);
        \draw (3L)--(AL)--(BL)--(CL)--(CR)--(BR)--(AR)--(3R);
        \draw (BL)--(FL);
        \draw (BR)--(FR);
        \draw [<->,red] (FL) to [out=130,in=50,looseness=1] (FR);
        \draw [<->,red] (CL) to [out=130,in=50,looseness=1] (CR);
        \draw [<->,red] (BL) to [out=-130,in=-50,looseness=1] (BR);
        \draw [<->,red] (AL) to [out=-130,in=-50,looseness=1] (AR);
        \draw [dashed,blue] (0,-1.8)--(0,2.5);
\end{tikzpicture}}}}
Our task is to implement a procedure which allows us to derive a well-defined index for wreathed quivers of type \eref{Z2wrmiddlenode} and \eref{Z2wrnomiddlenode}. In order to do so, let us ungauge every node that appears in quivers \eref{Z2wrmiddlenode} and \eref{Z2wrnomiddlenode} and, subsequently, let us analyse the various subquivers consisting of hypermultiplets beginning and ending at flavour nodes. Once the $\BZ_2$ wreathing on such matter contributions is implemented, we restore the original gauge nodes and we study how the $\BZ_2$ wreathing acts on 3d $\CN = 4$ vector multiplets.
\subsubsection*{Bifundamental hypermultiplets contribution}
Let us consider the following contribution to the ungauged quivers \eref{Z2wrmiddlenode} and \eref{Z2wrnomiddlenode}, coming from two identical subquivers consisting of two flavour nodes connected by a bifundamental hypermultiplet, one on the left and one on the right of the axis of symmetry: 
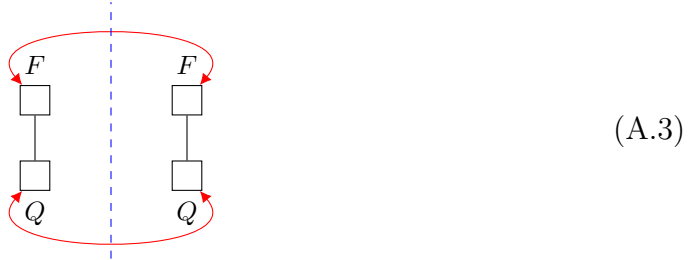
\bes{ \label{gaugeflavourZ2}
\vcenter{\hbox{\begin{tikzpicture}
        \node[flavour,label=below:
        {\footnotesize $Q$}] (BL) at (-1,0) {};
        \node[flavour,label=above:{\footnotesize $F$}] (FL) at (-1,1) {};
        \node[flavour,label=below:
        {\footnotesize $Q$}] (BR) at (1,0) {};
        \node[flavour,label=above:{\footnotesize $F$}] (FR) at (1,1) {};
        \draw (BL)--(FL);
        \draw (BR)--(FR);
        \draw [<->,red] (FL) to [out=130,in=50,looseness=1.3] (FR);
        \draw [<->,red] (BL) to [out=-130,in=-50,looseness=1.3] (BR);
        \draw [dashed,blue] (0,-1.1)--(0,2.3);
\end{tikzpicture}}}}
Let us denote with $q_{i,l}$ and $q_{i,r}$ the fugacities associated with the $\U(Q)$ nodes on the left-hand side and on the right-hand side of \eref{gaugeflavourZ2} respectively, with $i = 1, \ldots, Q$, and with $f_{j,l}$ and $f_{j,r}$ the ones associated with the $\U(F)$ nodes, with $j = 1, \ldots, F$. We can associate with the chiral field in the $(\mathbf{Q},\bar{\mathbf{F}})$ representation of $\U(Q) \times \U(F)$ on the left-hand side of \eref{gaugeflavourZ2} the matrix
\bes{ \label{Pimatrix}
\Pi_l  =  \left(
\begin{array}{cccccccccc}
 \frac{q_{1,l}}{f_{1,l}} & \ldots & 0 & 0 & 0 & 0 & 0 & 0 & 0 & 0 \\
 0 & \ddots & 0 & 0 & 0 & 0 & 0 & 0 & 0 & 0 \\
 0 & \ldots & \frac{q_{Q,l}}{f_{1,l}} & \ldots & 0 & 0 & 0 & 0 & 0 & 0 \\
 0 & 0 & \ldots & \frac{q_{1,l}}{f_{2,l}} & \ldots & 0 & 0 & 0  & 0 & 0 \\
 0 & 0 & 0 & 0 & \ddots & 0 & 0 & 0  & 0 & 0 \\
 0 & 0 & 0 & 0 & \ldots & \frac{q_{Q,l}}{f_{2,l}} & \ldots & 0  & 0 & 0 \\
 0 & 0 & 0 & 0 & 0 & 0 & \ddots & 0  & 0 & 0 \\
 0 & 0 & 0 & 0 & 0 & 0 & \ldots & \frac{q_{1,l}}{f_{F,l}}  & \ldots & 0 \\
 0 & 0 & 0 & 0 & 0 & 0 & 0 & 0 & \ddots & 0 \\
 0 & 0 & 0 & 0 & 0 & 0 & 0 & 0 & \ldots & \frac{q_{Q,l}}{f_{F,l}} \\
\end{array}
\right)
}
and, similarly, we can define the matrix $\Pi_r$, associated with the chiral field in the $(\mathbf{Q},\bar{\mathbf{F}})$ representation of $\U(Q) \times \U(F)$ on the right-hand side of \eref{gaugeflavourZ2}, by simply substituting the subscript $l$ with $r$ in \eref{Pimatrix}. We can therefore combine the chiral fields in the $(\mathbf{Q},\bar{\mathbf{F}})$ representation of $\U(Q) \times \U(F)$ appearing on both sides of \eref{gaugeflavourZ2} into the matrix
\bes{ 
\Pi = \left(
\begin{array}{c|c}
 \Pi_l & \ZM \\
 \hline
 \ZM & \Pi_r \\
\end{array}
\right)~,
}
whereas the matrix $\Pi^{-1}$ is associated with the chiral fields, both on the left-hand side and right-hand side of \eref{gaugeflavourZ2}, in the $(\bar{\mathbf{Q}},\mathbf{F})$ representation of $\U(Q) \times \U(F)$. It follows that $\Pi$ and $\Pi^{-1}$, taken together, describe both bifundamental hypermultiplets in \eref{gaugeflavourZ2}. Now, we can act on $\Pi$ and $\Pi^{-1}$ with the matrix representation of the elements of $\BZ_2$
\bes{ \label{Z2mat}
\gamma_{\langle(12)\rangle} = \left\{\gamma_{\ID} =\left(
 \begin{array}{c|c}
 \ID & \ZM \\
 \hline
 \ZM & \ID \\
\end{array}
\right) ~, \, \gamma_{(12)} = \left(
\begin{array}{c|c}
 \ZM & \ID \\
 \hline
 \ID & \ZM \\
\end{array}
\right)\right\}~,
}
and, at the level of the Higgs branch Hilbert series, we perform the $\BZ_2$ wreathing depicted in \eref{gaugeflavourZ2} as follows:
\bes{ 
\frac{1}{2} \left[\frac{1}{\det{\left(\ID - \Pi t \right)} \det{\left(\ID - \Pi^{-1} t \right)}} + \frac{1}{\det{\left(\ID - \gamma_{(12)} \Pi t \right)} \det{\left(\ID - \gamma_{(12)} \Pi^{-1} t \right)}}\right]~,
}
where the contribution associated with the identity element of $\BZ_2$ is the usual one for two bifundamental hypermultiplets, \ie
\bes{ 
\scalebox{0.93}{$
\begin{split}
&\frac{1}{\det{\left(\ID - \Pi t \right)} \det{\left(\ID - \Pi^{-1} t \right)}} \\ &= \PE\left[\left\{\sum_{i=1}^Q q_{i,l} \sum_{j=1}^F f_{j,l}^{-1} + \sum_{i=1}^Q q_{i,l}^{-1} \sum_{j=1}^F f_{j,l} + \sum_{i=1}^Q q_{i,r} \sum_{j=1}^F f_{j,r}^{-1} + \sum_{i=1}^Q q_{i,r}^{-1} \sum_{j=1}^F f_{j,r}\right\} t\right]~,
 \end{split}
$}}
whereas the contribution coming from the element $(1 2)$ of $\BZ_2$ reads
\bes{ \label{HSZ2wr12}
&\frac{1}{\det{\left(\ID - \gamma_{(12)} \Pi t \right)} \det{\left(\ID - \gamma_{(12)} \Pi^{-1} t \right)}} \\
& = \PE\left[\left\{\sum_{i=1}^Q q_{i,l} q_{i,r} \sum_{j=1}^F \left(f_{j,l} f_{j,r}\right)^{-1} + \sum_{i=1}^Q \left(q_{i,l} q_{i,r}\right)^{-1} \sum_{j=1}^F f_{j,l} f_{j,r}\right\} t^2\right]~.
}
We recognise \eref{HSZ2wr12} to be the Hilbert series arising from a $\U(Q) \times \U(F)$ bifundamental hypermultiplet, with the fugacities associated with the nodes $\U(Q)$ and $\U(F)$ being $q_i \equiv q_{i,l} q_{i,r}$ and $f_j \equiv f_{j,l} f_{j,r}$ respectively, but with $t \rightarrow t^2$. From \eqref{CBHBlimits} with $h = t$, this is equivalent to sending $(x, a) \rightarrow (x^2, a^2)$, leading us to identify \eref{gaugeflavourZ2} with
\bes{ \label{gaugeflavourZ2wr}
\eref{gaugeflavourZ2} = \frac{1}{2} \left[
\vcenter{\hbox{\begin{tikzpicture}
        \node[flavour,label=below:
        {\footnotesize $Q$},label=left:{\blue{\scriptsize $\vec{q}_l$}}] (BL) at (-0.5,0) {};
        \node[flavour,label=above:{\footnotesize $F$},label=left:{\blue{\scriptsize $\vec{f}_l$}}] (FL) at (-0.5,1) {};
        \node[flavour,label=below:
        {\footnotesize $Q$},label=right:{\blue{\scriptsize $\vec{q}_r$}}] (BR) at (0.5,0) {};
        \node[flavour,label=above:{\footnotesize $F$},label=right:{\blue{\scriptsize $\vec{f}_r$}}] (FR) at (0.5,1) {};
        \draw (BL)--(FL);
        \draw (BR)--(FR);
\end{tikzpicture}}}
\quad + \quad 
\vcenter{\hbox{\begin{tikzpicture}
        \node[flavour,thick,draw=red,label=below:
        {\footnotesize $Q$},label=left:{\red{\scriptsize $\vec{q}$}}] (B) at (0,0) {};
        \node[flavour,thick,draw=red,label=above:{\footnotesize $F$},label=left:{\red{\scriptsize $\vec{f}$}}] (F) at (0,1) {};
        \draw[thick,red] (B) to node[midway,right] {\textcolor{red}{\scriptsize $x^2, a^2$}} (F);
\end{tikzpicture}}}
\right]
}
where we highlight in {\blue blue} the flavour fugacities associated with each node in the unwreathed quiver, associated with the identity element of $\BZ_2$, and in {\red red} the flavour fugacities and the modification to the definition of the fugacities $x$ and $a$ in the quiver associated with the element $(12)$ of $\BZ_2$. Therefore, the index of \eref{gaugeflavourZ2} reads
\bes{ \label{indexgaugeflavourZ2wr}
&\CI_{\eqref{gaugeflavourZ2}}(\vec{q}_l, \vec{m}^q_l|\vec{q}_r, \vec{m}^q_r|\vec{f}_l, \vec{m}^f_l|\vec{f}_r, \vec{m}^f_r| a, n_a; x) \\ &  = \frac{1}{2} \left\{ \prod_{i=1}^Q \prod_{j=1}^F \left[ \prod_{s_l=\pm 1} \CZ^{1/2}_{\chi} \left(q_{i,l}^{s_l} f_{j,l}^{-s_l} a; s_l m^q_{i,l} - s_l m^f_{j,l} + n_a; x\right) \right.\right. \\ & \left. \left. \qquad \qquad \quad \, \times \prod_{s_r=\pm 1} \CZ^{1/2}_{\chi} \left(q_{i,r}^{s_r} f_{j,r}^{-s_r} a; s_r m^q_{i,r} - s_r m^f_{j,r} + n_a; x\right) \right] \right. \\ & \left.
\quad \, \, \, \, + \prod_{i=1}^Q \prod_{j=1}^F \prod_{s=\pm 1} \CZ^{1/2}_{\chi} \left(q_i^s f_j^{-s} a^2; s m^q_i - s m^f_j + 2 n_a; x^2\right)\right\}~.
}
\subsubsection*{Common node on the axis of symmetry}
Let us now deal with the contribution coming from two bifundamental hypermultiplets, one on the left and one on the right of the axis of symmetry of quiver \eref{Z2wrmiddlenode}, in the case in which the axis of symmetry passes through a common node:
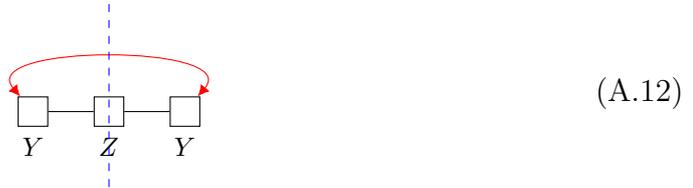
\bes{ \label{gaugeflavourZ2axis}
\vcenter{\hbox{\begin{tikzpicture}
        \node[flavour,label=below:
        {\footnotesize $Y$}] (CL) at (-1,0) {};
        \node[flavour,label=below:
        {\footnotesize $Z$}] (D) at (0,0) {};
        \node[flavour,label=below:
        {\footnotesize $Y$}] (CR) at (1,0) {};
        \draw (CL)--(D)--(CR);
        \draw [<->,red] (CL) to [out=130,in=50,looseness=1] (CR);
        \draw [dashed,blue] (0,-1)--(0,1.5);
\end{tikzpicture}}}}
If we denote with $y_{i,l}$ and $y_{i,r}$ the fugacities associated with the $\U(Y)$ nodes on the left-hand side and on the right-hand side of \eref{gaugeflavourZ2axis} respectively, where $i = 1, \ldots, Y$, and with $z_j$, with $j = 1, \ldots, Z$, the fugacities associated with the common $\U(Z)$ node, then we can construct the following matrix
\bes{ \label{Upsmatrix}
\Upsilon_l  =  \left(
\begin{array}{cccccccccc}
 \frac{y_{1,l}}{z_1} & \ldots & 0 & 0 & 0 & 0 & 0 & 0 & 0 & 0 \\
 0 & \ddots & 0 & 0 & 0 & 0 & 0 & 0 & 0 & 0 \\
 0 & \ldots & \frac{y_{Y,l}}{z_1} & \ldots & 0 & 0 & 0 & 0 & 0 & 0 \\
 0 & 0 & \ldots & \frac{y_{1,l}}{z_2} & \ldots & 0 & 0 & 0  & 0 & 0 \\
 0 & 0 & 0 & 0 & \ddots & 0 & 0 & 0  & 0 & 0 \\
 0 & 0 & 0 & 0 & \ldots & \frac{y_{Y,l}}{z_2} & \ldots & 0  & 0 & 0 \\
 0 & 0 & 0 & 0 & 0 & 0 & \ddots & 0  & 0 & 0 \\
 0 & 0 & 0 & 0 & 0 & 0 & \ldots & \frac{y_{1,l}}{z_Z}  & \ldots & 0 \\
 0 & 0 & 0 & 0 & 0 & 0 & 0 & 0 & \ddots & 0 \\
 0 & 0 & 0 & 0 & 0 & 0 & 0 & 0 & \ldots & \frac{y_{Y,l}}{z_Z} \\
\end{array}
\right)
}
and, analogously, the matrix $\Upsilon_r$ is defined by substituting $y_{i,l}$ with $y_{i,r}$ in \eref{Upsmatrix}. The matrices $\Upsilon_l$ and $\Upsilon_r$ are associated with the chiral fields in the $(\mathbf{Y},\bar{\mathbf{Z}})$ representation of $\U(Y) \times \U(Z)$ on the left-hand side and on the right-hand side of \eref{gaugeflavourZ2axis} respectively. They can be combined in a single matrix
\bes{ \label{Upsmatrixlr}
\Upsilon = \left(
\begin{array}{c|c}
 \Upsilon_l & \ZM \\
 \hline
 \ZM & \Upsilon_r \\
\end{array}
\right)~,
}
from which we can also derive $\Upsilon ^{-1}$. The two matrices $\Upsilon $ and $\Upsilon ^{-1}$ together describe the two bifundamental hypermultiplets of $\U(Y) \times \U(Z)$ both on the left and on the right of the axis of symmetry in \eref{gaugeflavourZ2axis}. Acting on them with the matrix representation of the $\BZ_2$ elements defined in \eref{Z2mat}, we obtain the Higgs branch Hilbert series corresponding to the $\BZ_2$ wreathing described in \eref{gaugeflavourZ2axis} as
\bes{ 
\frac{1}{2} \left[\frac{1}{\det{\left(\ID - \Upsilon t \right)} \det{\left(\ID - \Upsilon^{-1} t \right)}} + \frac{1}{\det{\left(\ID - \gamma_{(12)} \Upsilon t \right)} \det{\left(\ID - \gamma_{(12)} \Upsilon^{-1} t \right)}}\right]~.
}
The first contribution is the one associated with two bifundamental hypermultiplets with a common node, \ie
\bes{ 
\scalebox{0.94}{$
\begin{split}
&\frac{1}{\det{\left(\ID - \Upsilon t \right)} \det{\left(\ID - \Upsilon^{-1} t \right)}} \\ &= \PE\left[\left\{\sum_{i=1}^Y y_{i,l} \sum_{j=1}^Z z_{j}^{-1} + \sum_{i=1}^Y y_{i,l}^{-1} \sum_{j=1}^Z z_{j} + \sum_{i=1}^Y y_{i,r} \sum_{j=1}^Z z_{j}^{-1} + \sum_{i=1}^Y y_{i,r}^{-1} \sum_{j=1}^Z z_{j}\right\} t\right]~,
 \end{split}
$}}
and the second contribution, which is the interesting one, yields
\bes{ \label{HSmiddle12}
&\frac{1}{\det{\left(\ID - \gamma_{(12)} \Upsilon t \right)} \det{\left(\ID - \gamma_{(12)} \Upsilon^{-1} t \right)}} \\
& = \PE\left[\left\{\sum_{i=1}^Y y_{i,l} y_{i,r} \sum_{j=1}^Z z_{j}^{-2} + \sum_{i=1}^Y \left(y_{i,l} y_{i,r}\right)^{-1} \sum_{j=1}^Z z_{j}^2\right\} t^2\right]~.
}
The latter contribution coincides with the Higgs branch Hilbert series of a $\U(Y) \times \U(Z)$ bifundamental hypermultiplet, with fugacities $y_i \equiv y_{i,l} y_{i,r}$ associated with the $\U(Y)$ node and $z_j^2$ associated with the common $\U(Z)$ node. Note that, again, the variable $t$ is squared, which is equivalent to sending $(x, a) \rightarrow (x^2, a^2)$ according to \eref{CBHBlimits}. This leads us to propose that \eref{gaugeflavourZ2axis} can be identified with
\bes{ \label{gaugeflavourZ2wraxis}
\eref{gaugeflavourZ2axis} = \frac{1}{2} \left[
\vcenter{\hbox{\begin{tikzpicture}
        \node[flavour,label=below:
        {\footnotesize $Y$},label=left:{\blue{\scriptsize $\vec{y}_{l}$}}] (CL) at (-1,-0.5) {};
        \node[flavour,label=below:
        {\footnotesize $Z$},label=above:{\blue{\scriptsize $\vec{z}$}}] (D) at (0,0.5) {};
        \node[flavour,label=below:
        {\footnotesize $Y$},label=right:{\blue{\scriptsize $\vec{y}_{r}$}}] (CR) at (1,-0.5) {};
        \draw (CL)--(D)--(CR);
\end{tikzpicture}}}
\quad + \quad 
\vcenter{\hbox{\begin{tikzpicture}
        \node[flavour,thick,draw=red,label=below:
        {\footnotesize $Y$},label=left:{\red{\scriptsize $\vec{y}$}}] (C) at (0,0) {};
        \node[flavour,thick,draw=red,label=above:{\footnotesize $Z$},label=left:{\red{\scriptsize $\vec{z}^2$}}] (D) at (0,1) {};
        \draw[thick,red] (C) to node[midway,right] {\textcolor{red}{\scriptsize $x^2, a^2$}} (D);
\end{tikzpicture}}}
\right]
}
It follows that the index of \eref{gaugeflavourZ2axis} can be expressed as
\bes{ \label{indexgaugeflavourZ2wraxis}
&\CI_{\eqref{gaugeflavourZ2axis}}(\vec{y}_l, \vec{m}^y_l|\vec{y}_r, \vec{m}^y_r|\vec{z}, \vec{m}^z| a, n_a; x) \\ &  = \frac{1}{2} \left\{ \prod_{i=1}^Y \prod_{j=1}^Z \left[ \prod_{s_l=\pm 1} \CZ^{1/2}_{\chi} \left(y_{i,l}^{s_l} z_{j}^{-s_l} a; s_l m^y_{i,l} - s_l m^z_{j} + n_a; x\right) \right.\right. \\ & \left. \left. \qquad \qquad \quad \, \times \prod_{s_r=\pm 1} \CZ^{1/2}_{\chi} \left(y_{i,r}^{s_r} z_{j}^{-s_r} a; s_r m^y_{i,r} - s_r m^z_{j} + n_a; x\right) \right] \right. \\ & \left.
\quad \, \, \, \, + \prod_{i=1}^Y \prod_{j=1}^Z \prod_{s=\pm 1} \CZ^{1/2}_{\chi} \left(y_i^s z_{j}^{-2 s} a^2; s m^y_{i} - s m^z_{j} + 2 n_a; x^2\right)\right\}~.
}
Let us anticipate a comment that will be explained below. In the last line of \eref{indexgaugeflavourZ2wraxis}, even if the fugacities $z_j$ are rescaled to $z_j^2$, the associated magnetic fluxes do not get rescaled to $2 m^z_j$. This is due to the fact that, when we restore the original $\U(Z)$ gauge node, since this is intersected by the vertical axis of symmetry of quiver \eref{Z2wrmiddlenode}, the associated 3d $\CN = 4$ vector multiplet contribution is not affected by the $\BZ_2$ wreathing. A more detailed explanation will be provided around \eref{gaugeZ2middle} and \eref{indexgaugeZ2wrmiddle}.
\subsubsection*{Bifundamental hypermultiplet cut by the axis of symmetry}
In the case in which the axis of symmetry cuts the bifundamental hypermultiplet connecting two nodes, instead of passing through a node, we have to analyse the following contribution appearing in quiver \eref{Z2wrnomiddlenode}:
\bes{ \label{gaugegaugeZ2axis}
\vcenter
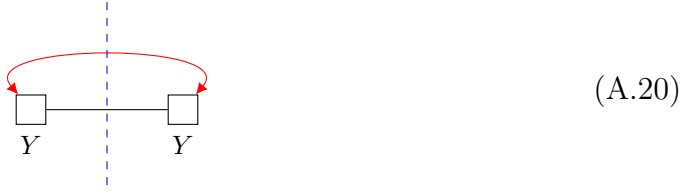
{\hbox{\begin{tikzpicture}
        \node[flavour,label=below:
        {\footnotesize $Y$}] (CL) at (-1,0) {};
        \node[flavour,label=below:
        {\footnotesize $Y$}] (CR) at (1,0) {};
        \draw (CL)--(CR);
        \draw [<->,red] (CL) to [out=130,in=50,looseness=1] (CR);
        \draw [dashed,blue] (0,-1)--(0,1.5);
\end{tikzpicture}}}}
If we still denote with $y_{i,l}$ and $y_{i,r}$ the fugacities associated with the two $\U(Y)$ nodes of \eref{gaugegaugeZ2axis}, where $i = 1, \ldots, Y$, then the contribution associated with the chiral field in the fundamental representation of the node on the left and in the antifundamental representation of the node on the right of the axis of symmetry can be encoded in the following matrix
\bes{ \label{Thetamatrix}
\Theta  =  \left(
\begin{array}{cccccccccc}
 \frac{y_{1,l}}{y_{1,r}} & \ldots & 0 & 0 & 0 & 0 & 0 & 0 & 0 & 0 \\
 0 & \ddots & 0 & 0 & 0 & 0 & 0 & 0 & 0 & 0 \\
 0 & \ldots & \frac{y_{Y,l}}{y_{1,r}} & \ldots & 0 & 0 & 0 & 0 & 0 & 0 \\
 0 & 0 & \ldots & \frac{y_{1,l}}{y_{2,r}} & \ldots & 0 & 0 & 0  & 0 & 0 \\
 0 & 0 & 0 & 0 & \ddots & 0 & 0 & 0  & 0 & 0 \\
 0 & 0 & 0 & 0 & \ldots & \frac{y_{Y,l}}{y_{2,r}} & \ldots & 0  & 0 & 0 \\
 0 & 0 & 0 & 0 & 0 & 0 & \ddots & 0  & 0 & 0 \\
 0 & 0 & 0 & 0 & 0 & 0 & \ldots & \frac{y_{1,l}}{y_{Y,r}}  & \ldots & 0 \\
 0 & 0 & 0 & 0 & 0 & 0 & 0 & 0 & \ddots & 0 \\
 0 & 0 & 0 & 0 & 0 & 0 & 0 & 0 & \ldots & \frac{y_{Y,l}}{y_{Y,r}} \\
\end{array}
\right)~,
}
whereas the matrix $\Theta^{-1}$ describes the chiral field in the antifundamental representation of the node on the left and in the fundamental representation of the node on the right of the axis of symmetry. Therefore, the matrices $\Theta$ and $\Theta^{-1}$ describe the bifundamental hypermultiplet connecting the two $\U(Y)$ nodes in \eref{gaugegaugeZ2axis}. Upon gauging the $\BZ_2$ action corresponding to the left-right symmetry, the Higgs branch Hilbert series of \eref{gaugegaugeZ2axis} reads
\bes{ 
\frac{1}{2} \left[\frac{1}{\det{\left(\ID - \Theta t \right)} \det{\left(\ID - \Theta^{-1} t \right)}} + \frac{1}{\det{\left(\ID - \gamma_{(12)} \Theta t \right)} \det{\left(\ID - \gamma_{(12)} \Theta^{-1} t \right)}}\right]~.
}
The first contribution yields
\bes{ 
\frac{1}{\det{\left(\ID - \Theta t \right)} \det{\left(\ID - \Theta^{-1} t \right)}} = \PE\left[\left\{\sum_{i,j=1}^Y y_{i,l} y_{j,r}^{-1} + \sum_{i,j=1}^Y y_{i,l}^{-1} y_{j,r}\right\} t\right]~,
}
which is the Hilbert series of a bifundamental hypermultiplet. On the other hand, the second contribution coincides with the one of a chiral field in the adjoint representation of $\U(Y)$, namely
\bes{ 
\frac{1}{\det{\left(\ID - \gamma_{(12)} \Theta t \right)} \det{\left(\ID - \gamma_{(12)} \Theta^{-1} t \right)}} = \PE\left[\left\{\sum_{i,j=1}^Y y_{i,l} y_{i,r} \left(y_{j,l} y_{j,r}\right)^{-1}\right\} t^2\right]~,
}
where, again, since the variable $t^2$ appears, we use the map \eref{CBHBlimits} and we send $(x, a) \rightarrow (x^2, a^2)$, with the $R-$charge being $1/2$. It follows that \eref{gaugegaugeZ2axis} can be represented as
\bes{ \label{gaugegaugeZ2wraxis}
\eref{gaugegaugeZ2axis} = \frac{1}{2} \left[
\vcenter{\hbox{\begin{tikzpicture}
        \node[flavour,label=below:
        {\footnotesize $Y$},label=left:{\blue{\scriptsize $\vec{y}_{l}$}}] (CL) at (-1,0) {};
        \node[flavour,label=below:
        {\footnotesize $Y$},label=right:{\blue{\scriptsize $\vec{y}_{r}$}}] (CR) at (1,0) {};
        \draw (CL)--(CR);
\end{tikzpicture}}}
\quad + \quad 
\vcenter{\hbox{\begin{tikzpicture}
        \node[flavour,thick,draw=red,label=below:
        {\footnotesize $Y$},label=left:{\red{\scriptsize $\vec{y}$}}] (C) at (0,0) {};
        \draw[dashed,red,thick] (C) edge [out=-45,in=45,looseness=10] node[midway,right] {\textcolor{red}{\scriptsize $x^2, a^2$}} (C);
\end{tikzpicture}}}
\right]
}
where $y_i \equiv y_{i,l} y_{i,r}$ and we use the dashed line to indicate an adjoint chiral, instead of an adjoint hyper. The corresponding index is given by
\bes{ \label{indexgaugegaugeZ2wraxis}
&\CI_{\eqref{gaugegaugeZ2axis}}(\vec{y}_l, \vec{m}^y_l|\vec{y}_r, \vec{m}^y_r| a, n_a; x) \\ &  = \frac{1}{2} \left\{ \prod_{i,j=1}^Y  \prod_{s=\pm 1} \CZ^{1/2}_{\chi} \left(y_{i,l}^{s} y_{j,r}^{-s} a; s m^y_{i,l} - s m^y_{j,r} + n_a; x\right) \right. \\ & \left.
\quad \, \, \, \, \, + \prod_{i,j=1}^Y \CZ^{1/2}_{\chi} \left(y_{i} y_{j}^{-1} a^2; s m^y_{i} - s m^y_{j} + 2 n_a; x^2\right)\right\}~.
}
\subsubsection*{Adjoint hypermultiplets contribution}
Finally, we can analyse the last contribution appearing in the ungauged quivers \eref{Z2wrmiddlenode} and \eref{Z2wrnomiddlenode}, that is the one associated with two identical adjoint hypermultiplets, one on the left and one on the right of the axis of symmetry:
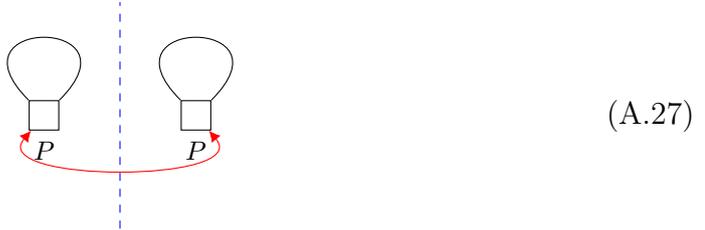
\bes{ \label{adjointhyperZ2}
\vcenter{\hbox{\begin{tikzpicture}
        \node[flavour,label=below:{\footnotesize $P$}] (AL) at (-1,0) {};
        \node[flavour,label=below:{\footnotesize $P$}] (AR) at (1,0) {};
        \path (AL) edge [out=45,in=135,looseness=10] (AL);
        \path (AR) edge [out=45,in=135,looseness=10] (AR);
        \draw [<->,red] (AL) to [out=-130,in=-50,looseness=1] (AR);
        \draw [dashed,blue] (0,-1.5)--(0,1.5);
\end{tikzpicture}}}}
Let us denote with $p_{i,l}$ and $p_{i,r}$ the fugacities associated with the $\U(P)$ nodes on the left-hand side and on the right-hand side of \eref{adjointhyperZ2} respectively, with $i = 1, \ldots, P$. Then, the matrix associated with a chiral field in the adjoint representation of the $\U(P)$ node on the left of \eref{adjointhyperZ2} is given by
\bes{ \label{Phimatrix}
\Phi_l  =  \left(
\begin{array}{ccccccccc}
 1 & \ldots & 0 & 0 & 0 & 0 & 0 & 0 & 0 \\
 0 & \frac{p_{2,l}}{p_{1,l}} & \ldots & 0 & 0 & 0 & 0 & 0 & 0 \\
 0 & 0 & \ddots & 0 & 0 & 0 & 0 & 0 & 0 \\
 0 & 0 & \ldots & \frac{p_{P,l}}{p_{1,l}} & \ldots & 0 & 0 & 0 & 0 \\
 0 & 0 & 0 & 0 & \ddots & 0 & 0 & 0 & 0\\
 0 & 0 & 0 & 0 & \ldots & \frac{p_{1,l}}{p_{P,l}} & \ldots & 0 & 0  \\
 0 & 0 & 0 & 0 & 0 & 0 & \ddots & 0 & 0\\
 0 & 0 & 0 & 0 & 0 & 0 & \ldots & \frac{p_{P-1,l}}{p_{P,l}} & 0 \\
 0 & 0 & 0 & 0 & 0 & 0 & 0 & \ldots & 1\\
\end{array}
\right)~,
}
from which we can also derive $\Phi_r$, which corresponds to a chiral field in the adjoint representation of the $\U(P)$ node on the right of \eref{adjointhyperZ2}, by substituting $p_{i,l}$ with $p_{i,r}$. The two matrices $\Phi_l$ and $\Phi_r$ can be combined into the matrix
\bes{ \label{PhimatrixLR}
\Phi = \left(
\begin{array}{c|c}
 \Phi_l & \ZM \\
 \hline
 \ZM & \Phi_r \\
\end{array}
\right)
}
and the contribution associated with the two hypermultiplets in the adjoint representation of the two $\U(P)$ nodes of \eref{adjointhyperZ2} is described by both $\Phi$ and $\Phi^{-1}$. Combining them with the action of the $\BZ_2$ elements defined in \eref{Z2mat}, the Higgs branch Hilbert series of the $\BZ_2$ wreathing described in \eref{adjointhyperZ2} reads
\bes{ 
\frac{1}{2} \left[\frac{1}{\det{\left(\ID - \Phi t \right)} \det{\left(\ID - \Phi^{-1} t \right)}} + \frac{1}{\det{\left(\ID - \gamma_{(12)} \Phi t \right)} \det{\left(\ID - \gamma_{(12)} \Phi^{-1} t \right)}}\right]~.
}
The first term reproduces the usual contribution coming from two adjoint hypermultiplets, \ie
\bes{ 
\frac{1}{\det{\left(\ID - \Phi t \right)} \det{\left(\ID - \Phi^{-1} t \right)}} = \PE\left[2 \left\{\sum_{i,j=1}^P p_{i,l} p_{j,l}^{-1} + \sum_{i,j=1}^P p_{i,r} p_{j,r}^{-1}\right\} t\right]~,
}
whereas the second term is given by
\bes{
\frac{1}{\det{\left(\ID - \gamma_{(12)} \Phi t \right)} \det{\left(\ID - \gamma_{(12)} \Phi^{-1} t \right)}} = 
\scalebox{0.98}{$
\PE\left[2 \left\{\sum\limits_{i,j=1}^P p_{i,l} p_{i,r} \left(p_{j,l} p_{j,r}\right)^{-1}\right\} t^2\right]~.
$}
}
The latter coincides with the contribution coming from a single hypermultiplet in the adjoint representation of $\U(P)$, whose fugacities are denoted by $p_{i} \equiv p_{i,l} p_{i,r}$, with $(x, a) \rightarrow (x^2, a^2)$, according to \eref{CBHBlimits}. Hence, we can depict \eref{adjointhyperZ2} as follows:
\bes{ \label{adjointhyperZ2wr}
\eref{adjointhyperZ2} = \frac{1}{2} \left[
\vcenter{\hbox{\begin{tikzpicture}
        \node[flavour,label=below:{\footnotesize $P$},label=left:{\blue{\scriptsize $\vec{p}_{l}$}}] (AL) at (-1,0) {};
        \node[flavour,label=below:{\footnotesize $P$},label=right:{\blue{\scriptsize $\vec{p}_{r}$}}] (AR) at (1,0) {};
        \path (AL) edge [out=45,in=135,looseness=10] (AL);
        \path (AR) edge [out=45,in=135,looseness=10] (AR);
\end{tikzpicture}}}
\quad + \quad 
\vcenter{\hbox{\begin{tikzpicture}
        \node[flavour,thick,draw=red,label=below:
        {\footnotesize $P$},label=left:{\red{\scriptsize $\vec{p}$}}] (A) at (0,0) {};
        \path[thick,red] (A) edge [out=-45,in=45,looseness=10] node[midway,right] {\textcolor{red}{\scriptsize $x^2, a^2$}} (A);
\end{tikzpicture}}}
\right]
}
This leads us to propose the following expression for the index of \eqref{adjointhyperZ2}:
\bes{ \label{indexadjointhyperZ2wr}
&\CI_{\eqref{adjointhyperZ2}}(\vec{p}_l, \vec{m}^p_l|\vec{p}_r, \vec{m}^p_r| a, n_a; x) \\ &  = \frac{1}{2} \left\{ \prod_{i,j=1}^P \left[ \prod_{s=\pm 1} \CZ^{1/2}_{\chi} \left(p_{i,l}^{s} p_{j,l}^{-s} a; s m^p_{i,l} - s m^p_{j,l} + n_a; x\right) \right. \right. \\ & \left. \left. \qquad \qquad \, \times \prod_{s=\pm 1} \CZ^{1/2}_{\chi} \left(p_{i,r}^{s} p_{j,r}^{-s} a; s m^p_{i,r} - s m^p_{j,r} + n_a; x\right) \right] \right. \\ & \left.
\quad \, \, \, \, \, + \prod_{i,j=1}^P \prod_{s=\pm 1} \CZ^{1/2}_{\chi} \left(p_{i}^s p_{j}^{-s} a^2; s m^p_{i} - s m^p_{j} + 2 n_a; x^2\right)\right\}~.
}
\subsubsection*{Restoring the original gauge nodes}
Now, we can restore the original gauge nodes of quivers \eref{Z2wrmiddlenode} and \eref{Z2wrnomiddlenode}. For instance, let us focus on the two $\U(Y)$ gauge nodes in \eref{Z2wrmiddlenode} and \eref{Z2wrnomiddlenode}.
\bes{ \label{gaugeZ2}
\vcenter{\hbox{\begin{tikzpicture}
        \node[gauge,label=below:{\footnotesize $Y$}] (AL) at (-1,0) {};
        \node[gauge,label=below:{\footnotesize $Y$}] (AR) at (1,0) {};
        \draw [<->,red] (AL) to [out=130,in=50,looseness=1] (AR);
        \draw [dashed,blue] (0,-1)--(0,1);
\end{tikzpicture}}}}
In doing so, we have to take into account the contribution coming from the adjoint chiral fields in the 3d $\CN = 4$ vector multiplets. The two adjoint chiral multiplets coming from the two $\U(Y)$ gauge nodes are associated with the matrix $\Phi$ defined in \eref{PhimatrixLR}, but with entries $y_{i,l}$ and $y_{i,r}$, where $i = 1, \ldots, Y$, instead of $p_{j,l}$ and $p_{j,r}$. Acting on $\Phi$ with the matrix representation of the $\BZ_2$ elements reported in \eref{Z2mat}, we obtain the following contribution to the Higgs branch Hilbert series of the quiver wreathed by $\BZ_2$:
\bes{ \label{HSZ2wradjchiral}
\frac{1}{2} \left[\det{\left(\ID - \Phi t^2 \right)} + \det{\left(\ID - \gamma_{(12)} \Phi t^2 \right)}\right]~,
}
where the first term corresponds to the usual contribution coming from two adjoint chiral fields, namely
\bes{ 
\det{\left(\ID - \Phi t^2 \right)} = \PE\left[- \left\{\sum_{i,j=1}^Y y_{i,l} y_{j,l}^{-1} + \sum_{i,j=1}^Y y_{i,r} y_{j,r}^{-1}\right\} t^2\right]~,
}
whereas the second term is given by
\bes{ \label{secondtermadjchiral}
\det{\left(\ID - \gamma_{(12)} \Phi t^2 \right)} = \PE\left[- \left\{\sum_{i,j=1}^Y y_{i,l} y_{i,r} \left(y_{j,l} y_{j,r}\right)^{-1}\right\} t^4\right]~.
}
Observe that the latter reproduces the contribution coming from a single chiral field in the adjoint representation of the $\U(Y)$ gauge node, with fugacities $y_{i} \equiv y_{i,l} y_{i,r}$, but with $t^4$ appearing instead of $t^2$, meaning that we have to send $(x, a) \rightarrow (x^2, a^2)$ according to \eref{CBHBlimits}.

When we restore the original gauge nodes, in addition to \eref{HSZ2wradjchiral}, we also have to analyse how the Haar measure appearing in the Hilbert series is affected by the $\BZ_2$ wreathing. For this purpose, let us define the $m \times m$ matrix, with $m = Y (Y - 1)/2$, whose diagonal elements are $y_{i,l} y_{j,l}$, with $i < j$, namely
\bes{ \label{Mumatrix}
\mu_l  =  \left(
\begin{array}{ccccccccc}
 \frac{y_{1,l}}{y_{2,l}} & \ldots & 0 & 0 & 0 & 0 & 0 & 0 & 0\\
 0 & \frac{y_{1,l}}{y_{3,l}} & \ldots & 0 & 0 & 0 & 0 & 0 & 0 \\
 0 & 0 & \ddots & 0 & 0 & 0 & 0 & 0 & 0 \\
 0 & 0 & \ldots & \frac{y_{1,l}}{y_{Y,l}} & \ldots & 0 & 0 & 0 & 0 \\
 0 & 0 & 0 &\ldots & \frac{y_{2,l}}{y_{3,l}} & \ldots & 0 & 0 & 0 \\
 0 & 0 & 0 & 0 & 0 & \ddots & 0 & 0 & 0\\
 0 & 0 & 0 & 0 & 0 & \ldots & \frac{y_{2,l}}{y_{Y,l}} & \ldots & 0 \\
 0 & 0 & 0 & 0 & 0 & 0 & 0 & \ddots & 0\\
 0 & 0 & 0 & 0 & 0 & 0 & 0 & \ldots & \frac{y_{Y-1,l}}{y_{Y,l}} \\
\end{array}
\right)~.
}
Analogously, we define the matrix $\mu_r$ by substituting $y_{i,l}$ with $y_{i,r}$ in \eref{Mumatrix}. These two can then by combined in the following $2 m \times 2 m$ matrix
\bes{ \label{Mumatrixlr}
\mu = \left(
\begin{array}{c|c}
 \mu_l & \ZM \\
 \hline
 \ZM & \mu_r \\
\end{array}
\right)~,
}
thanks to which the Haar measure for the two $\U(Y)$ gauge groups, the one on the left-hand side and the one on the right-hand side of \eref{gaugeZ2}, can be expressed as
\bes{ \label{HaarmeasUAUA}
&\int_{\U(Y) \times \U(Y)} d \mu_{\U(Y) \times \U(Y)} \\ &= \frac{1}{(2 \pi i)^{2 Y}} \left(\prod_{i=1}^Y \oint_{|y_{i,l}|=1} \frac{d y_{i,l}}{y_{i,l}} \oint_{|y_{i,r}|=1} \frac{d y_{i,r}}{y_{i,r}}\right) \det{\left(\ID - \mu\right)} \\ & = \frac{1}{(2 \pi i)^{2 Y}} \left(\prod_{i=1}^Y \oint_{|y_{i,l}|=1} \frac{d y_{i,l}}{y_{i,l}} \oint_{|y_{i,r}|=1} \frac{d y_{i,r}}{y_{i,r}}\right) \prod_{i < j}^Y \left(1 - y_{i,l} y_{j,l}^{-1}\right) \left(1 - y_{i,r} y_{j,r}^{-1}\right)~.
}
In order to perform the $\BZ_2$ wreathing, we act on $\mu$ with the $\BZ_2$ matrices \eref{Z2mat} and the Haar measure gets modified as
\bes{
\frac{1}{(2 \pi i)^{2 Y}} \left(\prod_{i=1}^Y \oint_{|y_{i,l}|=1} \frac{d y_{i,l}}{y_{i,l}} \oint_{|y_{i,r}|=1} \frac{d y_{i,r}}{y_{i,r}}\right) \frac{1}{2} \left[\det{\left(\ID - \mu\right)} + \det{\left(\ID - \gamma_{(12)} \mu\right)}\right]~.
}
The first contribution is \eref{HaarmeasUAUA}, whereas the determinant in the second term yields
\bes{ \label{secondtermHaar}
\det{\left(\ID - \gamma_{(12)} \mu\right)} = \prod_{i < j}^Y \left[1 - y_{i,l} y_{i,r} \left(y_{j,l} y_{j,r}\right)^{-1}\right]~.
}
Note that this is the contribution entering in the Haar measure of a single $\U(Y)$ group, with gauge fugacities $y_{i} \equiv y_{i,l} y_{i,r}$.

Taking into account both \eref{secondtermadjchiral} and \eref{secondtermHaar}, and the discussion below them, it follows that \eref{gaugeZ2} can be traded for
\bes{ \label{gaugeZ2wr}
\eref{gaugeZ2} = \frac{1}{2} \left[
\vcenter{\hbox{\begin{tikzpicture}
        \node[gauge,label=below:{\footnotesize $Y$},label=left:{\blue{\scriptsize $\vec{y}_{l}$}}] (AL) at (-0.5,0) {};
        \node[gauge,label=below:{\footnotesize $Y$},label=right:{\blue{\scriptsize $\vec{y}_{r}$}}] (AR) at (0.5,0) {};
\end{tikzpicture}}}
\quad + \quad 
\vcenter{\hbox{\begin{tikzpicture}
        \node[gauge,thick,draw=red,label=below:
        {\footnotesize $Y$},label=left:{\red{\scriptsize $\vec{y}$}},label=right:{\red{\scriptsize $x^2, a^{-4}$}}] (A) at (0,0) {};
\end{tikzpicture}}}
\right]
}
where we remark that the adjoint chiral coming from the second term in the sum \eref{gaugeZ2wr} is refined with fugacities $x^2$ and $a^{-4}$.\footnote{The adjoint chiral sitting in the 3d $\CN=4$ vector multiplet has charge $-2$ under the $\U(1)$ axial symmetry. Hence, if we send $a \rightarrow a^2$, the adjoint chiral coming from the second term in \eref{gaugeZ2wr} has charge $-4$ under the axial symmetry.\label{foot:axial}} This means that, at the level of the index, \eref{gaugeZ2} reads
\bes{ \label{indexgaugeZ2wr}
\scalebox{0.88}{$
\begin{split}
&\CI_{\eqref{gaugeZ2}}(w_{y,l}, n_{y,l}|w_{y,r}, n_{y,r}| a, n_a; x)  \\ & = \frac{1}{2} \left\{\left[ \frac{1}{(Y!)^2} \sum_{(\vec{m}^y_l, \vec{m}^y_r) \in \BZ^{2 Y}} \oint \left(\prod_{j = 1}^Y \frac{d y_{j,l}}{2 \pi i y_{j,l}} \frac{d y_{j,r}}{2 \pi i y_{j,r}} y_{j,l}^{n_{y,l}} y_{j,r}^{n_{y,r}} w_{y,l}^{m^y_{j,l}} w_{y,r}^{m^y_{j,r}}\right) \right. \right. \\ & \left.\left. \qquad \quad \times  \CZ^{\U(Y)}_{\text{vec}}\left(\vec{y}_l; \vec{m}^y_l; x\right) \CZ^{\U(Y)}_{\text{vec}}\left(\vec{y}_r; \vec{m}^y_r; x\right) \right. \right. \\ & \left.\left. \qquad \times \prod_{i,j = 1}^Y \CZ^{1}_{\chi} \left(y_{i,l} y_{j,l}^{-1} a^{-2}; m^y_{i,l} - m^y_{j,l} -2 n_a; x\right) \CZ^{1}_{\chi} \left(y_{i,r} y_{j,r}^{-1} a^{-2}; m^y_{i,r} - m^y_{j,r} -2 n_a; x\right) \right] \right. \\ & \left.\quad +\left[ \frac{1}{Y!} \sum_{\vec{m}^y \in \BZ^Y} \oint \left(\prod_{j = 1}^Y \frac{d y_{j}}{2 \pi i y_{j}} y_{j}^{n_y} w_y^{m^y_{j}}\right) \CZ^{\U(Y)}_{\text{vec}}\left(\vec{y}; \vec{m}^y; x^2\right) \right. \right. \\ & \left. \left. \qquad \times \prod_{i,j = 1}^Y \CZ^{1}_{\chi} \left(y_{i} y_{j}^{-1} a^{-4}; m^y_{i} - m^y_{j} -4 n_a; x^2\right)\right]\right\}~,
\end{split}
$}
}
where $(w_{y,l}, n_{y,l})$ and $(w_{y,r}, n_{y,r})$ are the fugacities and background fluxes associated with the topological symmetries of the two $\U(Y)$ gauge nodes appearing in the first term of the sum \eref{gaugeZ2wr}. Moreover, $(w_y, n_y)$ are the topological fugacity and the corresponding background flux related to the single $\U(Y)$ gauge node in the second term of the sum \eref{gaugeZ2wr}, with $w_y \equiv w_{y,l} w_{y,r}$.

Let us also provide more details on the comment anticipated below \eref{indexgaugeflavourZ2wraxis}, regarding the restoration of the original $\U(Z)$ gauge node, which is intersected by the vertical axis of symmetry in the wreathed quiver \eref{Z2wrmiddlenode}. 
\bes{ \label{gaugeZ2middle}
\vcenter{\hbox{\begin{tikzpicture}
        \node[gauge,label=below:{\footnotesize $Z$}] (Z) at (0,0) {};
        \draw [<->,red] (-1,0) to [out=130,in=50,looseness=1] (1,0);
        \draw [dashed,blue] (0,-1)--(0,1);
\end{tikzpicture}}}}
The 3d $\CN = 4$ vector multiplet associated with this node is not affected by the $\BZ_2$ wreathing, \ie its contribution to the index reads
\bes{ \label{indexgaugeZ2wrmiddle}
&\CI_{\eqref{gaugeZ2middle}}(w_z, n_z| a, n_a; x)  \\ & = \frac{1}{Z!} \sum_{\vec{m}^z \in \BZ^Z} \oint \left(\prod_{j = 1}^Z \frac{d z_{j}}{2 \pi i z_{j}} z_{j}^{n_z} w_z^{m^z_{j}} \right)  \CZ^{\U(Z)}_{\text{vec}}\left(\vec{z}; \vec{m}^z; x\right) \\& \qquad \times \prod_{i,j = 1}^Z \CZ^{1}_{\chi} \left(z_{i} z_{j}^{-1} a^{-2}; m^z_{i} - m^z_{j} -2 n_a; x\right) ~,
}
where $(w_z, n_z)$ are the fugacity and background flux associated with the topological symmetry of the $\U(Z)$ gauge node. Importantly, since \eref{indexgaugeZ2wrmiddle} is just the standard vector multiplet contribution with gauge fugacities $z_j$ and associated magnetic fluxes $m^z_j$, then the latter do not get rescaled to $2 m^z_j$ in the last line of \eref{indexgaugeflavourZ2wraxis}, as one might naively expect. Note, however, that the gauge fugacities get rescaled to $z_j^2$ in the last line of \eref{indexgaugeflavourZ2wraxis}, since this follows from the discussion around \eref{HSmiddle12} involving the matter contribution \eref{gaugeflavourZ2axis}.

Putting everything together, the contributions \eref{gaugeflavourZ2wr}, \eref{gaugeflavourZ2wraxis}, \eref{gaugegaugeZ2wraxis}, \eref{adjointhyperZ2wr} and \eref{gaugeZ2wr} imply that the wreathed quivers \eref{Z2wrmiddlenode} and \eref{Z2wrnomiddlenode} admit the following descriptions:
\begin{subequations}  
\begin{align}
\begin{split} \label{Z2wrQmiddlenode}
\eref{Z2wrmiddlenode} = \frac{1}{2} \left[\eref{Z2wrmiddlenode}_{\ID} + \eref{Z2wrmiddlenode}_{(12)}\right]~,
\end{split} \\
\begin{split} \label{Z2wrQnomiddlenode}
\eref{Z2wrnomiddlenode} = \frac{1}{2} \left[\eref{Z2wrnomiddlenode}_{\ID} + \eref{Z2wrnomiddlenode}_{(12)}\right]~,
\end{split}
\end{align}
\end{subequations}
with the various quivers appearing in the sums above which can be depicted as follows:
\bes{ \label{quivermiddlenode}
\eref{Z2wrmiddlenode}_{\ID} =
\vcenter{\hbox{\begin{tikzpicture}
        \node[label=below:{}] (3L) at (-4,0) {$\ldots$};
        \node[gauge,label=below:{\footnotesize $P$}] (AL) at (-3,0) {};
        \node[gauge,label=below:
        {\footnotesize $Q$}] (BL) at (-2,0) {};
        \node[flavour,label=above:{\footnotesize $F$}] (FL) at (-2,1) {};
        \node[gauge,label=below:
        {\footnotesize $Y$}] (CL) at (-1,0) {};
        \node[gauge,label=below:
        {\footnotesize $Z$}] (D) at (0,0) {};
        \node[gauge,label=below:
        {\footnotesize $Y$}] (CR) at (1,0) {};
        \node[gauge,label=below:
        {\footnotesize $Q$}] (BR) at (2,0) {};
        \node[flavour,label=above:{\footnotesize $F$}] (FR) at (2,1) {};
        \node[gauge,label=below:{\footnotesize $P$}] (AR) at (3,0) {};
        \node[label=below:{}] (3R) at (4,0) {$\ldots$};
        \path (AL) edge [out=45,in=135,looseness=10] (AL);
        \path (AR) edge [out=45,in=135,looseness=10] (AR);
        \draw (3L)--(AL)--(BL)--(CL)--(D)--(CR)--(BR)--(AR)--(3R);
        \draw (BL)--(FL);
        \draw (BR)--(FR);
\end{tikzpicture}}}
}
\bes{ \label{quivermiddlenode12}
\eref{Z2wrmiddlenode}_{(12)} = 
\vcenter{\hbox{\begin{tikzpicture}
        \node[label=below:{}] (3D) at (-4,0) {\red{$\ldots$}};
        \node[gauge,thick,draw=red,label=below:{\footnotesize $P$}] (A) at (-3,0) {};
        \node[gauge,thick,draw=red,label=below:
        {\footnotesize $Q$}] (B) at (-2,0) {};
        \node[flavour,thick,draw=red,label=above:{\footnotesize $F$}] (F) at (-2,1) {};
        \node[gauge,thick,draw=red,label=below:
        {\footnotesize $Y$}] (C) at (-1,0) {};
        \node[gauge,label=below:
        {\footnotesize $Z$}] (D) at (0,0) {};
        \path[thick, red] (A) edge [out=45,in=135,looseness=10] (A);
        \draw[thick, red] (3D)--(A)--(B)--(C)--(D);
        \draw[thick, red] (B)--(F);
        \node (nn) at (0.5,0.75) {\red{\scriptsize $(x,a,z) \rightarrow (x^2, a^2,z^2)$}};
\end{tikzpicture}}}
}
\bes{ \label{quivernomiddlenode}
\eref{Z2wrnomiddlenode}_{\ID} =
\vcenter{\hbox{\begin{tikzpicture}
        \node[label=below:{}] (3L) at (-3.5,0) {$\ldots$};
        \node[gauge,label=below:{\footnotesize $P$}] (AL) at (-2.5,0) {};
        \node[gauge,label=below:
        {\footnotesize $Q$}] (BL) at (-1.5,0) {};
        \node[flavour,label=above:{\footnotesize $F$}] (FL) at (-1.5,1) {};
        \node[gauge,label=below:
        {\footnotesize $Y$}] (CL) at (-0.5,0) {};
        \node[gauge,label=below:
        {\footnotesize $Y$}] (CR) at (0.5,0) {};
        \node[gauge,label=below:
        {\footnotesize $Q$}] (BR) at (1.5,0) {};
        \node[flavour,label=above:{\footnotesize $F$}] (FR) at (1.5,1) {};
        \node[gauge,label=below:{\footnotesize $P$}] (AR) at (2.5,0) {};
        \node[label=below:{}] (3R) at (3.5,0) {$\ldots$};
        \path (AL) edge [out=45,in=135,looseness=10] (AL);
        \path (AR) edge [out=45,in=135,looseness=10] (AR);
        \draw (3L)--(AL)--(BL)--(CL)--(CR)--(BR)--(AR)--(3R);
        \draw (BL)--(FL);
        \draw (BR)--(FR);
\end{tikzpicture}}}
}
\bes{ \label{quivernomiddlenode12}
\eref{Z2wrnomiddlenode}_{(12)} = 
\vcenter{\hbox{\begin{tikzpicture}
        \node[label=below:{}] (3D) at (-4,0) {\red{$\ldots$}};
        \node[gauge,thick,draw=red,label=below:{\footnotesize $P$}] (A) at (-3,0) {};
        \node[gauge,thick,draw=red,label=below:
        {\footnotesize $Q$}] (B) at (-2,0) {};
        \node[flavour,thick,draw=red,label=above:{\footnotesize $F$}] (F) at (-2,1) {};
        \node[gauge,thick,draw=red,label=below:
        {\footnotesize $Y$}] (C) at (-1,0) {};
        \path[thick,red] (A) edge [out=45,in=135,looseness=10] (A);
        \draw[dashed,red,thick] (C) edge [out=-45,in=45,looseness=10] node[midway,right] {} (C);
        \draw[thick,red] (3D)--(A)--(B)--(C);
        \draw[thick,red] (B)--(F);
        \node (nn) at (0,0.75) {\red{\scriptsize $(x,a) \rightarrow (x^2, a^2)$}};
\end{tikzpicture}}}
}
where we denote with $(x,a,\vec{z}) \rightarrow (x^2, a^2, \vec{z}^2)$ and $(x,a) \rightarrow (x^2, a^2)$  the rescaling of the fugacities in the contribution of each red line and node. Using \eref{indexgaugeflavourZ2wr}, \eref{indexgaugeflavourZ2wraxis}, \eref{indexgaugegaugeZ2wraxis}, \eref{indexadjointhyperZ2wr}, \eref{indexgaugeZ2wr} and \eref{indexgaugeZ2wrmiddle}, the corresponding indices are given explicitly by the following expressions.
\bi
\item For \eref{quivermiddlenode}, we have
\bes{ \label{indexquivermiddlenode}
\scalebox{0.9}{$
\begin{split}
&\CI_{\eref{quivermiddlenode}} = \CI_{\eref{Z2wrmiddlenode}_{\ID}}(\vec{f}_{lr}, \vec{m}^f_{lr}| w_{p,lr}, n_{p,lr}|w_{q,lr}, n_{q,lr}| w_{y,lr}, n_{y,lr}|w_{z}, n_{z}| a, n_a; x) \\
& = \frac{1}{(P!)^2 (Q!)^2 (Y!)^2 Z!} \sum_{\vec{m}^p_{lr} \in \BZ^{2 P}} \, \sum_{\vec{m}^q_{lr} \in \BZ^{2 Q}} \, \sum_{\vec{m}^y_{lr} \in \BZ^{2 Y}} \, \sum_{\vec{m}^z \in \BZ^Z} \oint \,\, \CI^{\text{vec}}_{\eref{Z2wrmiddlenode}_{\ID}} \,\,\CI^{\text{adj}}_{\eref{Z2wrmiddlenode}_{\ID}} \,\,\CI^{\text{chir}}_{\eref{Z2wrmiddlenode}_{\ID}}~,
\end{split}
$}
}
where
\bes{ \label{indexquivermiddlenodevec}
\CI^{\text{vec}}_{\eref{Z2wrmiddlenode}_{\ID}}  =& \left(\prod_{j = 1}^P \prod_{d= l,r} \frac{d p_{j,d}}{2 \pi i p_{j,d}} p_{j,d}^{n_{p,d}} w_{p,d}^{\vec{m}^p_{j,d}}\right) \prod_{d= l,r}\CZ^{\U(P)}_{\text{vec}}\left(\vec{p}_d; \vec{m}^p_d; x\right) \\ \times & \left(\prod_{j = 1}^Q \prod_{d= l,r} \frac{d q_{j,d}}{2 \pi i q_{j,d}} q_{j,d}^{n_{q,d}} w_{q,d}^{\vec{m}^q_{j,d}}\right) \prod_{d= l,r} \CZ^{\U(Q)}_{\text{vec}}\left(\vec{q}_d; \vec{m}^q_d; x\right) \\ \times & \left(\prod_{j = 1}^Y \prod_{d= l,r} \frac{d y_{j,d}}{2 \pi i y_{j,d}} y_{j,d}^{n_{y,d}} w_{y,d}^{\vec{m}^y_{j,d}}\right) \prod_{d= l,r} \CZ^{\U(Y)}_{\text{vec}}\left(\vec{y}_d; \vec{m}^y_d; x\right) \\ \times & \left(\prod_{j = 1}^Z \frac{d z_{j}}{2 \pi i z_{j}} z_{j}^{n_z} w_z^{\vec{m}^z_j}\right) \CZ^{\U(Z)}_{\text{vec}}\left(\vec{z}; \vec{m}^z; x\right) \times \ldots~,
}
\bes{ \label{indexquivermiddlenodeadj}
\CI^{\text{adj}}_{\eref{Z2wrmiddlenode}_{\ID}} =& \prod_{i,j = 1}^P \prod_{d= l,r} \CZ^{1}_{\chi} \left(p_{i,d} p_{j,d}^{-1} a^{-2}; m^p_{i,d} - m^p_{j,d} -2 n_a; x\right) \\ 
\times & \prod_{i,j = 1}^Q \prod_{d= l,r} \CZ^{1}_{\chi} \left(q_{i,d} q_{j,d}^{-1} a^{-2}; m^q_{i,d} - m^q_{j,d} -2 n_a; x\right) \\ \times & \prod_{i,j = 1}^Y \prod_{d= l,r} \CZ^{1}_{\chi} \left(y_{i,d} y_{j,d}^{-1} a^{-2}; m^y_{i,d} - m^y_{j,d} -2 n_a; x\right) \\ \times & \prod_{i,j = 1}^Z \CZ^{1}_{\chi} \left(z_{i} z_{j}^{-1} a^{-2}; m^z_i - m^z_j -2 n_a; x\right) \times \ldots~,
}
\bes{ \label{indexquivermiddlenodechir}
\CI^{\text{chir}}_{\eref{Z2wrmiddlenode}_{\ID}} =&
 \prod_{i=1}^P \prod_{j=1}^Q \prod_{d= l,r} \prod_{s=\pm 1} \CZ^{1/2}_{\chi} \left(p_{i,d}^s q_{j,d}^{-s} a; s m^p_{i,d} - s m^q_{j,d} + n_a; x\right) \\ 
 \times & \prod_{i=1}^Q \prod_{j=1}^F \prod_{d= l,r} \prod_{s=\pm 1} \CZ^{1/2}_{\chi} \left(q_{i,d}^s f_{j,d}^{-s} a; s m^q_{i,d} - s m^f_{j,d} + n_a; x\right) \\ \times & \prod_{i=1}^Q \prod_{j=1}^Y \prod_{d= l,r} \prod_{s=\pm 1} \CZ^{1/2}_{\chi} \left(q_{i,d}^s y_{j,d}^{-s} a; s m^q_{i,d} - s m^y_{j,d} + n_a; x\right) \\ \times & \prod_{i=1}^Y \prod_{j=1}^Z \prod_{d= l,r} \prod_{s=\pm 1} \CZ^{1/2}_{\chi} \left(y_{i,d}^s z_{j}^{-s} a; s m^y_{i,d} - s m^z_j + n_a; x\right) \\ \times & \prod_{i,j=1}^P \prod_{d= l,r} \prod_{s=\pm 1} \CZ^{1/2}_{\chi} \left(p_{i,d}^s p_{j,d}^{-s} a; s m^p_{i,d} - s m^p_{j,d} + n_a; x\right) \times \ldots~.
}
\item For \eref{quivermiddlenode12}, we have
\bes{ \label{indexquivermiddlenode12}
\scalebox{0.98}{$
\begin{split}
&\CI_{\eref{quivermiddlenode12}} = \CI_{\eref{Z2wrmiddlenode}_{(12)}}(\vec{f}, \vec{m}^f| w_{p}, n_{p}|w_{q}, n_{q}| w_{y}, n_{y}|w_{z}, n_{z}| a, n_a; x) \\
& = \frac{1}{P! Q! Y! Z!} \sum_{\vec{m}^p \in \BZ^P} \, \sum_{\vec{m}^q \in \BZ^Q} \, \sum_{\vec{m}^y \in \BZ^Y} \, \sum_{\vec{m}^z \in \BZ^Z} \oint \,\, \CI^{\text{vec}}_{\eref{Z2wrmiddlenode}_{(12)}}\,\, \CI^{\text{adj}}_{\eref{Z2wrmiddlenode}_{(12)}}\,\, \CI^{\text{chir}}_{\eref{Z2wrmiddlenode}_{(12)}}~,
\end{split}
$}
}
where
\bes{ \label{indexquivermiddlenodevec12}
\CI^{\text{vec}}_{\eref{Z2wrmiddlenode}_{(12)}} =& \left(\prod_{j = 1}^P \frac{d p_{j}}{2 \pi i p_{j}} p_{j}^{n_p} w_p^{\vec{m}^p_j}\right) \CZ^{\U(P)}_{\text{vec}}\left(\vec{p}; \vec{m}^p; x^2\right) \\ \times & \left(\prod_{j = 1}^Q \frac{d q_{j}}{2 \pi i q_{j}} q_{j}^{n_q} w_q^{\vec{m}^q_j}\right) \CZ^{\U(Q)}_{\text{vec}}\left(\vec{q}; \vec{m}^q; x^2\right) \\ \times & \left(\prod_{j = 1}^Y \frac{d y_{j}}{2 \pi i y_{j}} y_{j}^{n_y} w_y^{\vec{m}^y_j}\right) \CZ^{\U(Y)}_{\text{vec}}\left(\vec{y}; \vec{m}^y; x^2\right) \\ \times & \left(\prod_{j = 1}^Z \frac{d z_{j}}{2 \pi i z_{j}} z_{j}^{n_z} w_z^{\vec{m}^z_j}\right) \CZ^{\U(Z)}_{\text{vec}}\left(\vec{z}; \vec{m}^z; x\right) \times \ldots~,
}
\bes{ \label{indexquivermiddlenodeadj12}
\CI^{\text{adj}}_{\eref{Z2wrmiddlenode}_{(12)}}=& \prod_{i,j = 1}^P \CZ^{1}_{\chi} \left(p_{i} p_{j}^{-1} a^{-4}; m^p_i - m^p_j -4 n_a; x^2\right) \\ \times & \prod_{i,j = 1}^Q \CZ^{1}_{\chi} \left(q_{i} q_{j}^{-1} a^{-4}; m^q_i - m^q_j -4 n_a; x^2\right) \\ \times & \prod_{i,j = 1}^Y \CZ^{1}_{\chi} \left(y_{i} y_{j}^{-1} a^{-4}; m^y_i - m^y_j -4 n_a; x^2\right) \\ \times & \prod_{i,j = 1}^Z \CZ^{1}_{\chi} \left(z_{i} z_{j}^{-1} a^{-2}; m^z_i - m^z_j -2 n_a; x\right) \times \ldots~,
}
\bes{ \label{indexquivermiddlenodechir12}
\CI^{\text{chir}}_{\eref{Z2wrmiddlenode}_{(12)}} =& \prod_{i=1}^P \prod_{j=1}^Q \prod_{s=\pm 1} \CZ^{1/2}_{\chi} \left(p_{i}^s q_{j}^{-s} a^2; s m^p_i - s m^q_j + 2 n_a; x^2\right) \\
\times & \prod_{i=1}^Q \prod_{j=1}^F \prod_{s=\pm 1} \CZ^{1/2}_{\chi} \left(q_{i}^s f_{j}^{-s} a^2; s m^q_i - s m^f_j + 2 n_a; x^2\right) \\ \times & \prod_{i=1}^Q \prod_{j=1}^Y \prod_{s=\pm 1} \CZ^{1/2}_{\chi} \left(q_{i}^s y_{j}^{-s} a^2; s m^q_i - s m^y_j + 2 n_a; x^2\right) \\ \times & \prod_{i=1}^Y \prod_{j=1}^Z \prod_{s=\pm 1} \CZ^{1/2}_{\chi} \left(y_{i}^s z_{j}^{-2 s} a^2; s m^y_i - s m^z_j + 2 n_a; x^2\right) \\ \times & \prod_{i,j=1}^P \prod_{s=\pm 1} \CZ^{1/2}_{\chi} \left(p_{i}^s p_{j}^{-s} a^2; s m^p_i - s m^p_j + 2 n_a; x^2\right) \times \ldots~.
}
\item For \eref{quivernomiddlenode}, we have
\bes{ \label{indexquivernomiddlenode}
&\CI_{\eref{quivernomiddlenode}} = \CI_{\eref{Z2wrnomiddlenode}_{\ID}}(\vec{f}_{lr}, \vec{m}^f_{lr}| w_{p,lr}, n_{p,lr}|w_{q,lr}, n_{q,lr}| w_{y,lr}, n_{y,lr}| a, n_a; x) \\
& = \frac{1}{(P!)^2 (Q!)^2 (Y!)^2} \sum_{\vec{m}^p_{lr} \in \BZ^{2 P}} \, \sum_{\vec{m}^q_{lr} \in \BZ^{2 Q}} \, \sum_{\vec{m}^y_{lr} \in \BZ^{2 Y}} \oint 
\,\, \CI^{\text{vec}}_{\eref{Z2wrnomiddlenode}_{\ID}}\,\, \CI^{\text{adj}}_{\eref{Z2wrnomiddlenode}_{\ID}}\,\, \CI^{\text{chir}}_{\eref{Z2wrnomiddlenode}_{\ID}} ~,
}
where
\bes{ \label{indexquivernomiddlenodevec}
\CI^{\text{vec}}_{\eref{Z2wrnomiddlenode}_{\ID}} =&
\left(\prod_{j = 1}^P \prod_{d= l,r} \frac{d p_{j,d}}{2 \pi i p_{j,d}} p_{j,d}^{n_{p,d}} w_{p,d}^{\vec{m}^p_{j,d}}\right) \prod_{d= l,r}\CZ^{\U(P)}_{\text{vec}}\left(\vec{p}_d; \vec{m}^p_d; x\right) \\ 
\times & \left(\prod_{j = 1}^Q \prod_{d= l,r} \frac{d q_{j,d}}{2 \pi i q_{j,d}} q_{j,d}^{n_{q,d}} w_{q,d}^{\vec{m}^q_{j,d}}\right) \prod_{d= l,r} \CZ^{\U(Q)}_{\text{vec}}\left(\vec{q}_d; \vec{m}^q_d; x\right) \\ \times & \left(\prod_{j = 1}^Y \prod_{d= l,r} \frac{d y_{j,d}}{2 \pi i y_{j,d}} y_{j,d}^{n_{y,d}} w_{y,d}^{\vec{m}^y_{j,d}}\right) \prod_{d= l,r} \CZ^{\U(Y)}_{\text{vec}}\left(\vec{y}_d; \vec{m}^y_d; x\right) \times \ldots~,
}
\bes{ \label{indexquivernomiddlenodeadj}
\CI^{\text{adj}}_{\eref{Z2wrnomiddlenode}_{\ID}} =& \prod_{i,j = 1}^P \prod_{d= l,r} \CZ^{1}_{\chi} \left(p_{i,d} p_{j,d}^{-1} a^{-2}; m^p_{i,d} - m^p_{j,d} -2 n_a; x\right) \\ \times & \prod_{i,j = 1}^Q \prod_{d= l,r} \CZ^{1}_{\chi} \left(q_{i,d} q_{j,d}^{-1} a^{-2}; m^q_{i,d} - m^q_{j,d} -2 n_a; x\right) \\ \times & \prod_{i,j = 1}^Y \prod_{d= l,r} \CZ^{1}_{\chi} \left(y_{i,d} y_{j,d}^{-1} a^{-2}; m^y_{i,d} - m^y_{j,d} -2 n_a; x\right)\times \ldots~,
}
\bes{ \label{indexquivernomiddlenodechir}
\CI^{\text{chir}}_{\eref{Z2wrnomiddlenode}_{\ID}} = & \prod_{i=1}^P \prod_{j=1}^Q \prod_{d= l,r} \prod_{s=\pm 1} \CZ^{1/2}_{\chi} \left(p_{i,d}^s q_{j,d}^{-s} a; s m^p_{i,d} - s m^q_{j,d} + n_a; x\right) \\ \times & \prod_{i=1}^Q \prod_{j=1}^F \prod_{d= l,r} \prod_{s=\pm 1} \CZ^{1/2}_{\chi} \left(q_{i,d}^s f_{j,d}^{-s} a; s m^q_{i,d} - s m^f_{j,d} + n_a; x\right) \\ \times & \prod_{i=1}^Q \prod_{j=1}^Y \prod_{d= l,r} \prod_{s=\pm 1} \CZ^{1/2}_{\chi} \left(q_{i,d}^s y_{j,d}^{-s} a; s m^q_{i,d} - s m^y_{j,d} + n_a; x\right) \\ \times & \prod_{i,j=1}^Y  \prod_{s=\pm 1} \CZ^{1/2}_{\chi} \left(y_{i,l}^{s} y_{j,r}^{-s} a; s m^y_{i,l} - s m^y_{j,r} + n_a; x\right) \\ \times & \prod_{i,j=1}^P \prod_{d= l,r} \prod_{s=\pm 1} \CZ^{1/2}_{\chi} \left(p_{i,d}^s p_{j,d}^{-s} a; s m^p_{i,d} - s m^p_{j,d} + n_a; x\right) \times \ldots~.
}
\item For \eref{quivernomiddlenode12}, we have
\bes{ \label{indexquivernomiddlenode12}
&\CI_{\eref{quivernomiddlenode12}} = \CI_{\eref{Z2wrnomiddlenode}_{(12)}}(\vec{f}, \vec{m}^f| w_{p}, n_{p}|w_{q}, n_{q}| w_{y}, n_{y}| a, n_a; x) \\
& = \frac{1}{P! Q! Y! Z!} \sum_{\vec{m}^p \in \BZ^P} \, \sum_{\vec{m}^q \in \BZ^Q} \, \sum_{\vec{m}^y \in \BZ^Y} \oint \,\, \CI^{\text{vec}}_{\eref{Z2wrnomiddlenode}_{(12)}}\,\, \CI^{\text{adj}}_{\eref{Z2wrnomiddlenode}_{(12)}} \,\,
\CI^{\text{chir}}_{\eref{Z2wrnomiddlenode}_{(12)}} ~,
}
where 
\bes{ \label{indexquivernomiddlenodevec12}
\CI^{\text{vec}}_{\eref{Z2wrnomiddlenode}_{(12)}} =& \left(\prod_{j = 1}^P \frac{d p_{j}}{2 \pi i p_{j}} p_{j}^{n_p} w_p^{\vec{m}^p_j}\right) \CZ^{\U(P)}_{\text{vec}}\left(\vec{p}; \vec{m}^p; x^2\right) \\ 
\times & \left(\prod_{j = 1}^Q \frac{d q_{j}}{2 \pi i q_{j}} q_{j}^{n_q} w_q^{\vec{m}^q_j}\right) \CZ^{\U(Q)}_{\text{vec}}\left(\vec{q}; \vec{m}^q; x^2\right) \\ \times & \left(\prod_{j = 1}^Y \frac{d y_{j}}{2 \pi i y_{j}} y_{j}^{n_y} w_y^{\vec{m}^y_j}\right) \CZ^{\U(Y)}_{\text{vec}}\left(\vec{y}; \vec{m}^y; x^2\right) \times \ldots~,
}
\bes{ \label{indexquivernomiddlenodeadj12}
\CI^{\text{adj}}_{\eref{Z2wrnomiddlenode}_{(12)}} =& \prod_{i,j = 1}^P \CZ^{1}_{\chi} \left(p_{i} p_{j}^{-1} a^{-4}; m^p_i - m^p_j -4 n_a; x^2\right) \\ \times & \prod_{i,j = 1}^Q \CZ^{1}_{\chi} \left(q_{i} q_{j}^{-1} a^{-4}; m^q_i - m^q_j -4 n_a; x^2\right) \\ \times & \prod_{i,j = 1}^Y \CZ^{1}_{\chi} \left(y_{i} y_{j}^{-1} a^{-4}; m^y_i - m^y_j -4 n_a; x^2\right) \times \ldots~,
}
\bes{ \label{indexquivernomiddlenodechir12}
\CI^{\text{chir}}_{\eref{Z2wrnomiddlenode}_{(12)}} = & \prod_{i=1}^P \prod_{j=1}^Q \prod_{s=\pm 1} \CZ^{1/2}_{\chi} \left(p_{i}^s q_{j}^{-s} a^2; s m^p_i - s m^q_j + 2 n_a; x^2\right) \\ 
\times & \prod_{i=1}^Q \prod_{j=1}^F \prod_{s=\pm 1} \CZ^{1/2}_{\chi} \left(q_{i}^s f_{j}^{-s} a^2; s m^q_i - s m^f_j + 2 n_a; x^2\right) \\ \times & \prod_{i=1}^Q \prod_{j=1}^Y \prod_{s=\pm 1} \CZ^{1/2}_{\chi} \left(q_{i}^s y_{j}^{-s} a^2; s m^q_i - s m^y_j + 2 n_a; x^2\right) \\ \times &\prod_{i,j=1}^Y \CZ^{1/2}_{\chi} \left(y_{i} y_{j}^{-1} a^2; s m^y_{i} - s m^y_{j} + 2 n_a; x^2\right) \\ \times & \prod_{i,j=1}^P \prod_{s=\pm 1} \CZ^{1/2}_{\chi} \left(p_{i}^s p_{j}^{-s} a^2; s m^p_i - s m^p_j + 2 n_a; x^2\right) \times \ldots~.
 }
\ei
For compactness, given a generic fugacity $F$ and magnetic flux $M$, we use the notation $F_{lr} \equiv (F_l, F_r), M_{lr} \equiv (M_l, M_r)$. In case there is a special unitary gauge/flavour node $\SU(S)$, with gauge/flavour fugacities $s_j$ and corresponding magnetic fluxes $m^s_j$, with $j = 1, \ldots, S$, we set 
\bes{ \label{SUnode}
s_S = \left(\prod_{j = 1}^{S - 1} s_j\right)^{-1} ~,~ m^s_S = - \sum_{j = 1}^{S - 1} m^s_j~.
}
Upon setting the background magnetic fluxes associated with the flavour and topological symmetries to zero, the indices of the wreathed quivers $\eref{Z2wrmiddlenode}$ and $\eref{Z2wrnomiddlenode}$ are then given by
\begin{subequations}  
\begin{align}
\begin{split} \label{indZ2wrQmiddlenode}
&\CI_{\eref{Z2wrmiddlenode}}(\vec{f}| w_{p}|w_{q}| w_{y}|w_{z}| a; x) \\ &= \frac{1}{2} \left[\CI_{\eref{Z2wrmiddlenode}_{\ID}}(\vec{f}_{lr} = \vec{f}| w_{p,lr} = w_{p}|w_{q,lr} = w_{q}| w_{y,lr} = w_{y}|w_{z}| a; x) \right. \\ & \left. \quad \, \, \, + \CI_{\eref{Z2wrmiddlenode}_{(12)}}(\vec{f}^2| w_{p}^2| w_{q}^2| w_{y}^2|w_{z}| a; x)\right]~,
\end{split} \\
\begin{split} \label{indZ2wrQnomiddlenode}
&\CI_{\eref{Z2wrnomiddlenode}}(\vec{f}| w_{p}|w_{q}| w_{y}| a; x) \\ &= \frac{1}{2} \left[\CI_{\eref{Z2wrnomiddlenode}_{\ID}}(\vec{f}_{lr} = \vec{f}| w_{p,lr} = w_{p}|w_{q,lr} = w_{q}| w_{y,lr} = w_{y} a; x) \right. \\ & \left. \quad \, \, \, + \CI_{\eref{Z2wrnomiddlenode}_{(12)}}(\vec{f}^2| w_{p}^2| w_{q}^2| w_{y}^2| a; x)\right]~,
\end{split}
\end{align}
\end{subequations}
where the notation $F_{lr} = F$ indicates that we set both figacities $F_l$ and $F_r$ to be equal to $F$. Such a redefinition of the fugacities in \eref{indZ2wrQmiddlenode} and \eref{indZ2wrQnomiddlenode} is employed in order to obtain well-defined indices, whose series expansion does not contain half-integer coefficients in front of the various fugacities, which might otherwise arise due to the presence of the overall factor $1/2$.

\subsection{\texorpdfstring{Quivers wreathed by $S_n$ and subgroups thereof}{Quivers wreathed by Sn and subgroups thereof}}
Let us now generalise the procedure presented in the previous subsection to analyse quivers wreathed by (subgroups of) $S_n$. Our starting point is a 3d $\CN=4$ quiver gauge theory $\mathsf{Q}$, with (special) unitary gauge and flavour nodes, consisting of $n$ identical building blocks which give rise to an $S_n$ symmetry. A prototypical example is a star-shaped quiver with $n$ identical legs glued together via a common gauge node
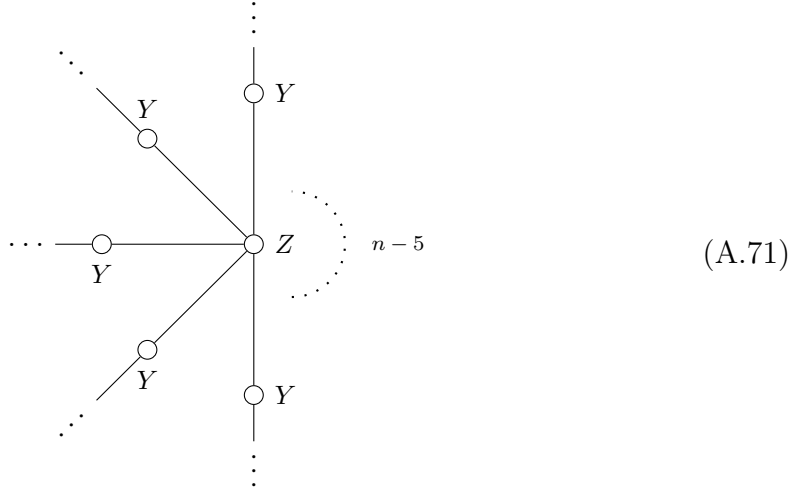
\bes{ \label{starshapedQ}
\vcenter{\hbox{\begin{tikzpicture}
        \node[gauge,label=right:{\footnotesize $Z$}] (0) at (0,0) {};
        \node[gauge,label=right:{\footnotesize $Y$}] (1) at (0,2) {};
        \node[gauge,label=above:{\footnotesize $Y$}] (2) at (-1.4,1.4) {};
        \node[gauge,label=below:{\footnotesize $Y$}] (3) at (-2,0) {};
        \node[gauge,label=below:{\footnotesize $Y$}] (4) at (-1.4,-1.4) {};
        \node[gauge,label=right:{\footnotesize $Y$}] (5) at (0,-2) {};
        \node[label=below:{}] (1D) at (0,3) {\rotatebox{90}{$\ldots$}};
        \node[label=below:{}] (2D) at (-2.4,2.4) {\rotatebox{-45}{$\ldots$}};
        \node[label=below:{}] (3D) at (-3,0) {$\ldots$};
        \node[label=below:{}] (4D) at (-2.4,-2.4) {\rotatebox{45}{$\ldots$}};
        \node[label=below:{}] (5D) at (0,-3) {\rotatebox{90}{$\ldots$}};
        \draw (0)--(1);
        \draw (0)--(2);
        \draw (0)--(3);
        \draw (0)--(4);
        \draw (0)--(5);
        \draw (1)--(1D);
        \draw (2)--(2D);
        \draw (3)--(3D);
        \draw (4)--(4D);
        \draw (5)--(5D);
        \draw[loosely dotted, thick, black] (0.5,-0.7) arc (-90:90:0.7) node[midway,right=0.2] {{\scriptsize $n - 5$}};
\end{tikzpicture}}}
}
The natural generalisation of formulae \eref{Z2wrQmiddlenode} and \eref{Z2wrQnomiddlenode} for the quiver $\mathsf{Q}$ wreathed by $H$, with $H \subseteq S_n$, reads
\bes{ \label{Snwrgeneralformula}
\mathsf{Q}/H = \frac{1}{|H|} \sum_{j = 1}^{|H|} \mathsf{Q}_{h_j}~,
}
where $|H|$ is the order of $H$ and $h_j$ are the elements of $H$. Given that a generic element of $H$ can be represented as a cycle or a product of disjoint cycles, let us focus on the element
\bes{ \label{elementH}
[b^u, c^v, \ldots, d^r] \equiv (b_1 b_2 \ldots b_u) (c_1 c_2 \ldots c_v) \ldots (d_1 d_2 \ldots d_r)~,
}
with $0 \le u, v, \ldots, r \le n$ and $u + v + \ldots + r = n$, which is a product of $p \le n$ cycles. This corresponds to the contribution $\mathsf{Q}_{[b^u, c^v, \ldots, d^r]}$ appearing in the sum \eref{Snwrgeneralformula}. Let us take $\mathsf{Q} = \eref{starshapedQ}$, \ie we want to study the contribution $\eref{starshapedQ}_{[b^u, c^v, \ldots, d^r]}$, and proceed as we did in the previous subsection by ungauging every node, analysing the subquivers consisting of hypermultiplets connecting flavour nodes and, finally, restoring the original gauge nodes.

\subsubsection*{Bifundamental hypermultiplets contribution}
Upon ungauging every node, we are interested in the following quiver
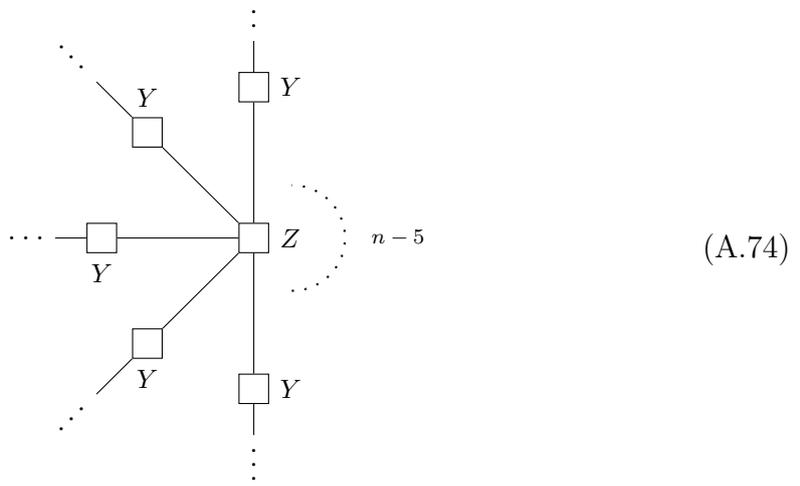
\bes{ \label{starshapedQungauged}
\vcenter{\hbox{\begin{tikzpicture}
        \node[flavour,label=right:{\footnotesize $Z$}] (0) at (0,0) {};
        \node[flavour,label=right:{\footnotesize $Y$}] (1) at (0,2) {};
        \node[flavour,label=above:{\footnotesize $Y$}] (2) at (-1.4,1.4) {};
        \node[flavour,label=below:{\footnotesize $Y$}] (3) at (-2,0) {};
        \node[flavour,label=below:{\footnotesize $Y$}] (4) at (-1.4,-1.4) {};
        \node[flavour,label=right:{\footnotesize $Y$}] (5) at (0,-2) {};
        \node[label=below:{}] (1D) at (0,3) {\rotatebox{90}{$\ldots$}};
        \node[label=below:{}] (2D) at (-2.4,2.4) {\rotatebox{-45}{$\ldots$}};
        \node[label=below:{}] (3D) at (-3,0) {$\ldots$};
        \node[label=below:{}] (4D) at (-2.4,-2.4) {\rotatebox{45}{$\ldots$}};
        \node[label=below:{}] (5D) at (0,-3) {\rotatebox{90}{$\ldots$}};
        \draw (0)--(1);
        \draw (0)--(2);
        \draw (0)--(3);
        \draw (0)--(4);
        \draw (0)--(5);
        \draw (1)--(1D);
        \draw (2)--(2D);
        \draw (3)--(3D);
        \draw (4)--(4D);
        \draw (5)--(5D);
        \draw[loosely dotted, thick, black] (0.5,-0.7) arc (-90:90:0.7) node[midway,right=0.2] {{\scriptsize $n - 5$}};
\end{tikzpicture}}}
}
where we denote with $y_{i,A}$ the fugacities associated with the $\U(Y)$ nodes, with $i = 1, \ldots, Y$ and $A = [b^u, c^v, \ldots, d^r]$, and with $z_j$ the fugacities associated with the common $\U(Z)$ node, with $j = 1, \ldots, Z$. We can therefore associate with the chiral fields in the $(\mathbf{Y},\bar{\mathbf{Z}})$ representation of each $\U(Y) \times \U(Z)$ contribution the matrices $\Upsilon_A$, which are defined as \eref{Upsmatrix} by substituting $y_{i,l}$ with $y_{i,A}$, and the generalisation of matrix \eref{Upsmatrixlr} reads
\bes{ 
\Upsilon = \left(
\begin{array}{c c c| c c c| c| c c c}
 \Upsilon_{b_1} & \ldots & \ZM & \ZM & \ldots & \ZM & \ldots & \ZM & \ldots & \ZM \\
 \vdots & \ddots & \vdots & \vdots & \ddots & \vdots & \ddots & \vdots & \ddots & \vdots \\
 \ZM & \ldots & \Upsilon_{b_u}& \ZM & \ldots & \ZM & \ldots & \ZM & \ldots & \ZM \\
 \hline
 \ZM & \ldots & \ZM & \Upsilon_{c_1} & \ldots & \ZM & \ldots & \ZM & \ldots & \ZM \\
 \vdots & \ddots & \vdots & \vdots & \ddots & \vdots & \ddots & \vdots & \ddots & \vdots \\
 \ZM & \ldots & \ZM & \ZM & \ldots & \Upsilon_{c_v} & \ldots & \ZM & \ldots & \ZM \\
 \hline
 \vdots & \ddots & \vdots & \vdots & \ddots & \vdots & \ddots & \vdots & \ddots & \vdots \\
 \hline
 \ZM & \ldots & \ZM & \ZM & \ldots & \ZM & \ldots & \Upsilon_{d_1} & \ldots & \ZM \\
 \vdots & \ddots & \vdots & \vdots & \ddots & \vdots & \ddots & \vdots & \ddots & \vdots \\
 \ZM & \ldots & \ZM & \ZM & \ldots & \ZM & \ldots & \ZM & \ldots & \Upsilon_{d_r} \\
\end{array}
\right)~,
}
from which we can also derive $\Upsilon^{-1}$. The two matrices $\Upsilon$ and $\Upsilon^{-1}$ describe the $n$ hypermultiplets in the bifundamental representation of $\U(Y) \times \U(Z)$ of quiver \eref{starshapedQungauged}. We can act on them with the matrix representation $\gamma_{[b^u, c^v, \ldots, d^r]}$ of the group element $[b^u, c^v, \ldots, d^r]$, whose entries are
\bes{ \label{matrepcycles}
\left(\gamma_{[b^u, c^v, \ldots, d^r]}\right)_{i,j} = \begin{cases}
    \ID \quad \text{if $[b^u, c^v, \ldots, d^r]$ exchanges $i$ and $j$} \\
    \ZM \quad \text{otherwise}
\end{cases}~,
}
where both $\ID$ and $\ZM$ are of the same dimension of the matrices they act upon.\footnote{In the present case, both $\ID$ and $\ZM$ are $Y Z \times Y Z$ matrices, since $\Upsilon_{A}$ are $Y Z \times Y Z$ matrices.} This yields the following Higgs branch Hilbert series
\bes{ \label{HSgenericcycle}
&\frac{1}{\det{\left(\ID - \gamma_{[b^u, c^v, \ldots, d^r]} \Upsilon t \right)} \det{\left(\ID - \gamma_{[b^u, c^v, \ldots, d^r]} \Upsilon^{-1} t \right)}} \\
& = \PE\left[\left\{\left(\sum_{i=1}^Y \prod_{b = b_1}^{b_u}y_{i,b} \right) \sum_{j=1}^Z z_{j}^{-u} + \left(\sum_{i=1}^Y \prod_{b = b_1}^{b_u}y_{i,b}^{-1} \right) \sum_{j=1}^Z z_{j}^u\right\} t^u \right. \\& \left. \qquad \, \, + \left\{\left(\sum_{i=1}^Y \prod_{c = c_1}^{c_v }y_{i,c} \right) \sum_{j=1}^Z z_{j}^{-v} + \left(\sum_{i=1}^Y \prod_{c = c_1}^{c_v }y_{i,c}^{-1} \right) \sum_{j=1}^Z z_{j}^v\right\} t^v + \ldots \right. \\& \left. \qquad \, \,+\left\{\left(\sum_{i=1}^Y \prod_{d = d_1}^{d_r} y_{i,d} \right) \sum_{j=1}^Z z_{j}^{-r} + \left(\sum_{i=1}^Y \prod_{d = d_1}^{d_r}y_{i,d}^{-1} \right) \sum_{j=1}^Z z_{j}^r\right\} t^r\right]~,
}
which coincides with the contribution coming from a star-shaped quiver consisting of a common $\U(Z)$ node linked to $p \le n$ $\U(Y)$ nodes via bifundamental hypermultiplets. If we denote each of the matter contributions connecting such $\U(Y)$ nodes and the $\U(Z)$ node with a subscript $1, 2, \ldots, p$, then the definition of the fugacities and the rescaling of the variable $t$ associated with each of them are as follows:
\bes{ \label{rescalefugacitiesstarshaped}
\begin{array}{lll}
\U(Y)_1: y_{i,1} \equiv \prod\limits_{b = b_1}^{b_u}y_{i,b}~, &\quad  \U(Z)_1: z_j^u~, &\quad  t_1: t \rightarrow t^u~,\\  \U(Y)_2: y_{i,2} \equiv \prod\limits_{c = c_1}^{c_v}y_{i,c} ~, & \quad \U(Z)_2: z_j^v~, &\quad  t_2: t \rightarrow t^v~, \\ & \quad \vdots & \quad \\  \U(Y)_p: y_{i,p} \equiv \prod\limits_{d = d_1}^{d_r}y_{i,d} ~, & \quad \U(Z)_p: z_j^r~, &\quad  t_p: t \rightarrow t^r ~.
\end{array}
}
Note that, according to \eref{CBHBlimits}, rescaling the variable $t$ into $t^m$ for some $m$ is equivalent to sending $(x, a) \rightarrow (x^m, a^m)$. This observation, together with \eref{rescalefugacitiesstarshaped}, leads us to identify $\eref{starshapedQungauged}_{[b^u, c^v, \ldots, d^r]}$ with
\bes{ \label{starshapedQf}
\eref{starshapedQungauged}_{[b^u, c^v, \ldots, d^r]} = \vcenter{\hbox{\begin{tikzpicture}
        \node[flavour,label=right:{\footnotesize $Z$}] (0) at (0,0) {};
        \node[flavour,thick,draw=red,label=below:{\footnotesize $Y$},label=above:{\red{\scriptsize $\vec{y}_{1}$}}] (1) at (-2,0) {};
        \node[flavour,thick,draw=blue,label=right:{\footnotesize $Y$},label=left:{\blue{\scriptsize $\vec{y}_{2}$}}] (2) at (0,2) {};
        \node[flavour,thick,draw=violet,label=right:{\footnotesize $Y$},label=left:{\violet{\scriptsize $\vec{y}_{p}$}}] (3) at (0,-2) {};
        \node[label=below:{}] (1D) at (-3,0) {{\red{$\ldots$}}};
        \node[label=below:{}] (2D) at (0,3) {\rotatebox{90}{\blue$\ldots$}};
        \node[label=below:{}] (3D) at (0,-3) {\rotatebox{90}{{\violet$\ldots$}}};
        \draw[thick,red] (0) to node[midway,below] {\textcolor{red}{\scriptsize $x^u, a^u, z^u$}} (1);
        \draw[thick,blue] (0) to node[midway,right] {\textcolor{blue}{\scriptsize $x^v, a^v, z^v$}} (2);
        \draw[thick,violet] (0) to node[midway,right] {\textcolor{violet}{\scriptsize $x^r, a^r, z^r$}} (3);
        \draw[thick,red] (1)--(1D);;
        \draw[thick,blue] (2)--(2D);
        \draw[thick,violet] (3)--(3D);
        \draw[loosely dotted, thick, black] (0.5,-0.7) arc (-90:90:0.7) node[midway,right=0.2] {{\scriptsize $p - 3$}};
\end{tikzpicture}}}
}
where we highlight in {\red red}, {\blue blue} and {\violet violet} the flavour fugacities and the modification to the definition of the fugacities $x$ and $a$ associated with the {\red first}, {\blue second} and {\violet $p$-th} matter contribution respectively. Consequently, the index of theory $\eref{starshapedQungauged}_{[b^u, c^v, \ldots, d^r]}$ reads
\bes{ \label{indexwrSnmatter}
&\CI_{\eref{starshapedQungauged}_{[b^u, c^v, \ldots, d^r]}}(\vec{y}_1, \vec{m}^y_1|\vec{y}_2, \vec{m}^y_2| \ldots |\vec{y}_p, \vec{m}^y_p|\vec{z}, \vec{m}^z| a, n_a; x) \\ &  = \prod_{i=1}^Y \prod_{j=1}^Z \prod_{s=\pm 1} \CZ^{1/2}_{\chi} \left(y_{i,1}^s z_{j}^{-u s} a^u; s m^{y}_{i,1} - s m^z_{j} + u n_a; x^u\right) \\ &  \times \prod_{i=1}^Y \prod_{j=1}^Z \prod_{s=\pm 1} \CZ^{1/2}_{\chi} \left(y_{i,2}^s z_{j}^{-v s} a^v; s m^{y}_{i,2} - s m^z_{j} + v n_a; x^v\right) \times \ldots \\ &  \times \prod_{i=1}^Y \prod_{j=1}^Z \prod_{s=\pm 1} \CZ^{1/2}_{\chi} \left(y_{i,p}^s z_{j}^{-r s} a^r; s m^{y}_{i,p} - s m^z_{j} + r n_a; x^r\right) ~,
}
where the magnetic fluxes $m^z_{j}$ do not get rescaled, even if the associated fugacities $z_j$ are rescaled. The reason for this is that, when we restore the original $\U(Z)$ node in the middle, this is not affected by wreathing, as already expalained below \eref{indexgaugeflavourZ2wraxis} and \eref{indexgaugeZ2wrmiddle}.

\subsubsection*{Restoring the original gauge nodes}
Next, we can restore the original gauge nodes of quiver \eref{starshapedQ} and we can analyse how the element $[b^u, c^v, \ldots, d^r]$ acts on the adjoint chiral fields appearing in the $n$ identical 3d $\CN = 4$ vector multiplets and on the Haar measure. For our purpose, in order to emphasise the action of the group element, let us consider a string of $u + v + \ldots + r = n$ $\U(Y)$ gauge nodes grouped into $p \le n$ sets containing $u, v, \ldots, r$ nodes
\bes{ \label{ngaugenodes}
\scalebox{0.97}{$
\left(\vcenter{\hbox{\begin{tikzpicture}
    \node[gauge,label=below:{\footnotesize $Y$}] (B1) at (-1,1) {};
    \node[gauge,label=below:{\footnotesize $Y$}] (B2) at (0,1) {};
    \node[label=above:{\scriptsize $u-3$}] (BD) at (1,1) {$\ldots$};
    \node[gauge,label=below:{\footnotesize $Y$}] (Bu) at (2,1) {};
\end{tikzpicture}}}\right)_1
\left(\vcenter{\hbox{\begin{tikzpicture}
    \node[gauge,label=below:{\footnotesize $Y$}] (C1) at (-1,0) {};
    \node[gauge,label=below:{\footnotesize $Y$}] (C2) at (0,0) {};
    \node[label=above:{\scriptsize $v-3$}] (CD) at (1,0) {$\ldots$};
    \node[gauge,label=below:{\footnotesize $Y$}] (Cv) at (2,0) {};
\end{tikzpicture}}}\right)_2
\ldots 
\left(\vcenter{\hbox{\begin{tikzpicture}
    \node[gauge,label=below:{\footnotesize $Y$}] (D1) at (-1,-2) {};
    \node[gauge,label=below:{\footnotesize $Y$}] (D2) at (0,-2) {};
    \node[label=above:{\scriptsize $r-3$}] (DD) at (1,-2) {$\ldots$};
    \node[gauge,label=below:{\footnotesize $Y$}] (Dr) at (2,-2) {};
\end{tikzpicture}}}\right)_p
$}
}
where each round bracket with a subscript $m$ contains the nodes belonging to the $m$-th set, with $m = 1, \ldots, p$.

Let us deal with the adjoint chiral multiplets first, whose contribution is encoded in the matrix
\bes{ 
\Phi = \left(
\begin{array}{c c c| c c c| c| c c c}
 \Phi_{b_1} & \ldots & \ZM & \ZM & \ldots & \ZM & \ldots & \ZM & \ldots & \ZM \\
 \vdots & \ddots & \vdots & \vdots & \ddots & \vdots & \ddots & \vdots & \ddots & \vdots \\
 \ZM & \ldots & \Phi_{b_u}& \ZM & \ldots & \ZM & \ldots & \ZM & \ldots & \ZM \\
 \hline
 \ZM & \ldots & \ZM & \Phi_{c_1} & \ldots & \ZM & \ldots & \ZM & \ldots & \ZM \\
 \vdots & \ddots & \vdots & \vdots & \ddots & \vdots & \ddots & \vdots & \ddots & \vdots \\
 \ZM & \ldots & \ZM & \ZM & \ldots & \Phi_{c_v} & \ldots & \ZM & \ldots & \ZM \\
 \hline
 \vdots & \ddots & \vdots & \vdots & \ddots & \vdots & \ddots & \vdots & \ddots & \vdots \\
 \hline
 \ZM & \ldots & \ZM & \ZM & \ldots & \ZM & \ldots & \Phi_{d_1} & \ldots & \ZM \\
 \vdots & \ddots & \vdots & \vdots & \ddots & \vdots & \ddots & \vdots & \ddots & \vdots \\
 \ZM & \ldots & \ZM & \ZM & \ldots & \ZM & \ldots & \ZM & \ldots & \Phi_{d_r} \\
\end{array}
\right)~,
}
which generalises \eref{PhimatrixLR}, where the matrices $\Phi_A$, with $A = [b^u, c^v, \ldots, d^r]$, are defined as \eref{Phimatrix} by substituting $p_{i,l}$ with $y_{i,A}$. We can then act on $\Phi$ with the matrix representation of element $[b^u, c^v, \ldots, d^r]$, defined in \eref{matrepcycles}, and obtain the following contribution to the Higgs branch Hilbert series:
\bes{ \label{Snwradjchirals}
\scalebox{0.94}{$
\begin{split}
&\det{\left(\ID - \gamma_{[b^u, c^v, \ldots, d^r]} \Phi t^2 \right)} \\ &= \PE\left[-\left\{\sum_{i,j = 1}^Y y_{i,1} y_{j,1}^{-1}\right\} t^{2 u} -\left\{\sum_{i,j = 1}^Y y_{i,2} y_{j,2}^{-1}\right\} t^{2 v} - \ldots -\left\{\sum_{i,j = 1}^Y y_{i,p} y_{j,p}^{-1}\right\} t^{2 r} \right]~.
\end{split}
$}
}
We recognise \eref{Snwradjchirals} to be the contribution coming from $p$ $\U(Y)$ adjoint chiral fields, whose associated gauge fugacities $y_{i,1}, y_{i,2}, \ldots, y_{i,p}$ are defined as in \eref{rescalefugacitiesstarshaped}, with $t^2$ rescaled into $t^{2 m}$, with $m = u, v, r, \ldots$, meaning that we have to send $(x, a)$ into $(x^m, a^m)$.

Finally, let us study the action of element $[b^u, c^v, \ldots, d^r]$ on the Haar measure for the $\U(Y)$ gauge groups in quiver \eref{starshapedQ}. Let us take the matrices $\mu_A$, which are defined as \eref{Mumatrix} by substituting $y_{i,l}$ with $y_{i,A}$, and combine them into a single matrix $\mu$, which generalises \eref{Mumatrixlr}, given by
\bes{ 
\mu = \left(
\begin{array}{c c c| c c c| c| c c c}
 \mu_{b_1} & \ldots & \ZM & \ZM & \ldots & \ZM & \ldots & \ZM & \ldots & \ZM \\
 \vdots & \ddots & \vdots & \vdots & \ddots & \vdots & \ddots & \vdots & \ddots & \vdots \\
 \ZM & \ldots & \mu_{b_u}& \ZM & \ldots & \ZM & \ldots & \ZM & \ldots & \ZM \\
 \hline
 \ZM & \ldots & \ZM & \mu_{c_1} & \ldots & \ZM & \ldots & \ZM & \ldots & \ZM \\
 \vdots & \ddots & \vdots & \vdots & \ddots & \vdots & \ddots & \vdots & \ddots & \vdots \\
 \ZM & \ldots & \ZM & \ZM & \ldots & \mu_{c_v} & \ldots & \ZM & \ldots & \ZM \\
 \hline
 \vdots & \ddots & \vdots & \vdots & \ddots & \vdots & \ddots & \vdots & \ddots & \vdots \\
 \hline
 \ZM & \ldots & \ZM & \ZM & \ldots & \ZM & \ldots & \mu_{d_1} & \ldots & \ZM \\
 \vdots & \ddots & \vdots & \vdots & \ddots & \vdots & \ddots & \vdots & \ddots & \vdots \\
 \ZM & \ldots & \ZM & \ZM & \ldots & \ZM & \ldots & \ZM & \ldots & \mu_{d_r} \\
\end{array}
\right)~.
}
If we consider the determinant
\bes{ \label{SnwrHaar}
\scalebox{0.99}{$
\det{\left(\ID - \gamma_{[b^u, c^v, \ldots, d^r]} \mu\right)} = \prod\limits_{i < j}^Y \left(1 - y_{i,1} y_{j,1}^{-1}\right) \left(1 - y_{i,2} y_{j,2}^{-1}\right) \times \ldots \times \left(1 - y_{i,p} y_{j,p}^{-1}\right) ~,
$}
}
we observe that this coincides with the contribution entering in the Haar measure of $p$ $\U(Y)$ gauge groups, with gauge fugacities $y_{i,1}, y_{i,2}, \ldots, y_{i,p}$.

Taking into account both \eref{Snwradjchirals} and \eref{SnwrHaar}, it follows that $\eref{ngaugenodes}_{[b^u, c^v, \ldots, d^r]}$ can be depicted as
\bes{ \label{Snwrgauge}
\eref{ngaugenodes}_{[b^u, c^v, \ldots, d^r]} =
\vcenter{\hbox{\begin{tikzpicture}
    \node[gauge,thick,draw=red,label=below:{\footnotesize $Y$},label=left:{\red{\scriptsize $\vec{y}_1$}},label=above:{\red{\scriptsize $x^u, a^{-2 u}$}}] (B) at (-2,0) {};
    \node[gauge,thick,draw=blue,label=below:{\footnotesize $Y$},label=left:{\blue{\scriptsize $\vec{y}_2$}},label=above:{\blue{\scriptsize $x^v, a^{-2 v}$}}] (C) at (0,0) {};
    \node[label=above:{\scriptsize $p-3$}] (d) at (1.5,0) {$\ldots$};
    \node[gauge,thick,draw=violet,label=below:{\footnotesize $Y$},label=left:{\violet{\scriptsize $\vec{y}_p$}},label=above:{\violet{\scriptsize $x^r, a^{-2 r}$}}] (D) at (3,0) {};
\end{tikzpicture}}}
}
where we highlight in {\red red}, {\blue blue} and {\violet violet} the gauge fugacities and the rescaling of the fugacities $x$ and $a$ associated with the gauge contributions coming from the {\red first}, {\blue second} and {\violet $p$-th} set of nodes in \eref{ngaugenodes} respectively.\footnote{Recall that the adjoint chiral belonging to the 3d $\CN = 4$ vector multiplet has charge $2$ under the $\U(1)$ axial symmetry, see Footnote \ref{foot:axial}. Hence, if we send $a \rightarrow a^m$, the adjoint chirals in \eref{Snwrgauge} are refined with fugacity $a^{-2 m}$.} The associated index reads
\bes{ \label{indexgaugeSnwr}
&\CI_{\eref{ngaugenodes}_{[b^u, c^v, \ldots, d^r]}}(w_{y,1}, n_{y,1}|w_{y,2}, n_{y,2}| \ldots|w_{y,p}, n_{y,p}|a, n_a; x) \\&= \frac{1}{(Y!)^p} \sum_{(\vec{m}^y_1, \vec{m}^y_2, \ldots, \vec{m}^y_p) \in \BZ^l} \oint \left(\prod_{j = 1}^Y \frac{d y_{j,1}}{2 \pi i y_{j,1}} y_{j,1}^{n_{y,1}} w_{y,1}^{m^y_{j,1}} \right) \CZ^{\U(Y)}_{\text{vec}}\left(\vec{y}_1; \vec{m}^y_1; x^u\right) \\ & \qquad \qquad \qquad \qquad \qquad \, \, \, \times \left(\prod_{j = 1}^Y \frac{d y_{j,2}}{2 \pi i y_{j,2}} y_{j,2}^{n_{y,2}} w_{y,2}^{m^y_{j,2}} \right) \CZ^{\U(Y)}_{\text{vec}}\left(\vec{y}_2; \vec{m}^y_2; x^v\right) \times \ldots \\ & \qquad \qquad \qquad \qquad \qquad \, \, \, \times \left(\prod_{j = 1}^Y \frac{d y_{j,p}}{2 \pi i y_{j,p}} y_{j,p}^{n_{y,p}} w_{y,p}^{m^y_{j,p}} \right) \CZ^{\U(Y)}_{\text{vec}}\left(\vec{y}_p; \vec{m}^y_p; x^r\right) \\ & \qquad \qquad \times \prod_{i,j = 1}^Y \CZ^{1}_{\chi} \left(y_{i,1} y_{j,1}^{-1} a^{-2 u}; m^y_{i,1} - m^y_{j,1} -2 u n_a; x^u\right) \\ & \qquad \qquad \times \prod_{i,j = 1}^Y \CZ^{1}_{\chi} \left(y_{i,2} y_{j,2}^{-1} a^{-2 v}; m^y_{i,2} - m^y_{j,2} -2 v n_a; x^v\right) \times \ldots \\ & \qquad \qquad \times \prod_{i,j = 1}^Y \CZ^{1}_{\chi} \left(y_{i,p} y_{j,p}^{-1} a^{-2 r}; m^y_{i,p} - m^y_{j,p} -2 r n_a; x^r\right) ~,
}
where $(w_{y,1}, n_{y,1}), (w_{y,2}, n_{y,2}), \ldots, (w_{y,p}, n_{y,p})$ are the fugacities and background fluxes associated with the topological symmetries of the $p$ $\U(Y)$ gauge nodes appearing in \eref{Snwrgauge}. In particular, if we denote with $ w_{j,m}$ the topological fugacities of the $\U(Y)$ nodes in \eref{ngaugenodes}, with $m = 1, \ldots, p$ and $j = 1, \ldots l(m) \equiv$ number of $\U(Y)$ nodes in the $m$-th set, then we define $w_{y,m} \equiv \prod_{j = 1}^{l(m)} w_{j,m}$.

On the other hand, the 3d $\CN = 4$ vector multiplet associated with the $\U(Z)$ node in \eref{starshapedQ} is not affected by the action of the element $[b^u, c^v, \ldots, d^r]$ and its contribution to the index is the usual one for a vector multiplet, namely it is given by \eref{indexgaugeZ2wrmiddle}.

In conclusion, taking into account \eref{starshapedQf} and \eref{Snwrgauge}, we find that the contribution coming from the action of the group element $[b^u, c^v, \ldots, d^r]$ on the star-shaped quiver \eref{starshapedQ} can be depicted as
\bes{ \label{starshapedQg}
\eref{starshapedQ}_{[b^u, c^v, \ldots, d^r]} = \vcenter{\hbox{\begin{tikzpicture}
        \node[gauge,label=right:{\footnotesize $Z$}] (0) at (0,0) {};
        \node[gauge,thick,draw=red,label=below:{\footnotesize $Y$},label=above:{\red{\scriptsize $\vec{y}_{1}$}}] (1) at (-2,0) {};
        \node[gauge,thick,draw=blue,label=right:{\footnotesize $Y$},label=left:{\blue{\scriptsize $\vec{y}_{2}$}}] (2) at (0,2) {};
        \node[gauge,thick,draw=violet,label=right:{\footnotesize $Y$},label=left:{\violet{\scriptsize $\vec{y}_{p}$}}] (3) at (0,-2) {};
        \node[label=below:{}] (1D) at (-3,0) {{\red{$\ldots$}}};
        \node[label=below:{}] (2D) at (0,3) {\rotatebox{90}{\blue$\ldots$}};
        \node[label=below:{}] (3D) at (0,-3) {\rotatebox{90}{{\violet$\ldots$}}};
        \draw[thick,red] (0) to node[at end,below=0.5] {\textcolor{red}{\scriptsize $(x, a, z) \rightarrow (x^u, a^u, z^u)$}} (1);
        \draw[thick,blue] (0) to node[midway,right] {\textcolor{blue}{\scriptsize $(x, a, z) \rightarrow (x^v, a^v, z^v)$}} (2);
        \draw[thick,violet] (0) to node[midway,right] {\textcolor{violet}{\scriptsize $(x, a, z) \rightarrow (x^r, a^r, z^r)$}} (3);
        \draw[thick,red] (1)--(1D);;
        \draw[thick,blue] (2)--(2D);
        \draw[thick,violet] (3)--(3D);
        \draw[loosely dotted, thick, black] (0.5,-0.7) arc (-90:90:0.7) node[midway,right=0.2] {{\scriptsize $p - 3$}};
\end{tikzpicture}}}
}
where the wreathed quiver $\eref{starshapedQ}/H$ can be obtained from equation \eref{Snwrgeneralformula} by summing over all possible contributions of the form \eref{starshapedQg} corresponding to elements of $H$. Using \eref{indexgaugeZ2wrmiddle}, \eref{indexwrSnmatter} and \eref{indexgaugeSnwr}, the index of \eref{starshapedQg} reads
\bes{ \label{indexstarshapedQSnwr}
\scalebox{0.97}{$
\begin{split}
&\CI_{\eref{starshapedQg}}=\CI_{\eref{starshapedQ}_{[b^u, c^v, \ldots, d^r]}}(w_{y,1}, n_{y,1}|w_{y,2}, n_{y,2}| \ldots|w_{y,p}, n_{y,p}|w_z, n_z| a, n_a; x) \\&= \frac{1}{(Y!)^p Z!} \sum_{(\vec{m}^y_1, \ldots, \vec{m}^y_p) \in \BZ^{p Y}} \, \sum_{\vec{m}^z \in \BZ^Z} \CI^{\text{vec}}_{\eref{starshapedQ}_{[b^u, c^v, \ldots, d^r]}} \,\,
\CI^{\text{adj}}_{\eref{starshapedQ}_{[b^u, c^v, \ldots, d^r]}} \,\,
\CI^{\text{chir}}_{\eref{starshapedQ}_{[b^u, c^v, \ldots, d^r]}}~,
\end{split}
$}
}
where
\bes{
\CI^{\text{vec}}_{\eref{starshapedQ}_{[b^u, c^v, \ldots, d^r]}} =&
\left(\prod_{j = 1}^Y \frac{d y_{j,1}}{2 \pi i y_{j,1}} y_{j,1}^{n_{y,1}} w_{y,1}^{\vec{m}^y_{j,1}} \right) \CZ^{\U(Y)}_{\text{vec}}\left(\vec{y}_1; \vec{m}^y_1; x^u\right) \\ \times & \left(\prod_{j = 1}^Y \frac{d y_{j,2}}{2 \pi i y_{j,2}} y_{j,2}^{n_{y,2}} w_{y,2}^{\vec{m}^y_{j,2}} \right) \CZ^{\U(Y)}_{\text{vec}}\left(\vec{y}_2; \vec{m}^y_2; x^v\right) \times \ldots \\ \times & \left(\prod_{j = 1}^Y \frac{d y_{j,p}}{2 \pi i y_{j,p}} y_{j,p}^{n_{y,p}} w_{y,p}^{\vec{m}^y_{j,p}} \right) \CZ^{\U(Y)}_{\text{vec}}\left(\vec{y}_p; \vec{m}^y_p; x^r\right) \\ \times & \left(\prod_{j = 1}^Z \frac{d z_{j}}{2 \pi i z_{j}} z_{j}^{n_{z}} w_{z}^{\vec{m}^z_{j}} \right) \CZ^{\U(Z)}_{\text{vec}}\left(\vec{z}; \vec{m}^z; x\right) \times \ldots ~,
}
\bes{
\CI^{\text{adj}}_{\eref{starshapedQ}_{[b^u, c^v, \ldots, d^r]}} =&  \prod_{i,j = 1}^Y \CZ^{1}_{\chi} \left(y_{i,1} y_{j,1}^{-1} a^{-2 u}; m^y_{i,1} - m^y_{j,1} -2 u n_a; x^u\right) \\ 
\times & \prod_{i,j = 1}^Y \CZ^{1}_{\chi} \left(y_{i,2} y_{j,2}^{-1} a^{-2 v}; m^y_{i,2} - m^y_{j,2} -2 v n_a; x^v\right) \times \ldots \\ \times & \prod_{i,j = 1}^Y \CZ^{1}_{\chi} \left(y_{i,p} y_{j,p}^{-1} a^{-2 r}; m^y_{i,p} - m^y_{j,p} -2 r n_a; x^r\right) \\ \times & \prod_{i,j = 1}^Z \CZ^{1}_{\chi} \left(z_{i} z_{j}^{-1} a^{-2}; m^z_{i} - m^z_{j} -2 n_a; x\right) \times \ldots~,
}
\bes{
\CI^{\text{chir}}_{\eref{starshapedQ}_{[b^u, c^v, \ldots, d^r]}} =& \prod_{i=1}^Y \prod_{j=1}^Z \prod_{s=\pm 1} \CZ^{1/2}_{\chi} \left(y_{i,1}^s z_{j}^{-u s} a^u; s m^{y}_{i,1} - s m^z_{j} + u n_a; x^u\right) \\ 
\times & \prod_{i=1}^Y \prod_{j=1}^Z \prod_{s=\pm 1} \CZ^{1/2}_{\chi} \left(y_{i,2}^s z_{j}^{-v s} a^v; s m^{y}_{i,2} - s m^z_{j} + v n_a; x^v\right) \times \ldots \\ \times & \prod_{i=1}^Y \prod_{j=1}^Z \prod_{s=\pm 1} \CZ^{1/2}_{\chi} \left(y_{i,p}^s z_{j}^{-r s} a^r; s m^{y}_{i,p} - s m^z_{j} + r n_a; x^r\right) \times \ldots~.
}
In case there are special unitary nodes, we have to set the corresponding fugacities and magnetic fluxes as specified in \eref{SUnode}. Upon setting the background magnetic fluxes associated with the topological symmetries to zero, the index of the wreathed quiver $\eref{starshapedQ}/H$ can then be obtained by summing over the indices of all possible contributions corresponding to elements of $H$ and dividing by the order of $H$:
\bes{ \label{SnwrgeneralformulaIndex}
\scalebox{0.9}{$
\begin{split}
&\CI_{\eref{starshapedQ}/H}(\vec{w}|w_z| a; x) \\&= \frac{1}{|H|} \sum_{[b^u, c^v, \ldots, d^r] \in H} \CI_{\eref{starshapedQ}_{[b^u, c^v, \ldots, d^r]}}(w_{y,1} = w_{b_i}^{l(1)}|w_{y,2} = w_{c_j}^{l(2)}| \ldots|w_{y,p} = w_{d_k}^{l(p)}|w_z| a; x)~,
\end{split}
$}
}
where we denote with $l(m)$ the length of the $m$-th cycle appearing in the group element $[b^u, c^v, \ldots, d^r]$, with $m = 1, \ldots, p$, and with $b_i, c_j, \ldots, d_k$ generic entries of the $m$-th cycle of such element, as specified by \eref{elementH}. In order to obtain a well-defined index, we have to set $w_m = w_{m'}$ whenever, throughout the sum over all the elements of $H$, two different group elements act on the same $\U(Y)$ node of quiver \eref{starshapedQ} via two distinct cycles $m$ and $m'$, namely the same $\U(Y)$ node appears in two different sets labelled by subscripts $m$ and $m'$ in \eref{ngaugenodes}.\footnote{For the sake of explicitness, let us take $H = S_3$ and $n = 4$ in quiver \eref{starshapedQ}. We can redefine the topological fugacities associated to each contribution $\eref{starshapedQ}_{[b^u, c^v, \ldots, d^r]}$ participating in the sum over all elements of $S_3$ as follows:
\bes{ 
\begin{array}{lll}
\ID \equiv (1)(2)(3)(4)~ &:\quad  w_{y,1} = w_1, w_{y,2} = w_2, w_{y,3} = w_3, w_{y,4} = w_4~,\\  (12) \equiv (12)(3)(4) ~ &: \quad w_{y,1} = w_1^2, w_{y,2} = w_3, w_{y,3} = w_4~,\\ (13) \equiv (13)(2)(4) ~ &: \quad w_{y,1} = w_1^2, w_{y,2} = w_2, w_{y,3} = w_4~,\\ (23) \equiv (23)(1)(4) ~ &: \quad w_{y,1} = w_2^2, w_{y,2} = w_1, w_{y,3} = w_4~,\\ (123) \equiv (123)(4) ~ &: \quad w_{y,1} = w_1^3, w_{y,2} = w_4~,\\ (132) \equiv (132)(4) ~ &: \quad w_{y,1} = w_1^3, w_{y,2} = w_4,~.
\end{array}
}
Since the node $\U(Y)_2$ associated with the fugacity $w_2$ appears both in the first set and in the second set corresponding to the first and second cycles, we have to set $w_2 = w_1$. A similar consideration holds also for the node $\U(Y)_3$ associated with the fugacity $w_3$, meaning that we also have to set $w_3 = w_1$.} This observation generalises the comment below \eref{indZ2wrQnomiddlenode} and ensures that the series expansion of the index does not contain half-integer coefficients, which might otherwise arise due to the presence of the overall factor $1/|H|$.

\bibliographystyle{JHEP}
\bibliography{bibli.bib}

\end{document}